\documentclass[a4paper]{report}
\usepackage{latexsym}
\usepackage{amsfonts}
\usepackage{amssymb}
\usepackage{amsmath}
\usepackage{amscd}
\usepackage{amsthm}
\usepackage{feynmf}
\input xy
\xyoption{all}

\newcommand{\doublespaced}{\renewcommand{\baselinestretch}{2}\normalfont}
\newcommand{\singlespaced}{\renewcommand{\baselinestretch}{1}\normalfont}
\newcommand{\draftspaced}{\singlespaced} 

\newtheorem{theorem}{Theorem}[section]
\newtheorem {thm}{Theorem}[section]
\newtheorem{cor}[theorem]{Corollary}
\newtheorem{lem}[theorem]{Lemma}
\newtheorem{prop}[theorem]{Proposition}

\theoremstyle{defn}
\newtheorem{defn}[theorem]{Definition}

\theoremstyle{rem}
\newtheorem{rem}[theorem]{Remark}
\newtheorem{example}[theorem]{Example}

\numberwithin{equation}{section}

\doublespaced
\def\thetitle{{\bf Riemann-Hilbert Problem and Quantum Field Theory:\\
Integrable Renormalization, Dyson-Schwinger Equations
}}
\def\theauthor{\textsf{Ali Shojaei-Fard}}
\def\theyear{2010}

\begin{document}

\pagenumbering{roman}
\doublespaced
\large\newlength{\oldparskip}\setlength\oldparskip{\parskip}\parskip=.3in
\thispagestyle{empty}

\begin{center}

\vspace*{\fill}

\thetitle

\theauthor

\end{center}

\noindent\singlespaced\large

\singlespaced

\vspace{5in}

\doublespaced\large
\begin{center}
November \theyear
\end{center}


\normalsize\parskip=\oldparskip

\newpage
\singlespaced






\emph{\small \textit{... \textbf{We are here, because we have a dream.\\
But this is not a reason and this is not a why.\\
The dream itself is a means and it is not an end ...}}}







\newpage
\singlespaced

\chapter*{\textsf{Acknowledgments}}

First, it is my pleasure to thank Prof. Dr. Matilde Marcolli
because of her helpful ideas, advices and also, her great scientific supports during the Ph.D. program specially, my thesis. Second, I would like to thank Prof. Vida Milani because of her valuable encouragements and scientific supports during the graduate period. Third, I would also like to thank Prof. Kurusch Ebrahimi-Fard and Prof. Dominique Manchon because of helpful and useful discussions at MPI and ESI.

I acknowledge with thanks the financial supports from Hausdorff
Research Institute for Mathematics (HIM) for the program
Geometry-Physics (May 2008 - August 2008), Max Planck Institute for
Mathematics (MPIM) as a member of IMPRS program (September 2008 -
February 2009), Erwin Schrodinger International Institute for
Mathematical Physics (ESI) for the program Number Theory and Physics
(March 2009 - April 2009) and also, partially research fellowship
from M.S.R.T. (May 2008 - September 2008).

Finally, I would like to thank Institute for Studies in Theoretical Physics and Mathematics (IPM) for giving time to me for talks \cite{S10, S14}.

\newpage


\begin{center}
  \textbf{ABSTRACT}\\
\vspace{.15in}
\end{center}

{\it Finding a practical formalism for eliminating ultraviolet divergences in quantum field theory was developed on the basis of perturbation theory.
Kreimer could interpret the combinatorics of BPHZ perturbative renormalization based on a Hopf algebraic structure on Feynman diagrams. It was applied to determine an infinite dimensional complex Lie group connected with a renormalizable theory underlying minimal subtraction scheme in dimensional regularization.

Practically, Connes and Kreimer reformulated the BPHZ method based on the extraction of finite values with respect to the Birkhoff factorization on elements of the mentioned Lie group and then Connes and Marcolli provided a new geometric interpretation from physical information based on the Riemann-Hilbert correspondence. Moreover, they could introduce a categorical configuration to describe renormalizable theories such that as the result, formulating a universal treatment is an important contribution in this direction.

This Hopf algebraic approach improved immediately in several fields. On the one hand, with attention to the multiplicativity of renormalization and the theory of Rota-Baxter algebras, Connes-Kreimer theory determined an attractive procedure to consider integrable systems underlying renormalizable theories.
On the other hand, people reconsidered Quantum Chromodynamics, Quantum Electrodynamics and
(non-)abelian gauge theories such that in this process, giving a comprehensive interpretation from
non-perturbative theory is known as one important expected challenge in the Connes-Kreimer-Marcolli theory. With applying Hochschild cohomology theory, Kreimer introduced a new  combinatorial formulation from Dyson-Schwinger equations where it provides an interesting intelligent strategy to analyze non-perturbative situations. Furthermore, Connes and Marcolli could discover a meaningful connection between quantum field theory and theory of mixed Tate motives.

This work is written based on this Hopf algebraic modeling in the study of quantum field theories. We focus on some essential anticipated questions around this formalism for instance theory of integrable systems and non-perturbative theory.

In the first purpose, we concentrate on the theory of quantum integrable systems underlying the Connes-Kreimer approach. We introduce a new family of Hamiltonian systems depended on the perturbative renormalization process (i.e. renormalization and regularization prescriptions) in renormalizable theories. It is observed that the renormalization group can
determine an infinite dimensional integrable system such that this fact provides a link between this proposed class of motion integrals and
renormalization flow. Moreover, with help of the integral renormalization theorems, we study motion integrals underlying Bogoliubv character and BCH series to obtain a new family of fixed point equations.

In the second goal, we consider the combinatorics of Connes-Marcolli approach to provide a Hall rooted tree type reformulation from one particular object in this theory namely,  universal Hopf algebra of renormalization $H_{\mathbb{U}}$. As the consequences, interesting relations between this Hopf algebra and some well-known combinatorial Hopf algebras are obtained and also, one can make a new Hall polynomial representation from universal singular frame such that based on the universal nature of this special loop, one can expect a Hall tree type scattering formula
for physical information such as counterterms.

In the third aim, with attention to the given rooted tree version of
$H_{\mathbb{U}}$ and by applying the
Connes-Marcolli's universal investigation, we are going to improve the
notion of an intrinsic
geometrical interpretation from non-perturbative theory.
In this process, at the first step we consider combinatorial Dyson-Schwinger equations at
the level of the universal Hopf algebra of renormalization. At the
second step, with respect to factorization of Feynman diagrams into primitive components,
the universality of $H_{\mathbb{U}}$ at the level of these equations is discussed. And finally, we find a bridge between
these equations and objects of the universal category of flat
equi-singular vector bundles such that by this way, the universal
property of this category at the level of these equations will be
observed.}
\newpage

\textbf{Keywords.} Combinatorial Hopf Algebras; Connes-Kreimer Renormalization Group; Connes-Kreimer-Marcolli Perturbative Renormalization; Dyson-Schwinger Equations; Hall Rooted Trees; Quantum Integrable Systems; Renormalizable Quantum Field Theory; Riemann-Hilbert Correspondence; Universal Hopf Algebra of Renormalization.

\doublespaced

\textbf{MSC 2000.} 05C05; 83C47; 18D50; 05E05; 22E65; 37K25; 37K30; 35Q15; 81T15

\doublespaced

\textbf{PACS.} 02.10.De; 11.10.Gh; 11.10.Hi; 2.30.Ik; 02.10.Ox; 02.20.Sv; 02.20.Tw; 11.10.-z

\doublespaced

\textbf{Email Address:}  shojaeifa@yahoo.com

\noindent


\vspace*{\fill}

\newpage
\tableofcontents
\newpage
\draftspaced
\pagenumbering{arabic}
\include{introdept}
\include{back}
\include{finitedept}
\include{infinitedept}


\chapter{\textsf{Introduction}}

\emph{\small \textit{\textbf{ ... " You see, one thing is that I can live with doubt and \\
uncertainty and not knowing. I think it's much more interesting \\
to live not knowing than to have the answers that might be wrong ", \\
Richard Feynman ...}}}

\vspace{.4in}

Quantum Field Theory (QFT) is the most important profound formulated manifestation
in modern physics for the description of occurrences at the smallest length
scales with highest energies.
In fact, this mysterious theoretic hypothesis is the fundamental result of merging the two
crucial achievements in physics namely, Quantum Mechanics and
Special Relativity such that its essential target can be summarized in finding an
unified interpretation from interactions between elementary particles. The development of
QFT is done in several formalizations such that Wightman's axiomatic
configuration (i.e. constructive QFT) together with Haag's algebraic
formulation in terms of von Neumann algebras are the most
influenced efforts \cite{H7, S13, WS1}. Perturbation approach is also another successful and useful point of view
to QFT. It is based on perturbative expansions related to
Feynman graphs where in these expansions one can find some
ill-defined iterated Feynman integrals such that they should be removed in a physical sensible procedure. This problem can be considered with a well-known analytic algorithm namely, renormalization. \cite{M5, S13}

The apparatus of renormalization developed in the perturbation theory led to attractive success in quantum field theory where it has in fact an analytic meaning. But Kreimer could find the appearance of a combinatorial nature inside of this analytic method and he showed that one can explain the renormalization procedure (as a recursive formalism for the elimination of
(sub-)divergences from diagrams) based on one particular Hopf algebra structure $H_{FG}$ on Feynman diagrams. The discovery of this smooth mathematical construction encapsulated in the renormalization process covered the lack of a modern practical mechanism
for the principally description of this technique and also,
it introduced a new rich relationship between this essential physical
method and modern mathematics. It should be mentioned that although the Bogoliubov recursion performs renormalization
without using any Hopf algebra structure but when we go to the
higher loop orders, the advantages of this Hopf algebraic reconstruction in computations will be observed decisively. \cite{BK1, BK2, EGGV, EK1, K6, K2, K3, KD1, W2, W1}

There are different approaches to renormalization. For instance, in the Bogoliubov method, the renormalization is done without regularization and counterterms but in the Bogoliubov-Parasiuk-Hepp-Zimmermann (BPHZ) method, we work on dimensional regularization (as the regularization scheme) and minimal subtraction (as the renormalization map).
Originally, the regularization can parametrize ultraviolet
divergences appearing in amplitudes to reduce them formally finite
together with a special subtraction of ill-defined expressions
(associated with physical principles). But this
procedure determines some non-physical parameters and indeed,
it changes the nature of Feynman rules (identified by the given physical theory) to algebra morphisms from the
renormalization Hopf algebra $H_{FG}$ to the commutative
algebra $A_{dr}$ of Laurent series with finite pole part in dimensional regularization. It is obviously seen that this
commutative algebra is characterized with the given regularization
method and it means that by changing the regularization scheme, we
should work on its associated algebra. In general, interrelationship between Feynman diagrams and Feynman integrals can be described by the Feynman rules of the theory and Kreimer could interpret this rules on the basis of characters of $H_{FG}$.  \cite{CM1, EM1}

Soon thereafter, Connes and Kreimer introduced a new practical reformulation for the BPHZ perturbative renormalization. They could associate an infinite dimensional complex Lie group $G(\mathbb{C})$ to each
renormalizable physical theory $\Phi$ and then they investigated that this BPHZ scheme is in fact an example of a general mathematical procedure namely, the extraction
of finite values on the basis of the Riemann-Hilbert problem in the sense that one can calculate some important physical information for instance counterterms and related renormalized values
with applying the Birkhoff decomposition on elements of this Lie group. In other words,
according to their programme, based on the regularization scheme, the bare unrenormalizaed theory produces a meromorphic loop $\gamma_{\mu}$ on $C = \partial \bold{\Delta}$ with values in the space of characters (i.e. the Lie group $G(\mathbb{C})$) where $\bold{\Delta}$ is an infinitesimal disk centered at $z=0$. For each $z \in C$, $\gamma_{\mu}(z)=\phi^{z}$ is called dimensionally regularized Feynman rules character. It means that after application of the special character $\phi$ on an arbitrary Feynman diagram $\Gamma$, one can get an iterated Feynman integral such that regularized version of this character namely, $\phi^{z}$ maps each Feynman diagram $\Gamma$ to a Laurent series  $U_{\mu}^{z}(\Gamma)$ in $z$ with finite pole part (i.e. regularized unrenormalized version of the related Feynman integral). Then they showed that the renormalized theory is just the evaluation at the integer dimension $\textsc{D}$ of space-time of the holomorphic positive part of the Birkhoff decomposition of $\gamma_{\mu}$. As the result, they could reformulate physical information such as counterterms and related renormalized values, renormalization group and its infinitesimal generator (i.e. $\beta-$function) based on components of this decomposition of $\gamma_{\mu}$ with values in the group of formal diffeomorphisms of the space of coupling constants. \cite{CK1, CK2, CM2, CM1, G3}

The existence and the uniqueness of the Connes-Kreimer's Birkhoff decomposition are connected with the Rota-Baxter property of the chosen regularization and renormalization couple (i.e. multiplicativity of renormalization). This fact reports about some interesting relations between the theory of Rota-Baxter type algebras and the Riemann-Hilbert correspondence such that as a consequence one can expect to study quantum integrable systems in this Hopf algebraic language. \cite{EG2, EGK1, EGK2, S2}

Connes and Marcolli widely improved this mathematical machinery from
perturbative renormalization by giving a categorical algebro-geometric
dictionary for the analyzing of physical information in the
minimal subtraction scheme in dimensional regularization underlying
the Riemann-Hilbert correspondence. According to their approach,
the dimensional regularization parameter $z \in \bold{\Delta}$ determines a principal $\mathbb{C}^{*}-$bundle $B$ over the infinitesimal disk $\bold{\Delta}$ (i.e. $p: \bold{\Delta} \times \mathbb{C}^{*} \longrightarrow \bold{\Delta}$). Letting $B^{0}:=B-p^{-1}\{0\}$ and $P^{0}:=B^{0} \times G(\mathbb{C})$. Each arbitrary equivalence class $\overline{\omega}$ of flat connections on $P^{0}$ denotes a differential equation
$\bold{D} \gamma = \overline{\omega}$ (i.e. $\bold{D}: G(\mathbb{C}) \longrightarrow \Omega^{1}(\mathfrak{g}(\mathbb{C})), \  \  \phi \longmapsto \phi^{-1} d \phi$)
such that it has a unique solution. Equi-singularity condition on $\overline{\omega}$ describes the independency of the type of singularity of $\gamma$ at $z=0$ from sections of $B$. It means that for sections $\sigma_{1}, \sigma_{2}$ of $B$, $\sigma^{*}_{1}(\gamma)$ and $\sigma^{*}_{2}(\gamma)$ have the same singularity at $z=0$.
Connes and Marcolli firstly could reformulate components of the Birkhoff factorization of loops $\gamma_{\mu} \in Loop(G(\mathbb{C}),\mu)$ based on time-ordered exponential and elements of the Lie algebra $\mathfrak{g}(\mathbb{C})$. Secondly, they found a bijective correspondence between minus parts of this kind of decomposition on elements in $Loop(G(\mathbb{C}), \mu)$ (which determine counterterms) and elements in $\mathfrak{g}(\mathbb{C})$ and finally, they proved that each element of this Lie algebra determines a class of flat equi-singular connections on $P^{0}$. As the conclusion, one can see that this family of connections encode geometrically counterterms such that the independency of counterterms from the mass parameter $\mu$ is equivalent to the equi-singularity condition on connections.

In addition, they showed that these classes of connections can play the role of objects of a category $\mathcal{E}^{\Phi}$ such that it can be recovered by the category $\mathcal{R}_{G^{*}}$ of finite dimensional representations of the affine group scheme $G^{*}$.
In the next step and in a general configuration, they introduced the universal category of flat
equi-singular vector bundles $\mathcal{E}$ such that its universality
comes from this interesting notion that for each renormalizable
theory $\Phi$, one can put its related category of flat equi-singular
connections $\mathcal{E}^{\Phi}$ as a full subcategory in $\mathcal{E}$.
Because of the neutral Tannakian nature of this universal
category, one important Hopf algebra can be determined from the
procedure. That is universal Hopf algebra of renormalization $H_{\mathbb{U}}$.
By this special Hopf algebra and its associated affine group scheme $\mathbb{U}^{*}$,
renormalization groups and counterterms of renormalizable physical
theories have universal and canonical lifts. Connes and
Marcolli developed this strong mathematical treatment to the motivic
Galois theory. \cite{CM2, CM4, CM3, CM1}

The systematic extension of this Hopf algebraic modeling to different kinds of (local) quantum field theories would be
an attractive topic for people and we can observe the improvement
of this aspect for example in the reformulation of Quantum Electrodynamics (QED) (i.e. describes the interaction of charged particles such as electrons with photons), Quantum Chromodynamics (QCD) (i.e. describes the strong interaction between quarks and gluons) and quantum (non-)abelian gauge theories. \cite{BF, K4, K15, RDQ, V2, V3, V6, VP1}

Furthermore, finding a comprehensive
description from non-perturbative theory based on the Riemann-Hilbert problem
is also known as an important and interesting
challenge in this Hopf algebraic viewpoint. In \cite{CM2} the authors
suggest a procedure by the
Birkhoff factorization and the effective couplings of the
(renormalized) perturbative theories. On the other hand, there is a
specific class of equations in physics for the analytic studying of
non-perturbative situations, namely Dyson-Schwinger equations (DSEs).
With attention to the combinatorial nature of the Hopf
algebra of renormalization (as the guiding structure) and also, with the help of Hochschild
cohomology theory, Kreimer introduced a new significant combinatorial version from
these equations. Working on classification and
also calculating explicit solutions for these equations eventually lead to a constructive achievement
for the much better understanding of non-perturbative theory. One can refer to \cite{BK1, BK2, K7, K14, K2, K3} for more details and considerable advances.
The study of this kind of
equations on different Hopf algebras of rooted trees can help us to improve
our knowledge for the identification of combinatorial nature of non-perturbative events and
its reason comes back to this essential note that
Hopf algebras of renormalizable theories are representable with a
well-known Hopf algebra on rooted trees, namely Connes-Kreimer Hopf
algebra. Some interesting results about the study of DSEs on rooted
trees are collected in \cite{F1, H3, H2}.
In conclusion, with combining the perturbation theory, the combinatorics of renormalization, the geometry of dimensional regularization, the Connes-Marcolli categorification method and combinatorial Dyson-Schwinger equations, people hope to find a perfect conceptual understanding of quantum field theory.

Finally, let us sketch the outline of the present dissertation. Here we are going to improve our knowledge about the applications of this Hopf algebraic formalism in the study of quantum field theory. This research is on the basis of two general purposes namely, the theory of quantum integrable Hamiltonian systems with respect to the Connes-Kreimer framework \cite{S11, S12} and the geometric interpretation of non-perturbative theory with respect to the Connes-Marcolli approach and Dyson-Schwinger equations \cite{S5}.
At the basis of these scopes, after reviewing some preliminary facts about Hopf
algebras (in the next section), in chapter three we familiar with the
Connes-Kreimer Hopf algebra and some others important combinatorial Hopf algebra
structures. The fourth section contains an overview from the Hopf algebraic perturbative renormalization. In chapter five we consider our new
point of view to study quantum integrable systems underlying the Connes-Kreimer theory \cite{S11, S10, S12}. This work yields a deep conceptional relation between the  theory of Rota-Baxter algebras and the Riemann-Hilbert problem. In the sixth part we study the
geometry of dimensional regularization and its categorical results. The consideration of a Hall rooted tree type
representation from universal Hopf algebra of renormalization is done in chapter seven. It makes possible to study interesting relations between this special Hopf algebra and some important well-known combinatorial Hopf algebras which have essential roles to study the combinatorics of renormalization. In addition, we extend this new formulation of $H_{\mathbb{U}}$ to the level of its associated complex Lie group (as a motivic Galois group) and Lie algebra. With attention to this procedure, we provide a Hall rooted tree representation from one important loop namely, universal singular frame $\gamma_{\mathbb{U}}$ and then we will discuss about the application of this new reconstruction of $\gamma_{\mathbb{U}}$ in the study of physical information \cite{S6}.  In the eight section, with
notice to the given rooted tree version of $H_{\mathbb{U}}$, we study
combinatorial DSEs at the level of this specific Hopf algebra. And
also, with attention to this class of equations and factorization of Feynman diagrams into primitive components, we show that how one can extend the universality of $H_{\mathbb{U}}$
to the level of non-perturbative theory. At last, with extending the universality of the category of equi-singular vector bundles to the level of DSEs, we improve the geometric knowledge about this important class of equations in modern physics such that this process implies a categorical configuration in the study of DSEs. \cite{S7, S5, S14}


\chapter{\textsf{Theory of Hopf algebras}}

The concept of Hopf algebra was introduced based on the work of Hopf on manifolds and then its widely applications in topology, representation theory, combinatorics, quantum groups and noncommutative geometry displayed the power of this structure in different branches of mathematics \cite{A1, K13, S4}. Indeed, it is important to know that Hopf algebras
provide generalizations for group theory and Lie theory. As a well-known example, it can be seen that the dual of the universal enveloping algebra of a simple Lie
algebra determines a Hopf algebra. Moreover, this powerful mathematical
construction can provide a new opportunity to find useful interrelationships between
the pure world of mathematics and some complicate techniques in modern physics
such as perturbative renormalization. Additionally, with
the help of a special class of Hopf algebras namely, quantum groups,
one can observe the developments of this theory in mathematical physics and theoretical physics
\cite{K11, K5, M3, MM1, W2}.

Since Hopf algebras play a skeleton key for this work, therefore it is essential to have enough information about them.
With attention to our future requirements,
in this chapter we familiar with the concept of Hopf algebra and then we will have a short overview from its basic
properties.


\section{{\it \textsf{Elements of Hopf algebras}}}

Let $\mathbb{K}$ be a field with characteristic zero. A
$\mathbb{K}-$vector space $A$ together with an associative bilinear
map $m:A\otimes A\to A$ and a unit $\bold{1}$  is called {\it unital
algebra} such that its associativity and its unit are expressed respectively by
the following commutative diagrams:
\begin{equation} \label{algebra1}
\xymatrix{ A\otimes A\otimes A \ar[d]^{id \otimes m}\ar[r]^{m\otimes
id}
    &A\otimes A \ar[d]^m    \\
A\otimes A \ar[r]^m & A}
\end{equation}

\begin{equation} \label{algebra2}
\xymatrix{ \mathbb{K} \otimes A \ar[r]^{\mu \otimes id} \ar[dr]^\sim
& A\otimes A \
 \ar[d]_m   &A\otimes \mathbb{K} \ar[l]_{id \otimes \mu}\ar[dl]_\sim \\
&A&}
\end{equation}
where the map $\mu: \mathbb{K} \longrightarrow A$ is defined by
$\mu(\lambda)=\lambda \bold{1}$. This algebra is {\it commutative}, if $m\circ \tau=m$ such that
$\tau:A\otimes A\to A\otimes A$ is the {\it flip map} defined by
\begin{equation} \label{flipmap1}
\tau(a\otimes b)=b\otimes a.
\end{equation}

With reversing all arrows in the above diagrams, one can define
the dual structure of algebra. A $\mathbb{K}-$vector space $B$
together with a co-associative bilinear map $\Delta:B\to B\otimes B$
and a counit $\varepsilon: B \longrightarrow \mathbb{K}$ is called
{\it coalgebra}, if we have the following commutative diagrams.

\begin{equation} \label{coalgebra1}
\xymatrix{ B\otimes B\otimes B
    &B \otimes B \ar[l]_{\Delta \otimes id}    \\
B\otimes B \ar[u]_{id \otimes \Delta} & B
\ar[l]_{\Delta}\ar[u]_{\Delta}}
\end{equation}

\begin{equation} \label{coalgebra2}
\xymatrix{ \mathbb{K} \otimes B      & B \otimes B
\ar[l]_{\varepsilon\otimes id}\
    \ar[r]^{id \otimes \varepsilon} & B \otimes \mathbb{K}   \\
& B \ar[u]^\Delta \ar[ul]_\sim \ar[ur]^\sim &}
\end{equation}
This coalgebra is {\it co-commutative}, if
$\tau\circ\Delta=\Delta$. With using the {\it Sweedler's notation}
for the coproduct namely,
\begin{equation}
\Delta x=\sum_{(x)}x_1\otimes x_2,
\end{equation}
the co-associativity and the co-commutativity conditions will be
written with
\begin{equation} \label{coassociativity1}
(\Delta \otimes id)\circ \Delta(x)=\sum_{(x)}x_{1,1}\otimes
x_{1,2}\otimes x_2=
    \sum_{(x)}x_1\otimes x_{2,1}\otimes x_{2,2}
    =(id \otimes \Delta)\circ\Delta(x),
\end{equation}

\begin{equation} \label{cocommutativity1}
\sum_{(x)}x_1\otimes x_2=\sum_{(x)}x_2\otimes x_1.
\end{equation}

\begin{rem} \label{substr.}
It is clear that sub-structures (i.e. sub-algebra, sub-coalgebra) are defined in a natural way.
\end{rem}

For given algebra $A$ and coalgebra $B$, one can define a
product namely, {\it convolution product} on the space $L(B,A)$ of all
linear maps from $B$ to $A$. For each $\varphi, \psi$ in
$L(B,A)$, it is given by
\begin{equation} \label{convolutionpro1}
\varphi*\psi:= m_{A}\circ(\varphi\otimes\psi)\circ\Delta_{B}.
\end{equation}

\begin{defn} \label{bialgebra}
A $\mathbb{K}-$vector space $H$ together with the unital algebra
structure $(m, \mu)$ and the counital coalgebra structure $(\Delta,
\varepsilon)$ is called {\it bialgebra}, if $\Delta$ and $
\varepsilon$ are algebra morphisms and $\mu$ is a coalgebra
morphism. These conditions are determined with the following
commutative diagrams.
\begin{equation}
\xymatrix{ H\otimes H \otimes H \otimes  H \ar[rr]^{\tau_{23}} && H
\otimes  H \otimes  H \otimes H
\ar[d]^{m\otimes m} \\
H \otimes  H    \ar[u]_{\Delta\otimes\Delta} \ar[r]_m &  H
\ar[r]_{\Delta} & H \otimes  H}
\end{equation}

\begin{equation}
\xymatrix{ H \otimes  H \ar[d]^m
\ar[r]^{\varepsilon\otimes\varepsilon}
    & \mathbb{K} \otimes \mathbb{K} \ar[d]^\sim &&& H\otimes H &\mathbb{K} \otimes \mathbb{K} \ar[l]_{\mu \otimes \mu}\\
H \ar[r]^\varepsilon & \mathbb{K} &&&  H \ar[u]_\Delta & \mathbb{K}
\ar[l]_ \mu \ar[u]_\sim}
\end{equation}
\end{defn}

\begin{defn} \label{Hopfalgebra1}
A bialgebra $H$ together with a linear map $S: H \longrightarrow H$
is called {\it Hopf algebra}, if there is a compatibility between $S$ and bi-algebraic structure given by the following commutative diagram.
\begin{equation}
\xymatrix{& H \otimes  H  \ar[rr]^{S \otimes id}
                && H \otimes  H \ar[dr]^{m}   & \\
H\ar[rr]^\varepsilon \ar[dr]^\Delta \ar[ur]^\Delta
                &&  \mathbb{K} \ar[rr]^ \mu  && H\\
& H \otimes  H    \ar[rr]^{id \otimes S}
                && H \otimes H \ar[ur]^{m}   &}
\end{equation}
The map $S$ is called {\it antipode}.
\end{defn}

There are many examples of Hopf algebras and with attention to our scopes in continue at first we familiar with some important Hopf algebras and reader can find more other samples in \cite{A1, K11, S4}.

\begin{example}
For a fixed invertible element $q \in \mathbb{K}$, Hopf algebra
$H_{q}$ is generated by $1$ and elements $a, b, b^{-1}$ together with
relations
$$ bb^{-1} = b^{-1}b = 1, \  \  ba = q ab. $$
Its structures are determined by
$$ \Delta a = a \otimes 1 + b \otimes a, \  \  \Delta b = b \otimes b, \ \ \Delta b^{-1} = b^{-1} \otimes b^{-1}, $$
$$\varepsilon(a) = 0, \ \ \varepsilon(b) = \varepsilon(b^{-1}) = 1, \ \ S(a)=-b^{-1}a, \ \  S(b)=b^{-1}, \ \ S(b^{-1})=b.$$
\cite{M3}
\end{example}

\begin{example}
Let $G$ be a finite group and $\mathbb{K}G$ the vector space
generated by $G$. Identify a Hopf algebra structure on
$\mathbb{K}G$ such that its product is given by the group structure of
$G$ and
$$1=e, \ \ \Delta(g) = g \otimes g, \ \ \varepsilon(g)=1, \ \ S(g)=g^{-1}.$$ \cite{M3}
\end{example}

\begin{example} \label{universalenv}
Let $\mathfrak{g}$ be a finite dimensional Lie algebra over
$\mathbb{K}$. The universal enveloping algebra $U(\mathfrak{g})$ is
the noncommutative algebra generated by $1$ and elements of the
Lie algebra with respect to the relation
$$ [x,y]=xy - yx.$$
It introduces a Hopf algebra structure such that
$$ \Delta(x) = x \otimes 1 + 1 \otimes x, \ \ \varepsilon(x)=0, \ \ S(x)=-x.$$
One can show that $U(\mathfrak{g})$ is the quotient of the tensor
algebra $T(\mathfrak{g})$ modulo an ideal generated by the commutator. \cite{CM1}
\end{example}

\begin{example} \label{positivenumHopf}
Consider the algebra of multiple semigroup $H$ of natural positive
integers $\mathbb{N}$ such that $(e_{n})_{n \in \mathbb{N}}$ is its
basis as a vector space. With the help of decomposition of numbers
into the prime factors, one can define a commutative cocommutative
connected graded (with the number of prime factors (including
multiplicities)) Hopf algebra such that its coproduct and its
antipode are determined by
$$\Delta(e_{p_1\cdots p_k})=\sum_{I\amalg J=\{1,\ldots ,k\}}e_{p_I}\otimes e_{p_J},$$
$$S(e_n)=(-1)^{|n|}e_n,$$
where $p_I$ denotes the product of the primes $p_j,j\in I$. \cite{M2}
\end{example}

By adding some additional structures such as grading and filtration, one can apply Hopf algebras in physics. For example it helps us to classify all Feynman diagrams with respect to loop numbers or number of internal edges and
it will be useful when we do renormalization.

\begin{defn} \label{graded-hopf}
(i) A Hopf algebra $H$ over $\mathbb{K}$ is called graded, if it is
a graded $\mathbb{K}-$vector space $H=\bigoplus_{n\ge 0} H_n$ such
that
$$H_p. H_q \subset  H_{p+q}, \ \ \Delta(H_n)\subset \bigoplus_{p+q=n}
H_p\otimes H_q, \ \ S(H_n)\subset H_n.$$

(ii) A connected filtered Hopf algebra $H$ is a
$\mathbb{K}$-vector space together with an increasing
$\mathbb{Z}_{+}$-filtration:
$$H^0 \subset H^1 \subset \cdots \subset H^n \subset \cdots, \  \
\bigcup_n H^n = H$$ such that $H^{0}$ is one dimension and
$$H^p.H^q \subset  H^{p+q},
\ \ \Delta(H^n) \subset \sum_{p+q=n} H^p \otimes H^q, \ \ S(H^n)
\subset H^n.$$
\cite{EG2, M2}
\end{defn}

\begin{lem}  \label{increasingfilt3}
A graded bialgebra determines an increasing filtration $
H^n=\bigoplus_{p=0}^n H_p$. \cite{M2}
\end{lem}

The convolution product together with a
filtration structure determine an antipode on a bialgebra.

\begin{lem} \label{antipode1}
Any connected filtered bialgebra $H$ is a filtered Hopf algebra. Its
antipode structure is given by
$$S(x)=\sum_{k\ge 0}(\mu \varepsilon-Id_{H})^{*k}(x).$$ \cite{M2}
\end{lem}

For a fixed connected filtered bialgebra $H$ and an algebra $A$,
set a map $e:=\mu_{A}\circ\varepsilon_{H}$ such that
$e(\bold{1})=\bold{1}_{A}$ and $e(x)=0$ for any $x\in
 Ker \ \varepsilon_{H}$. It is easy to see that $e$ plays the role of unit
 for the convolution product $*$ on the set $L(H,A)$. Set
\begin{equation} \label{characters1}
G:=\{\varphi \in  L(H,A),\ \varphi(\bold{1})=\bold{1}_{A}\},
\end{equation}

\begin{equation} \label{deriv1}
\mathfrak{g}:= \{ \alpha \in L(H,A), \ \alpha(\bold{1})=0 \}.
\end{equation}
It will be shown that elements of these sets make possible to find a new prescription from physical
information of a quantum field theory.
\begin{thm} \label{char-der1}
(i) $(G,*)$ is a group such that for each $\varphi \in  G$, its
inverse is given by
$$ \varphi^{*-1}(x)= \bigl(e-(e-\varphi)\bigr)^{*-1}(x)
                =\sum_{k\ge 0}(e-\varphi)^{*k}(x).$$

(ii) $\mathfrak{g}$ is a subalgebra of $(L(H,A),*)$ such that
commutator with respect to the convolution product introduces a Lie
algebra structure on it.

(iii) $G=e + \mathfrak{g}$.

(iv) For any $x\in H^n$, the {\it exponential map} is defined by
$$e^{*\alpha}(x)=\sum_{k\ge 0}{\alpha^{*k}(x)\over k!}.$$
It determines a bijection map from $\mathfrak{g}$ onto $G$ such that
its inverse namely, the {\it logarithmic map} is given by
$$Log (1+\alpha)(x)=\sum_{k\ge 1}{(-1)^{k-1}\over k}\alpha^{*k}(x).$$

(v) The above sums have just finite terms. \cite{EG2, K11, M2, MM1}
\end{thm}

It is also interesting to know that the increasing
filtration on $H$ can identify a complete metric structure on $L(H,A)$. Set
\begin{equation} \label{metr1}
L^n:=\{\alpha\in L(H,A), \alpha|_{H^{n-1}}=0\}.
\end{equation}
It can be seen that for each positive
integer numbers $p$ and $q$,
\begin{equation} \label{metr2}
L^{p} * L^{q} \subset L^{p+q}.
\end{equation}
It gives a decreasing filtration on $L(H,A)$ such that $ L^0=
L(H,A)$ and $L^1= \mathfrak{g}$.
For each element $\varphi \in L(H,A)$, the value $val \ \varphi$ is
defined as the biggest integer $k$ such that $\varphi$ is in $L^k$.
The map
\begin{equation} \label{metr3}
d(\varphi,\psi)=2^{- val(\varphi-\psi)}
\end{equation}
gives us a complete metric on $L(H,A)$.

We close this section with introducing one
important technique for constructing Hopf algebras from Lie algebras
namely, Milnor-Moore theorem. Example \ref{universalenv} leads to a closed relation between Lie algebras and Hopf
algebras but in general, it is impossible to reconstruct a Hopf algebra
from a Lie algebra. By adding some conditions, one can find very
interesting process to recover Hopf algebras.

\begin{defn} \label{prim1}
(i) An element $p$ in the Hopf algebra $H$ is called {\it
primitive}, if
$$\Delta(p)= p \otimes 1 + 1 \otimes p.$$
(ii) A graded Hopf algebra is called {\it finite type}, if each of
the homogenous components $H_{i}$ are finite dimensional vector
spaces.
\end{defn}

One should mark that if the graded Hopf algebra $H$ (of finite
type) is an infinite dimensional vector space, its graded dual
\begin{equation}
H^{*}= \bigoplus_{n \ge 0} H_{n}^{\star}
\end{equation}
is strictly contained in the space
of linear functionals $H^{\star}:= L (H, \mathbb{K})$.

\begin{rem} \label{gradeddual1}
Let $H$ be a commutative (cocommutative) connected graded finite
type Hopf algebra.

(i)  $Ker \ \varepsilon \simeq \bigoplus_{i >0} H_{i}$ is an ideal
in $H$. It is called {\it augmentation ideal}.

(ii) A linear map $f \in H^{\star}$ belongs to $H^{*}$ if and only
if $f|_{H_{i}}=0$, for each component $H_{i}$ but for a finite
number.

(iii) If $H$ is finite dimensional vector space, then $H^{*} =
H^{\star}$.

(iv) There is a cocommutative (commutative) connected graded Hopf
algebra structure on $H^{*}$. \cite{EG2, K11, M2, MM1}
\end{rem}

For a given Hopf algebra $H$, an element $x$ in the augmentation
ideal is called {\it indecomposable}, if it can not been written as
a linear combination of products of elements in $Ker \ \varepsilon$.
The set of all indecomposable elements is denoted by $In(H)$.

\begin{thm} \label{gradeddual2}
For a given connected, graded and finite type Hopf algebra $H$,

(i) There is a correspondence between the set of all primitive
elements of $H$ namely, $Prim(H)$ and $In(H^{*})$.

(ii) There is a correspondence between $In(H)$ and $Prime(H^{*})$.
\cite{K11, MM1}
\end{thm}

Milnor and Moore proved that the reconstruction of a given Hopf algebra (under some conditions) on the basis of primitive elements is possible. In fact, this result plays an essential role in the Connes-Marcolli categorical configuration in the study of renormalizable quantum field theories.

\begin{thm} \label{milnor-moore}
Let $H$ be a connected graded commutative finite type Hopf
algebra. It can be reconstructed with the Lie subalgebra $Prim(H)$ of $L(H,
\mathbb{K})$ and it means that $H \simeq U(Prim(H))^{*}$. \cite{CM1, MM1}
\end{thm}

Two classes of elements in $H^{\star}$ have particular roles namely, {\it
characters} and {\it infinitesimal characters}. It is discussed by Kreimer that Feynman rules of a given quantum field theory can be capsulated in characters where this ability provides a new reformulation from counterterms, renormalized values, elements of the renormalization group and its related infinitesimal generator ($\beta-$function).

\begin{defn} \label{deriv2}
(i)
An element $f \in H^{\star}$ is called character, if $f(1)=1$ and for each $x,y \in H$,
$$f(xy)=f(x)f(y).$$

(ii) An element $g
\in H^{\star}$ is called derivation (or infinitesimal character), if
for each $x,y \in H$,
$$ g(xy) = g(x) \varepsilon(y) + g(y) \varepsilon(x).$$
\end{defn}

\begin{rem}
It is important to note that each primitive element of the Hopf
algebra $H^{*}$ determines a derivation.
\end{rem}


\chapter{\textsf{Combinatorial Hopf algebras}}

Hopf algebra of renormalization is introduced on the set of all Feynman diagrams of a given renormalizable physical theory such that its structures completely related to the renormalization process on these graphs. Since we want to have a general framework to consider perturbation theory, so it is necessary to identify a Hopf algebra structure independent of physical theories and further, it is reasonable to provide a universal simplified toy model for this Hopf algebraic formalization to apply it in computations. Fortunately, investigation of the combinatorics of the renormalization can help us to find a solution for this problem. Indeed, Kreimer applied decorated version of non-planar rooted trees (as combinatorial objects) to represent Feynman diagrams such that labels could help us to restore divergent sub-diagrams and their positions (i.e. nested loops) in origin graphs. Then with respect to the recursive mechanism for removing sub-divergences, he introduced a coproduct structure on these labeled rooted trees such that as the result one can produce a combinatorial Hopf algebra independent of physical theories. It is called Connes-Kreimer Hopf algebra of rooted trees  \cite{K6, K12}. Even more in a categorical configuration, this rooted tree type model is equipped with a universal property with respect to Hochschild cohomology theory such that the grafting operator can determine its related  Hochschild one cocycles \cite{EGK1, EGK2, EK1}. At this level one can expect the applications of combinatorial techniques in the study of perturbative renormalization \cite{CL1, EGGV, F2, H3, H2, H5, H6, H4, K13, KW1}.

In this chapter we focus on combinatorial objects and review the structures of some important defined combinatorial Hopf algebras which are connected with the Connes-Kreimer Hopf algeba.


\section{{\it \textsf{Connes-Kreimer Hopf algebra}}}

Rooted trees allow us to investigate the combinatorial basement of renormalization programme
such that as one expected result, it determines a universal simplification for explaining the removing of sub-divergences procedure. Here we consider the most important Hopf algebra structure on non-planar rooted trees in the study of QFTs.

\begin{defn} \label{rootedtree1}
A non-planar rooted tree $t$ is an oriented, connected and simply connected
graph together with one distinguished vertex with no incoming edge
namely, root. A monomial in rooted trees (that
commuting with each other) is called {\it forest}.
\end{defn}

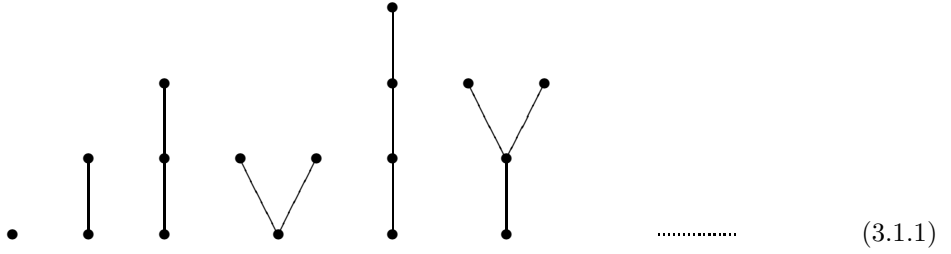
\begin{equation}
\xy
(-75,0)*{\bullet},
(-65,0)*{\bullet},
(-65,0);(-65,10) **@{-},
(-65,10)*{\bullet},
(-55,0)*{\bullet},
(-55,0);(-55,10) **@{-},
(-55,10)*{\bullet},
(-55,10);(-55,20) **@{-},
(-55,20)*{\bullet},
(-40,0)*{\bullet},
(-45,10);(-40,0) **@{-},
(-45,10)*{\bullet},
(-40,0);(-35,10) **@{-},
(-35,10)*{\bullet},
(-25,0)*{\bullet},
(-25,0);(-25,10) **@{-},
(-25,10)*{\bullet},
(-25,10);(-25,20) **@{-},
(-25,20)*{\bullet},
(-25,20);(-25,30) **@{-},
(-25,30)*{\bullet},
(-10,0)*{\bullet},
(-10,0);(-10,10) **@{-},
(-10,10)*{\bullet},
(-15,20);(-10,10) **@{-},
(-15,20)*{\bullet},
(-5,20)*{\bullet},
(-10,10);(-5,20) **@{-},
(10,0);(20,0) **@{.},
\endxy
\end{equation}

A rooted tree $t$ with a given embedding in the plane
is called planar rooted tree. For example,

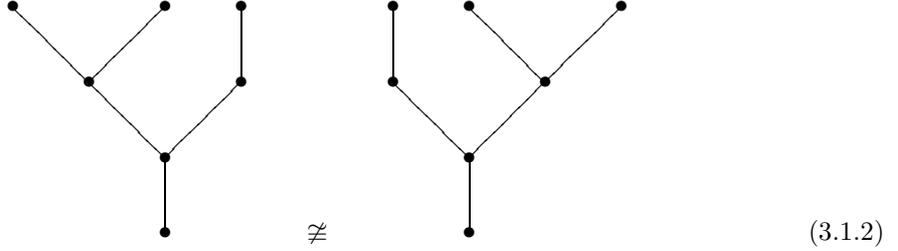
\begin{equation}
\xy
(20,0)*{\bullet},
(20,0);(20,10) **@{-},
(20,10)*{\bullet},
(10,20);(20,10) **@{-},
(10,20)*{\bullet},
(10,20),(0,30) **@{-},
(0,30)*{\bullet},
(10,20);(20,30) **@{-},
(20,30)*{\bullet},
(20,10);(30,20) **@{-},
(30,20)*{\bullet},
(30,20);(30,30) **@{-},
(30,30)*{\bullet},
(40,0)*{\ncong},
(60,0)*{\bullet},
(60,0);(60,10) **@{-},
(60,10)*{\bullet},
(50,20);(60,10)**@{-},
(50,20)*{\bullet},
(50,20);(50,30) **@{-},
(50,30)*{\bullet},
(60,10);(70,20) **@{-},
(70,20)*{\bullet},
(60,30);(70,20) **@{-},
(60,30)*{\bullet},
(70,20);(80,30) **@{-},
(80,30) *{\bullet},
\endxy
\end{equation}

Let $\textbf{T}$ be the set of all non-planar rooted trees and
$\mathbb{K} \textbf{T}$ be the vector space over the field
$\mathbb{K}$ (with characteristic zero) generated by $\textbf{T}$.
It is graded by the number of non-root vertices of rooted trees and
it means that
\begin{gather} \label{gradedtree}
\bold{T_{n}}:= \{ t \in \bold{T}: |t|=n+1 \}, \   \   \
\bold{\mathbb{K}T}:= \bigoplus_{n \ge 0} \mathbb{K} \bold{T_{n}}.
\end{gather}
Consider graded free unital
commutative symmetric algebra $H(\textbf{T})$ containing $\mathbb{K} \textbf{T}$ such that the
empty tree is its unit. We equip this
space with the counit $\epsilon: H(\textbf{T}) \longrightarrow
\mathbb{K}$ given by
\begin{gather} \label{counittree}
\epsilon(\mathbb{I})=1, \   \    \epsilon(t_{1}...t_{n})=0, \  \
t_{1}...t_{n}\neq \mathbb{I}.
\end{gather}
With respect to the BPHZ renormalization process, some edges and vertices from rooted trees should be removed step by step and this can be formulated with a special family of cuts.
\begin{defn} \label{admisiblecut1}
An \textit{admissible cut} $c$ of a rooted tree $t$ is a collection
of its edges with this condition that along any path from the root
to the other vertices, it meets at most one element of $c$. By
removing the elements of an admissible cut $c$ from a rooted tree
$t$, we will have a rooted tree $R_{c}(t)$ with the original root
and a forest $P_{c}(t)$ of rooted trees.
\end{defn}

For instance,
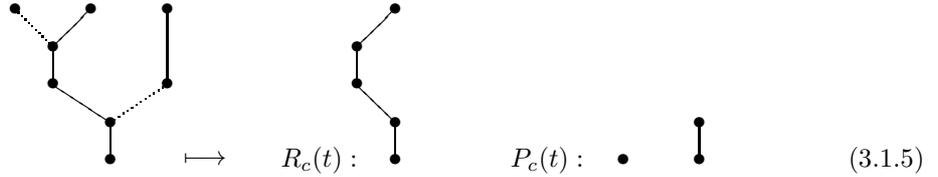
\begin{equation}
\xy
(10,20)*{\bullet},
(10,20);(15,15) **@{.},
(15,15)*{\bullet},
(20,20);(15,15) **@{-},
(20,20)*{\bullet},
(15,15);(15,10) **@{-},
(15,10)*{\bullet},
(30,20);(30,10) **@{-},
(30,10)*{\bullet},
(30,20)*{\bullet},
(22.5,0);(22.5,5) **@{-},
(22.5,0)*{\bullet},
(22.5,5);(15,10) **@{-},
(22.5,5)*{\bullet},
(22.5,5);(30,10) **@{.},
(35,0)*{\longmapsto},
(50,0)*{R_{c}(t):},
(60,0)*{\bullet},
(60,0);(60,5)**@{-},
(60,5)*{\bullet},
(55,10);(60,5) **@{-},
(55,10)*{\bullet},
(55,10);(55,15) **@{-},
(55,15)*{\bullet},
(55,15);(60,20) **@{-},
(60,20)*{\bullet},
(80,0)*{P_{c}(t):},
(90,0)*{\bullet},
(100,0)*{\bullet},
(100,0);(100,5) **@{-},
(100,5)*{\bullet},
\endxy
\end{equation}
shows an admissible cut but the cut
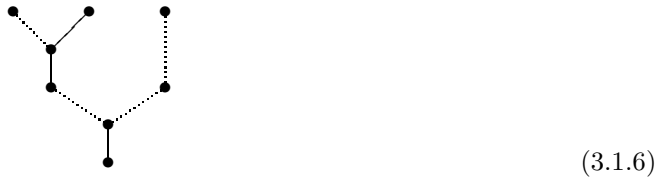
\begin{equation}
\xy
(10,20)*{\bullet},
(10,20);(15,15) **@{.},
(15,15)*{\bullet},
(20,20);(15,15) **@{-},
(20,20)*{\bullet},
(15,15);(15,10) **@{-},
(15,10)*{\bullet},
(30,20);(30,10) **@{.},
(30,10)*{\bullet},
(30,20)*{\bullet},
(22.5,0);(22.5,5) **@{-},
(22.5,0)*{\bullet},
(22.5,5);(15,10) **@{.},
(22.5,5)*{\bullet},
(22.5,5);(30,10) **@{.},
\endxy
\end{equation}
can not be admissible. This concept determines a coproduct structure on $H(\textbf{T})$
given by
\begin{gather} \label{cop-ck-1}
\Delta: H(\textbf{T}) \longrightarrow H(\textbf{T}) \otimes
H(\textbf{T}),  \   \  \Delta (t) = t \otimes \mathbb{I} +
\mathbb{I} \otimes t + \sum_{c} P_{c} (t) \otimes R_{c} (t)
\end{gather}
where the sum is over all possible non-trivial admissible cuts on
$t$. As an example,
$$
\xy
(-35,0)*{\Delta(},
(-30,0)*{\bullet},
(-30,0);(-30,5) **@{-},
(-30,5)*{\bullet},
(-30,5);(-30,10) **@{-},
(-30,10)*{\bullet},
(-25,15);(-30,10) **@{-},
(-25,15)*{\bullet},
(-35,15);(-30,10) **@{-},
(-35,15)*{\bullet},
(-25,0)*{)=},
(-15,0)*{\bullet},
(-15,0);(-15,5) **@{-},
(-15,5)*{\bullet},
(-15,5);(-15,10) **@{-},
(-15,10)*{\bullet},
(-20,15);(-15,10) **@{-},
(-20,15)*{\bullet},
(-15,10);(-10,15) **@{-},
(-10,15)*{\bullet},
(-5,0)*{\otimes},
(0,0)*{\mathbb{I}},
(5,0)*{+},
(10,0)*{\mathbb{I}},
(15,0)*{\otimes},
(20,0)*{\bullet},
(20,0);(20,5) **@{-},
(20,5)*{\bullet},
(20,5);(20,10) **@{-},
(20,10)*{\bullet},
(15,15);(20,10) **@{-},
(15,15)*{\bullet},
(20,10);(25,15) **@{-},
(25,15)*{\bullet},
(30,0)*{+},
(35,0)*{\bullet},
(40,0)*{\otimes},
(45,0)*{\bullet},
(45,0);(45,5) **@{-},
(45,5)*{\bullet},
(40,10);(45,5) **@{-},
(40,10)*{\bullet},
(45,5);(50,10) **@{-},
(50,10)*{\bullet},
(55,0)*{+},
(60,0)*{\bullet},
(60,0);(60,5) **@{-},
(60,5)*{\bullet},
(65,0)*{\otimes},
(70,0)*{\bullet},
(65,5);(70,0) **@{-},
(65,5)*{\bullet},
(70,0);(75,5) **@{-},
(75,5)*{\bullet},
(80,0)*{+},
\endxy
$$
\begin{equation}
\xy
(0,0)*{2};
(5,0)*{\bullet},
(5,0);(5,5) **@{-},
(5,5)*{\bullet},
(5,5);(5,10) **@{-},
(5,10)*{\bullet},
(5,10);(5,15) **@{-},
(5,15)*{\bullet},
(10,0)*{\otimes};
(15,0)*{\bullet},
(20,0)*{+},
(25,0)*{\bullet},
(25,0);(25,5) **@{-},
(25,5)*{\bullet},
(25,5);(25,10) **@{-},
(25,10)*{\bullet},
(30,0)*{\otimes},
(35,0)*{\bullet},
(40,0)*{\bullet},
\endxy
\end{equation}

It should be remarked that this coproduct can be rewritten in a
recursive way. Let $B^{+}:H(\textbf{T}) \longrightarrow
H(\textbf{T})$ be a linear operator that mapping a forest to a
rooted tree by connecting the roots of rooted trees in the forest to
a new root.
$$
\xy
(0,5)*{r},
(0,0)*{\bullet},
(0,0);(-15,-10) **@{-},
(0,0);(-10,-10) **@{-},
(0,0);(10,-10) **@{-},
\endxy
$$
\begin{equation}
\xy
(0,-5)*{t_{1}},
(0,0)*{\bullet},
(5,-5)*{t_{2}},
(5,0)*{\bullet},
(10,0);(20,0) **@{.},
(25,-5)*{t_{n}},
(25,0)*{\bullet},
\endxy
\end{equation}
Operator $B^{+}$ is an isomorphism of graded vector spaces and
for the rooted tree $t=B^{+}(t_{1}...t_{n})$, we have
\begin{equation} \label{cop-ck-2}
\Delta B^{+}(t_{1}...t_{n})=t \otimes \mathbb{I} + (id \otimes
B^{+}) \Delta (t_{1}...t_{n}).
\end{equation}
$\Delta$ is extended linearity to define it as an algebra
homomorphism. On the other hand, with the help of admissible cuts one can define recursively an
antipode on $H(\textbf{T})$ given by
\begin{equation} \label{antip-ck1}
S(t)=-t- \sum_{c} S(P_{c}(t))R_{c}(t).
\end{equation}

\begin{thm}
The symmetric algebra $H(\textbf{T})$ together with the coproduct (\ref{cop-ck-1}) and the
antipode (\ref{antip-ck1}) is a finite type connected graded
commutative noncocommutative Hopf algebra. It is called {\it
Connes-Kreimer Hopf algebra} and denoted by $H_{CK}$. \cite{CK4, EG2, EGK1, EGK2}
\end{thm}


The study of Hopf subalgebras of $H_{CK}$ can be useful. For instance one can consider the cocommutative Hopf subalgebra of {\it
ladder trees} (i.e. rooted trees without any side-branchings)
$H(\textbf{LT})$ such that it is applied to work on
the relations between perturbative QFTs and representation theory
of Lie algebras. For this case, $H(\textbf{LT})$ is reduced to a
polynomial algebra freely generated by ladder trees and with the
help of increasing or decreasing the degree of generators, one can
induce insertion and elimination operators. \cite{MD1, MD2}

By theorem \ref{char-der1}, the convolution product $*$ determines a
group structure on the space $\mathbf{char H_{CK}}$ of all
characters and a graded Lie algebra structure on the space $\mathbf{
\partial char H_{CK}}$ of all derivations where naturally, there is a bijection map $\exp^{*}$ from $\mathbf{ \partial char H_{CK}}$ to $\mathbf{char H_{CK}}$
(which plays an essential role in the representation of components of
the Birkhoff decomposition of characters). \cite{EGK1, EGK2}

Finally one should mark to the {\it universal property} of this
Hopf algebra such that it is the essential result of the universal problem in
Hochschild cohomology.

\begin{thm} \label{universality-ck}
Let $\mathcal{C}$ be a category with objects $(H,L)$ consisting of a
commutative Hopf algebra $H$ and a Hochschild one cocycle $L:H
\longrightarrow H$. It means that for each $x \in H$,
$$\Delta L(x) = L(x) \otimes \mathbb{I} + (id \otimes L)\Delta(x).$$
And also Hopf algebra homomorphisms, that commute with cocycles, are
morphisms in this category. $(H_{CK},B^{+})$ is the universal
element in $\mathcal{C}$. In other words, for each object $(H,L)$
there exists a unique morphism of Hopf algebras $\phi: H_{CK}
\longrightarrow H$ such that $L \circ \phi = \phi \circ B^{+}$.
$H_{CK}$ is unique up to isomorphism. \cite{CK4}
\end{thm}


\section{{\it \textsf{Rooted trees and (quasi-)symmetric functions}}}

There are different Hopf algebra structures on (non-)planar rooted trees and in fact, Connes-Kreimer Hopf algebra is one particular choice. Here we try to familiar with some important combinatorial Hopf algebras and then with using (quasi-)symmetric functions, their relations with $H_{CK}$ will be considered.

\begin{defn}
Let $t,s$ be rooted trees such that  $t=B^{+}(t_{1},...t_{n})$ and
$|s|=m$. The new product $t \bigcirc s$ is defined with the sum of rooted trees given by
attaching each of $t_{i}$ to a vertex of $s$.
\end{defn}

One can define a
coproduct compatible with $\bigcirc$ on $\bold{\mathbb{K}T}$. It is
given by
\begin{equation}
\Delta_{GL} B^{+}(t_{1},...t_{k}) = \sum_{I \cup J = \{1,2, ...,k\}}
B^{+}(t(I)) \otimes B^{+}(t(J)).
\end{equation}

\begin{thm} \label{GL-1}
$H_{GL}:=(\bold{\mathbb{K}T}, \bigcirc, \Delta_{GL})$ is a connected
graded noncommutative cocommutative Hopf algebra and it is called
{\it Grossman-Larson Hopf algebra}. $H_{GL}$ is the graded dual of
$H_{CK}$ and it is the universal enveloping algebra of its Lie
algebra of primitives. \cite{H3, P1}
\end{thm}

Let $\bold{P}$ be the set of all planar rooted trees and
$\bold{\mathbb{K}P}$ be its graded vector space. Tensor algebra
$T(\bold{\mathbb{K}P})$ is an algebra of ordered forests of planar
rooted trees and $B^{+}:T(\bold{\mathbb{K}P}) \longrightarrow
\bold{\mathbb{K}P}$ is an isomorphism of graded vector spaces. There
are two interesting Hopf algebra structures on $\bold{P}$.

\begin{defn} \label{bbr1}
A balanced bracket representation (BBR) of a planar rooted tree
contains symbols $<$ and $>$ satisfying in the following rules:

- For a planar rooted tree of weight $n$, the symbol $<$ and the
symbol $>$ occur $n$ times,

- In reading from left to right, the count of $<$'s is agree with
the count of $>$'s,

- The empty BBR is a tree with just one vertex.
\end{defn}

For example, one represents planar rooted trees
\begin{equation}
\xy
(-55,0)*{\bullet},
(-55,0);(-55,10) **@{-},
(-55,10)*{\bullet},
(-55,10);(-55,20) **@{-},
(-55,20)*{\bullet},
(-40,0)*{\bullet},
(-45,10);(-40,0) **@{-},
(-45,10)*{\bullet},
(-40,0);(-35,10) **@{-},
(-35,10)*{\bullet},
(-25,0)*{\bullet},
(-25,0);(-25,10) **@{-},
(-25,10)*{\bullet},
(-25,10);(-25,20) **@{-},
(-25,20)*{\bullet},
(-25,20);(-25,30) **@{-},
(-25,30)*{\bullet},
(-10,0)*{\bullet},
(-10,0);(-10,10) **@{-},
(-10,10)*{\bullet},
(-15,20);(-10,10) **@{-},
(-15,20)*{\bullet},
(-5,20)*{\bullet},
(-10,10);(-5,20) **@{-},
(10,0)*{\bullet},
(0,10);(10,0) **@{-},
(0,10)*{\bullet},
(10,0);(10,10) **@{-};
(10,10)*{\bullet},
(10,0);(20,10) **@{-},
(20,10)*{\bullet},
\endxy
\end{equation}
with
\begin{equation}
<<>>, \  \  \   <><>,  \  \  \  <<<>>>,  \  \  \  <<><>>,  \  \  \  <><><>,
\end{equation}
respectively.

\begin{defn} \label{bbr2}
A BBR $F$ is called irreducible, if $F=<G>$ for some BBR $G$ and
otherwise it can be written by a juxtaposition $F_{1}...F_{k}$ of
irreducible BBRs. These components correspond with the branches of the
root in the associated planar rooted tree.
\end{defn}

\begin{defn}
Let $t,s$ be two planar rooted trees with BBR representations
$F_{t}, F_{s}$ such that $F_{t}=F^{1}_{t}...F^{k}_{t}$. Define a new
product $t \diamond s$ by a sum of planar rooted trees such that
their BBRs are given by shuffling the components of $F_{t}$ into the
$F_{s}$.
\end{defn}

Moreover, with help of the decomposition of elements into
their irreducible components, one can modify a compatible coproduct
$\Delta_{\diamond}$ on $\bold{\mathbb{K}P}$.

\begin{thm} \label{bbr3} \label{bbr4}

(i) Based on the balanced bracket representation, there is a connected graded noncommutative Hopf
algebra structure on $\bold{\mathbb{K}P}$ and it is denoted by
$H_{\bold{P}}:= (\bold{\mathbb{K}P}, \diamond, \Delta_{\diamond})$.

(ii) Based on the coproduct (\ref{cop-ck-1}), there is a graded connected
noncommutative Hopf algebra. It is called {\it Foissy Hopf algebra}
and denoted by $H_{F}$. $H_{F}$ is self-dual and isomorphic to
$H_{\bold{P}}$. \cite{F1, H2, H4}

\end{thm}

Relation between rooted trees and noncommutative geometry can be clear when the reconstruction of one important Hopf algebra in computations of transverse index theory,
based on rooted trees, is done \cite{CK4, K13}. Consider a Hopf algebra $H$ with the generators $x,y,\delta_{n}$ $ \
(n \in \mathbb{N})$, together with the following relations
\begin{gather} \label{cm-relations}
[x,y]=-x, \ \  [x,\delta_{n}]=\delta_{n+1}, \ \  [y,\delta_{n}]=n
\delta_{n}, \ \ [\delta_{n},\delta_{m}]=0,
\end{gather}
such that its coproduct structure on the generators is given by
$$\Delta(x)=x \otimes 1 + 1 \otimes x + \delta_{1},$$
$$\Delta(y)=y \otimes 1 + 1 \otimes y, $$
\begin{equation} \label{cm-cop}
\Delta(\delta_{1})=\delta_{1} \otimes 1 + 1 \otimes \delta_{1}.
\end{equation}
It is easy to present the generators $\delta_{n}$ with rooted trees. Define
a linear operator $N$ on rooted trees such that its application on a
rooted tree $t$ is a sum of rooted trees given by adding an edge to
each vertex of $t$. Now identify $\delta_{1}$ with $\bullet$ and
$\delta_{n}=N^{n}(\mathbb{I})$.

\begin{lem} \label{cm-hopf}
The set of generators $\{\delta_{n}\}_{n}$ introduces a Hopf subalgebra of $H$ such that it
is equivalent to the Connes-Moscovici Hopf algebra
$H_{CM}$. \cite{CK4}
\end{lem}

It will be shown that how one can reduce $H_{CM}$ based on Dyson-Schwinger
equations and additionally, a copy of this Hopf algebra related to the
universal Hopf algebra of renormalization will be determined. These results report the importance of this Hopf algebra in the study of quantum field theory.

\begin{defn}
Let $\mathbb{K}[[x_{1},x_{2},...]]$ be the ring of formal power
series. A formal series $f$ is called

(i) symmetric, if for any sequence of distinct positive integers
$n_{1},...,n_{k}$, the coefficients in $f$ of the monomials
$x^{i_{1}}_{n_{1}} ... x^{i_{k}}_{n_{k}}$ and $x^{i_{1}}_{1} ...
x^{i_{k}}_{k}$ equal.

(ii) quasi-symmetric, if for any increasing sequence $n_{1} < ... <
n_{k}$, the coefficients in $f$ of the monomials
$x^{i_{1}}_{n_{1}} ... x^{i_{k}}_{n_{k}}$ and $x^{i_{1}}_{1} ...
x^{i_{k}}_{k}$ equal.

(iii) Let $SYM$ ($QSYM$) be the set
of all symmetric (quasi-symmetric) functions. It is easy to see that
$SYM \subset QSYM$.
\end{defn}

For better understanding, it can be seen that for each $n$ the symmetric group
$\mathbb{S}_{n}$ acts on $\mathbb{K}[[x_{1},x_{2},...]]$ by permuting
the variables and a symmetric function is invariant under these
actions and it means that after each permutation coefficients of its
monomials remain without any change.

\begin{lem} \label{gen-qusi-sym}
(i) As a vector space, $QSYM$ is generated by the monomial
quasi-symmetric functions $M_{I}$ such that $I=(i_{1},...,i_{k})$
and $M_{I}:=\sum_{n_{1} < n_{2} < ... < n_{k}} x^{i_{1}}_{n_{1}} ...
x^{i_{k}}_{n_{k}}$.

(ii) If we forget order in a composition, then the generators $m_{\lambda} := \sum_{\phi(I)= \lambda}
M_{I}$ of $SYM$ (viewed as a vector space) can be determined. \cite{GKLLRT, H2}
\end{lem}

\begin{thm} \label{main10}
(i) There is a graded connected commutative cocommutative self-dual Hopf
algebra structure on $SYM$.

(ii) There is a graded connected
commutative non-cocommutative Hopf algebra structure on $QSYM$ such
that its graded dual is denoted by $NSYM$. As an algebra, $NSYM$ is
the noncommutative polynomials on the variables $z_{n}$ of degree
$n$. \cite{GKLLRT, H2}
\end{thm}

Hoffman could find new important relations between rooted trees and (quasi-)symmetric
functions such that we will extend his results to the level of the universal Hopf
algebra of renormalization.

\begin{thm} \label{rel-tree-symm}
There are following commutative diagrams of Hopf algebra
homomorphisms.  \cite{H2}

\begin{gather}
\begin{CD}
NSYM @>{\alpha_{1}}>> H_{F}\\
@V{\alpha_{3}}VV @V{\alpha_{2}}VV\\
SYM @>{\alpha_{4}}>> H_{CK}
\end{CD}
\ \ \ \ \
\begin{CD}
SYM @<{\alpha_{4}^{\star}}<< H_{GL}\\
@V{\alpha_{3}^{\star}}VV @V{\alpha_{2}^{\star}}VV\\
QSYM @<{\alpha_{1}^{\star}}<< H_{\bold{P}}
\end{CD}
\end{gather}
\end{thm}

\begin{proof}

It is enough to define homomorphisms on generators. With attention to the definitions of Hopf algebras, we have

- $\alpha_{1}$ sends each variable $z_{n}$ to the ladder tree
$l_{n}$ of degree $n$.

- $\alpha_{2}$ maps each planar rooted tree to its corresponding
rooted tree without notice to the order in products.

- $\alpha_{3}$ sends each $z_{n}$ to the symmetric function
$m_{\underbrace{(1,...,1)}_{n}}$.

- $\alpha_{4}$ maps $m_{\underbrace{(1,...,1)}_{n}}$ to the ladder
tree $l_{n}$.

- For the composition $I=(i_{1},...,i_{k})$ define a planar
rooted tree $t_{I}:=B^{+}(l_{i_{1}},...,l_{i_{k}})$. For each planar
rooted tree $t$, if $t=t_{I}$, then define
$\alpha_{1}^{\star}(t):=M_{I}$ and otherwise
$\alpha_{1}^{\star}(t):=0$.

- For each rooted tree $t$, $\alpha_{2}^{\star}(t):=|sym(t)| \sum_{s
\in \alpha_{2}^{-1}(t)} s$.

- $\alpha_{3}^{\star}$ is the inclusion map.

- For the partition $J=(j_{1},...,j_{k})$, define a rooted tree
$t_{J}:=B^{+}(l_{j_{1}},...,l_{j_{k}})$. For each rooted tree $t$,
if $t=t_{J}$ (for some partition $J$), then define
$\alpha_{4}^{\star}(t):=|sym(t_{J})|m_{J}$ and otherwise
$\alpha_{4}^{\star}(t):=0$.

\end{proof}

\begin{defn} \label{zhao-def} \label{zhao} \label{zhao1} \label{zhao2}
Recursively define the following morphism
$$
Z: NSYM \longrightarrow H_{GL},  \  \  Z(z_{n}) = \epsilon_{n}
$$
such that rooted trees $\epsilon_{n}$ are given by
$$\epsilon_{0}:= \bullet$$
$$
\epsilon_{n}:=k_{1} \bigcirc \epsilon_{n-1} - k_{2} \bigcirc
\epsilon_{n-2} + ... + (-1)^{n-1} k_{n}
$$
where
$$
k_{n}:= \sum_{|t|=n+1} \frac{t}{|sym(t)|} \in H_{GL}.
$$
It is called {\it Zhao's homomorphism}.
\end{defn}

From definition it is clear that $Z$ is an injective homomorphism of Hopf algebras.

\begin{lem}  \label{zhao3} \label{zhao4}
Dual of Zhao's homomorphism exists uniquely. \cite{H5, H6}
\end{lem}

\begin{proof}
Suppose
$$
A^{+}: QSYM \longrightarrow QSYM, \   \   M_{I} \longmapsto M_{I
\sqcup (1)}.
$$
It is a linear map with the cocycle property. For each ladder tree $l_{n}$ and monomial $u$ of rooted trees, define a morphism
$Z^{\star}: H_{CK} \longrightarrow QSYM$ such that
$$l_{n} \longmapsto M_{\underbrace{(1,...,1)}_{n}},$$
$$
B^{+}(u) \longmapsto A^{+} (Z^{\star}(u)).
$$
One can see that
$Z^{\star}$ is the unique homomorphism with respect to the map $A^{+}$.
\end{proof}

We show that it is possible to lift the
Zhao's homomorphism and its dual to the level of Hall rooted trees and Lyndon words. Roughly, this process provides an extension of this homomorphism to the level of the universal Hopf
algebra of renormalization.


\section{{\it \textsf{Incidence Hopf algebras}}}

On the one hand, rooted trees introduce one important class of Hopf algebras namely, combinatorial type and on the other hand, incidence Hopf algebras, induced in operad theory, provide another general class of Hopf algebras such that rooted trees (as kind of posets) characterize interesting examples in this procedure.
The essential part of this story is that incidence Hopf algebra related to one special family of operads introduces the Connes-Kreimer Hopf algebra.

The story of operad theory was begun with the study of loop spaces and then its applications in different branches of mathematics were found very soon. There is a closed relation between operads and objects of symmetric monoidal categories such as category of sets, category of topological spaces, category of vector spaces, ... . Additionally, operads can determine interesting source of Hopf algebras namely, incidence Hopf algebras \cite{CL1, S1, V1}. In this part we are going to consider this important family of Hopf algebras related to posets.

\begin{defn} \label{poset}
A partially ordered set (poset) is a set with a partial order
relation. A growing sequence of the elements of a poset is called
chain. A poset is pure, if for any $x \le y$ the maximal chains
between $x$ and $y$ have the same length. A bounded and pure poset
is called graded poset.
\end{defn}

\begin{example}
One can define a graded partial order on the set
$[n]=\{1,2,...,n\}$ by the refinement of partitions and it is called
partition poset.
\end{example}

\begin{defn} \label{operad}
(i) An operad $(P, \bold{co}, \bold{u})$ is a monoid in the monoidal category
$\mathbb{S}-Mod$ of
$\mathbb{S}-$modules (i.e. a collection $\{P(n)\}_{n}$ of (right)
$\mathbb{S}_{n}-$modules). It means that the composition morphism
$\bold{co}: P \circ P \longrightarrow P$ is associative and the morphism
$\bold{u}: \mathbb{I}\longrightarrow P$ is unit.

(ii) This operad is called augmented, if there exists a morphism of
operads $\psi_{P}: P \longrightarrow \mathbb{I}$ such that $\psi_{P} \circ
\bold{u}=id$.
\end{defn}

\begin{example}
A $\mathbb{S}-$set is a collection $\{P_{n}\}_{n}$
of sets $P_{n}$ equipped with an action of the group
$\mathbb{S}_{n}$. A monoid $(P, \bold{co}, \bold{u})$ in the monoidal category
of $\mathbb{S}-$sets is called a set operad.
\end{example}

\begin{defn} \label{basic-operad}
For a given set operad $P$ and for each $(x_{1},...,x_{t}) \in P_{i_{1}} \times ... \times
P_{i_{t}}$, define a map
$$\lambda_{(x_{1},...,x_{t})}: P_{t} \longrightarrow P_{i_{1}+...+i_{t}}, \
x \longmapsto \bold{co}(x \circ (x_{1},...,x_{t})).$$
A set opeard $P$ is called basic, if each
$\lambda_{(x_{1},...,x_{t})}$ be injective.
\end{defn}

Incidence Hopf algebras are introduced based on special partition posets associated to set operads. At first we should know the structure of this class of posets.

\begin{defn} \label{diagonal-action}
Suppose $(P, \bold{co}, \bold{u})$ be a set operad and for the given set $A$ with $n$
elements, let $\mathbb{A}$ be the set of ordered sequences of the
elements of $A$ such that each element appearing once. For each $n$, there is an action of
the group $\mathbb{S}_{n}$ on $P_{n}$ such that for each
element $x_{n} \times (a_{i_{1}},...,a_{i_{n}})$ in $P_{n} \times
\mathbb{A}$, its image under an element $\sigma$ of $\mathbb{S}_{n}$
is given by
$$\sigma(x_{n}) \times
(a_{\sigma^{-1}(i_{1})},...,a_{\sigma^{-1}(i_{n})}).$$
It is called diagonal action and its orbit is denoted by $\overline{x_{n}
\times (a_{i_{1}},...,a_{i_{n}})}$.
\end{defn}

\begin{defn} \label{p-partition} \label{diagonal2}
Let $\mathfrak{P}_{n}(A):=P_{n}
\times_{\mathbb{S}_{n}} \mathbb{A}$ be the set of all orbits under the
diagonal action. Set
$$
P(A):=(\bigsqcup_{f:[n] \longrightarrow A} P_{n})_{\sim}
$$
where $f$ is a bijection and $(x_{n},f) \sim (\sigma(x_{n}),f \circ \sigma^{-1})$ is an
equivalence relation. A $P-$partition of $[n]$ is a set of components $B_{1},...,B_{t}$
such that

- Each $B_{j}$ belongs to $\mathfrak{P}_{i_{j}} (I_{j})$
where $i_{1}+...+i_{t}=n$,

- Family $\{I_{j}\}_{1 \le j \le t}$ is a
partition of $[n]$.
\end{defn}

\begin{lem}
One can extend maps $\lambda_{(x_{1},...,x_{t})}$ to
$\lambda^{\sim}$ at the level of $P(A)$. \cite{V1}
\end{lem}

\begin{proof}
Define
$$ \lambda^{\sim}: P_{t} \times (\mathfrak{P}_{i_{1}}(I_{1}) \times ... \times \mathfrak{P}_{i_{t}}(I_{t})) \longrightarrow   \mathfrak{P}_{i_{1}+...+i_{t}}(A)$$
$$x \times (c_{1},...,c_{t}) \longmapsto \overline{\bold{co}( x \circ
(x_{1},...,x_{t})) \times (a^{1}_{1},...,a^{t}_{i_{t}})}$$
such that $\{I_{j}\}_{1 \le j \le t}$ is a partition of $A$ and each
$c_{r}$ is represented by $\overline{x_{r} \times
(a^{r}_{1},...,a^{r}_{i_{r}})}$ where $x_{r} \in P_{i_{r}}$,
$I_{r}=\{ a^{r}_{1},...,a^{r}_{i_{r}} \}$.
\end{proof}

\begin{defn} \label{part-poset1}
For the set operad $P$ and $P-$partitions $\mathfrak{B} =
\{B_{1},...,B_{r}\}$, $\mathfrak{C} = \{C_{1},...,C_{s}\}$ of $[n]$
such that $B_{k} \in \mathfrak{P}_{i_{k}} (I_{k})$ and $C_{l} \in
\mathfrak{P}_{j_{l}} (J_{l})$, we say that the $P-$partition
$\mathfrak{C}$ is larger than $\mathfrak{B}$, if for any $k \in
\{1,2,...,r\}$ there exists $\{p_{1},...,p_{t}\} \subset
\{1,2,...,s\}$ such that

- Family $\{J_{p_{1}},...,J_{p_{t}}\}$ is a
partition of $I_{k}$,

- There exists an element $x_{t} \in
P_{t}$ such that $B_{k}=\lambda^{\sim} (x_{t} \times
(C_{p_{1}},...,C_{p_{t}}))$.

This poset is called operadic
partition poset associated to the operad $P$ and denoted by
$\Pi_{P}([n])$.
\end{defn}

One can develop the notion of this poset to each locally finite set
$A= \bigsqcup A_{n}$ such that in this case a $P-$partition of $[A]$
is a disjoint union (composition) of $P-$partitions of $[A_{n}]$s
and therefore the operadic partition poset associated to the operad
$P$ will be a composition of posets $\Pi_{P}([A_{n}])$ and denoted
by $\Pi_{P}([A])$.

\begin{defn} \label{part-poset2}
A collection $(\mathfrak{p}_{i})_{i \in I}$ of posets is called good
collection, if

- Each poset $\mathfrak{p}_{i}$ has a minimal element $\bold{0}$ and
a maximal element $\bold{1}$ (an interval),

- For all $i \in I,$ $x \in \mathfrak{p}_{i}$, the interval
$[\bold{0},x]$ (or $[x,\bold{1}]$) is isomorphic to a product of posets
$\prod_{j} \mathfrak{p}_{j}$ (or $\prod_{k} \mathfrak{p}_{k}$).
\end{defn}

\begin{rem} \label{part-poset3}
For a given good collection $\mathcal{A}:= (\mathfrak{p}_{i})_{i \in
I}$, it is possible to make a new good collection $\mathcal{A}^{-}$
of all finite products $\prod_{i} \mathfrak{p}_{i}$ of elements such
that it is closed under products and closed under taking
subintervals. \cite{CL1}
\end{rem}

Let $[\mathcal{A}]$ ($[\mathcal{A}^{-}]$) be the set of isomorphism
classes of posets in $\mathcal{A}$ ($\mathcal{A}^{-}$) such that
elements in these sets denoted by $[i], [j], ...$ and
$H_{\mathcal{A}}$ be a vector space generated by the family
$\{F_{[i]}\}_{[i] \in [\mathcal{A}^{-}]}$. It is equipped with a
commutative product (i.e. direct product of posets) $F_{[i]}F_{[j]}
= F_{[i \times j]}$ such that $F_{[e]}$ is the unit (where $[e]$ is
the isomorphism class of the singleton interval).

\begin{rem}
As an algebra $H_{\mathcal{A}}$ may not be free.
\end{rem}

\begin{lem}  \label{inc-co}
Based on subintervals, there is a coproduct structure on $H_{\mathcal{A}}$
given by
$$\Delta (F_{[i]}) = \sum_{x \in \mathfrak{p}_{i}} F_{[\bold{0},x]}
\otimes F_{[x,\bold{1}]}.$$
It determines a commutative Hopf algebra.
\end{lem}

\begin{thm} \label{inci-hopf}
Let $\Pi_{P}$ be a family of the operadic partition posets
associated to the set operad $P$. One can find a good collection of
posets $(\mathfrak{p}_{i})$ (depended upon $\Pi_{P}$) such that
its related Hopf algebra $H_{P}$ is called incidence Hopf
algebra. \cite{CL1, S1}
\end{thm}

\begin{rem}
Incidence Hopf algebra $H_{P}$ has a basis indexed by the
isomorphism classes of intervals in the posets $\Pi_{P}(I)$ (for all
sets $I$) and this identification makes the sets $I$ disappear and
it means that the construction of this Hopf algebra is independent
of any label.
\end{rem}

A rooted tree looks like a poset with a unique minimal element
(root) such that for any element $v$, the set of elements descending
$v$ forms a chain (i.e. the graph has no loop) and maximal elements
are called {\it leaves}. There is an interesting basic set operad on
rooted trees such that its incidence Hopf algebra
determines a well known object.

\begin{defn} \label{compo-tree}
For the set $I$ with the partition $\{J_{i}\}_{i \ge 1 }$, suppose
$NAP(I)$ be the set of rooted trees with vertices labeled by $I$.
For $s_{i} \in NAP(J_{i})$ and $t \in NAP(I)$, we consider the
disjoint union of the rooted trees $s_{i}$ such that for each edge
of $t$ between $i_{1}, i_{2}$ in $I$, add an edge between the root
of $s_{i_{1}}$ and the root of $s_{i_{2}}$. The resulting graph is a
rooted tree labeled by $\bigsqcup_{i} J_{i}$ and its root is the
root of $s_{k}$ such that $k$ is the label of the root of $t$. It
defines the composition $t ((s_{i})_{i \in I})$.
\end{defn}

\begin{thm}
Operad $NAP$ is a functor from the groupoid of sets to the
category of sets. \cite{CL1, S1, V1}
\end{thm}

The operadic partition poset $\Pi_{NAP} (I)$ is a
set of forests of $I-$labeled rooted trees such that a forest $X$ is
covered by a forest $Y$, if $Y$ is obtained from $X$ by grafting the
root of one component of $X$ to the root of another component of
$X$. Or $X$ is obtained from $Y$ by removing an edge incident to the
root of one component of $Y$.

\begin{rem} \label{nap2}
Any interval in $\Pi_{NAP} (I)$ is a product of intervals of the
form $[\bold{0},t_{i}]$ such that $t_{i} \in NAP(J_{i})$. If
$t=B^{+}(t_{1},...,t_{k})$, then the poset $[\bold{0},t]$ is
isomorphic to the product of the posets $[\bold{0}, B^{+}(t_{i})]$
for $i \in \{1,2,...,k\}$.
\end{rem}

\begin{lem} \label{nap1}
The incidence Hopf algebra $H_{NAP}$ is a free commutative algebra
on unlabeled rooted trees of root-valence $1$ such that elements
$F_{[t]}$ (where $t$ is a rooted tree) form a basis at the vector
space level.
\end{lem}
According to the theorem \ref{universality-ck} and the structure of
$H_{NAP}$, one can obtain the next important result.

\begin{thm} \label{nap-ck}
$H_{NAP}$ is isomorphic to $H_{CK}$ by the unique Hopf algebra
isomorphism $\rho: F_{[B^{+} (t_{1},...,t_{k})]} \longmapsto
t_{1}...t_{k}$. \cite{CL1}
\end{thm}

This theorem allows us to discover an operadic partition poset formalism for the Connes-Kreimer Hopf algebra
of rooted trees such that after finding a rooted tree reformulation for $H_{U}$, one can apply theorem \ref{nap-ck} to recognize an
operadic source for this specific universal Hopf algebra in Connes-Marcolli treatment.


\chapter{\textsf{Connes-Kreimer theory of the perturbative renormalization}}

The initial motivation in collaboration between the theory of Hopf algebras
and the perturbation theory in renormalizable QFT was determined carefully with the description of perturbative renormalization underlying dimensional regularization in minimal subtraction scheme
in an algebro-geometric framework. In other words, Connes and Kreimer
discovered an interesting revolutionary bridge between the BPHZ prescription in renormalization and the Riemann-Hilbert correspondence. They
proved that perturbative renormalization is in fact one special case of the general mathematical
process of the extraction of finite values based on the
Riemann-Hilbert problem in the reconstruction of differential
equations from data of their monodromy representation such that
for the algebraic reformalization of the BPHZ method,
one can look at to the local regular-singular version of this
problem where at this level the application of the Birkhoff
factorization in the study of QFTs can be investigated. Because in fact,
negative part of this decomposition can be applied to correct
the behavior of solutions near singularities without introducing
new singularities. \cite{CK1, CK2, CM4, CM3, CM1, G3}

According to this mathematical mechanism, for a given renormalizable QFT $\Phi$ one can associate an infinite
dimensional complex Lie group $G(\mathbb{C})$ (i.e.
Lie group of diffeographisms of the theory)
determined with the Hopf algebra $H_{FG}$ of Feynman diagrams of the theory
and depended on the chosen regularization method (i.e. a commutative algebra $A$).
It should be noticed that since Hopf algebra $H_{FG}$ is graded and finite type therefore the group $G(\mathbb{C})$
is pro-unipotent.

With working on the dimensional regularization in the minimal
subtraction scheme, one can find an algebro-geometric machinery to consider perturbative renormalization.
It means that each character carries a geometric meaning in the sense that instead of working on characters
one can reproduce physical information of a given theory $\Phi$ from
factorization of loops (which are depended on the mass parameter $\mu$
and the dimensional regularization parameter $z$) with values in the
Lie group $G(\mathbb{C})$.

For instance in \cite{CK1, CK2}, authors show that passing from
unrenormalized value to the renormalized value is equivalent to the
replacement of a given loop $z \longmapsto \gamma_{\mu}(z) \in G(\mathbb{C})$
on the infinitesimal punctured disk $\bold{\Delta}^{*}$ (identified by the regularization parameter) with the value of its
positive component of the Birkhoff decomposition at the
critical integral dimension $\textsc{D}$. In addition, one can recover the related
counterterm from the negative part of this decomposition.

These results strongly depend on this essential fact that each regularized unrenormalized value
$U_{\mu}^{z}(\Gamma(p_{1},...,p_{n}))$ determines a loop
$\gamma_{\mu}(z)$ on $\bold{\Delta}^{*}$ around the origin and with values in $G(\mathbb{C})$ such that
with the minimal subtraction this unrenormalized value for different
values $z$ will be subtracted. \cite{CM2, CM4}

In this chapter, with a pedagogical intention, we are going to consider the Connes-Kreimer approach to renormalization to provide enough knowledge about this new Hopf algebraic interpretation from physical information. We start with the definition of the Hopf algebra of Feynman diagrams and then consider some of its properties such as grading structures, gluing operator, its rooted tree type representation. Finally, Hopf algebraic renormalization will be studied.


\section{{\it \textsf{Hopf algebra of Feynman diagrams}}}

A renormalizable perturbative quantum field theory can be introduced based on a family of graphs namely, Feynman diagrams which describe possible circumstances
between different types of elementary particles. In these graphs vertices report interactions and edges indicate propagators.
Here one can see some examples of different types of vertices and edges in 3-dimensional scalar field theory, QCD, QED and Gravity:
$$
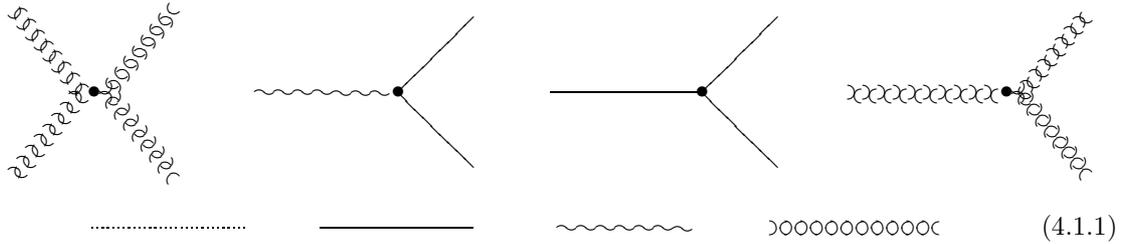
\xy
(-40,0)*{\bullet},
(-50,10);(-42,0) **@{)},
(-50,10.75);(-42,0.75) **@{(},
(-50,-10);(-42,0) **@{)},
(-50,-11);(-42,-1) **@{(},
(-38,0);(-30,10) **@{)},
(-38,0.75);(-30,10.75) **@{(},
(-38,0);(-30,-10) **@{)},
(-38,-1);(-30,-11) **@{(},
(-20,0);(0,0) **@{~},
(0,0)*{\bullet},
(0,0);(10,10) **@{-},
(0,0);(10,-10) **@{-},
(20,0);(40,0) **@{-},
(40,0)*{\bullet},
(40,0);(50,10) **@{-},
(40,0);(50,-10) **@{-},
(60,0);(78,0) **@{)},
(60,-0.5);(78,-0.5) **@{(},
(80,0)*{\bullet},
(82,0);(90,10) **@{)},
(82,-0.5);(90,9.5) **@{(},
(82,0);(90,-10) **@{)},
(82,-0.5);(90,-10.5) **@{(},
\endxy
$$

\begin{equation}
\xy
(-40,0);(-20,0) **@{.},
(-10,0);(10,0) **@{-},
(20,0);(40,0) **@{~},
(50,0);(70,0) **@{)},
(50.5,0);(70.5,0) **@{(},
\endxy
\end{equation}

Consider a theory with the set $R_{V}$ (consists of all possible interactions) and the set $R_{E}$ (consists of all propagators).

\begin{defn}
A Feynman diagram $\Gamma$ is an oriented graph that contains a finite set $\Gamma^{0}$ of vertices and a finite set $\Gamma^{1}$ of edges such that

- For each vertex $v$, its type is determined by the set
$$f_{v}:= \{e \in \Gamma^{1}: e \cap v \neq \emptyset \}.$$

- The set $\Gamma^{1}$ decomposes into two different subsets

(Int) $\Gamma^{1}_{int}$ consists of all internal edges (i.e. an edge together with begin and end vertices),

(Ext) $\Gamma^{1}_{ext}$ consists of all external edges (i.e. an edge with an open end).

- Based on Feynman rules of a theory, all edges are labeled with physical parameters (i.e. momenta of particles).

- If $p_{1}, ..., p_{k}$ are momenta of external edges, then $\sum_{i} p_{i} = 0$. (conservation law)
\end{defn}

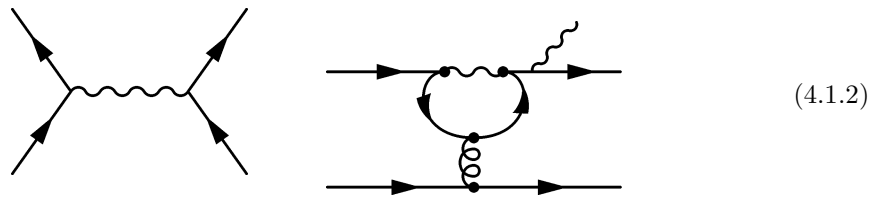
\begin{equation}
\begin{tabular}{cccccccccccccccc}
\begin{fmffile}{one}
  \fmfframe(1,7)(1,7){
   \begin{fmfgraph*}(110,62)
    \fmfleft{i1,i2}
    \fmfright{o1,o2}
    \fmflabel{$e^-$}{i1}
    \fmflabel{$e^+$}{i2}
    \fmflabel{${\ensuremath{\erlpm}}$}{o1}
    \fmflabel{${\ensuremath{\erlpm}}$}{o2}
    \fmf{fermion}{i1,v1,i2}
    \fmf{fermion}{o1,v2,o2}
    \fmf{photon,label=$\gamma/Z^0$}{v1,v2}
   \end{fmfgraph*}
  }
\end{fmffile}




\begin{fmffile}{three}
  \fmfframe(1,2)(1,2){
\begin{fmfgraph*}(110,62)
  \fmfipair{Vtb,Vts,b,s,ep,em,p,p',ga,gb,tm,bm}
  \fmfiequ{tm}{.5[nw,ne]}
  \fmfiequ{bm}{.5[sw,se]}
  \fmfiequ{.5[Vtb,Vts]}{.7[bm,tm]}
  \fmfiequ{Vts}{Vtb+(.2w,0)}
  \fmfiequ{b}{.7[sw,nw]}
  \fmfiequ{s}{.7[se,ne]}
  \fmfiequ{ga}{Vts+(.1w,0)}
  \fmfiequ{gb}{ne-(.15w,0)}
  \fmfiequ{p'}{bm}
  \fmfiequ{p}{p'+(0,.3h)}
  \fmfiequ{em}{se}
  \fmfiequ{ep}{sw}
  \fmfi{photon,lab=$W$,
    lab.sid=right}{Vtb--Vts}
  \fmfi{fermion}{b--Vtb}
  \fmfi{fermion}{ga--s}
  \fmfi{plain}{Vts--ga}
  \fmfi{fermion,lab=$t/c/u$}{Vtb{b-Vtb}
    .. tension 1 .. {right}p}
  \fmfi{fermion,lab=$t/c/u$}{p{right}
    .. tension 1 .. {Vts-s}Vts}
  \fmfi{gluon,label=$g$, lab.sid=right }{p--p'}
  \fmfi{photon,label=$\gamma$}{gb--ga}
  \fmfi{fermion}{ep--p'}
  \fmfi{fermion}{p'--em}
  \fmfiv{d.siz=3thin,lab=$b$}{b}
  \fmfiv{d.siz=3thin,lab=$s$}{s}
  \fmfiv{d.siz=3thin,lab=$\overline{u}$}{ep}
  \fmfiv{d.siz=3thin,lab=$\overline{u}$}{em}
  \fmfiv{d.sh=circle,d.siz=3thin}{Vtb}
  \fmfiv{d.sh=circle,d.siz=3thin}{Vts}
  \fmfiv{d.sh=circle,d.siz=3thin}{p}
  \fmfiv{d.sh=circle,d.siz=3thin}{p'}
 \end{fmfgraph*}
  }
\end{fmffile}
\end{tabular}
\end{equation}

A special class of these diagrams together with an algebraic operation (i.e. insertion) are enough to construct the whole theory. They are one particle irreducible (1PI)
Feynman graphs without any sub-divergences which play the role of building blocks for defining a mathematical structure (i.e. Hopf algebra).

\begin{defn}
An n-particle irreducible (n-PI) graph is a Feynman diagram $\Gamma$ with this property that upon removal of $n$ internal edges, it is still connected. It is clear that for $n \ge 2$, each n-PI graph is a (n-1)-PI.
\end{defn}

\begin{equation}
\xy
(0,0);(20,0) **@{-},
(10,0)*{>},
(10,2.5)*{p},
(20,0)*{\bullet},
(20,0);(40,0) **\crv{(20,7)&(40,7)},
(30,5)*{>},
(30,8)*{p+k},
(20,0);(40,0) **\crv{(20,-7)&(40,-7)},
(30,-5.25)*{<},
(30,-8)*{k},
(40,0)*{\bullet},
(40,0);(60,0) **@{-},
(50,0)*{>},
(50,2.5)*{p},
(80,0);(90,0) **@{-},
(90,0)*{\bullet},
(90,0);(100,0) **\crv{(90,5)&(100,5)},
(90,0);(100,0) **\crv{(90,-5)&(100,-5)},
(100,0)*{\bullet},
(100,0);(110,0) **@{-},
(110,0)*{\bullet},
(110,0);(120,0) **\crv{(110,5)&(120,5)},
(110,0);(120,0) **\crv{(110,-5)&(120,-5)},
(120,0)*{\bullet},
(120,0);(130,0) **@{-},
(105,-15)*{not},
(112,-15)*{1PI},
(30,-15)*{1PI},
\endxy
\end{equation}

\begin{defn}
For each arbitrary Feynman diagram $\Gamma$,

(i) $res(\Gamma)$ is a new graph as the result of shrinking all of the internal edges and vertices of $\Gamma$ into one vertex. The resulting graph consists of a vertex together with all of the external edges of $\Gamma$.

(ii) For each Feynman subgraph $\gamma$ of $\Gamma$, the graph $\Gamma / \gamma$ is defined by shrinking $\gamma$ into a vertex. The resulting diagram is called quotient graph.
\end{defn}

\begin{equation}
\xy
(0,0);(10,0) **@{~},
(10,0)*{\bullet},
(10,0);(30,20) **@{-},
(10,0);(30,-20) **@{-},
(20,-10);(20,10) **@{-},
(35,0)*{\longmapsto},
(35,2.5)*{res},
(40,0);(50,0) **@{~},
(50,0)*{\bullet},
(50,0);(60,10) **@{-},
(50,0);(60,-10) **@{-},
\endxy
\end{equation}

$$
\xy
(65,0)*{\Gamma=},
(70,0);(80,0) **@{-},
(80,0)*{\bullet},
(80,0);(120,0) **@{~},
(85,0)*{<},
(80,0);(120,0) **\crv{(80,10)&(120,10)},
(90,0);(110,0) **\crv{(90,5)&(110,5)},
(90,0)*{\bullet},
(110,0)*{\bullet},
(100,0)*{<},
(115,0)*{<},
(100,7.5)*{>},
(120,0)*{\bullet},
(120,0);(130,0) **@{~},
(140,0)*{\gamma=},
(145,0);(155,0) **@{-},
(150,0)*{<},
(155,0)*{\bullet},
(155,0);(165,0) **@{-},
(165,0)*{\bullet},
(160,0)*{<},
(155,0);(160.5,5) **@{~},
(159.25,5);(165,0) **@{~},
(165,0);(175,0)**@{-},
(170,0)*{<},
\endxy
$$

\begin{equation}
\xy
(-10,0)*{\frac{\Gamma}{\gamma}=},
(0,0);(20,0) **@{~},
(20,0)*{\bullet},
(20,0);(40,0) **\crv{(20,5)&(40,5)},
(40,0)*{\bullet},
(30,3.5)*{>},
(20,0);(40,0) **\crv{(20,-5)&(40,-5)},
(30,-3.5)*{<},
(40,0);(60,0) **@{~},
\endxy
\end{equation}

\begin{defn} \label{prelie}
Define a bilinear operation
$\star$ on the set of 1PI graphs given by
$$\Gamma_{1} \star \Gamma_{2} := \sum_{\Gamma} n(\Gamma_{1},
\Gamma_{2}; \Gamma) \Gamma$$
where the sum is over 1PI graphs $\Gamma$ and $n(\Gamma_{1},
\Gamma_{2}; \Gamma)$ counts the number of ways that a subgraph
$\Gamma_{2}$ can be reduced to a point in $\Gamma$ such that
$\Gamma_{1}$ is obtained and also
$$|\Gamma| = |\Gamma_{1}| +
|\Gamma_{2}|, \  res(\Gamma)=res(\Gamma_{1}).$$
\end{defn}

\begin{equation}
\xy
(0,0);(5,0) **@{-},
(5,0);(10,0) **\crv{(5,2.5)&(10,2.5)},
(5,0);(10,0) **\crv{(5,-2.5)&(10,-2.5)},
(10,0);(15,0) **@{-},
(20,0)*{\star},
(25,0);(30,0) **@{-},
(30,0);(35,5) **@{-},
(35,5);(35,-5) **@{-},
(30,0);(35,-5) **@{-},
(35,5);(40,10) **@{-},
(35,-5);(40,-10) **@{-},
(45,0) *{=},
(50,0);(55,0) **@{-},
(55,0);(65,10) **@{-},
(58,3);(63,8) **\crv{(58,10)&(63,8)},
(65,10);(65,-10) **@{-},
(55,0);(65,-10) **@{-},
(65,10);(67,12) **@{-},
(65,-10);(67,-12) **@{-},
(70,0)*{+},
(75,0);(80,0) **@{-},
(80,0);(90,10) **@{-},
(90,10);(90,-10) **@{-},
(90,5);(90,-5) **\crv{(95,5)&(95,-5)},
(80,0);(90,-10) **@{-},
(90,-10);(92,-12) **@{-},
(90,10);(92,12) **@{-},
(97,0)*{+},
(100,0);(105,0) **@{-},
(105,0);(115,10) **@{-},
(105,0);(115,-10) **@{-},
(115,10);(115,-10) **@{-},
(115,10);(117,12) **@{-},
(115,-10);(117,-12) **@{-},
(108,-3);(113,-8) **\crv{(108,-10)&(113,-8)},
\endxy
\end{equation}

\begin{equation}
\xy
(50,0);(55,0) **@{-},
(55,0);(65,10) **@{-},
(65,10);(65,-10) **@{-},
(55,0);(65,-10) **@{-},
(65,10);(70,15) **@{-},
(65,-10);(70,-15) **@{-},
(75,0)*{\star},
(80,0);(85,0) **@{-},
(85,0);(90,0) **\crv{(85,5)&(90,5)},
(85,0);(90,0) **\crv{(85,-5)&(90,-5)},
(90,0);(95,0)**@{-},
(100,0)*{=},
(103,0)*{2},
(105,0);(110,0) **@{-},
(110,0);(120,0) **\crv{(110,10)&(120,10)},
(110,0);(120,0) **\crv{(110,-10)&(120,-10)},
(120,0);(125,0) **@{-},
(115,7.5);(115,-7.5) **@{-},
\endxy
\end{equation}

\begin{rem} \label{prop-prelie}

(i) Finitely of Feynman diagrams show that the above sum is finite.

(ii) $res(\Gamma_{1} \star \Gamma_{2}) = res(\Gamma_{1})$,

(iii) The operation $\star$ is pre-Lie, namely
$$ [\Gamma_{1} \star \Gamma_{2}] \star \Gamma_{3} - \Gamma_{1} \star [\Gamma_{2} \star \Gamma_{3}] =
[\Gamma_{1} \star \Gamma_{3}] \star \Gamma_{2} - \Gamma_{1} \star
[\Gamma_{3} \star \Gamma_{2}], $$

(iv) For some integers $r$ and $k_{j}$ that $j=1,...,r$, any
non-primitive 1PI graph $\Gamma$ can be written at most in $r$
different forms
$$ \Gamma = \prod_{i=1}^{k_{j}} \gamma_{j} \star_{j,i} \Gamma_{j,i} $$
such that $\gamma_{j}$s are primitive graphs. When $r>1$, the graph
$\Gamma$ is called {\it overlapping}. \cite{K10, K2, K3}
\end{rem}

So this operator determines a Lie algebra structure on Feynman diagrams such that the Lie bracket is the commutator with respect to the $\star$. From physical point of view, this insertion operator technically can be expounded by the gluing of Feynman graphs based on types of edges. The reader interested in this quest of a deeper level of understanding should consult \cite{CK4, CK3, K10}.

\begin{thm}  \label{cop-feynman}
Graded dual of the universal
enveloping algebra of the Lie algebra $\mathcal{L}$ on 1PI graphs
(determined with the definition \ref{prelie}) is a graded connected
commutative non-cocommutative Hopf algebra. It is called Hopf algebra of
Feynman diagrams of the theory $\Phi$ and denoted by $H_{FG}=H(\Phi)$. \cite{CM2, CM1, K6, K12}
\end{thm}

\begin{proof}
It is the immediate result of the Milnor-Moore theorem. Based on the gluing information, one can determine sub-diagrams of Feynman graphs such that it leads to the coproduct structure. For each
Feynman diagram $\Gamma$, its coproduct can be written by
$$
\Delta(\Gamma) = \Gamma \otimes \mathbb{I} + \mathbb{I} \otimes
\Gamma + \sum_{\gamma \subset \Gamma} \gamma \otimes \Gamma / \gamma
$$
such that the sum is over all disjoint unions of 1PI superficially divergent
proper subgraphs with residue in $R_{V} \cup R_{E}$ where the associated amplitudes of their residues
need renormalization. Now expand it to the free products of 1PI graphs.
\end{proof}

\begin{rem}
There are different choices for grading structure on the Hopf algebra of Feynman diagrams such as
number of vertices, number of internal edges, number of independent
loops, .... Grading with the number of internal edges determines finite type property for this Hopf algebra. \cite{CM2, CM1}
\end{rem}

It is remarkable to know that one can reformulate this Hopf algebra by a certain decorated version of the
Connes-Kreimer Hopf algebra of rooted trees such that decorations
conserve some physical information such as (sub-)divergences (i.e. nested loops) of
Feynman diagrams. As an example, the diagram
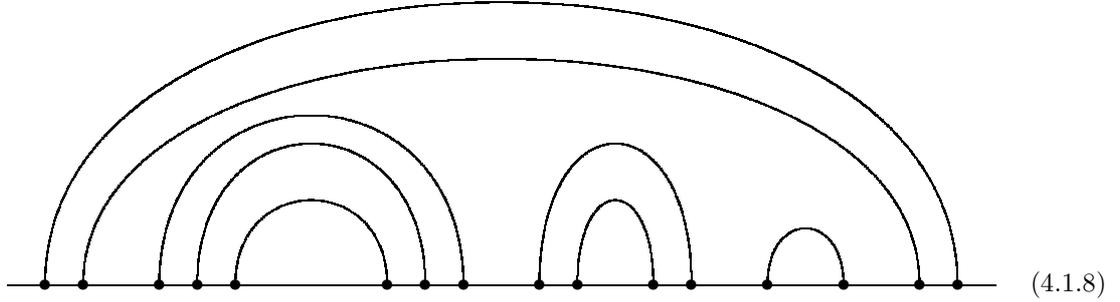
\begin{equation}
\xy
(0,0);(10,0) **@{-},
(10,0)*{\bullet},
(10,0);(120,0) **@{-},
(120,0)*{\bullet},
(120,0);(130,0) **@{-},
(10,0);(120,0) **\crv{(10,40)&(120,40)},
(20,0);(60,0) **\crv{(20,30)&(60,30)},
(20,0)*{\bullet},
(60,0)*{\bullet},
(75,0);(85,0) **\crv{(75,15)&(85,15)},
(75,0)*{\bullet},
(85,0)*{\bullet},
(70,0)*{\bullet},
(90,0)*{\bullet},
(70,0);(90,0) **\crv{(70,25)&(90,25)},
(30,0);(50,0) **\crv{(30,15)&(50,15)},
(30,0)*{\bullet},
(50,0)*{\bullet},
(5,0)*{\bullet},
(5,0);(125,0) **\crv{(5,50)&(125,50)},
(25,0)*{\bullet},
(25,0);(55,0) **\crv{(25,25)&(55,25)},
(55,0)*{\bullet},
(100,0)*{\bullet},
(110,0)*{\bullet},
(100,0);(110,0) **\crv{(100,10)&(110,10)},
(125,0)*{\bullet},
\endxy
\end{equation}
can be represented by the labeled rooted tree
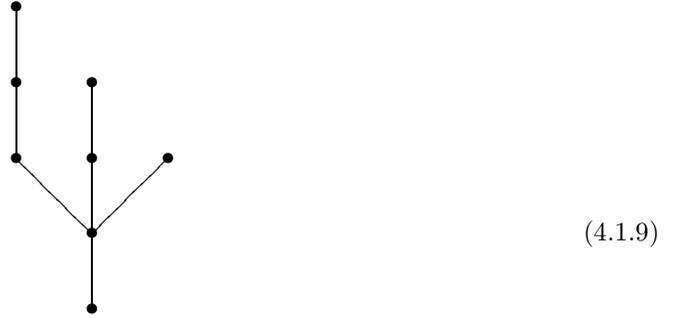
\begin{equation}
\xy
(0,-10)*{\bullet},
(0,-10);(0,0) **@{-},
(0,0)*{\bullet},
(0,0);(-10,10) **@{-},
(0,0);(10,10) **@{-},
(-10,10)*{\bullet},
(10,10)*{\bullet},
(-10,10);(-10,20) **@{-},
(-10,20)*{\bullet},
(0,10)*{\bullet},
(0,0);(0,10) **@{-},
(0,20)*{\bullet},
(0,10);(0,20) **@{-},
(-10,30)*{\bullet},
(-10,20);(-10,30) **@{-},
\endxy
\end{equation}
such that each vertex reports the divergent primitive sub-diagram
$
\xy
(5,0);(10,0) **@{-},
(10,0)*{\bullet},
(10,0);(20,0) **@{-},
(10,0);(20,0) **\crv{(10,5)&(20,5)},
(20,0)*{\bullet},
(20,0);(25,0) **@{-},
\endxy
$
and edges show the locations of sub-diagrams in the main graph. Representation of Feynman diagrams together with overlapping divergences based on rooted trees is also studied. In \cite{K1, KW1} authors show that how these kinds of nested sub-divergences can be reduced to linear combinations of rooted trees.

The Lie algebra $\mathcal{L}$ gives rise to two representations
acting as derivations on $H(\Phi)$. They are

\begin{equation}
 < Z_{\Gamma_{1}}^{+}, \Gamma_{2}> := \Gamma_{2} \star \Gamma_{1}
\end{equation}

\begin{equation}
 <Z_{\Gamma_{1}}^{-}, \Gamma_{2}> := \sum_{i} < Z_{\Gamma_{1}}^{+}, (\Gamma_{2})^{'}_{i}> (\Gamma_{2})^{"}_{i}
\end{equation}
such that
\begin{equation}
\Delta (\Gamma_{2})= \mathbb{I} \otimes \Gamma_{2} + \Gamma_{2}
\otimes \mathbb{I} +  \sum_{i} (\Gamma_{2})^{'}_{i} \otimes
(\Gamma_{2})^{"}_{i}.
\end{equation}

\begin{rem}
If $\Gamma_{2}$ be a 1PI graph, then for each term in the above sum,
there is a unique gluing data $G_{i}$ that describes how one can
reach to the graph $\Gamma_{2}$ by gluing of the components
$(\Gamma_{2})^{'}_{i}$ into $(\Gamma_{2})^{"}_{i}$. \cite{CK3, F2, K10}
\end{rem}

There is also another interesting and useful grading where it can be applied to explain the relation between the Connes-Kreimer Hopf algebra of rooted tress and the Hopf algebra of Feynman diagrams, proof of locality of counterterms and also in the study of Dyson-Schwinger equations.

\begin{defn}  \label{bidegree3}
Let $Ker \ \epsilon$ be
the augmentation ideal of $H(\Phi)$ and
$$P: H(\Phi) \longrightarrow ker \ \epsilon, \  \  P:=id-\mathbb{I}
\epsilon$$
be the projection on to this ideal. Define a new map
$$Aug^{(m)}:= (\underbrace {P \otimes ... \otimes P}_{m})
\Delta^{m-1}: H(\Phi) \longrightarrow \{ker \ \epsilon\}^{\otimes m}$$
and set
$$H(\Phi)^{(m)} := \frac{KerAug^{(m+1)}}{KerAug^{(m)}},  \  \ m \ge 1.$$
It is called bidegree.
\end{defn}

\begin{lem} \label{bidegree4} \label{bidegree5}
One can show that
$$H(\Phi)= \bigoplus_{m \ge 0} H(\Phi)_{m} = \bigoplus_{m \ge 0}
H(\Phi)^{(m)}$$
such that for each $m \ge 1$,
$$H(\Phi)_{m} \subset \bigoplus^{m}_{j=1} H(\Phi)^{(j)}, \   \
H(\Phi)_{0} \simeq H(\Phi)^{(0)} \simeq \mathbb{K}.$$
\cite{K2, M2}
\end{lem}

\begin{rem}
(i) All Feynman graphs that contain (sub-)divergences (i.e. nested loops)
belong to the augmentation ideal and it means that $H_{aug}(\Phi):=
\bigoplus_{i \ge 1} H(\Phi)_{i}$ stores quantum information.

(ii) For each 1PI graph $\Gamma$, one can identify a linear
generator $\delta_{\Gamma}$ and set $H_{lin}(\Phi):= span \
\{\delta_{\Gamma}\}_{\Gamma}$. It is observed that $H_{lin}(\Phi)
\subset H_{aug}(\Phi)$.
\end{rem}

The grafting operator $B^{+}$ is defined on rooted trees but with attention to the decorations one can lift it to the level of Feynman diagrams. For much better understanding of this translation, letting $H_{CK}(\Phi)$ be a labeled version of the
Connes-Kreimer Hopf algebra of rooted trees (decorated by primitive
1PI Feynman graphs of the renormalizable theory $\Phi$). By choosing a
primitive element $\gamma$, the operator $B^{+}_{\gamma}$ is an
homogeneous operator of degree one such that after its application to
a forest, it connects the roots in the forest to a new root
decorated by $\gamma$. As an example, one can see that
\begin{equation}
\xy
(0,5)*{B^{+}},
(0,0);(20,0) **@{-},
(5,0);(15,0) **\crv{(5,5)&(15,5)},
(10,0)*{>},
(20,5)*{(},
(22,5);(42,5) **@{-},
(27,5);(36,5) **\crv{(27,10)&(36,10)},
(32,5)*{>},
(44,5)*{)=},
(50,5);(80,5) **@{-},
(55,5);(75,5) **\crv{(55,15)&(75,15)},
(60,5);(70,5) **\crv{(60,10)&(70,10)},
(65,5)*{>},
\endxy
\end{equation}
In chapter eight, it will be discussed that the grafting operator determines Hochchild one cocycles of a chain complex connected with the renormalization coproduct. Now bidegree (as the grading factor) and the operator $B^{+}$ can provide a decomposition of diagrams which contain primitive components such that it helps us to have a practical instruction for studying Feynman diagrams.

\begin{thm} \label{ck-feynmangraphs}
Define a homomorphism $\Psi: H(\Phi) \longrightarrow H_{CK}(\Phi)$ given by
$$\Psi(\Gamma) = \sum_{j=1}^{r}
B^{+}_{\gamma_{j},G_{j,i}}[\prod_{i=1}^{k_{j}} \Psi(\Gamma_{j,i})].$$
One can show that

(i) $\Psi$ is defined by induction on bidegree,

(ii) It is a Hopf algebra homomorphism,

(iii) Its image is a closed Hopf subalgebra,

(iv) $G_{j,i}$s are the gluing data,

(v) $B^{+}_{\gamma_{j},G_{j,i}}$s are one-cocycles. \cite{CK3, F2, K10, M2}

\end{thm}

Feynman diagrams of a theory are equipped with physical
information. For example external edges have external momenta and
because of that it should be reasonable to pay attention to theses
physical properties of graphs in the structure of the Hopf algebra. Therefore
people prefer to consider this story at two levels: discrete part (i.e.
Feynman diagrams without notice to the external structures) for studying the toy model and full
structure (i.e. Feynman diagrams together with external data).

\begin{thm} \label{hop-Feyn-dis-fu}
For a given renormalizable quantum field theory $\Phi$,

(i) The discrete Hopf algebra of Feynman diagrams $H_{D}(\Phi)$ is
made on the free commutative algebra over $\mathbb{C}$ generated by
pairs $(\Gamma,w)$ such that $\Gamma$ is a 1PI graph and $w$ is a
monomial with degree agree with the number of external edges of
graph.

(ii) The full Hopf algebra $H_{F}(\Phi)$ is made on the symmetric
algebra of the linear space of distributions defined by the external
structures.

(iii) $H(\Phi)$ is isomorphic to $H_{D}(\Phi)$  (in the case:
without external structure) or $H_{F}(\Phi)$ (in the case: with
external structures). \cite{CK1, CK2, CM1, K2}
\end{thm}

It is discussed that how one can arrange Feynman diagrams of given physical theory into a Hopf algebra based on the recognizing of sub-divergences of diagrams.
Now it is important to have an explicit understanding from the concept of "{\it renormalizability}" underlying this Hopf algebra structure and so at the final part of this section we consider this essential concept.

Start with the Largrangian of a given physical theory $\Phi$. We know that
each monomial in the Lagrangian corresponds to an amplitude. Letting
$\mathcal{A}$ be the set of all amplitudes.

\begin{defn}
The physical theory $\Phi$ is renormalizable if it has
a finite subset $\mathcal{R}_{+} \subset \mathcal{A}$ as the set of
amplitudes which need renormalization.
\end{defn}

For example, it can be easily seen that $\phi^{3}$
in dimension $\textsc{D} \le 6$ is a renormalizable theory.

Each amplitude $a\in \mathcal{A}$ specifies an integer $n=n(a)$ which gives the number of external
edges. Let $\mathcal{M}_{a}$ be the set of all 1PI graphs
contributing to the amplitude $a$, $|\Gamma|$ be the number of
independent loops in $\Gamma$ and $|sym(\Gamma)|$ be the rank of the
automorphism group of the graph.

\begin{defn} \label{dse3}
Let $\phi$ be the Feynman rules character associated to the theory
$\Phi$. The Green function related to an amplitude $a$ and the
character $\phi$ is given by
$$
G_{\phi}^{a}=1 \pm \sum_{\Gamma \in \mathcal{M}_{a}}
\alpha^{|\Gamma|} \frac{\phi(\Gamma)}{|sym(\Gamma)|} =  1 \pm
\sum_{k \ge 1} \alpha^{k} \phi(c^{a}_{k})$$
where
$$
c^{a}_{k}= \sum_{\Gamma \in \mathcal{M}_{a}, |\Gamma|=k}
\frac{\Gamma}{|sym(\Gamma)|}
$$
such that the sum is over all 1PI graphs of order $k$ contributing
to the amplitude $a$. The plus sign is taken if $n(a) \ge 3$ and the
minus sign for $n(a)=2$.
\end{defn}

Since we are interested to study graphs together with
(sub-)divergences, therefore it is not necessary to consider all graphs. Because for instance in $\phi^{3}$ theory in six dimension (as a toy model), one investigates that superficial divergent diagrams are those with the number of external edges $\le 3$ and further {\it
tadpole amplitudes} (i.e. $n(a)=1$) and {\it vacuum amplitudes}
(i.e. $n(a)=0$) have vanished Green functions.
\begin{equation}
\xy
(0,0)*{\bullet},
(20,0)*{\bullet},
(0,0);(20,0) **\crv{(0,10)&(20,10)},
(0,0);(20,0) **\crv{(0,-10)&(20,-10)},
(10,-12)*{Vacuum},
(40,0);(50,0) **@{-},
(50,0)*{\bullet},
(50,0);(70,0) **\crv{(50,10)&(70,10)},
(50,0);(70,0) **\crv{(50,-10)&(70,-10)},
(70,0)*{\bullet},
(60,-12)*{Tadpole},
\endxy
\end{equation}
Indeed, it is enough to study Green functions
depended on 1PI graphs with 2 or 3 external legs. Because with the help of these graphs, one can
build a basis for the Hopf algebra $H_{FG}$. Now by indicating the
structure of the sum
\begin{equation}
\Gamma^{a}= \mathbb{I} \pm \sum_{\Gamma \in \mathcal{M}_{a}}
\alpha^{|\Gamma|} \frac{\Gamma}{|sym(\Gamma)|}
\end{equation}
in the definition \ref{dse3}, it is implicitly observed that after the
application of the Feynman rules character $\phi$ on this sum, we
will obtain its related Green function and it means that
\begin{equation}
G_{\phi}^{a}= \phi (\Gamma^{a}).
\end{equation}

\begin{lem} \label{dse4}
For a given renormalizable theory $\Phi$, one can decompose the set of amplitudes $\mathcal{A}$
into two disjoint subsets $\mathcal{R}_{+},$
$\mathcal{R}_{-}$ such that for each $r \in \mathcal{R}_{+}$, we
have
$$
\Gamma^{r} =  \mathbb{I} \pm \sum_{k \ge 1} \alpha ^{k} B^{+}_{k;r}
(\Gamma^{r} Q^{n_{r}k})
$$
where $B^{+}_{k;r}$ are Hochschild 1-cocycles and $Q^{n_{r}}$ is a
monomial in $\Gamma^{r}$ or its inverse. Since the amplitudes from
the set $\mathcal{R}_{-}$ are determined with the knowledge of
elements in $\mathcal{R}_{+}$, therefore for the study of
renormalizable theories, it is enough to focus on the elements in
$\mathcal{R}_{+}$. \cite{BK1, BK2, K4, K2, K3, KY1}
\end{lem}

Recognizing Feynman diagrams together with (sub-)divergences can help us to consider more precisely their related Feynman integrals.
There is an important factor (connected with the dimension of theory) to identify divergences in
graphs namely, {\it superficial degree of divergency}
$\omega$.

\begin{rem}
Parameter $\omega$ has the following properties:

(i) All Feynman diagrams with the same external structure have the
same superficial degree of divergence. It means that
$$
res(\Gamma_{1})=res(\Gamma_{2}) \Longrightarrow \omega
(\Gamma_{1})=\omega (\Gamma_{2}).
$$

(ii) This degree shows the existence of divergency only for a finite
number of distinct external structures $r$ in $\mathcal{R}_{+}$. \cite{F2}
\end{rem}

Actually, the superficial degree of divergency of a 1PI graph
$\Gamma$ measures the degree of singularity of the integral in the
amplitude $a_{\Gamma}$ with respect to the integrated variables
$q_{1}, q_{2}, ....$. Under the transformation of momenta $q_{i}
\longmapsto tq_{i}$ ($ t \in \mathbb{R}$), we have $a_{\Gamma}
\longmapsto t^{\omega(\Gamma)} a_{\Gamma}$.
With notice to this factor, a classification of amplitudes of a
given renormalizable theory is possible.

\begin{lem}
The amplitude of a diagram
$\Gamma$ with just one loop

(i) is {\it convergent} if $\omega(\Gamma)<0$,

(ii) has a {\it logarithmic divergency} if $\omega(\Gamma)=0$,

(iii) has a {\it polynomial divergency} if $\omega(\Gamma)>0$.
\end{lem}

The computation of the superficial degree of divergency of a Feynman graph $\Gamma$
with more than one loop is also possible with starting from 1PI
subgraphs with one loop and continue the process by enlarging
subgraphs until to reach to the main graph.

\begin{rem} \label{dse5}
If we look at to the above procedure, the existence of a
self-similar recursive way determines the formal sums $\Gamma^{r} \
(r \in \mathcal{R}_{+})$ in terms of themselves and the action of
suitable maps (i.e. one cocycles) $B^{+}_{k;r}$.
\end{rem}

For each Feynman diagram $\Gamma$,
$B^{+}_{n}(\Gamma)$ is defined by the insertion of a collection of
(sub-)divergences $\Gamma$ into the identified $n-$loop primitive
graphs. Kreimer could introduce a new measure to translate the combinatorics of Feynman diagrams to the normal analytic picture which physicists familiar with it.

\begin{lem} \label{dse6}
There is a measure $\mu_{+}$ such that for each Feynman rules
character $\phi$, it is determined by $\phi (B^{+}_{n} (\mathbb{I}))
:= \int d \mu_{+}$ where the expression $\phi (B^{+}_{n}(\Gamma)) =
\int \phi(\Gamma) d \mu_{+}$ shows that subgraphs become
subintegrals under the Feynman rules. \cite{K7, K15, KY1}
\end{lem}

In summary, it is considered that for each arbitrary renormalizable perturbative physical theory one can
introduce a graded connected commutative non-cocommutative Hopf
algebra of finite type such that 1PI Feynman graphs play the role of its
building blocks. And also, with the help of a decorated version of the Connes-Kreimer Hopf algebra of rooted trees (such that primitive divergent sub-graphs are put in labels), one can display a toy model from this Hopf algebra where it will be useful to do renormalization in a simpler procedure for different theories. This toy model can provide a universal framework in the study of different theories.


\section{{\it \textsf{Algebraic perturbative renormalization}}}

Attempts at eliminating divergences had been started from the birth of quantum field theory.
In experimentally calculation of amplitudes, we have encountered two different
kinds of divergences: {\it infra-red}  (i.e. an amplitude which
becomes infinite for vanishingly small values of some momenta) and
{\it ultra-violet} (i.e. an amplitude which becomes infinite for
arbitrary large values of momenta in a loop integration). The
applied structure in interpretation of ultra-violet divergences will also
allow to cancel infra-red divergences. The main idea of
renormalization is to correct the original Lagrangian of a quantum field theory by an
infinite series of counterterms, labeled with Feynman graphs of the
theory. By these counterterms we can disappear ultra-violet
divergences. \cite{CM1, LP1, M5}

Feynman diagrams together with some special rules have the ability of presenting all of the possible happenings in a renormalizable theory. For example in a Feynman graph,  external edges are the symbols for particles with assigned
momenta, vertices show interactions and internal edges for creation
and annihilation of virtual particles are applied.

With applying Feynman rules of the given theory, one can associate a Feynman iterated integral to
each Feynman diagram. Generally since these integrals have (sub-)divergences, therefore they are complex and ill-defined. This problem is in fact the most conceptional difficulty in quantum field theory such that with the help of various approaches, people try to find suitable solutions for this problem. As a consequence, nowadays renormalization is one understandable powerful technique to consider the behavior of diagrams together with (sub-)divergences. Furthermore, we studied that how one can connect a Hopf algebra structure to a given theory and now it surely very important to know that how this new mathematical structure leads to a new algebraic reformulation from the perturbative renormalization process underlying BPHZ method. It should be remarked that the wide advantage of this point of view (in the study of
renormalization) than the Bogoliubov recursive standard process can be observed more clearly, when we
want to work on the Feynman diagrams with high loop orders.
\cite{CK3, EGGV, K1, K2, K15, KD1, W2, W1}

Here we are going to have a short overview from renormalization process with respect to this Hopf algebraic modeling. This formalism is discovered by Connes and Kreimer and they could find very closed relation between this well-known physical technique and one important general instruction in mathematics, namely Riemann-Hilbert problem.

There are several mechanisms for renormalization. Connes-Kreimer approach to the perturbative renormalization is certainly
a practical reformulation from the BPHZ method on the basis of the Hopf algebra $H_{FG}$
connected with a given renormalizable theory
$\Phi$ such that Feynman rules can be
determined with special characters of this Hopf algebra. One should
point out that because of the universal property of the Connes-Kreimer
Hopf algebra $H_{CK}$ of rooted tree (with respect to the Hochschild cohomology), the using
of a decorated version of this Hopf algebra in computations helps us to find a simplified toy
model for studying. \cite{EGK1, EGK2}

So we should concentrate on the renormalization procedure depended upon one regularization parameter. In this kind, the first stage of renormalization is done by regularization such that with the help of regularization parameter, all divergences
appearing in amplitudes can be parameterized to reach to the formally finite expansions
together with a subtraction of ill-defined expressions. In the
regularization process some non-physical parameters will be created
such that this fact could be changed the nature of Feynman rules to
algebra homomorphisms from the Hopf algebra $H_{FG}$ to a commutative algebra
$A$ where this algebra is characterized by the regularization
prescription. Connes and Kreimer proved that minimal subtraction scheme in dimensional regularization can be rewritten based on the Birkhoff decomposition of characters of $H_{FG}$ with values in the algebra $A_{dr}$ of Laurent series with finite pole part such that components of this factorization of very special characters (identified
with the Feynman rules of a given theory) provide counterterms and renormalized values. Moreover, one can redefine the renormalization group and its infinitesimal generator ($\beta-$function) by using the negative component. In this procedure, one important algebraic property plays an essential role. It is the Rota-Baxter property of the pair $(A_{dr},R_{ms})$ such that it reports the multiplicativity of renormalization and it leads to apply the Riemann-Hilbert correspondence in the study of perturbative QFT.  \cite{EG2, EGK1, EGK2, W2}

\begin{defn} \label{char1}  \label{conv1}
Supposing that $H$ be the Hopf algebra of Feynman diagrams of a renormalizable QFT $\Phi$ and $A$ be a commutative algebra with respect to the regularization scheme.
Set
$$G(A):=Hom(H,A) = \{\phi: H\longrightarrow A:
\phi(xy)=\phi(x)\phi(y), \ \phi(1)=1\} \subset L(H,A)$$
and consider the convolution product on $G(A)$ such that for each
$\phi_{1},\phi_{2} \in G(A)$, it is given by
$$\phi_{1}*\phi_{2}:=m_{A} \circ (\phi_{1}\otimes \phi_{2}) \circ
\bigtriangleup_{H}.$$
\end{defn}

\begin{thm}

(i) The convolution product $*$ determines a group structure on $G(A)$.

(ii) For a fixed commutative Hopf algebra $H$,
there is a representable covariant functor $\bold{G}$  (represented by $H$)
from the category of unital commutative $\mathbb{K}-$algebras to the
category of groups.

(iii) Every covariant representable functor
between two above categories is determined by an affine group scheme
$\bold{G}$. For each commutative algebra $A$, $\bold{G}(A)$ is called affine group scheme. \cite{CK1, CK2, CM5, CM2, CM4, CM3, CM1}

\end{thm}

One can improve this categorical configuration to the Lie algebra level.

\begin{thm}
For a given affine group scheme $\bold{G}$ (viewed as a functor), one can
extend it to the covariant functor
$$\mathfrak{g}=Lie \ \bold{G}, \ \  A \longmapsto \mathfrak{g}(A)$$
from the category of commutative unital $\mathbb{K}-$algebras to the
category of Lie algebras. $\mathfrak{g}(A)$ is the Lie algebra of linear maps
$l:H\longrightarrow A$ such that for each $x,y\in H$,
$$l(xy)=l(x)\epsilon(y)+ \epsilon(x)l(y)$$
where $\epsilon$ is the augmentation of $H$. \cite{CM5, CM2, CM4, CM3, CM1}
\end{thm}

\begin{rem} \label{deriv3}
Equivalently there is another picture from the elements of $\mathfrak{g}(A)$ given by
linear maps $t:H\longrightarrow H$ with the properties
$$t(xy)=xt(y)+t(x)y, \  \  \bigtriangleup t=(Id\otimes
t)\bigtriangleup.$$
In this case, the Lie bracket is defined by the commutator with
respect to the composition. \cite{CM4, CM1}
\end{rem}

Since characters include Feynman rules of a theory therefore it is important to have enough information about the dual space $L(H,A)$ of all linear maps from $H$ to $A$.  We know that filtration can determine a concept of distance related to Hopf algebras.
Loop number of Feynman diagrams induces an increasing filtration on $H$ such that in the dual level, it defines
 a decreasing filtration on $L(H,A)$. With the help of
(\ref{metr1}), (\ref{metr2}) and (\ref{metr3}), one can have a
complete metric $\textbf{d}_{\Phi}$ on the space $L(H,A)$. In future it will be seen that the dual space plays an essential role in the study of quantum integrable systems.

It is good place to familiar with another additional algebraic structure namely, Rota-Baxter maps.

\begin{defn} \label{rb1}
Let $\mathbb{K}$ be a field with characteristic zero and $A$ an
associative unital $\mathbb{K}-$algebra. A $\mathbb{K}-$linear map
$R:A\longrightarrow A$ is called {\it Rota-Baxter operator of weight
$\lambda\in \mathbb{K}$}, if for all $x,y \in A$, it satisfies
\begin{equation}
R(x)R(y) + \lambda R(xy)=R(R(x)y+xR(y)).
\end{equation}
The pair $(A,R)$ is called Rota-Baxter algebra.
\end{defn}

\begin{lem} \label{adkin}

(i) For $\lambda \neq 0$, the standard form
$$R(x)R(y) + R(xy)=R(R(x)y+xR(y))$$
is given by the transformation $R  \longmapsto  \lambda^{-1}R$.

(ii) If $R$ is a Rota-Baxter map, then $\widetilde{R}:= Id_{A}-R$
will be a Rota-Baxter map and also $Im \ R,$ $Im \ \widetilde{R}$
are subalgebras in $A$.

(iii) For a given Rota-Baxter algebra $(A,R)$, define a product
$$a \circ_{R} b:= R(a)b + aR(b) - ab$$
on $A$. $A_{R}:=(A, \circ_{R},R)$ is a Rota-Baxter algebra. It is called
double Rota-Baxter algebra and one can continue this process to obtain an
infinite sequence of doubles.

(iv) The pair $(A,R)$ has a unique Birkhoff decomposition
$(R(A),-\widetilde{R}(A))\subset A\times A$ if and only if it is a
Rota-Baxter algebra.

(v) There is a natural way to extend the Rota-Baxter property to the Lie
algebra level. For a given Rota-Baxter algebra $(A,R)$, let $(A, [.,.])$ be
its Lie algebra with respect to the commutator. One can show that
for each $x, y \in A$,
\begin{equation} \label{4}
[R(x),R(y)] + R([x,y])=R([R(x),y]+[x,R(y)]).
\end{equation}
Triple $(A,[.,.],R)$ is called Lie Rota-Baxter algebra. \cite{EG2, EGK1, EGK2, M2}

\end{lem}

Based on a given Rota-Baxter structure on the algebra $A$, it is possible to determine a new algebraic structure on the dual space $L(H,A)$ such that it will be applied to reformulate renormalization process.

\begin{thm} \label{information2}
Suppose regularization and renormalization schemes in a given theory are introduced by Rota-Baxter algebra
$(A,R)$. Another Rota-Baxter map $\Upsilon$ can be inherited on $L(H,A)$ given by
$\phi \longmapsto \Upsilon(\phi):=R\circ \phi.$

(i) Triple  $\widetilde{\Phi}:=(L(H,A),*,\Upsilon)$ is a complete filtered noncommutative associative
unital Rota-Baxter $\mathbb{K}-$algebra of weight one.

(ii) One can extend this Rota-Baxter map to the Lie algebra level. \cite{EG2, G3}
\end{thm}

Now consider one particular renormalization method in modern physics namely, minimal subtraction scheme in dimensional regularization. This interesting scheme is identified with the commutative algebra
of Laurent series with finite pole part $A_{dr}:=\mathbb{C}[[z, z^{-1}]$ and
the renormalization map $R_{ms}$ on $A_{dr}$ where it is given by
\begin{equation} \label{minimal-sub}
R_{ms}(\sum_{i \ge -m}^{\infty} c_{i} z^{i}):= \sum_{i \ge
-m}^{-1} c_{i} z^{i}.
\end{equation}
It is an idempotent Rota-Baxter map of weight one. Connes and
Kreimer proved that the Bogoliubov's recursive formula for
counterterms and renormalized values (depended on determined
characters) can be reconstructed with the Birkhoff decomposition on elements of $G(A_{dr})$ and the map $R_{ms}$ \cite{CK1, CK2, CM1}.

\begin{thm} \label{char2}

For a given renormalizable physical theory $\Phi$ underlying the BPHZ method with the related Hopf algebra $H_{D}(\Phi)$ (for the discrete level) and $H_{F}(\Phi)$ (for the full level),

(i) The groups of diffeographisms Diffg$(\Phi)_{D}$ (for the discrete
level) and Diffg$(\Phi)_{F}$ (for the full level) are the pro-unipotent affine group schemes of the Hopf algebras
$H_{D}(\Phi)$ and $H_{F}(\Phi)$. The relation between these two groups
is determined by a semidirect product as follows
$$
Diffg(\Phi)_{F}= Diffg(\Phi)_{ab} \rtimes Diffg(\Phi)_{D}.
$$

(ii) Set $H=H(\Phi)$. The graded dual Hopf algebra
$H^{*}$ contains all finite linear combinations of homogeneous
linear maps on $H$. If $L:=Lie \  Prim H^{*}$, then there is a
canonical isomorphism of Hopf algebras between $H$ and the graded
dual of the universal enveloping algebra of $L$ and moreover, $L=Lie
 \  G(A_{dr})$. \cite{CM2, CM1}
\end{thm}

The phrase "{\it diffeographism}" is motivated from this fact that the space Diffg$(\Phi)$ acts on coupling constants of the theory through formal diffeomorphisms tangent to the identity. These diffeographisms together with Birkhoff factorization provide enough information to calculate counterterms and renormalized values.

\begin{thm} \label{bphz2}
For the dimensionally regularized Feynman rules character $\phi \in
G(A_{dr})$, there is a unique pair $(\phi_{-}, \ \phi_{+})$ of characters in
$G(A_{dr})$ such that
$$ \phi = \phi_{-}^{-1} * \phi_{+}.$$
It determines an algebraic Birkhoff decomposition for the chosen
character.
\cite{CM4, EG2, G3}
\end{thm}

Based on the Riemann-Hilbert correspondence as a motivation, Connes and Kreimer proved that physical parameters of a given theory can reformulate with the characters $\phi_{-}$ and $\phi_{+}$. It was the first bridge between Birkhoff decomposition and theory of quantum fields.

\begin{thm} \label{bphz1}
For the dimensionally regularized Feynman rules character $\phi$ in
$G(A_{dr})$,

(i) Its related Birkhoff components are determined by

$$ \phi_{-}(\Gamma) = e_{A_{dr}} \circ \epsilon_{H} (\Gamma) -
\Upsilon(\phi_{-} * (\phi - e_{A_{dr}} \circ \epsilon_{H}))
(\Gamma)$$
$$ = -R_{ms}(\phi(\Gamma) + \sum_{\gamma \subset \Gamma} \phi_{-}(\gamma) \phi(\frac{\Gamma}{\gamma})), $$
$$\phi_{+}(\Gamma) = e_{A_{dr}} \circ \epsilon_{H}(\Gamma) -
\widetilde{\Upsilon}(\phi_{+} * (\phi^{-1}-e_{A_{dr}} \circ
\epsilon_{H})) (\Gamma)$$
$$= \phi(\Gamma) + \phi_{-}(\Gamma) + \sum_{\gamma \subset \Gamma} \phi_{-}(\gamma) \phi(\frac{\Gamma}{\gamma}),$$
such that $\Gamma \in ker \ \epsilon_{H}$ and the sum is over all disjoint
unions of superficially divergent 1PI proper subgraphs.

(ii) BPHZ renormalization (i.e. unrenormalized regularized value, counterterms, renormalized values) can be rewritten algebraically by
$$\phi(\Gamma) = U_{\mu}^{z}(\Gamma), \ \
\phi_{-}(\Gamma)=c(\Gamma), \ \
\phi_{+}(\Gamma)=rv(\Gamma).$$
\cite{CK1, CK2, CM4, EG2, G3}
\end{thm}

To remove divergences step by step in a perturbative expansion of Feynman diagrams is
the main idea of renormalization and in this language a theory is
renormalizable, if the disappearing of all divergences be possible by
such a finite recursive procedure. With attention to the algebraic reformulation of the BPHZ prescription, in the next step of this section, we consider the structure of the renormalization group and its related infinitesimal character. We observe that how geometric concepts (such as loop space) would be entered in the story to provide a complete algebro-geometric description from physical information.

Let $U_{\mu}^{z}(\Gamma (p_{1},...p_{N}))$ be a regularized
unrenormalized value. It determines a loop $\gamma$ on the infinitesimal punctured disk $\bold{\Delta^{*}}$ (connected with the regularization parameter) around the $z=0$ with values
in the group of diffeographisms such that this loop has a Birkhoff
factorization
\begin{equation} \label{birk-loop1}
\gamma(z)=\gamma_{-}(z)^{-1} \gamma_{+}(z).
\end{equation}
where $\gamma_{-}(z)$ (holomorphic in $\mathbb{P}(\mathbb{C}) - \{0\}$)
gives the counterterm and $\gamma_{+}(z)$ (regular at $z=0$)
determines the renormalized value. Since $U_{\mu}^{z}(\Gamma
(p_{1},...p_{N}))$ depends on the mass parameter $\mu$, its related
loop should have a dependence on this parameter. In summary,
\begin{equation} \label{birk-loop2}
U_{\mu}^{z}(\Gamma (p_{1},...p_{N}))  \Longleftrightarrow
\gamma_{\mu}: \bold{\Delta^{*}} \longrightarrow Diffg(\Phi).
\end{equation}
Now one can see that the space $Loop \ (Diffg(\Phi),\mu)$ contains physical information of the theory $\Phi$.

\begin{thm} \label{birk-loop3}
Let $G(\mathbb{C})$ be the pro-unipotent complex Lie group associated to the
positively graded connected commutative finite
type Hopf algebra $H$ of the renormalizable theory $\Phi$
underlying the minimal subtraction scheme in dimensional regularization.
Suppose $\gamma_{\mu}(z)$ be a loop with values in $G(\mathbb{C})$ that encodes
$U_{\mu}^{z}(\Gamma(p_{1},...,p_{n}))$. It has a unique Birkhoff
decomposition
$$ \gamma_{\mu}(z)= \gamma_{\mu-}(z)^{-1} \gamma_{\mu+}(z)$$
such that

(i) $\frac {\partial}{\partial \mu} \gamma_{\mu -}(z)=0,$

(ii) For each $\phi \in H^{\star}_{n}$, $t \in \mathbb{C}:$
$\theta_{t}(\phi)= e^{nt} \phi$ is a $1-$parameter group of
automorphisms on $G(\mathbb{C}),$

(iii) $\gamma_{e^{t}\mu}(z)= \theta_{tz}(\gamma_{\mu}(z)),$

(iv) The limit
$$F_{t}=lim_{z \longrightarrow 0} \gamma_{-}(z)
\theta_{tz}(\gamma_{-}(z)^{-1})$$ exists and it denotes a
$1-$parameter subgroup $F_{t}$ of $G(\mathbb{C})$. It means that for each $s,t$,
$$F_{s+t}=F_{s}*F_{t}.$$

(v) For each Feynman diagram $\Gamma$, $F_{t}(\Gamma)$ is a
polynomial in $t$.

(vi) $\forall t \in \mathbb{R}:$ $\gamma _{e^{t} \mu +}(0)= F_{t}
\gamma _{\mu +}(0)$. \cite{CK1, CK2, CM2, CM4, CM1}
\end{thm}

\begin{defn} \label{ren-beta}
The $1-$parameter subgroup $\{F_{t}\}_{t}$ of $G(\mathbb{C})$ identifies
the renormalization group of the theory such that its infinitesimal generator
$$
\beta= \frac {d}{dt} |_{t=0} F_{t}$$
determines the
$\beta-$function of the theory.
\end{defn}

\begin{rem}
$G(\mathbb{C})$ is a topological group with the topology of pointwise
convergence and it means that for each sequence $\{\gamma_{n}\}_{n}$ of
loops with values in $G(\mathbb{C})$ and for each Feynman diagram $\Gamma \in H$,
$$\gamma_{n} \longrightarrow \gamma \Longleftrightarrow
\gamma_{n}(\Gamma) \longrightarrow \gamma(\Gamma).$$
\end{rem}

Letting $\mathfrak{g}(\mathbb{C})$ be the Lie algebra of diffeographisms of $\Phi$. We know that
it contains all of the linear maps $Z:H \longrightarrow A$
satisfying in the Libniz's law namely, derivations such that its Lie
bracket is given by the commutator with respect to the convolution
product. There is a bijection between this Lie algebra and its
corresponding Lie group given by the exponential map. One can expand
the Lie algebra $\mathfrak{g}(\mathbb{C})$ with an additional generator $Z_{0}$
such that for each $Z \in \mathfrak{g}(\mathbb{C})$,
\begin{equation}
[Z_{0},Z]=Y(Z)
\end{equation}
where $Y$ is the grading operator.

\begin{lem} \label{sca1}

(i) $\frac{d \theta_{t}}{dt}|_{t=0} = Y$.

(ii) For each $\phi \in H^{*}$ and Feynman diagram $\Gamma \in H$, we have
$$<\theta_{t}(\phi),\Gamma> = <\phi,\theta_{t}(\Gamma)>.$$
It means that
$\theta_{t} = Ad_{exp^{*}(tZ_{0})}.$   \cite{CM2, CM4, EGK1}
\end{lem}

In a more practical point of view and for the simplicity in computations, one can have a scattering type formula for components of
the dimensionally regularized Feynman rules character $\phi$. It means that

\begin{lem}  \label{sca2}  \label{sca3} \label{sca4}

(i) $F_{t}=lim_{z \longrightarrow 0} \phi_{-} \theta_{tz} \phi_{-}^{-1},$

(ii)  $\beta(\phi) = \phi_{\pm} * Y(\phi_{\pm}^{-1}) = \phi_{\pm} *
[Z_{0},\phi_{\pm}^{-1}],$

(iii)
$exp^{*} (t (\beta(\phi) + Z_{0})) * exp^{*}(-tZ_{0}) = \phi_{\pm} *
\theta_{t}(\phi_{\pm}^{-1}) \longrightarrow^{t \rightarrow \infty} \phi_{\pm}.$

\cite{EGK1, EGK2, S2}
\end{lem}

Finally, let us recalculate physical information to emphasize more the role of the Connes-Kreimer Hopf algebra in this algebraic formalism. Indeed, this machinery works with the antipode map \cite{K15}. If $\phi$ be the Feynman rules character, then consider the undeformed character $\phi \circ S$ and deform this character by the renormalization map. So for each Feynman diagram $\Gamma$, the BPHZ renormalization can be summarized in the equations

\begin{equation}
S_{R_{ms}}^{\phi}(\Gamma) = - R_{ms}(\phi(\Gamma)) - R_{ms} (\sum_{\gamma \subset \Gamma} S_{R_{ms}}^{\phi}(\gamma) \phi(\frac{\Gamma}{\gamma})),
\end{equation}
\begin{equation}
\Gamma \longmapsto S_{R_{ms}}^{\phi} * \phi (\Gamma).
\end{equation}
Because it is easy to see that
\begin{equation}
S_{R_{ms}}^{\phi} * \phi (\Gamma) = \overline{R}(\Gamma) + S_{R_{ms}}^{\phi}(\Gamma)
\end{equation}
such that the Bogoliubov operation $\overline{R}$ is given by
\begin{equation}
\overline{R}(\Gamma) = U_{\mu}^{z}(\Gamma) + \sum_{\gamma \subset \Gamma} c(\gamma) U_{\mu}^{z}(\frac{\Gamma}{\gamma})
= \phi(\Gamma) + \sum_{\gamma \subset \Gamma} S_{R_{ms}}^{\phi}(\gamma) \phi(\frac{\Gamma}{\gamma}).
\end{equation}
It means that $S_{R_{ms}}^{\phi}(\Gamma)$ and $S_{R_{ms}}^{\phi} * \phi (\Gamma)$ are counterterm and renormalized value depended on the Feynman diagram $\Gamma$.

In summary, Connes-Kreimer perturbative renormalization introduces a new algebraic interpretation to calculate the renormalization group based on the loop space of characters and Birkhoff factorization. The geometric nature of this procedure will be shown more, when we consider the renormalization bundle and its related flat equi-singular connections (formulated by Connes and Marcolli). Moreover, we know that the
minimal subtraction scheme in dimensional regularization indicates a Rota-Baxter algebra such that it provides Birkhoff decomposition for characters. As we shall see later that how this algebraic property helps us to indicate a new approach to consider theory of integrable systems in renormalizable quantum field theories such that renormalization has a central role.


\chapter{\textsf{Integrable renormalization: Integrable quantum Hamiltonian systems based on the perturbative renormalization}}

It was shown that with the help of Hopf algebra of renormalization on Feynman diagrams of a renormalizable QFT and based on components of the Birkhoff decomposition of some particular elements in the loop space on diffeographisms, one can determine counterterms, renormalized values, the Connes-Kreimer renormalization group and its infinitesimal generator. Indeed, the identification of renormalization with the Riemann-Hilbert problem provides a new conceptual interpretation of physical information.

With applying Atkinson theorem, one can show that the existence and the uniqueness of this factorization is a direct consequence of
an important and interesting concept in physics namely, multiplicativity of renormalization which is prescribed mathematically with the
Rota-Baxter condition of the chosen renormalization scheme.
On the other hand and in a roughly speaking, one can find the importance of this class of equations in the study of theory of integrable systems specially, (modified) Yang-Baxter equations. This fact reports obviously a pure algebraic configuration of renormalization where it indeed seems wise to introduce a coherent ideology for considering quantum integrable systems with respect to the Connes-Kreimer theory. \cite{EG1, EGK1, EGK2,
K2, KS1}

With attention to the theory of integrable systems in classical
level \cite{C1, D1, KGT, KM1, KS1, S1, S3}, this (rigorous) algebraic machinery for the
description of renormalizable QFTs and the power of noncommutative differential forms \cite{G1}, we are going to find a new family of Hamiltonian systems which arise from  the Connes-Kreimer approach to perturbative renormalization
and moreover, we show that how integrability condition can be determined naturally based on Poisson brackets related to Rota-Baxter type algebras.
The beauty of this new viewpoint to integrable systems will be identified, when we consider the minimal subtraction underlying the dimensional
regularization (as the renormalization mechanism) and then study possible relation between introduced motion integrals and renormalization group. After that, based on the Rosenberg framework, we familiar with other group of quantum Poisson brackets and their related motion integrals. We close this chapter with introducing a new family of fixed point equations modified with motion integral condition and Bogoliubov character. In summary, it seems favorable to report about the appearance of a glimpse of one general relation between the theory of Rota-Baxter type algebras and the Riemann-Hilbert problem underlying quantum field theory. \cite{S11, S10, S12}


\section{{\it \textsf{What is an integrable system? From finite dimension (geometric approach) to infinite dimension (algebraic approach)}}}

Let $V$ be a $m$-dimensional real vector space and $\omega:V\times V
\longrightarrow \mathbb{R}$ a skew-symmetric bilinear map.

\begin{defn} \label{symplecticform2}
For a
map $\widetilde{\omega}:V\longrightarrow V^{*}$ given by
$$\widetilde{\omega}(v)(u):=\omega(v,u),$$
$\omega$ is called a symplectic form, if $\widetilde{\omega}$ be a
bijective map. The pair $(V,\omega)$ is called symplectic
vector space and it is clear that each symplectic vector space has
even dimension.
\end{defn}

\begin{defn} \label{symplecticform3}
A differential 2-form $\omega$ on a manifold $M$ is called
symplectic, if it is closed and for each $p\in M$, $\omega_{p}$ is a
symplectic form on $T_{p}M$.
\end{defn}

\begin{defn}
(i) A complex structure on $V$ is a linear map $J:V\longrightarrow
V$ such that $J^{2}=-Id_{V}$. $(V,J)$ is called a complex
vector space.

(ii) An almost complex structure on a manifold $M$ is a smooth
field of complex structures on the tangent spaces and it means that
for each $x \in M$,
$$
x \longmapsto J_{x}:T_{x}M \longrightarrow T_{x}M: Linear, \ \
J^{2}_{x}=-Id.
$$
The pair $(M,J)$ is called an almost complex manifold.

(iii) An almost complex structure $J$ is called integrable, if $J$ is
induced by a structure of complex manifold on $M$.

\end{defn}

There is a well known operator to characterize integrable almost complex structures on manifolds. It will be shown that this map is in fact a starting key for us to consider quantum integrable systems underlying an algebraic formalism.

\begin{defn} \label{nij1}
For the
almost complex manifold $(M,J)$, the {\it Nijenhuis tensor} is
defined by
$$
NT(v,w):=[Jv,Jw]-J[v,Jw]-J[Jv,w]-[v,w]
$$
such that $v,w$ are vector fields on $M$ and for each $f\in
C^{\infty}(M)$,
$$
[v,w].f=v.(w.f)-w.(v.f)
$$
where $v.f=df(v)$.
\end{defn}

\begin{thm} \label{nij2}
For the almost complex manifold $(M,J)$, the following facts are
equivalent:

(i) $M$ is a complex manifold,

(ii) $J$ is integrable,

(iii) $NT\equiv 0$,

(iv) $d=\partial + \overline{\partial}$. \cite{AKN, C1}
\end{thm}

Definition \ref{nij1} and theorem \ref{nij2} show that if $J$ is an
integrable structure, then for each $x\in M$ and $v,w\in T_{x}M$, we
have
\begin{equation}
[J_{x}v,J_{x}w]=J_{x}[v,J_{x}w]+J_{x}[J_{x}v,w]+[v,w].
\end{equation}

Equations of motion are the results of variational problems in
classical mechanics and in a system with $n$ particles in
$\mathbb{{R}}^{n}$, all of the physical trajectories follow from the
Newton's second law. In these paths the mean value of the difference
between kinetics and potential energies is minimum.

\begin{lem} \label{sympl.man.1}
Let $(M,\omega)$ be a symplectic manifold and $H:M \longrightarrow
\mathbb{R}$ a smooth function. There is a unique vector field
$X_{H}$ on $M$ such that $i_{X_{H}}\omega=dH$. One can identify a
1-parameter family of diffeomorphisms $\rho_{t}:M\longrightarrow
M$ such that

$$\rho_{0}=Id_{M}, \   \ \frac{d\rho_{t}}{dt}\circ
\rho_{t}^{-1}=X_{H}, \  \ \rho_{t}^{*}\omega=\omega.$$ \cite{AKN, C1, KGT}
\end{lem}

This lemma shows that each smooth function $H:M \longrightarrow
\mathbb{R}$ determines a family of {\it symplectomorphisms}. The
function $H$ is called {\it Hamiltonian function} such that the vector
field $X_{H}$ is its corresponding {\it Hamiltonian vector field}.

\begin{defn} \label{ham.sys.1}
The triple $(M,\omega,H)$ is called a classical Hamiltonian system.
\end{defn}

On the Euclidean space $\mathbb{R}^{2n}$ with the local coordinate
$(q_{1},...,q_{n},p_{1},...,p_{n})$ and the canonical symplectic
form $\omega=\sum dq_{i}\wedge dp_{i}$, the curve
$\alpha_{t}=(q(t),p(t))$ is an {\it integral curve} of the vector
field $X_{H}$ if and only if it determines the Hamiltonian equations of motion
\begin{gather}
\frac{dq_{i}}{dt}=\frac{\partial H}{\partial p_{i}}, \  \
\frac{dp_{i}}{dt}=-\frac{\partial H}{\partial q_{i}}
\end{gather}
on $\mathbb{R}^{2n}$.

\begin{thm} \label{ham.sys.2}
The Newton's second law on the configuration space $\mathbb{R}^{n}$
is equivalent to the Hamiltonian equations of motion on the phase
space $\mathbb{R}^{2n}$. \cite{AKN, C1, KGT}
\end{thm}
For a given symplectic manifold $M$  and $f,g\in C^{\infty}(M)$, letting
$X_{f},X_{g}$ be the Hamiltonian vector fields with respect to these
functions. From this class of vector fields, one expects a Poisson bracket on $C^{\infty}(M)$ given
by
\begin{equation}
\{f,g\}:=\omega(X_{f},X_{g}).
\end{equation}
The pair $(C^{\infty}(M),\{.,.\})$ is called {\it Poisson algebra related to
the configuration space $M$} and from now $M$ is called a {\it Poisson manifold}.

\begin{rem}
Dual space of a Lie algebra is one useful example of a Poisson
manifold but generally, these Poisson manifolds are not
symplectic.
\end{rem}

\begin{lem} \label{poisson1}
For any $H \in C^{\infty}(M)$, its related
Hamiltonian vector field which acts on the elements of
$C^{\infty}(M)$ (by the Poisson bracket) can be rewritten with the equation
$$
X_{H} f = \{H,f\}
$$
such that for each $x \in M$, the vectors $X_{H}(x)$ span a linear
subspace in $T_{x}M$. \cite{AKN, C1, KGT}
\end{lem}

We know that Hamiltonian vector fields are tangent to symplectic
leaves and it means that the Hamiltonian flows enable to preserve
each leaf separately. So for a given Hamiltonian system $(M,\omega,H)$,
equation
\begin{equation} \label{2}
\{f,H\}=0
\end{equation}
is equivalent with this fact that $f$ is constant along the integral curves of $X_{H}$.
The function $f$ is called {\it integral of motion}. \cite{C1, S1}

\begin{defn} \label{integrablesys.1}
The Hamiltonian system $(M, \omega, H)$ is called {\it integrable}, if there exist
$n=\frac{1}{2}dimM$ independent integrals of motion
$f_{1}=H,f_{2},...,f_{n}$ such that $\{f_{i},f_{j}\}=0$. \cite{AKN, C1}
\end{defn}

Suppose $\mathfrak{g}$ be a Lie algebra with the dual
space $\mathfrak{g}^{\star}$ and $P(\mathfrak{g}^{\star})$ be the
space of polynomials on the dual space. Elements of the Lie algebra can determine a bracket
on $P(\mathfrak{g}^{\star})$ such
that for each $ g_{1},g_{2} \in \mathfrak{g}$ and $h^{\star} \in
\mathfrak{g}^{\star},$ it is given by
\begin{equation}
\{g_{1},g_{2}\}(h^{\star}):= h^{\star}([g_{1},g_{2}]).
\end{equation}
Since $P(\mathfrak{g}^{\star})$ is dense in
$C^{\infty}(\mathfrak{g}^{\star})$, one can expand canonically this
bracket to obtain a {\it Lie Poisson bracket} such that for each $s_{1},
s_{2} \in C^{\infty}(\mathfrak{g}^{\star})$, we have
\begin{equation} \label{lie-poisson}
\{s_{1}, s_{2}\}(h^{\star})=h^{\star}([d s_{1}(h^{\star}), d
s_{2}(h^{\star})]).
\end{equation}
\begin{thm} \label{liepoison}
For a given Lie group $G$ with the Lie algebra $\mathfrak{g}$, the
symplectic leaves of (\ref{lie-poisson}) are $G-$orbits in
$\mathfrak{g}^{\star}$. \cite{AKN, KGT, S3}
\end{thm}

One can apply automorphisms on a Lie algebra to introduce other Lie Poisson brackets. This method provides a new favorable point of view to study infinite dimensional  integrable systems.

\begin{defn} \label{rmatrix}
For a fixed Lie
algebra $\mathfrak{g}$, an endomorphism $R \in End(\mathfrak{g})$ is
called R-matrix, if for each $g_{1},g_{2} \in \mathfrak{g}$,
the bracket
$$
[g_{1},g_{2}]_{R}= \frac{1}{2}([R (g_{1}),g_{2}] + [g_{1},R
(g_{2})])
$$
satisfies in the Jacobi identity and it means that $[.,.]_{R}$ is a
Lie bracket.
\end{defn}

Yang-Baxter equation is enough to introduce a new class of Poisson brackets on $\mathfrak{g}^{\star}$ which arise from this kind of Lie brackets.

\begin{thm} \label{cY.B}
Let $R \in End(\mathfrak{g})$. For each $g_{1}, g_{2} \in
\mathfrak{g}$, set
$$B_{R}(g_{1},g_{2}):=[R (g_{1}),R (g_{2})] - R([R (g_{1}),g_{2}] + [g_{1},R (g_{2})]).$$
The $R-$bracket defined in \ref{rmatrix} follows the Jacobi identity if and only if
for each $g_{1},g_{2},g_{3} \in \mathfrak{g}$, we have
$$ [B_{R}(g_{1},g_{2}),g_{3}] + [B_{R}(g_{2},g_{3}),g_{1}] + [B_{R}(g_{3},g_{1}),g_{2}]=0. $$
This condition is called {\it classical Yang-Baxter equation}. The
simplest sufficient condition is given by $\ B_{R}(g_{1},g_{2})=0$.
\cite{KGT, KM1, KS1, S3}
\end{thm}

Yang-Baxter equations make possible to consider integrable systems in an algebraic framework such that this ability together with the Hopf algebra of Feynman diagrams help us to study quantum Hamiltonian systems and integrability condition depended on the algebraic renormalization. The critical point in this procedure can be summarized in reinterpretation of the renormalization group on the basis of Rota-Baxter type algebras. We show that how one can find a new family of symplectic structures where its existence is strongly connected with the choosing of renormalization prescription.


\section{{\it \textsf{Rota-Baxter type algebras: Nijenhuis algebras}}}

In this part we dwell a moment on a special class of Rota-Baxter type algebras namely, Nijenhuis algebras to discuss that how one can
apply these maps to deform the initial product. It gives a family of new associative algebras together with related compatible Lie brackets.

For a linear map $N:A \longrightarrow A$, define a new product on
$A$. It is given by
\begin{equation} \label{circ1}
(x,y) \longmapsto x \circ_{N} y:= N(x)y+xN(y)-N(xy).
\end{equation}

\begin{rem}
If $e$ be the unit of $(A,m)$ and $N(e)=e$ then
$(A,\circ_{N})$ has the same unit.
\end{rem}

\begin{lem} \label{pro-new}

(i) $(A, \circ_{N})$ is an algebra.

(ii) The product $\circ_{N}$ is associative if and only if for each
$x, y \in A$,
$$T_{N}(x,y):=N(x \circ_{N} y) - N(x)N(y)$$
be a Hochschild 2-cocycle of the algebra $A$ (with respect to the
Hochschild coboundary operator $\textbf{b}$ connected with the product of
$A$).  It means that for each $x, y, z \in A$,
$$\textbf{b}T_{N}(x,y,z):=xT_{N}(y,z) - T_{N}(xy,z) + T_{N}(x,yz) - T_{N}(x,y)z = 0.$$

(iii) $N$ is a derivation in the original algebra if and only if $N$
is a 1-cocycle with respect to the coboundary operator $\textbf{b}$. In
this case, the new product $\circ_{N}$ is trivial. \cite{CGM1, D2}
\end{lem}

\begin{defn} \label{nij3}
The linear map $N$ is called {\it Nijenhuis tensor} supported by
$\circ_{N}$, if
\begin{equation}
N(x \circ_{N} y) = N(x)N(y).
\end{equation}
The pair $(A,N)$ is called Nijenhuis algebra.
\end{defn}

\begin{rem}

(i) If $N$ is a Nijenhuis tensor, then $\circ_{N}$
is an associative product on $A$ and also for each $\lambda \in
\mathbb{K}$, $m + \lambda \circ_{N}$ is an associative product
on $A$ (such that $m$ is the original product of $A$).

(ii) For a given Rota-Baxter algebra $(A,R)$ with the
idempotent Rota-Baxter map $R$ and for each $\lambda \in
\mathbb{K}$, the operator $N_{\lambda}:= R- \lambda \widetilde{R}$
has Nijenhuis property and it means that $(A,N_{\lambda})$ is a
Nijenhuis algebra. \cite{EG2, EGK1, EGK2}

\end{rem}

The classical Yang-Baxter equation is essentially governed by
the extension of this kind of operators to the Lie algebra level.

\begin{defn}  \label{2bra}
A Nijenhuis tensor for the Lie algebra $(A,[.,.])$ is a linear map
$N:A \longrightarrow A$ such that for each $x, y \in A$,
$$N([x,y]_{N})=[N(x),N(y)]$$
where

$$[x,y]_{N}:=[N(x),y]+[x,N(y)]-N([x,y]).$$

\end{defn}

\begin{rem} \label{n1}
The compatibility of this Lie bracket is strongly related to the
Nijenhuis property of $N$ and it is easy to see that in this case
$$
[N(x),N(y)]=N([N(x),y])+ N([x,N(y)])-N^{2}([x,y]).
$$
The triple $(A,[.,.],N)$ is called Nijenhuis Lie algebra. \cite{D2}
\end{rem}

\begin{thm} \label{nij4}
Let $(A,\circ_{N})$ be the associative algebra with respect to the
Nijenhuis tensor $N$ on $A$. $N$ is a Nijenhuis tensor for the Lie
algebra $(A,[.,.])$ (i.e. $[.,.]$ is the commutator with respect to the product $m$) and
for each $x,y \in A$, we will have
\begin{equation}
[x,y]_{N}=x \circ_{N} y - y \circ_{N} x.
\end{equation}
\cite{CGM1, D2, EG2, EGK1, EGK2}
\end{thm}

Because of the importance of Nijenhuis algebras, we are going to mention an algorithmic instruction for constructing this type of algebras from
each arbitrary commutative algebra. Let $(A,m)$
be a commutative $\mathbb{K}-$algebra with the tensor algebra
$T(A):= \oplus_{n \ge 0} A^{\otimes n}$. Elements of $A$ play
the role of {\it letters} and generators $U=a_{1} \otimes ...
\otimes a_{n}$ in $A^{\otimes n}$ (where $a_{i} \in A$) are
identified with {\it words} $a_{1}...a_{n}$.

\begin{lem} \label{quasi-shufle1}
For each $a, b \in A,$ $U \in A^{\otimes n},$  $V \in A^{\otimes m}$
and $\lambda \in \mathbb{K}$,  one can define recursively an
associative commutative quasi-shuffle product on $T(A)$ given
by
$$
aU \star bV:=a(U \star bV)+b(aU \star V)- \lambda m(a,b)(U \star V)
$$
such that the empty word plays the role of its unit. \cite{E1, EG1, H1}
\end{lem}

The product given by the lemma \ref{quasi-shufle1} determines
another kind of shuffle product.

\begin{lem} \label{quasi-shufle2}
An augmented quasi-shuffle product on the augmented tensor module
$\overline{T}(A):= \oplus_{n
> 0} A^{\otimes n}$ is given by
$$
aU \odot bV:= m(a,b)(U \star V).
$$
It is an associative commutative product such that the unit $e$
of the algebra $A$ is the unit for this product. \cite{E1, EG1, H1}
\end{lem}

One can define a modified version of the above products.

\begin{lem} \label{quasi-shufle3}
The modified quasi-shuffle product on $T(A)$ is defined by
$$
aU \ominus bV:= a(U \ominus bV) + b(aU \ominus V) - e m(a,b)(U
\ominus V)
$$
such that the empty word is the unit for this commutative
associative product. Its augmented version namely, augmented
modified quasi-shuffle product on $\overline{T}(A)$ is defined by
$$
aU \oslash bV:= m(a,b) (U \ominus V)
$$
such that the unit $e$ of $A$ is the unit for this new product. \cite{E1, EG1, H1}
\end{lem}

These products can be applied to make a special family of Rota-Baxter
and Nijenhuis algebras with the universal property.

\begin{thm} \label{univ-nij}
For a unital commutative associative $\mathbb{K}-$algebra $A$ with
the unit $e$, let $B^{+}_{e}$ be a linear operator on
$\overline{T}(A)$ given by
$$B^{+}_{e}(a_{1}...a_{n}):=ea_{1}...a_{n}.$$

(i) $(T(A), \star, \lambda, B^{+}_{e})$ is a RB algebra of weight
$\lambda$.

(ii) $(T(A), \ominus, B^{+}_{e})$ is a Nijenhuis algebra.

(iii) $(\overline{T}(A),\odot, \lambda, B^{+}_{e})$ is the universal
RB algebra of weight $\lambda$ generated by $A$.

(iv) $(\overline{T}(A), \oslash, B^{+}_{e})$ is the universal
Nijenhuis algebra generated by $A$.  \cite{E1}
\end{thm}

\begin{rem}
For instance, universal Nijenhuis algebra (generated by $A$) means
that for any Nijenhuis algebra $B$ and algebra homomorphism $f: A
\longrightarrow B$, there exists a unique Nijenhuis homomorphism
$\widetilde{f}:\overline{T}(A) \longrightarrow B$ such that
$\widetilde{f} \circ j_{A} = f$.
\end{rem}

Everything is prepared to introduce a new class of quantum Hamiltonian systems as the consequence of the Connes-Kreimer algebraic framework to perturbative theory. It will be shown that how the Hopf algebraic renormalization group can give us some examples of integrable systems from this class of Hamiltonian systems.


\section{{\it \textsf{Theory of quantum integrable systems}}}

In this part we want to focus on the algebraic basis of the Connes-Kreimer theory namely, the Rota-Baxter property induced from renormalization to improve theory of integrable systems to the level of renormalizable physical theories. With the help of noncommutative differential forms (associated with Hamiltonian derivations) and with attention to the chosen renormalization method (i.e. regularization algebra and renormalization map), we are going to introduce a new family of Hamiltonian systems based on the Connes-Kreimer Hopf algebra of Feynman diagrams. It is discussed that how integrability condition on these systems are strongly connected with the perturbative renormalization process. \cite{S10, S12}

Renormalization prescription makes possible two different types of deformations. In one class,
we consider deformed algebras which deformation process is performed by an idempotent renormalization map and in another class, with respect to regularization scheme (independent of the renormalization map), we turn to Ebrahimi-Fard's aspect in defining the universal Nijenhuis algebra (as the kind of deformation method) and then we will deform the initial algebra by this universal Nijenhuis tensor. Finally, with working on differential forms of these deformed algebras, we will illustrate (as the conclusion) Hamiltonian systems and also integrability condition.

Roughly speaking, Connes-Kreimer Hopf algebra and regularization algebra determines a new noncommutative algebra such that with attention to the renormalization scheme, one can provide new deformed algebras. Then with working on the noncommutative differential calculus with respect to these mentioned deformed algebras, we can obtain symplectic structures and so Hamiltonian systems. Because of applying this noncommutative differential formalism on the deformed algebras depended upon the renormalization process, it does make sense to use the phrase "{\it integrable renormalization}" for this new captured approach to quantum integrable systems.

\begin{center}
\textbf{First Class}
\end{center}

Suppose $\Phi$ be a renormalizable QFT with the associated Hopf algebra of Feynman diagrams $H$ and letting the perturbative renormalization is performed with the idempotent
renormalization map $R$ and the regularization scheme $A$. In summary, we denote its algebraic
reformulation with
\begin{equation}
\widetilde{\Phi}=(L(H,A),*,\Upsilon)
\end{equation}
such that
the idempotent Rota-Baxter map $\Upsilon$ on $L(H,A)$ is given by $R$.
It is easy to show that for each $\lambda \in \mathbb{K}$, the operator
$\Upsilon_{\lambda}:=\Upsilon - \lambda \widetilde{\Upsilon}$ (where
$\widetilde{\Upsilon}:=Id- \Upsilon$) has Nijenhuis property.

\begin{defn} \label{int.1}
By the formula (\ref{circ1}), for each $\lambda$ a new product
$\circ_{\lambda}$ on $L(H,A)$ can be defined such that for each
$\phi_{1},\phi_{2} \in L(H,A)$,
$$
\phi_{1} \circ_{\lambda} \phi_{2}:= \Upsilon_{\lambda} (\phi_{1}) *
\phi_{2} + \phi_{1} * \Upsilon_{\lambda} (\phi_{2}) -
\Upsilon_{\lambda} (\phi_{1} * \phi_{2}).
$$
\end{defn}

\begin{rem}

(i) One can show that
$$\Upsilon_{\lambda} (\phi_{1} \circ_{\lambda} \phi_{2}) =
\Upsilon_{\lambda} (\phi_{1}) * \Upsilon_{\lambda} (\phi_{2}).$$

(ii) Lemma \ref{pro-new} shows that $\circ_{\lambda}$ is an
associative product and one can identify the following compatible Lie
bracket
$$[\phi_{1},\phi_{2}]_{\lambda}:=[\Upsilon_{\lambda}(\phi_{1}),
\phi_{2}] + [\phi_{1}, \Upsilon_{\lambda} (\phi_{2})] -
\Upsilon_{\lambda} ([\phi_{1},\phi_{2}])$$
such that definition \ref{int.1} provides that
$$[\phi_{1},\phi_{2}]_{\lambda}= \phi_{1} \circ_{\lambda} \phi_{2} -
\phi_{2} \circ_{\lambda} \phi_{1}.$$
It means that one can extend the Nijenhuis property of
$\Upsilon_{\lambda}$ to the Lie algebra level (with the commutator with
respect to the product $*$).

\end{rem}

\begin{defn} \label{int.2} \label{family}
The new spectral information $(\widetilde{\Phi}_{\lambda}, [.,.]_{\lambda})
:=(L(H,A),\circ_{\lambda}, \Upsilon_{\lambda}, [.,.]_{\lambda})$
is called $\lambda-$information based on the theory $\Phi$
and with respect to the Nijenhuis map $\Upsilon_{\lambda}$.
\end{defn}

\begin{center}
\textbf{Second Class}
\end{center}

Generally, when we study the algebraic renormalization methods, the renormalization map might not have idempotent property and therefore we should apply another technique to receive Nijenhuis tensors.
In this situation one can focus on a commutative algebra (which reflects the
regularization scheme) and apply the universal Nijenhuis tensor based on this algebra.

Let $\Phi$ be a renormalizable theory with the related Hopf algebra
$H$ such that the regularization scheme is given by the associative commutative unital algebra $A$. Theorem \ref{univ-nij} shows
that $(\overline{T}(A), \oslash, B^{+}_{e})$ is the universal Nijenhuis algebra
based on $A$ and therefore one can define a Nijenhuis map
$\Upsilon^{+}_{e}$ on $L(H,\overline{T}(A))$ given by
\begin{equation} \label{int.3}
\Upsilon^{+}_{e}(\psi):= B^{+}_{e} \circ \psi.
\end{equation}

\begin{defn} \label{int.4} \label{circ}
With help of the operator $\Upsilon^{+}_{e}$, a new product
$\circ_{u}$ on $L(H,\overline{T}(A))$ is introduced such that for
each $\psi_{1},\psi_{2} \in L(H,\overline{T}(A))$,
$$\psi_{1} \circ_{u} \psi_{2}:= \Upsilon^{+}_{e} (\psi_{1})
*_{\oslash} \psi_{2} +  \psi_{1} *_{\oslash} \Upsilon^{+}_{e}
(\psi_{2}) - \Upsilon^{+}_{e} (\psi_{1} *_{\oslash} \psi_{2}),$$
where
$$\psi_{1} *_{\oslash} \psi_{2}:= \oslash (\psi_{1} \otimes \psi_{2})
\circ \Delta_{H}.$$
\end{defn}

\begin{rem}
(i) The Nijenhuis property of $\Upsilon^{+}_{e}$ provides that
$$
\Upsilon^{+}_{e} (\psi_{1} \circ_{u} \psi_{2}) = \Upsilon^{+}_{e}
(\psi_{1}) *_{\oslash} \Upsilon^{+}_{e} (\psi_{2}),
$$
such that it supports the associativity of the product $\circ_{u}$.

(ii) A compatible Lie bracket $[.,.]_{u}$ can be defined on $L(H,\overline{T}(A))$ by
$$[\psi_{1},\psi_{2}]_{u}:=[\Upsilon^{+}_{e}(\psi_{1}), \psi_{2}] +
[\psi_{1}, \Upsilon^{+}_{e} (\psi_{2})] - \Upsilon^{+}_{e}(
[\psi_{1},\psi_{2}]).$$
Moreover we have
$$[\psi_{1},\psi_{2}]_{u} = \psi_{1} \circ_{u} \psi_{2} - \psi_{2}
\circ_{u} \psi_{1}.$$
It means that Nijenhuis property of
$\Upsilon^{+}_{e}$ can be extended to the Lie algebra level (with the commutator with
respect to the product $*_{\oslash}$).
\end{rem}

\begin{defn} \label{int.5}
The spectral information $(\widetilde{\Phi}_{u}, [.,.]_{u}) :=(L(H,\overline{T}(A)),\circ_{u},
\Upsilon^{+}_{e},[.,.]_{u})$ is called $u-$information based on the theory $\Phi$ and
with respect to the Nijenhuis map $\Upsilon^{+}_{e}$.
\end{defn}

\begin{center}
\textbf{Symplectic Structures}
\end{center}

Here we introduce a certain family of symplectic structures related to this
algebraic preparation from renormalizable physical theories and for
this goal we need theory of noncommutative differential calculus
over an algebra based on the space of its derivations.

\begin{defn} \label{int.6}
Let $C$ be an associative unital algebra over the field $\mathbb{K}$
(with characteristic zero) with the center $Z(C)$. A derivation
$\theta :C \longrightarrow C$ is an infinitesimal automorphism of
$C$ such that it is a linear map satisfying the Leibniz rule
where if it could have been exponentiated, then the map
$exp \ \theta$ will be an automorphism of $C$.
\end{defn}

\begin{rem}

(i) Geometrically,
$\theta$ is a vector field on a noncommutative space and
$t\longmapsto \exp (t\theta)$ is the one parameter flow of
automorphisms (i.e. integral curves) of our noncommutative space
generated by $\theta$.

(ii) Suppose $Der(C)$ be the space of all derivations on $C$. It is a module
over $Z(C)$ such that it has a Lie algebra structure with the Lie
bracket given by the commutator with respect to the composition of
derivations. \cite{G1, V4, V5}
\end{rem}

\begin{defn} \label{int.7}
Letting $\Omega_{Der}^{n}(C)$ be the space of all $Z(C)-$multilinear
antisymmetric mappings from $Der(C)^{n}$ into $C$ such that
$\Omega_{Der}^{0}(C)=C$. A differential graded algebra
$\Omega^{\bullet}_{Der}(C)= \bigoplus_{n \ge 0} \Omega_{Der}^{n}(C)$ can be
defined where for each $\omega \in \Omega_{Der}^{n}(C)$ and
$\theta_{i} \in Der(C),$ its antiderivation differential operator
$d$ of degree one is given by
$$(d \omega) (\theta_{0},...,\theta_{n}) := $$
$$
\sum_{k=0}^{n} (-1)^{k} \theta_{k}
\omega(\theta_{0},...,\widehat{\theta_{k}},...,\theta_{n}) + \sum_{0
\le r<s \le n} (-1)^{r+s} \omega([\theta_{r},\theta_{s}],
\theta_{0},...,\widehat{\theta_{r}},...,\widehat{\theta_{s}},...,
\theta_{n}).$$
\end{defn}

\begin{rem} \label{dif-grad-Alg}
(i) $d^{2}=0$.

(ii) For a given derivation $\theta$ of $C$ and $\omega \in
\Omega_{Der}^{n}(C)$,  consider an anti-derivation operator $i_{\theta}$ of
degree $(-1)$ defined by
$$i_{\theta} \omega
(\theta_{1},...,\theta_{n-1})=\omega(\theta,\theta_{1},...,\theta_{n-1}).
$$
It is observed that
$$ i_{\theta_{1}} i_{\theta_{2}} + i_{\theta_{2}}i_{\theta_{1}}
= 0,$$
and if $L_{\theta}:=d \circ i_{\theta} + i_{\theta} \circ d$,
then
$$ L_{\theta_{1}} i_{\theta_{2}} - i_{\theta_{2}} L_{\theta_{1}} =
i_{[\theta_{1},\theta_{2}]},$$
$$ L_{\theta_{1}}L_{\theta_{2}} - L_{\theta_{2}}L_{\theta_{1}} =
L_{[\theta_{1},\theta_{2}]}.$$
\cite{G1}
\end{rem}

\begin{defn} \label{int.8}
The noncommutative deRham complex on $Der(C)$ is defined by
$$DR_{Der}^{\bullet}(C):=\frac{\Omega_{Der}^{\bullet}(C)}{[\Omega_{Der}^{\bullet}(C),\Omega_{Der}^{\bullet}(C)]}.$$

\end{defn}

We know that an algebra $C$ equipped with a bi-derivation Lie
bracket $\{.,.\}$ which satisfies in the Jacobi identity determines a {\it Poisson algebra}. There is a class of derivations on this algebra which contains an essential  geometric meaning.

\begin{defn} \label{int.9}
For each $c$ in the Poisson algebra $C$, derivation
$$ham(c): x \longmapsto \{c,x\}$$
is called a Hamiltonian derivation (vector field) corresponding to
$c$.
\end{defn}

\begin{defn} \label{int.10}
For a given Poisson algebra $C$, a $Z(C)-$bilinear antisymmetric map
$\omega$ in $\Omega^{2}_{Der}(C)$ is called non-degenerate, if for
any element $c \in C$, there exists a derivation $\theta_{c}$ of $C$
such that for each derivation $\theta$,
$$
\omega(\theta_{c},\theta)=\theta(c).
$$
In this case derivation $\theta_{c}=ham(c)$ is unique and the function
$\theta \longmapsto i_{\theta} \omega$ is linear and injective and
it is observed that
$$
(i_{ham(c)} \omega) (\theta) = \omega(ham(c),\theta)=
\theta(c)=:(dc)(\theta)
$$
such that $dc:Der(C) \longrightarrow C$ is a 1-form.
\end{defn}

\begin{defn} \label{int.11}
A closed non-degenerate element $\omega$ in $\Omega^{2}_{Der}(C)$ is
called a symplectic structure.
\end{defn}

With the help of this symplectic form, one can define an
antisymmetric bilinear bracket on $C$ such that for each $x, y \in
C$, it is given by
\begin{equation} \label{int.12}
\{x,y\}_{\omega}:= \omega(ham(x),ham(y)).
\end{equation}
It satisfies Leibniz law and Jacobi identity and therefore it
determines a Poisson bracket on $C$.

\begin{lem} \label{int.13}
(i) There is a Lie algebra homomorphism from $(C,\{.,.\}_{\omega})$
to $(Der(C),[.,.])$.

(ii) When $Z(C)-$module generated by the set
$$Ham(C):=\{ham(c): c \in C \}$$
be the entire of the space $Der(C)$, the Jacobi identity for the bracket
$\{.,.\}_{\omega}$ and closed condition for the symplectic structure
$\omega$ are equivalent. \cite{G1, G2, V4, V5}
\end{lem}

\begin{proof}

One can show that
$$[ham(x),ham(y)]=ham(\{x,y\}_{\omega}).$$
It means that {\it ham} plays the role of a Lie homomorphism.
\end{proof}

If a given associative algebra $C$ follows condition (ii) in the
lemma \ref{int.13}, then the Poisson bracket on $C$ is called {\it
non-degenerate} and otherwise it is called {\it degenerate}. It is
possible to reach to a symplectic structure from a non-degenerate
Poisson bracket.

\begin{lem} \label{sym-pois}
For a non-degenerate Poisson bracket $\{.,.\}$ on the associative
algebra $C$,  there exists a symplectic structure $\omega$ such that
its related Poisson bracket coincides with $\{.,.\}$. \cite{G1, G2}
\end{lem}
\begin{proof}
Let $\theta_{1}, \theta_{2}$ be derivations on $C$. From non-degeneracy
of the Poisson bracket, we know that
$$Z(C).Ham(C) = Der(C).$$
Therefore there exist $\{x_{1},...,x_{m},y_{1},...,y_{n} \} \subset
C$ and $\{u_{1},...,u_{m},v_{1},...,v_{n}\} \subset Z(C)$ such that
$$\theta_{1}= \sum_{i} u_{i} \ ham(x_{i}), \ \ \ \theta_{2}=\sum_{j}
v_{j} \ ham(y_{j}).$$
Define
$$\omega(\theta_{1},\theta_{2}):= \sum_{i,j} u_{i}v_{j} \{x_{i},y_{j}\}.$$
$\omega$ is our interesting symplectic structure.
\end{proof}

Now we are ready to combine the theory of noncommutative differential forms with the Connes-Kreiemr approach to obtain a new family of Hamiltonian systems completely depended upon the renormalization procedure.
At first suppose in theory
$\Phi$ one can perform renormalization with an
idempotent Rota-Baxter renormalization map $R$ and the related
$\lambda-$information $\widetilde{\Phi}_{\lambda}$ ($\lambda \in
\mathbb{K}$). The Lie bracket $[.,.]_{\lambda}$ determines a Poisson
bracket in a natural way where lemma \ref{sym-pois} can induce a symplectic structure $\omega_{\lambda}$ but for this result we need the non-degeneracy of
this Poisson bracket such that in general maybe it does not happen.
According to the proof of this lemma and since for the
identification of integrals of motion just we should concentrate on
Hamiltonian vector fields (derivations), therefore for removal this
problem it is enough to work on

-  $Z(\widetilde{\Phi}_{\lambda})-$module
$Der_{Ham}(\widetilde{\Phi}_{\lambda})$ generated by the set
$Ham(\widetilde{\Phi}_{\lambda})$ (i.e. all Hamiltonian derivations of
the algebra $\widetilde{\Phi}_{\lambda}$) instead of the set of all
derivations $Der(\widetilde{\Phi}_{\lambda})$.

- And restrict
the differential graded algebra
$\Omega^{\bullet}_{Der}(\widetilde{\Phi}_{\lambda})$ into the
differential graded algebra
$\Omega^{\bullet}_{Der_{Ham}}(\widetilde{\Phi}_{\lambda})$ such that
$\Omega^{0}_{Der_{Ham}}(\widetilde{\Phi}_{\lambda})=
\widetilde{\Phi}_{\lambda}$ and
$\Omega_{Der_{Ham}}^{n}(\widetilde{\Phi}_{\lambda})$ is the space of
all $Z(\widetilde{\Phi}_{\lambda})-$multilinear antisymmetric
mappings from $Der_{Ham}(\widetilde{\Phi}_{\lambda})^{n}$ into
$\widetilde{\Phi}_{\lambda}$.

\begin{cor} \label{int.14}
\textbf{Symplectic forms related to the first class}. The differential form
$$
\omega_{\lambda}: Der_{Ham}(\widetilde{\Phi}_{\lambda}) \times
Der_{Ham}(\widetilde{\Phi}_{\lambda}) \longrightarrow
\widetilde{\Phi}_{\lambda}$$
in $\Omega^{2}_{Der_{Ham}}(\widetilde{\Phi}_{\lambda})$ given by
$$\omega_{\lambda}(\theta,\theta'):= \sum_{i,j} u_{i} \circ_{\lambda}
v_{j} \circ_{\lambda} [f_{i},h_{j}]_{\lambda}$$
such that $
\{f_{1},...,f_{m},h_{1},...,h_{n} \} \subset L(H,A), \
\{u_{1},...,u_{m},v_{1},...,v_{n}\} \subset
Z(\widetilde{\Phi}_{\lambda})$ and
$\theta= \sum_{i} u_{i} \circ_{\lambda} ham(f_{i}), \  \theta'=\sum_{j} v_{j} \circ_{\lambda} ham(h_{j})$
is a $Z(\widetilde{\Phi}_{\lambda})-$bilinear, antisymmetric,
non-degenerate and closed element (i.e. a symplectic structure) .
\end{cor}

At second if renormalization map does not have idempotent property or
in general, then one can concentrate on the universal Nijenhuis
tensor to obtain a Poisson bracket $[.,.]_{u}$. Maybe it has not
non-degeneracy and so for determining a symplectic
structure $\omega_{u}$, it is enough to focus on

- $Z(\widetilde{\Phi}_{u})-$module $Der_{Ham}(\widetilde{\Phi}_{u})$
generated by the set $Ham(\widetilde{\Phi}_{u})$.

- And restrict the differential graded algebra
$\Omega^{\bullet}_{Der}(\widetilde{\Phi}_{u})$ into the differential graded algebra
$\Omega^{\bullet}_{Der_{Ham}}(\widetilde{\Phi}_{u})$ such that
$\Omega^{0}_{Der_{Ham}}(\widetilde{\Phi}_{u})= \widetilde{\Phi}_{u}$
and $\Omega_{Der_{Ham}}^{n}(\widetilde{\Phi}_{u})$ is the space of
all $Z(\widetilde{\Phi}_{u})-$multilinear antisymmetric mappings
from $Der_{Ham}(\widetilde{\Phi}_{u})^{n}$ into
$\widetilde{\Phi}_{u}$.

\begin{cor} \label{int.15}
\textbf{Symplectic forms related to the second class}. The differential form
$$
\omega_{u}: Der_{Ham}(\widetilde{\Phi}_{u}) \times
Der_{Ham}(\widetilde{\Phi}_{u}) \longrightarrow
\widetilde{\Phi}_{u}$$
in
$\Omega^{2}_{Der_{Ham}}(\widetilde{\Phi}_{u})$ given by
$$\omega_{u}(\rho,\mu):= \sum_{i,j}
w_{i} \circ_{u} z_{j} \circ_{u} [g_{i},k_{j}]_{u}
$$
such that $ \{g_{1},...,g_{m},k_{1},...,k_{n} \} \subset
L(H,\overline{T}(A)), \ \{w_{1},...,w_{m},z_{1},...,z_{n}\} \subset
Z(\widetilde{\Phi}_{u})$ and
$\rho= \sum_{i} w_{i} \circ_{u} ham(g_{i}), \ \mu=\sum_{j} z_{j} \circ_{u} ham(k_{j})$
is a symplectic structure.
\end{cor}

\begin{prop}
The Connes-Kreimer algebraic perturbative renormalization (namely, the couple renormalization map and regularization algebra) determines a family of symplectic structures.
\end{prop}

Now it is convenient to use the terminology "{\it symplectic space}" for this new information obtained form the geometric studying of renormalization.

\begin{defn} \label{int.16}

Symplectic spaces related to the Hopf algebraic reconstruction of the perturbative renormalization in a renormalizable QFT are introduced by
$$
\Gamma_{\lambda}:= (\widetilde{\Phi}_{\lambda},
Der_{Ham}(\widetilde{\Phi}_{\lambda}), \omega_{\lambda}),$$
and
$$
\Gamma_{u}:=(\widetilde{\Phi}_{u},
Der_{Ham}(\widetilde{\Phi}_{u}),\omega_{u}).
$$
\end{defn}

\begin{lem} \label{int.16}
Suppose $\theta$ be a derivation in $Der_{Ham}(\widetilde{\Phi}_{\lambda})$
and $\rho$ be a derivation in $Der_{Ham}(\widetilde{\Phi}_{u})$. One can show that
$$
L_{\theta} \omega_{\lambda} =i_{\theta}\circ d_{\lambda}
\omega_{\lambda} + d_{\lambda} \circ
i_{\theta}\omega_{\lambda}=d_{\lambda}(i_{\theta} \omega_{\lambda})=
0
$$
and
$$
L_{\rho} \omega_{u} =i_{\rho}\circ d_{u} \omega_{u} + d_{u} \circ
i_{\rho}\omega_{u}=d_{u}(i_{\rho} \omega_{u})=0
$$
such that $d_{\lambda}$ ($d_{u}$) is the anti-derivation differential
operator for the differential graded algebra
$\Omega^{\bullet}_{Der_{Ham}}(\widetilde{\Phi}_{\lambda})$
($\Omega^{\bullet}_{Der_{Ham}}(\widetilde{\Phi}_{u})$).
This derivation is called {\it $\lambda-$symplectic} ({\it
$u-$symplectic}) vector field with respect to the symplectic
structure $\omega_{\lambda}$ ($\omega_{u}$).
\end{lem}

\begin{cor} \label{int.17}
Maps
$$
i_{\omega_{\lambda}}: Der_{Ham}(\widetilde{\Phi}_{\lambda})
\longrightarrow DR^{1}_{Der_{Ham}}(\widetilde{\Phi}_{\lambda}),$$
$$
\theta \longmapsto i_{\theta}\omega_{\lambda}
$$
and
$$
i_{\omega_{u}}: Der_{Ham}(\widetilde{\Phi}_{u}) \longrightarrow
DR^{1}_{Der_{Ham}}(\widetilde{\Phi}_{u}),$$
$$\rho \longmapsto i_{\rho}\omega_{u}
$$
are well defined, linear and bijection. It means that one can find a
bijection between closed one forms in
$DR^{1}_{Der_{Ham}}(\widetilde{\Phi}_{\lambda})$ (or
$DR^{1}_{Der_{Ham}}(\widetilde{\Phi}_{u})$) and $\lambda-$symplectic
(or $u-$symplectic) vector fields.
\end{cor}

\begin{proof}
For each $\lambda \in \mathbb{K}$, it is enough to know that
$\omega_{\lambda}$ ($\omega_{u}$) is non-degenerate and closed.
\end{proof}

\begin{defn} \label{int.18}
If $\theta^{\lambda}_{f}$ be the $\lambda-$symplectic vector field
associated with $d_{\lambda}f$ such that $f\in
\Omega^{0}_{Der_{Ham}}(\widetilde{\Phi}_{\lambda})=\widetilde{\Phi}_{\lambda}$. Then
a new Poisson bracket on $\widetilde{\Phi}_{\lambda}$ is defined by
$$
\{f,g\}_{\lambda}:=i_{\theta^{\lambda}_{f}}(d_{\lambda}g).
$$
It is called $\lambda-$Poisson bracket.
\end{defn}

\begin{cor} \label{pois-sym1}

(i) For each $\lambda \in \mathbb{K}$, there is a new Lie algebra
structure on $\widetilde{\Phi}_{\lambda}$ such that the map
$$
(\widetilde{\Phi}_{\lambda}, \{.,.\}_{\lambda}) \longrightarrow
(Der_{Ham}(\widetilde{\Phi}_{\lambda}),[.,.])
$$
is a Lie algebra homomorphism.

(ii) $\lambda-$Poisson bracket $\{.,.\}_{\lambda}$ is characterized with the symplectic
structure $\omega_{\lambda}$ (and therefore by the Lie brackets
$[.,.]_{\lambda}$). It means that
$$
\{f,g\}_{\lambda} = i_{\theta^{\lambda}_{f}}(d_{\lambda}g) =
i_{\theta^{\lambda}_{f}} i_{\theta^{\lambda}_{g}} \omega_{\lambda},
\   \ i_{[\theta^{\lambda}_{f},\theta^{\lambda}_{g}]}
\omega_{\lambda} =
d_{\lambda}i_{\theta^{\lambda}_{f}}(d_{\lambda}g)=d_{\lambda}
\{f,g\}_{\lambda}.
$$
\end{cor}

\begin{cor}
Derivation $\theta^{\lambda}_{\{f,g\}_{\lambda}}$ is the
unique $\lambda-$symplectic vector field with respect to
$d_{\lambda} \{f,g\}_{\lambda}$ and it means that
$\theta^{\lambda}_{\{f,g\}_{\lambda}}=[\theta^{\lambda}_{f},\theta^{\lambda}_{g}]$.
\end{cor}

There is a similar process for the symplectic space $\Gamma_{u}$ and one
can obtain a {\it $u-$Poisson bracket} $\{.,.\}_{u}$ on
$\widetilde{\Phi}_{u}$ induced with the symplectic structure
$\omega_{u}$ (and therefore by the Lie bracket $[.,.]_{u}$).

\begin{rem}
If the Poisson bracket $[.,.]_{\lambda}$ (or $[.,.]_{u}$) is
non-degenerate, then we can define symplectic structure
$\omega_{\lambda}$ (or $\omega_{u}$) on the whole space of
derivations of $\widetilde{\Phi}_{\lambda}$ (or
$\widetilde{\Phi}_{u}$) and therefore lemma \ref{sym-pois} shows that the
Poisson bracket $\{.,.\}_{\lambda}$ (or $\{.,.\}_{u}$) will be
coincide with $[.,.]_{\lambda}$ (or $[.,.]_{u}$). But in general, might
be they are not the same.
\end{rem}

\begin{center}
\textbf{Hamiltonian Systems}
\end{center}

In classical mechanics a Hamiltonian system consists of a symplectic manifold (as a configuration space) and a Hamiltonian operator such that a Poisson bracket can be inherited from this information. Then naturally, related motion integrals would be determined while as the result we will enable to identify integrable systems. Here we want to carry out the same procedure to obtain a new family of infinite Hamiltonian systems depended upon renormalizable perturbative theories.

\begin{defn} \label{25}
(i) For each $\lambda \in \mathbb{K}$ and a fixed $F \in L(H,A)$, the pair
$(\Gamma_{\lambda},F)$ is called {\it $\lambda-$Hamiltonian system}.

(ii) For $G \in L(H,\overline{T}(A))$, the pair $(\Gamma_{u},G)$ is
called {\it $u-$Hamiltonian system with respect to the theory $\Phi$
and the regularization scheme $A$}.

(iii) $F$ and $G$ are called {\it Hamiltonian functions}.
\end{defn}

With applying the natural exponential map on the space
$Der(\widetilde{\Phi}_{\lambda})$  (or $Der(\widetilde{\Phi}_{u})$)
given by the composition of derivations, we could characterize integrals
of motion.

\begin{prop} \label{lambda-integral} \label{uni-integral}

(i) Let $\Phi$ be a renormalizable theory with the idempotent
renormalization map $R$ and  $(\Gamma_{\lambda},F)$ be a
$\lambda-$Hamiltonian system with respect to it.

$$\{f,F\}_{\lambda}=0 \Longleftrightarrow $$

The map $f$ is constant along the
integral curves of $\theta^{\lambda}_{F}$ (i.e. 1-parameter flow
$\alpha^{\lambda}_{t}: t \longmapsto exp(t \theta^{\lambda}_{F})$ of
the automorphisms generated by $\theta^{\lambda}_{F}$).

The function $f$ is called {\it $\lambda-$integral of motion} of
this system.

(ii) Let $\Phi$ be a renormalizable theory with the regularization
scheme $A$  and $(\Gamma_{u},G)$ be a $u-$Hamiltonian system with
respect to it.

$$\{g,G\}_{u}=0 \Longleftrightarrow $$

The map $g$ is constant along the
integral curves of $\rho^{u}_{G}$ (i.e. 1-parameter flow
$\alpha^{u}_{t}: t \longmapsto exp(t \rho^{u}_{G})$ of the
automorphisms generated by $\rho^{u}_{G}$).

The function $g$ is called {\it $u-$integral of motion} for this
system.

\end{prop}

\begin{proof}
With using the {\it Cartan magic formula} \cite{C1}, we have
$$\frac{d}{dt} (\alpha^{\lambda}_{t})^{*}(f) = (\alpha^{\lambda}_{t})^{*} L_{\theta^{\lambda}_{F}} f = (\alpha^{\lambda}_{t})^{*} i_{\theta^{\lambda}_{F}} d_{\lambda}f$$
$$= (\alpha^{\lambda}_{t})^{*} i_{\theta^{\lambda}_{F}} i_{\theta^{\lambda}_{f}} \omega_{\lambda} = (\alpha^{\lambda}_{t})^{*} \omega_{\lambda}(\theta^{\lambda}_{f},\theta^{\lambda}_{F}) = (\alpha^{\lambda}_{t})^{*} \{f,F\}_{\lambda} = 0.$$
\end{proof}

And finally, the concept of integrability (based on the behavior of motion integrals) for this class of quantum Hamiltonian systems can be determined in an usual procedure. It means that

\begin{defn} \label{27}

(i) A $\lambda-$Hamiltonian system $(\Gamma_{\lambda},F)$ is called
{\it $(n,\lambda)-$integrable}, if there exist $n$ linearly
independent $\lambda-$integrals of motion $f_{1}=F,f_{2},...,f_{n}$
such that $\{f_{i},f_{j}\}_{\lambda}=0$.

(ii) A $u-$Hamiltonian system $(\Gamma_{u},G)$ is called {\it
$(n,u)-$integrable}, if there exist $n$  linearly independent
$u-$integrals of motion $g_{1}=G,g_{2},...,g_{n}$ such that
$\{g_{i},g_{j}\}_{u}=0$.

(iii) If a $\lambda-$Hamiltonian (or $u-$Hamiltonian) system has infinite
linearly independent $\lambda-$(or $u-$)integrals of motion, then
it is called infinite dimensional $\lambda-$integrable ( or
infinite dimensional $u-$integrable) system.
\end{defn}

Briefly speaking, we could provide the concept of integrability
of Hamiltonian systems in renormalizable QFTs with respect to regularization
algebras or idempotent renormalization schemes such that integrals of
motion are introduced in consistency of determined Nijenhuis operators. This process enables to reflect the dependency of this style of (integrable) Hamiltonian systems upon the perturbative renormalization.
More precisely, there is an interesting chance to check the strong compatibility of this approach with the Connes-Kreimer formalism and it can be done by the renormalization group. Actually, we will show that using Connes-Kreimer renormalization group can guide us to characterize some examples of integrable quantum Hamiltonian systems.


\section{{\it \textsf{Infinite integrable quantum Hamiltonian systems on the basis of the renormalization group}}}

We saw that how one can reach to a fundamental concept of Hamiltonian formalism from deformation of Connes-Kreimer convolution algebras. It is the place to show further the compatibility of this approach with the Hopf algebraic machinery in terms of the renormalization group. So in this part, we work on the BPHZ prescription to recognize motion integrals depended on characters of the renormalization Hopf algebra and also, we apply the Connes-Kreimer Birkhoff factorization to determine some conditions for components of this factorization on a Feynman rules character $\phi$ when they are motion integrals of $\phi$.

Roughly, here we consider the relation between defined
Nijenhuis type motion integrals and physical information underlying minimal
subtraction in dimensional regularization \cite{S12}. Particularly, we further concentrate on Hamiltonian
systems such that their Hamiltonian functions are dimensionally
regularized Feynman rules characters or elements of the Connes-Kreimer
renormalization group.

Starting with a renormalizable physical theory $\Phi$ together with the dimensionally regularized Feynman rules character $\phi$ underlying the renormalization prescription
$(A_{dr},R_{ms})$. We know that $R_{ms}$ is a Nijenhuis tensor
and so for each $\lambda \in \mathbb{K}$, the map $\Upsilon_{ms,
\lambda}$ on $L(H,A_{dr})$ has also this property such that one can
extend it to the Lie algebra level.
Since we need to study the behavior of the renormalization group, so just we focus on the case
$\lambda = 0$ and note that there are similar calculations for other values of $\lambda$. Consider the Rota-Baxter map
$\Upsilon_{ms}$ where for each $\phi \in L(H,A_{dr})$,
\begin{equation}
\Upsilon_{ms}(\phi):= R_{ms} \circ \phi.
\end{equation}
Induce
a new associative product and a Lie bracket on $L(H,A_{dr})$ given by

\begin{equation} \label{prod}
\phi_{1} \circ_{0} \phi_{2}:= R_{ms} \circ \phi_{1} * \phi_{2} +
\phi_{1} * R_{ms} \circ \phi_{2} - R_{ms} \circ (\phi_{1} *
\phi_{2}).
\end{equation}
\begin{equation} \label{lie0}
[\phi_{1},\phi_{2}]_{0}:=[R_{ms} \circ \phi_{1},\phi_{2}] +
[\phi_{1},R_{ms} \circ \phi_{2}] - R_{ms} \circ
([\phi_{1},\phi_{2}]).
\end{equation}
Equation (\ref{2bra}) shows that
\begin{equation} \label{3bra}
R_{ms} \circ [\phi_{1},\phi_{2}]_{0} = [R_{ms} \circ \phi_{1},
R_{ms} \circ \phi_{2}].
\end{equation}

\begin{defn}
The $0-$information based on the theory $\Phi$ and with respect to
the map $\Upsilon_{ms}$ is defined by
$$
(\widetilde{\Phi}_{0},[.,.]_{0}):=(L(H,A_{dr}), \circ_{0},
\Upsilon_{ms},[.,.]_{0}).
$$
\end{defn}

Corollary \ref{int.14} determines a symplectic structure for this $0-$information. For arbitrary derivations $ \theta =
\sum_{i} u_{i} \circ_{0} ham(f_{i})$ and $\theta' = \sum_{j} v_{j}
\circ_{0} ham(h_{j})$ in $Der_{Ham}(\widetilde{\Phi}_{0})$ where
$f_{1},...,f_{m}, h_{1},...,h_{n}$ are in $L(H,A_{dr})$ and
$u_{1},...,u_{m},v_{1},...,v_{n}$ are in $Z(\widetilde{\Phi}_{0})$, we have
\begin{equation}
\omega_{0}(\theta, \theta'):= \sum_{i,j} u_{i} \circ_{0} v_{j}
\circ_{0} [f_{i},h_{j}]_{0}.
\end{equation}

\begin{defn}
The symplectic space related to the theory $\Phi$ and the
renormalization map $R_{ms}$ is given by
$$
\Gamma_{0}:=(\widetilde{\Phi}_{0}, Der_{Ham}(\widetilde{\Phi}_{0}),
\omega_{0}).
$$
\end{defn}
By choosing a Hamiltonian function $F \in L(H,A_{dr})$, one can have
a $0-$Hamiltonian system $(\Gamma_{0},F)$ with respect to the
theory and from proposition \ref{lambda-integral}, a $0-$integral of
motion for this system is an element $f \in L(H,A_{dr})$ such that
\begin{equation}
\{f,F\}_{0}=0.
\end{equation}
Let $\theta_{F}^{0}, \theta_{f}^{0}$ be the
$0-$symplectic vector fields with respect to $d_{0}F, d_{0}f$. Therefore
\begin{gather}
i_{\theta_{F}^{0}} \omega_{0}= d_{0}F, \ \ i_{\theta_{f}^{0}} \omega_{0}= d_{0}f,
\end{gather}
\begin{equation}
\{f,F\}_{0}= i_{\theta_{F}^{0}} i_{\theta_{f}^{0}} \omega_{0} = \omega_{0}(\theta_{f}^{0},\theta_{F}^{0})=[f,F]_{0}=0.
\end{equation}

\begin{cor} \label{cond}
Let $f \in L(H,A_{dr})$ be a $0-$integral of motion for the system
$(\Gamma_{0},F)$. Then we have
$$R_{ms} \circ f * R_{ms} \circ F = R_{ms} \circ F * R_{ms} \circ f.$$
\end{cor}

\begin{proof}
By the formula (\ref{3bra}),
$$R_{ms} \circ [f,F]_{0} = [R_{ms} \circ f, R_{ms} \circ F] \Longrightarrow $$
$$ [R_{ms} \circ f, R_{ms} \circ F] = R_{ms}(0)=0 \Longrightarrow$$
$$R_{ms} \circ f * R_{ms} \circ F - R_{ms} \circ F * R_{ms} \circ f=0.$$
\end{proof}

\begin{cor}  \label{co-main}
The equation
$$R_{ms} \circ f * F - F * R_{ms} \circ f + f * R_{ms} \circ F -
R_{ms} \circ F * f - R_{ms} \circ (f*F) + R_{ms} \circ (F*f) = 0$$
determines a necessary and sufficient condition to
characterize $0-$integrals of motion for the $0-$Hamiltonian system $(\Gamma_{0},F)$ with respect to the minimal subtraction
scheme in dimensional regularization.
\end{cor}

\begin{proof}
It is proved by the equation (\ref{lie0}). Because it shows that for a
$0-$integral of motion $f$ we have
$$[f,F]_{0}=0 \ \Longleftrightarrow $$
$$[R_{ms} \circ f, F] + [f, R_{ms} \circ F] - R_{ms} \circ [f,F] = 0.$$
\end{proof}

If the Hamiltonian function of this system is the dimensionally
regularized Feynman rules character  $\phi$ of the theory, then one
can obtain interesting relations between the components of the
Birkhoff factorization of $\phi$ and $0-$integrals of motion of this
system.

\begin{cor} \label{main1}  \label{relation1}
For the dimensionally regularized Feynman rules character $\phi \in
L(H,A_{dr})$, let $f$ be a $0-$integral of motion for the
$0-$Hamiltonian system $(\Gamma_{0},\phi)$. Then for each Feynman
diagram $\Gamma \in ker \ \epsilon_{H}$, $f$ satisfies in the
equation
$$
\sum_{\gamma} R_{ms}(f(\gamma))R_{ms}(\phi(\frac{\Gamma}{\gamma})) =
\sum_{\gamma} R_{ms}(\phi(\gamma)) R_{ms}(f(\frac{\Gamma}{\gamma})).
$$
\end{cor}

\begin{proof}
For each $\Gamma \in ker \ \epsilon_{H}$, its coproduct is
given by
$$
\Delta(\Gamma) = \Gamma \otimes 1 + 1 \otimes \Gamma + \sum_{\gamma
\subset \Gamma} \gamma \otimes \frac{\Gamma}{\gamma}
$$
such that the sum is over all disjoint unions of 1PI superficially divergent
proper subgraphs. Let $f$ be a $0-$integral of motion of this
system. Renormalization coproduct on Feynman diagrams shows us that
$$(R_{ms} \circ f * R_{ms} \circ \phi) (\Gamma) = $$
$$
R_{ms}(f(1))R_{ms}(\phi(\Gamma)) + R_{ms}(f(\Gamma))R_{ms}(\phi(1))
+ \sum_{\gamma} R_{ms}(f(\gamma))
R_{ms}(\phi(\frac{\Gamma}{\gamma})),
$$
$$(R_{ms} \circ \phi * R_{ms} \circ f) (\Gamma) =$$
$$
R_{ms}(\phi(\Gamma))R_{ms}(f(1)) + R_{ms}(\phi(1))R_{ms}(f(\Gamma))
+ \sum_{\gamma} R_{ms}(\phi(\gamma))
R_{ms}(f(\frac{\Gamma}{\gamma})).
$$
Therefore with help of the corollary \ref{cond}, the mentioned formula should be obtained
such that the sum has finite terms where each term contains pole parts of
Laurent series.
\end{proof}

Let $f$ be a $0-$integral of motion for the system
$(\Gamma_{0},\phi)$ such that $\phi$ is the Feynman rules character
of the theory. Theorem \ref{bphz1} shows that for each primitive 1PI
(superficially divergent) proper subgraph $\gamma$ of the Feynman
diagram $\Gamma$,
\begin{equation} \label{prim}
\phi_{-}(\gamma)= - R_{ms}(\phi(\gamma)).
\end{equation}
Therefore with the help of corollary \ref{relation1} and (\ref{prim}), the
relations between components of the factorization of $\phi$ and this
$0-$integral of motion can be available. We have

$$\sum_{\gamma} R_{ms}(f(\gamma))R_{ms}(\phi(\frac{\Gamma}{\gamma})) = \sum_{\gamma}  -\phi_{-}(\gamma) R_{ms}(f(\frac{\Gamma}{\gamma}))$$
$$\Longrightarrow$$
$$\sum_{\gamma} R_{ms}(f(\gamma))R_{ms}(\phi(\frac{\Gamma}{\gamma}))R_{ms}(f(\frac{\Gamma}{\gamma}))^{-1} + \phi_{-}(\gamma)=0 $$
$$\Longrightarrow$$

\begin{equation} \label{relation5}
\sum_{\gamma} \phi_{-}(\gamma)= - \sum_{\gamma}
R_{ms}(f(\gamma))R_{ms}(\phi(\frac{\Gamma}{\gamma}))R_{ms}(f(\frac{\Gamma}{\gamma}))^{-1},
\end{equation}

\begin{equation} \label{relation6}
\sum_{\gamma} \phi_{+}(\gamma)= \sum_{\gamma} \phi(\gamma)-
\sum_{\gamma}
R_{ms}(f(\gamma))R_{ms}(\phi(\frac{\Gamma}{\gamma}))R_{ms}(f(\frac{\Gamma}{\gamma}))^{-1}.
\end{equation}

\begin{cor}
(i) With applying (\ref{relation5}) and (\ref{relation6}) in the
equations in theorem \ref{bphz1}, new representations
from the factorization components of a Feynman rules character $\phi$ (based on the $0-$integral of motion $f$) can be determined.

(ii) With putting them in (\ref{sca2}) and with the
help of lemma \ref{sca1} and corollary
\ref{main1}, relations between this $0-$integral of
motion and renormalization group and also $\beta-$function of the
theory can be investigated.
\end{cor}

Now we have a chance to search more geometrical meanings in the Feynman rules character $\phi$ and it can be performed based on motion integral condition for components of Birkhoff factorization of this character. It is necessary to emphasize that since the renormalization method is fixed and the factorization is unique, therefore consideration of this possibility helps us to familiar with some more hidden geometrical structures in a quantum field theory which provide the required integral conditions.

\begin{cor}
Suppose the negative part $\phi_{-}$ of the Birkhoff decomposition of
the dimensionally regularized  Feynman rules character $\phi \in
L(H,A_{dr})$ of a given theory $\Phi$ be a $0-$integral of motion for
the $0-$Hamiltonian system $(\Gamma_{0},\phi)$. Then for each
Feynman diagram $\Gamma$, we have
$$R_{ms} (\sum_{\gamma} \sum_{\gamma' \subset \Gamma_{\gamma}}
\phi_{-}(\gamma') \phi(\frac{\Gamma_{\gamma}}{\gamma'})) = 0.$$
\end{cor}

\begin{proof}
If $\phi_{-}$ be a $0-$integral of motion, then by given conditions in corollary
\ref{relation1} for each Feynman diagram $\Gamma$ one should have
$$\sum_{\gamma} R_{ms}(\phi_{-}(\gamma))R_{ms}(\phi(\frac{\Gamma}{\gamma})) = \sum_{\gamma} R_{ms}(\phi(\gamma))R_{ms}(\phi_{-}(\frac{\Gamma}{\gamma})).$$
By (\ref{prim}) and since $R_{ms}$ is an idempotent linear map, it can be seen that

$$\sum_{\gamma} -R_{ms}(R_{ms}(\phi(\gamma)))R_{ms}(\phi(\frac{\Gamma}{\gamma}))= $$
$$\sum_{\gamma} -R_{ms}(\phi(\gamma))R_{ms}(\phi(\frac{\Gamma}{\gamma})) = \sum_{\gamma} R_{ms}(\phi(\gamma))R_{ms}(\phi_{-}(\frac{\Gamma}{\gamma}))$$
$$
\Longrightarrow  \sum_{\gamma} -R_{ms}(\phi(\frac{\Gamma}{\gamma}))
= \sum_{\gamma} R_{ms}(\phi_{-}(\frac{\Gamma}{\gamma})) =
\sum_{\gamma} \phi_{-}(\frac{\Gamma}{\gamma}).
$$
Set $\Gamma_{\gamma}:= \frac{\Gamma}{\gamma}$. By theorem
\ref{bphz1},
$$\sum_{\gamma} \phi_{-}(\Gamma_{\gamma}) = -R_{ms} \sum_{\gamma} (\phi(\Gamma_{\gamma}) + \sum_{\gamma' \subset \Gamma_{\gamma}} \phi_{-}(\gamma') \phi(\frac{\Gamma_{\gamma}}{\gamma'}))$$
$$ \Longrightarrow  \sum_{\gamma} R_{ms} (\phi(\Gamma_{\gamma}) + \sum_{\gamma' \subset \Gamma_{\gamma}} \phi_{-}(\gamma') \phi(\frac{\Gamma_{\gamma}}{\gamma'})) =  \sum_{\gamma} R_{ms}(\phi(\Gamma_{\gamma}))$$
such that the sum is over all unions of 1PI proper superficially
divergent subgraphs of $\Gamma_{\gamma}$.
\end{proof}

With the help of theorems \ref{bphz2}, \ref{bphz1} and the equation
(\ref{co-main}) and since $R_{ms}$ is an idempotent map, one can
obtain more explicitly conditions for the components of decomposition of
$\phi$ which play the role of $0-$integrals of motion for the system
$(\Gamma_{0},\phi)$.

\begin{prop} \label{negative-positive}
For the $0-$Hamiltonian system $(\Gamma_{0},\phi)$ such that $\phi$
is the dimensionally regularized Feynman rules character (in
$L(H,A_{dr})$) of the theory $\Phi$, the components of the Birkhoff
factorization of $\phi$ are $0-$integrals of motion for the system if and only if

(i) $\phi_{-}$ satisfies in the equation
$$\phi_{+} - \phi * \phi_{-} + \phi_{-} * R_{ms} \circ \phi - R_{ms}
\circ \phi * \phi_{-}+ R_{ms} \circ (\phi * \phi_{-}) = 0.$$

(ii) $\phi_{+}$ satisfies in the equation
$$\phi_{+} * R_{ms} \circ \phi - R_{ms} \circ \phi * \phi_{+} - R_{ms}
\circ (\phi_{+} * \phi) + R_{ms} \circ (\phi * \phi_{+}) = 0.$$
\end{prop}

\begin{proof}
For the negative part $\phi_{-}$ we have
$$R_{ms} \circ \phi_{-} * \phi - \phi * R_{ms} \circ \phi_{-} + \phi_{-} * R_{ms} \circ \phi - R_{ms} \circ \phi * \phi_{-} - R_{ms} \circ (\phi_{-} * \phi) + R_{ms} \circ (\phi * \phi_{-}) = 0$$
$$ \Longleftrightarrow $$
$$ \phi_{-} * \phi - \phi * \phi_{-} + \phi_{-} * R_{ms} \circ \phi - R_{ms} \circ \phi * \phi_{-} - R_{ms} \circ (\phi_{-} * \phi) + R_{ms} \circ (\phi * \phi_{-}) = 0, \  \  \phi = \phi_{-}^{-1} * \phi_{+} $$
$$ \Longleftrightarrow $$
$$ \phi_{+} - \phi * \phi_{-} + \phi_{-} * R_{ms} \circ \phi - R_{ms} \circ \phi * \phi_{-} - R_{ms} \circ \phi_{+} +  R_{ms} \circ (\phi * \phi_{-}) = 0.$$
For the positive part $\phi_{+}$ we have
$$R_{ms} \circ \phi_{+} * \phi - \phi * R_{ms} \circ \phi_{+} + \phi_{+} * R_{ms} \circ \phi - R_{ms} \circ \phi * \phi_{+} - R_{ms} \circ (\phi_{+} * \phi) + R_{ms} \circ (\phi * \phi_{+}) = 0.$$
Now since $R_{ms} \circ \phi_{+} = 0,$ the proof is completed.
\end{proof}

It is good place to consider the behavior of the
renormalization group with respect to the $0-$Hamiltonian system
$(\Gamma_{0},\phi)$. With help of this group (associated
to the Feynman rules character $\phi$), we are going to introduce an
infinite dimensional $0-$integrable system with respect to the
theory $\Phi$.

\begin{prop} \label{renn}
For each $t$, an element $F_{t}$ of the renormalization group is a
$0-$integral of motion for the $0-$Hamiltonian system
$(\Gamma_{0},\phi)$ if and only if it satisfies in
$$F_{t} * R_{ms} \circ \phi - R_{ms} \circ \phi * F_{t} - R_{ms} \circ
(F_{t} * \phi) + R_{ms} \circ (\phi * F_{t}) = 0.$$
\end{prop}

\begin{proof}
Since for each $t$ and Feynman diagram $\Gamma$, $F_{t}(\Gamma)$ is
a polynomial in $t$, therefore $R_{ms}(F_{t}(\Gamma)) = 0$. With
notice to (\ref{co-main}), we have
$$R_{ms} \circ F_{t} * \phi - \phi * R_{ms} \circ F_{t} + F_{t} * R_{ms} \circ \phi - R_{ms} \circ \phi * F_{t} - R_{ms} \circ (F_{t} * \phi) + R_{ms} \circ (\phi * F_{t}) = 0.$$
\end{proof}

For
arbitrary elements $F_{t}, \ F_{s}$ of the renormalization group of
each given renormalizable theory $\Phi$, one can observe that
\begin{equation}
 \{F_{t},F_{s}\}_{0} = [R_{ms} \circ F_{t} , F_{s}] + [F_{t}, R_{ms} \circ F_{s}] - R_{ms} \circ ([F_{t},F_{s}]).
\end{equation}
Since the renormalization group is a 1-parameter subgroup of
$G(\mathbb{C})$ (i.e. $F_{t}*F_{s}=F_{t+s}$), therefore it is easy to
see that
\begin{equation}
\{F_{t},F_{s}\}_{0} = 0.
\end{equation}
It shows that the
renormalization group can give us an integrable system and this fact turns out a strong relation between (the given
Nijenhuis type) integrals of motion and the Connes-Kreimer renormalization group.

\begin{cor}
For the renormalizable physical theory $\Phi$, let $\phi \in
G(A_{dr})$ be its dimensionally regularized Feynman rules character
and $\{F_{t}\}_{t}$ be the renormalization group with respect to
this character. For each arbitrary element $F_{t_{0}}$ of the
renormalization group, $0-$Hamiltonain system $(\Gamma_{0},
F_{t_{0}})$ is an infinite dimensional $0-$integrable system.
\end{cor}

\begin{proof}
The renormalization group contains infinite linearly independent
$0-$integrals of motion $F_{t}$ for the system $(\Gamma_{0},
F_{t_{0}})$.
\end{proof}


\section{{\it \textsf{Rosenberg's strategy: The continuation of the standard process}}}

In \cite{BR1} the authors (by working on the Lie group of diffeographisms and its related Lie algebra of infinitesimal characters) improve the study of infinite dimensional Lie algebras and factorization problem to the level of the Connes-Kreimer theory. In this part, we focus on their strategy and apply their results to consider factorization problem on the previously introduced noncommutative algebras (deformed by Nijenhuis maps) such that consequently, a new family of integrals of motion will be determined in a natural manner \cite{S11, S10}.

The Lie brackets $[.,.]_{\lambda}$ make available another procedure to study
integrable systems at this level namely, identifying equations of motion and integral curves
from a Lax pair equation. Set $C^{dr}_{\lambda}:= (L(H,A_{dr}), \circ_{\lambda})$ and supposing $\mathcal{C}_{\lambda} = \mathcal{C}^{dr}_{\lambda}
\oplus \mathcal{C}^{dr*}_{\lambda}$ be its related semisimple trivial Lie bialgebra such that
$\mathcal{C}^{dr}_{\lambda}:= (C^{dr}_{\lambda},[.,.]_{\lambda} )$.

\begin{cor}
$\mathcal{C}_{\lambda}$ is the associated Lie
algebra of the Lie group $\widetilde{\mathcal{C}}_{\lambda}:=
\mathcal{C}^{dr}_{\lambda} \rtimes_{\sigma}
\mathcal{C}^{dr*}_{\lambda}$ such that
$$
\sigma:\mathcal{C}^{dr}_{\lambda} \times \mathcal{C}^{dr*}_{\lambda}
\longrightarrow \mathcal{C}^{dr*}_{\lambda}, \ \ (f,X) \longmapsto
Ad^{*}(f)(X).
$$
\end{cor}

\begin{proof}
It is directly obtained from \cite{BR1}.
\end{proof}

\begin{defn}
The loop algebra of $\mathcal{C}_{\lambda}$ is defined by the
set
$$
L \mathcal{C}_{\lambda}:= \{ F(c)= \sum_{j=- \infty}^{\infty}
c^{j} F_{j}, F_{j} \in \mathcal{C}_{\lambda}\}
$$
such that naturally, one can define the Lie bracket
$$
[\sum c^{i} F_{i},\sum c^{j} G_{j}]:= \sum_{k} c^{k} \sum_{i+j=k}
[F_{i},G_{j}]_{\lambda}
$$
on it.
\end{defn}

Decompose this set of formal power series into two parts
\begin{gather}
L \mathcal{C}_{\lambda, +}= \{\sum_{j=0}^{\infty} c^{j} F_{j}\},
\ \ L \mathcal{C}_{\lambda, -} = \{\sum_{j=-\infty}^{-1} c^{j}
F_{j}\}
\end{gather}
and let $P_{\pm}$ are the natural projections on these components
where $P:=P_{+}-P_{-}$.

\begin{cor}
For a given Casimir
function $\upsilon$ on $L \mathcal{C}_{\lambda}$, integral
curve $\Lambda(t)$ of the Lax pair equation $\frac{d\Lambda}{dt}=[M,\Lambda]$ where for $F(\lambda)=F(0)(\lambda) \in L \mathcal{C}_{\lambda}$,
$M=\frac{1}{2}P(I(d \upsilon (F(c)))) \in L \mathcal{C}_{\lambda}$
is given by
$$
\Lambda(t)= Ad^{*}_{L \widetilde{\mathcal{C}}_{\lambda}}
\gamma_{\pm}(t).\Lambda(0)
$$
such that the smooth curves $\gamma_{\pm}$ are the answers of the Birkhoff
factorization
$$
exp(-tX)=\gamma_{-}^{-1}(t)\gamma_{+}(t)
$$
where $X=I(d \upsilon(F(c))) \in L \mathcal{C}_{\lambda}$.
\end{cor}

\begin{proof}
It is easily calculated from \cite{BR1, S1}.
\end{proof}

\begin{rem}
(i) It is
important to know that one can project the above Lax pair equation
to an equation on loop algebra of the original Lie algebra
$\mathcal{C}^{dr}_{\lambda}$.

(ii) Based on the induced Nijenhuis type symplectic structures, one can introduce a symplectic space on the loop algebra $L \mathcal{C}_{\lambda}$. Next, depended motion integrals and therefore integrable Hamiltonian systems on this loop algebra can be determined.
\end{rem}


\section{{\it \textsf{Fixed point equations}}}

We observed that the Connes-Kreimer recalculation of the BPHZ
renormalization depends strongly on components of the Birkhoff factorization of dimensionally regularized Feynman rules characters and on the other hand, the existence of this factorization is supported originally by Atkinson's theorem \cite{G3}.

\begin{thm}
Let $A$ be an associative algebra over the field $\mathbb{K}$ and $R: A \longrightarrow A$ be a linear map. The pair $(A,R)$ is a Rota-Baxter algebra if and only if the algebra $A$ has a Birkhoff factorization. It means that there is a Cartesian product
$$(R(A),\widetilde{R}(A)) \subset A \times A$$
such that

- It is a subalgebra of $A \times A$,

- Each $x \in A$ admits a unique decomposition $x = R(x) \oplus \widetilde{R}(x)$. \cite{A2, GM1}
\end{thm}

This theorem can determine explicit inductive formulaes for these components of the Birkhoff factorization of
characters on the Connes-Kreimer Hopf algebra of rooted trees such that at this level, the
notion of a decomposition for Lie algebras will be available. With attention to the BPHZ prescription, in this part we are going to apply these recursive equations to introduce a new family of fixed point
equations related to the motion integral condition on components of a Feynman rules character. We will show that this type of equations are formalized by Bogoliubov character and BCH series. Furthermore, the behavior of $\beta-$function and renormalization group underlying these equations will be considered \cite{S11, S10}.
It should be important to note that because of the universality of the Connes-Kreimer Hopf algebra (with respect to the Hochschild cohomology theory), the study of this family of fixed point equations at this level can be lifted to other Hopf algebras of renormalizable theories.

\begin{lem}
Consider the group $char_{A_{dr}} \ H_{x}$ ($x=lrt,rt$) of characters on ladder trees (rooted trees) with its related Lie algebra $\partial \
char_{A_{dr}} \ H_{x}$.
This Lie algebra is generated by derivations
$Z^{t}$ indexed by ladder tree (rooted tree) $t$ and defined by the
natural paring
$$<Z^{t},s> = \delta_{t,s}.$$
\end{lem}

Based on
the bijection
between $char_{A_{dr}} \ H_{x}$ and $\partial \ char_{A_{dr}} \
H_{x}$, for each character $g$ with the corresponding
derivation $Z_{g}$, we have
\begin{equation} \label{exp-map}
g=exp^{*}(Z_{g}) = \sum_{n=0}^{\infty} \frac{Z_{g}^{*n}}{n!}.
\end{equation}

\begin{lem} \label{dec-partial1}
(i) The Lie algebra $\partial \ char_{A_{dr}} \ H_{x}$
together with the map $\mathcal{R}: f \longmapsto R_{ms} \circ f$ define a Lie Rota-Baxter algebra.

(ii) Each derivation $Z$ of this Lie algebra has a unique
decomposition
$$Z = \mathcal{R} (Z) \oplus \widetilde{\mathcal{R}} (Z).$$

(iii) Idempotent property of $R$ provides a
decomposition $A_{dr} = A^{+}_{dr} \oplus A^{-}_{dr}$ such that it can be
extended to $\partial \ char_{A_{dr}} \ H_{x}$ and it means that
$$
\partial \ char_{A_{dr}} \ H_{x} = (\partial \ char_{A_{dr}} \ H_{x})_{+} \oplus (\partial \ char_{A_{dr}} \
H_{x})_{-}
$$
where
$$(\partial \ char_{A_{dr}} \ H_{x})_{+}:= \widetilde{\mathcal{R}}
(\partial \ char_{A_{dr}} \ H_{x}), \ \ (\partial \ char_{A_{dr}} \
H_{x})_{-}:= \mathcal{R}(\partial \ char_{A_{dr}} \ H_{x})
$$
\cite{EG2, EGK1, EGK2}.
\end{lem}

For a fixed character $g \in C^{x}_{\lambda}$, let $f$ be its
integral of motion. Product $\circ_{\lambda}$, its related $\lambda-$Poisson bracket and motion integral condition show that
\begin{equation} \label{int.ren.7}
\{f,g\}_{\lambda}= i_{\theta_{g}^{\lambda}} i_{\theta_{f}^{\lambda}}
\omega_{\lambda} =
\omega_{\lambda}(\theta_{f}^{\lambda},\theta_{g}^{\lambda})=[f,g]_{\lambda}=0.
\end{equation}
And so it is apparently observed that
$$[\mathcal{R}_{\lambda}(f), g] + [f, \mathcal{R}_{\lambda}(g)] - \mathcal{R}_{\lambda}([f,g]) = 0 \ \Longleftrightarrow $$
\begin{equation} \label{int.ren.8}
\mathcal{R}_{\lambda}(f) * g - g * \mathcal{R}_{\lambda}(f) + f *
\mathcal{R}_{\lambda}(g) - \mathcal{R}_{\lambda}(g) * f -
\mathcal{R}_{\lambda}(f*g) + \mathcal{R}_{\lambda}(g*f) = 0.
\end{equation}

\begin{prop} \label{int.ren.50}
For the given character $g \in C^{x}_{\lambda}$ with the Birkhoff
factorization $(g_{-},g_{+})$,

(i) $ g = g_{-}^{-1} * g_{+}, \ \ R_{ms} \circ g_{-} = g_{-}, \ \ R_{ms} \circ g_{+} = 0.$

(ii) The negative component $g_{-}$ is an integral of motion for $g$
iff
$$ g_{+} - g * g_{-} + (1+ \lambda)g_{-} * \mathcal{R}(g) - (1 + \lambda) \mathcal{R}(g) * g_{-} + (1 + \lambda) \mathcal{R}(g*g_{-})=0.$$

(iii) The positive component $g_{+}$ is an integral of motion for $g$
iff
$$ - \lambda g_{+} * g + \lambda g * g_{+} + (1+ \lambda) g_{+} * \mathcal{R}(g) - (1+ \lambda) \mathcal{R}(g) * g_{+}
- (1+ \lambda) \mathcal{R}(g_{+}*g)+(1+ \lambda)
\mathcal{R}(g*g_{+})=0.$$
\end{prop}

Atkinson theorem determines very interesting recursive representation from components of decomposition of characters on rooted trees such that because of its practical structure in calculating physical information, this representation is called {\it tree renormalization}.

\begin{thm} \label{int.ren.1} \label{int.ren.2}  \label{int.ren.4} \label{int.ren.5}

(i) \textbf{Ladder tree renormalization.} Each arbitrary character $\phi \in
char_{A_{dr}} \ H_{lrt}$ has a unique Birkhoff factorization $(\phi_{-}^{-1},\phi_{+})$ such that
$$
\phi = exp^{*} (\mathcal{R}(Z_{\phi}) +
\widetilde{\mathcal{R}}(Z_{\phi})) = exp^{*} (\mathcal{R}(Z_{\phi}))
* exp^{*} (\widetilde{\mathcal{R}}(Z_{\phi})),
$$
$$
\phi_{-} = exp^{*} (- \mathcal{R}(Z_{\phi})), \ \ \phi_{+}= exp^{*} (\widetilde{\mathcal{R}}(Z_{\phi})).
$$

(ii) \textbf{Rooted tree renormalization.} Each arbitrary character
$\psi \in char_{A_{dr}} \ H_{rt}$ has a unique Birkhoff factorization
$(\psi_{-}^{-1},\psi)$ such that
$$
\psi = exp^{*}(Z_{\psi}) = exp^{*}(\mathcal{R}(\chi (Z_{\psi}))) *
exp^{*} (\widetilde{\mathcal{R}}(\chi (Z_{\psi}))),
$$
$$
\psi_{-} = exp^{*} (- \mathcal{R}(\chi(Z_{\psi}))), \ \ \psi_{+}= exp^{*} (\widetilde{\mathcal{R}}(\chi(Z_{\psi})))
$$
where the infinitesimal character $\chi$ is characterized with the
BCH series. \cite{EG2, EGK1, EGK2, GM1}

\end{thm}

Now with applying representations given in the theorem \ref{int.ren.2} and conditions given in the proposition \ref{int.ren.50}, one can obtain new equations at the level of Lie algebra for while these components
are motion integrals of a given character in the algebra $C_{\lambda}^{x}$.

In addition, there is another procedure to deform the initial product (based on double Rota-Baxter structures) such that this new deformed product can induce another representation from components.

\begin{defn} \label{new-prod}
Rota-Baxter map $\mathcal{R}$ deforms the convolution product $*$ to define
a well known associative product on the set $L(H_{x},A_{dr})$ given by
$$
f *_{\mathcal{R}} g:= f * \mathcal{R}(g) + \mathcal{R}(f) * g - f * g.
$$
\end{defn}

\begin{lem} \label{new-lie} \label{new-rep}

(i) Information $C^{x}_{\mathcal{R}}:=(L(H_{x},A_{dr}),*_{\mathcal{R}}, \mathcal{R})$ is a Rota-Baxter algebra with the corresponding $\mathcal{R}-$bracket
$$
[f,g]_{\mathcal{R}} = [f,\mathcal{R}(g)] + [\mathcal{R}(f),g] - [f,g].
$$

(ii) For each infinitesimal character $Z$, it can be seen that
$$
exp^{*} (\mathcal{R}(Z)) = \mathcal{R}
(exp^{*_{\mathcal{R}}}(Z)), \ \
exp^{*}(\widetilde{\mathcal{R}}(Z)) = - \widetilde{\mathcal{R}}
(exp^{*_{\mathcal{R}}}(-Z)).
$$
\end{lem}

\begin{cor} \label{new-eq}
For the given character $g \in C^{x}_{\mathcal{R}}$ with the Birkhoff
factorization $(g_{-},g_{+})$, the components are integrals of motion for $g$ iff

(i) $[g_{-}, \mathcal{R}(g)] = 0$,

(ii) $[g_{+},\mathcal{R}(g)] - [g_{+},g] = 0$,
respectively.
\end{cor}

\begin{proof}
It is proved by theorem \ref{int.ren.7}, definition \ref{new-prod} and lemma \ref{new-lie}.
\end{proof}

\begin{cor} \label{fix-point1}  \label{fix-point2}
Lie algebra version of the given equations in the corollary \ref{new-eq} are reformulated by

$$
(i) \ \mathcal{R}(exp^{*_{\mathcal{R}}} (-
\chi(Z_{\psi}))) * \mathcal{R}(exp^{*}(Z_{\psi})) - \mathcal{R}(exp^{*}(Z_{\psi})) * \mathcal{R}(exp^{*_{\mathcal{R}}} (-
\chi(Z_{\psi}))) = 0,
$$
$$ (ii)  - \widetilde{\mathcal{R}} (exp^{*_{\mathcal{R}}} (-
\chi(Z_{\psi}))) * \mathcal{R}(exp^{*}(Z_{\psi})) + \mathcal{R}(exp^{*}(Z_{\psi})) * \widetilde{\mathcal{R}} (exp^{*_{\mathcal{R}}} (-
\chi(Z_{\psi})))$$
$$
+ \widetilde{\mathcal{R}} (exp^{*_{\mathcal{R}}} (-
\chi(Z_{\psi}))) *  exp^{*}(Z_{\psi}) - exp^{*}(Z_{\psi}) * \widetilde{\mathcal{R}} (exp^{*_{\mathcal{R}}} (-
\chi(Z_{\psi}))) = 0.
$$
\end{cor}

Now we want to relate a family of fixed point equations to the equations induced in the corollary \ref{fix-point1} and because of that we need a new operator.

\begin{defn}
The morphism
$$b[\psi]: = exp^{*_{\mathcal{R}}}(-\chi(Z_{\psi}))$$
is called Bogoliubov character.
\end{defn}

\begin{lem} \label{int.ren.23}  \label{int.ren.24}
One can approximate Bogoliubov character with
the formula
$$
\mathcal{R}(b[\psi]) = -R_{ms} \circ \{exp^{*}(Z_{\psi}) +
\alpha_{\psi}\}
$$
such that
$$
\alpha_{\psi}:= \sum_{n \ge 0} \frac{1}{n!} \sum_{j=1}^{n-1}
\frac{n!}{j!(n-j)!} \mathcal{R}(-
\chi(Z_{\psi}))^{*(n-j)}*Z_{\psi}^{*j}
.$$
\cite{EGK2, GM1}
\end{lem}

\begin{cor} \label{fix-point3}   \label{fix-point4}
One can rewrite equations in the corollary \ref{fix-point1} based on the given estimation in the lemma \ref{int.ren.23}. We have
$$
(i)' \  \   - \mathcal{R}(\psi + \alpha_{\psi}) * \mathcal{R}(exp^{*}(Z_{\psi})) + \mathcal{R}(exp^{*}(Z_{\psi})) * \mathcal{R}(\psi + \alpha_{\psi}) = 0
$$
and
$$ (ii)' \ \ \widetilde{\mathcal{R}} (\psi + \alpha_{\psi}) * \mathcal{R}(exp^{*}(Z_{\psi})) - \mathcal{R}(exp^{*}(Z_{\psi})) * \widetilde{\mathcal{R}} (\psi + \alpha_{\psi})- \widetilde{\mathcal{R}} (\psi + \alpha_{\psi}) *  exp^{*}(Z_{\psi}) $$
$$
+ exp^{*}(Z_{\psi}) * \widetilde{\mathcal{R}} (\psi + \alpha_{\psi}) = 0.
$$
\end{cor}


Furthermore, we saw that Birkhoff factorization of characters of
the Connes-Kreimer Hopf algebra are identified with the special
infinitesimal character $\chi$ such that for each infinitesimal
character $Z \in \partial char_{A_{dr}} \ H_{rt}$, it is given by
\begin{equation}
\chi(Z)=Z+
\sum_{k=1}^{\infty} \chi_{Z}^{(k)}.
\end{equation}
The sum is a finite linear
combination of infinitesimal characters and $\chi_{Z}^{(k)}$s are
determined by solution of the fixed point equation
\begin{equation}
E: \ \ \chi(Z) = Z - \sum_{k=1}^{\infty}
c_{k}K^{(k)}(\mathcal{R}(\chi(Z)),\widetilde{\mathcal{R}}(\chi(Z)))
\end{equation}
such that terms $K^{(k)}$s are calculated with the BCH series \cite{EG1, EG2, EGK1, EGK2}. With putting the equation $E$
in the Bogoliubov character and then applying the result \ref{fix-point3}, one can reformulate the motion integral condition for components of a given character with respect to the fixed point equation $E$.

\begin{prop}
For a given Feynman rules character $\psi \in C_{\mathcal{R}}^{rt}$, if each of its Birkhoff factorization's components is an integral of motion for $\psi$, then a class of fixed point equations
will be determined.
\end{prop}


Probably discussion about renormalization group and its infinitesimal generator (i.e. beta function) can be interested at this level.
We saw that these physical information are defined based on
the grading operator $Y$ (that providing the scaling evolution of
the coupling constant) such that this element exists from the extension
of the Lie algebra $\partial \ char_{A_{dr}} H_{rt}$ by an element
$Z^{0}$ where for each rooted tree $t$, we have
\begin{equation}
[Z^{0},Z^{t}] =
Y(Z^{t}) = |t|Z^{t}.
\end{equation}

\begin{lem} \label{int.ren.42}
For each character $\psi \in char_{A_{dr}}
H_{rt}$, its related $\beta-$function is given by
$$\beta (\psi) = \psi_{-} * [Z^{0},\psi_{-}^{-1}] = \psi_{-} * Z^{0} *
\psi_{-}^{-1} - Z^{0}.$$
\cite{EGK2}
\end{lem}

It is obvious that with applying the exponential map, its related renormalization group
is determined by
\begin{equation}
F_{t} = exp^{*}(t \beta).
\end{equation}
On the other hand, we know that for each $t \in \mathbb{R}$, $F_{t}$ is a character given by a polynomial of the
variable $t$ and therefore
\begin{equation}
R_{ms} \circ F_{t}=0
\end{equation}
(\cite{CM1, EGK1,
S2}). Proposition \ref{int.ren.8} shows that

\begin{cor} \label{int.ren.41}  \label{int.ren.42}
Each element of the
renormalization group plays the role of an integral of motion for
$\psi$ in the algebras $C^{x}_{\lambda}$ and $C^{x}_{\mathcal{R}}$ if and only if
$$
-\lambda [F_{t},\psi]+[F_{t},\mathcal{R}_{\lambda}(\psi)] -
\mathcal{R}_{\lambda}([\psi,F_{t}])= 0,
$$
$$
[F_{t},\mathcal{R}(\psi)] - [F_{t},\psi] = 0,
$$
respectively.
\end{cor}

The second condition in corollary \ref{int.ren.42} introduces a fixed point equation depended on
the Feynman rules character $\psi$ and its related $\beta-$function.

\begin{cor} \label{100}
For a given Feynman rules character $\psi \in
C_{\mathcal{R}}^{rt}$, an element $F_{t}$ of the related
renormalization group plays the role of integral of motion for $\psi
$ iff the $\beta-$function satisfies in the fixed point equation
$$ [exp^{*}(t \beta),\mathcal{R} \{exp^{*}(\mathcal{R}(E))*exp^{*}(\widetilde{\mathcal{R}}(E))\}] - [exp^{*}(t \beta),exp^{*}(\mathcal{R}(E))*exp^{*}(\widetilde{\mathcal{R}}(E))]= 0.$$
\end{cor}

And finally, it should be remarked that
because of the one parameter property of the renormalization group (i.e.
$F_{t}*F_{s}=F_{t+s}$), one can easily show that for a fixed $t_{0}$
in the cases $C^{x}_{0}$ and $C^{x}_{\mathcal{R}}$, each
$F_{t}$ is an integral of motion for $F_{t_{0}}$. So it is reasonable to expect infinite integrable systems related with algebra $C^{x}_{\mathcal{R}}$.


\chapter{\textsf{Connes-Marcolli approach}}

The Connes-Kreimer conceptional interpretation of renormalization theory could provide a new Hopf algebraic reconstruction from physical information. In continuation of this approach, Connes and Marcolli initiated a new categorical framework based on geometric objects to describe renormalizable physical theories and in fact, they showed that there is a fundamental mathematical construction hidden inside of divergences. According to their programme, the BPHZ renormalization (i.e. minimal subtraction scheme in dimensional
regularization) determines a principal bundle such that solutions of classes of differential equations related to some particular flat connections on this bundle can encode counterterms. Then with collecting all of these connections into a category, they introduced a new categorical formalism to consider physical theories such that it leaded to a universal treatment with respect to the Connes-Kreimer theory. \cite{CM5, CM2, CM4, CM3, CM1}

Universal singular frame is evidently one of the most important foundations of this machinery
where it contains a deep physical meaning in the sense that with working at this level, all of the divergences can be disappeared and so one can provide a finite theory. Moreover, one can find significance relations between this special frame and noncommutative geometry based on the local index theorem. \cite{CM5, CM3}

Nowadays a new theory of mixed Tate motives is developed on the basis of the Connes-Marcolli formalism which it reports about very desirable connections between theory of motives and quantum field theory. \cite{CM5}

In this chapter, we want to consider the basic elements of this categorical geometric foundation in the study of perturbative renormalization theory.


\section{{\it \textsf{Geometric nature of counterterms: Category of flat equi-singular connections}}}

Reformulation of the Birkhoff decomposition
in terms of classes of differential equations determines
some geometric objects with respect to physical information of
renormalizable theories. In fact, negative parts of
decomposition of loops (or characters) can be applied to correct the
behavior of solutions of these differential systems by flat
connections (together with a special singularity, namely
equi-singularity) near the singularities, without making more
singularities elsewhere.
The importance of this
possibility can be investigated in making a geometrically encoding from
counterterms (divergences) \cite{CK1, CK2, CM2, CM4, CM3, CM1}. In this part an overview from this story is done.

\begin{defn} \label{time-ordered}
For a given connected graded commutative Hopf algebra $H$ of finite type with the related complex Lie group $G(\mathbb{C})$ and Lie algebra $\mathfrak{g}(\mathbb{C})$,  let $\alpha: I=[a,b] \subset \mathbb{R} \longrightarrow
\mathfrak{g}(\mathbb{C})$ be a smooth curve. Its associated time
ordered exponential is defined by
$$
Te^{\int ^{b}_{a} \alpha (t)dt}:=1 + \sum_{n \ge 1} \int_{a \le
s_{1} \le ... \le s_{n} \le b} \alpha (s_{1})...\alpha (s_{n})
ds_{1}...ds_{n}
$$
such that the product is taken in the graded dual space $H^{*}$ and $1
\in H^{*}$ is the unit corresponding to the counit $\epsilon$.
\end{defn}

\begin{rem} \label{43}

(i) This integral only depends on 1-form $\alpha(t)dt$ and because of
finite type property of Hopf algebra, for each element in $H$, it is
finite with values in the affine group scheme.

(ii) Time ordered exponential is the value $g(b)$ of the unique
solution $g(t) \in G(\mathbb{C})$ of the
differential equation
$$dg(t)=g(t) \alpha(t)dt$$
with the initial condition $g(a)=1$.

(iii) It is
multiplicative over the sum of paths.

(iv) Let $\omega$ be a flat $\mathfrak{g}(\mathbb{C})-$valued
connection on $\Omega \subset \mathbb{R}^{2}$ and $\alpha: [0,1]
\subset \mathbb{R} \longrightarrow \Omega$ be a curve. Then the time
ordered exponential $Te^{\int ^{1}_{0} \alpha^{\star} \omega}$ only
depends on the homotopy class $[\alpha]$ of paths such that
$\alpha(0)=a,$ $\alpha(1)=b$.

(v) Let $\omega \in \mathfrak{g}(\mathbb{C}(\{z\}))$ has a trivial
monodromy $M(\omega)=1$. Then there exists a solution $g \in
G(\mathbb{C}(\{z\}))$ for the equation $\bold{D}(g)=\omega$ such that the
logarithmic derivative $\bold{D}: G(K) \longrightarrow
\Omega^{1}(\mathfrak{g})$ is given by $\bold{D}(g):=g^{-1}dg$. \cite{CM5,
CM2, CM4, CM3, CM1}
\end{rem}

Time ordered exponential necessarily and sufficiently provides an useful representation from loops on the set
$Loop(G(\mathbb{C}),\mu)$ and their Birkhoff components such that it contains a reformulation of loops in terms
of a class of differential systems associated to a family of
equi-singular flat connections. As the consequence, this machinery delivers us a
geometric meaningful from divergences.

\begin{thm} \label{bir-dec-loops}
Let $\gamma_{\mu}(z)$ be a loop in the class
$Loop(G(\mathbb{C}),\mu)$. Then

(i) There exists a unique $\beta \in \mathfrak{g}(\mathbb{C})$ and a
loop $\gamma_{reg}(z)$ regular at $z=0$ such that
$$\gamma_{\mu}(z)=Te^{-\frac{1}{z} \int^{-zlog \mu}_{\infty} \theta_{-t}(\beta)dt} \theta_{zlog \mu} (\gamma_{reg}(z)).$$

(ii) There is a representation from components of the Birkhoff
factorization of $\gamma_{\mu}(z)$ based on time ordered exponential. We have
$$ \gamma_{\mu +}(z) = Te^{-\frac{1}{z} \int^{-zlog \mu}_{0} \theta_{-t}(\beta)dt} \theta_{zlog \mu} (\gamma_{reg}(z)),$$
$$ \gamma_{-}(z)= Te^{-\frac{1}{z} \int^{\infty}_{0} \theta_{-t}(\beta)dt}. $$

(iii) For each element $\beta \in \mathfrak{g}(\mathbb{C})$ and
regular loop $\gamma_{reg}(z)$, one unique element of
$Loop(G(\mathbb{C}),\mu)$ is identified. \cite{CM5,
CM2, CM4, CM3, CM1}
\end{thm}

\begin{defn}  \label{equival-rel}
Define an equivalence relation on connections (i.e.
$\mathfrak{g}(\mathbb{C})-$valued one forms). $\omega_{1},
\omega_{2}$ are equivalent if there exists $h \in
G(\mathbb{C}\{z\})$ such that
$$\omega_{2}=\bold{D}h+h^{-1}\omega_{1}h.$$
\end{defn}

\begin{rem} \label{equisingular1}
There is a correspondence between two equivalence
$\mathfrak{g}(\mathbb{C})-$valued connections $\omega_{1},
\omega_{2}$ with trivial monodromies and negative parts of the
Birkhoff decomposition of solutions of the equations
$$
\bold{D}f_{1}=\omega_{1}, \  \  \bold{D}f_{2}=\omega_{2}
$$
where $f_{1}, f_{2} \in
G(\mathbb{C}(\{z\}))$. It means that
$$\omega_{1} \sim \omega_{2} \Longleftrightarrow f^{-}_{1}=f^{-}_{2}.$$
\end{rem}

{\it Dimensional regularization} is an usual
regularization technique in perturbative renormalization. It is based on an
analytic continuation of Feynman integrals to the complex dimension
$d \in \mathbb{C}$ in a neighborhood of the integral (critical)
dimension $\textsc{D}$ at which ultra-violet divergences occur. Under this
regularization prescription, the procedure of renormalization with a
special renormalization scheme namely, {\it minimal subtraction}
(i.e. the subtraction of singular part of the Laurent series in
$z=d-\textsc{D}$ at each order in the loop expansion) can be performed.
In the BPHZ renormalization, all of the (sub-)divergences, in a recursive
procedure, disappear such that it can be
understood in terms of the extraction of finite values (i.e.
Birkhoff decomposition of loops with values in the space of diffeographisms).

\begin{thm}
There is a principal bundle connected with a given theory under
the Dim. Reg. + Min. Sub. scheme such that one special class of its related connections provides
a geometric reinterpretation from physical
information. This bundle is called renormalization bundle. \cite{CM2, CM4, CM3, CM1}
\end{thm}

\begin{proof} A sketch of proof:
Let $\bold{\Delta}$ be an infinitesimal disk corresponds with the
complexified dimension $\textsc{D}-z \in \bold{\Delta}$ of dimensional
regularization and $\mathbb{G}_{m}(\mathbb{C})=\mathbb{C}^{*}$
be the possible choices for the normalization of an integral in
dimension $\textsc{D}-z$. For the principal $\mathbb{C}^{*}-$bundle
$\mathbb{G}_{m} \longrightarrow B \longrightarrow \bold{\Delta}$, let $P=B
\times G$ be a trivial principal $G-$bundle. Set
$$
V:=p^{-1}(\{0\}) \subset B, \  \ B^{0}:=B-V, \ \  P^{0}=B^{0}
\times G.
$$
This bundle, together with connections on it, can store some geometrical meanings related to physical information.
\end{proof}

\begin{rem} \label{geo-dse5}
The choice of the unit of mass $\mu$ is the same as the choice of a
section $\sigma: \bold{\Delta} \longrightarrow B$ and indeed, the concept of equi-singularity turns to this fact.
\end{rem}

\begin{defn} \label{geo-dse6}
Fix a base point $y_{0} \in V$ and identify $B$ with $\bold{\Delta} \times
\mathbb{G}_{m}(\mathbb{C})$. A flat connection $\omega$ on $P^{0}$
is called equi-singular, if

- It is
$\mathbb{G}_{m}(\mathbb{C})-$invariant,

- For any solution $f$ of
the equation $\bold{D}f=\omega$, the restrictions of $f$ to the sections
$\sigma: \bold{\Delta} \longrightarrow B$ with $\sigma(0)=y_{0}$ have the
same singularity (namely, the pullbacks of a solution have the same
negative parts of the Birkhoff decomposition, independent of the
choice of the section and therefore the mass parameter).

\end{defn}

It is remarkable that one can expand equivalence relation given by definition \ref{equival-rel} to
this kind of connections. With working on classes of equi-singular flat connections on renormalization principal bundle, one can redisplay infinitesimal characters.

\begin{thm} \label{geo-dse7}
Equivalence classes of flat equi-singular $G-$connections on $P^{0}$
are represented by elements of $\mathfrak{g}(\mathbb{C})$ and
also, each element of the Lie algebra of affine group scheme of the
Hopf algebra associated to the renormalizable theory $\Phi$
identifies a specific class of equi-singular connections. The above
process is independent of the choice of a local regular section
$\sigma: \bold{\Delta} \longrightarrow B$ with $\sigma(0)=y_{0}$.
\cite{CM2, CM4, CM3, CM1}
\end{thm}

Theorems \ref{bir-dec-loops}, \ref{geo-dse7} provide a bijective
correspondence between negative parts of the Birkhoff
decomposition of loops with values in $G(\mathbb{C})$ and
classes of equi-singular flat connections on the renormalization bundle
(which stores the regularization parameter). Since
counterterms of the theory are identified with these negative parts (i.e. Connes-Kreimer formalization),
therefore these classes of connections are presenting the
counterterms.

In a categorical configuration, one can introduce a category such that equi-singular flat connections are its objects. The construction of this category
and its properties are completely analyzed by Connes and Marcolli and just we review their main result.

\begin{thm} \label{cat.1}
For a given renormalizable QFT $\Phi$ underlying Dim. Reg. + Min. Sub. with the associated Lie group $G(\mathbb{C})$ and vector bundle $B \longrightarrow \bold{\Delta}$, flat equi-singular connections on this bundle
introduce an abelian tensor category with a specific fiber functor. Additionally, this category is a neutral Tannakian category and therefore it is equivalent to the category of finite dimensional representations of affine group scheme $G^{*}:=G \rtimes \mathbb{G}_{m}$. \cite{CM5, CM2, CM4, CM3, CM1}
\end{thm}


\section{{\it \textsf{The construction of a universal Tannakian category}}}

The Riemann-Hilbert correspondence conceptually consists of describing a certain
category of equivalence classes of differential systems though a
representation theoretic datum. For a given renormalizable QFT
$\Phi$ with the related Hopf algebra $H$ and the space of diffeographisms
$G(\mathbb{C})$, it was shown that how one can identify a category of classes of flat equi-singular
$G-$connections. Categorification of the Connes-Kreimer theory allows us to have a reasonable idea to formulate this story
in a universal setting (underlying the Riemann-Hilbert problem) by constructing the universal category
$\mathcal{E}$ of equivalence classes of all flat equi-singular vector
bundles.

This category has the ability of covering the corresponding categories
of all renormalizable theories and it means that when we are working on
the theory $\Phi$, it is possible to consider the subcategory
$\mathcal{E}^{\Phi}$ of those flat equi-singular vector bundles which
is equivalent to the finite dimensional linear representations of
$G^{*}$.
In this part we try to consider some general features of
this very special category and its universality.

Start with a filtered vector bundle $(E,W)$ over $B$ with an increasing
filtration
\begin{gather}
W^{-n-1}(E) \subset W^{-n}(E), \  \ (W^{-n}(E)= \bigoplus_{m \ge n}
E_{m})
\end{gather}
and set
\begin{equation}
Gr^{W}_{n}= \frac{W^{-n}(E)}{W^{-n-1}(E)}.
\end{equation}

\begin{defn} \label{cat.2}
A $W-$connection on $E$ is a connection $\bigtriangledown$  on the
vector bundle $E^{0}=E|_{B^{0}}$ such that

- $\bigtriangledown$ is compatible with the filtration,

- $\bigtriangledown$ is the trivial connection on $Gr^{W}_{n}$.
\end{defn}

\begin{rem} \label{cat.3}
One can extend the mentioned equivalence relation in the definition
\ref{equival-rel} and the concept of equi-singularity to the level of $W-$connections. It means that

(i) Two $W-$connections $\bigtriangledown_{1}, \bigtriangledown_{2}$
on $E^{0}$ are $W-$equivalent if there exists an automorphism $T$ of
$E$ that preserves the filtration, where is identity on $Gr^{W}_{n}$
and $T \circ \bigtriangledown_{1} = \bigtriangledown_{2} \circ T$.

(ii) A flat $W-$connection $\bigtriangledown$ on $E$ is equi-singular
if it is $\mathbb{G}_{m}(\mathbb{C})-$invariant and the pullback
along different sections $\sigma$ of $B$ (such that
$\sigma(0)=y_{0}$) of a solution of the equation $\bigtriangledown
\eta = 0$ have the same type of singularity.
\end{rem}

\begin{defn} \label{cat.4}
The couple $(E,\bigtriangledown)$ is called a flat equi-singular vector bundle.
\end{defn}

These pairs introduce a category $\mathcal{E}$ such that its objects $Obj(\mathcal{E})$
are data $\Theta=[V,\bigtriangledown]$ where
$V$ is a $\mathbb{Z}-$graded finite dimensional vector space and
$\bigtriangledown$ is an equi-singular $W-$connection on the filtered
bundle $E^{0}=B^{0} \times V$ (with attention to the classes of
connections).

Each morphism $T \in Hom(\Theta, \Theta^{'})$ is a linear map $T: E
\longrightarrow E^{'}$ compatible with the grading such that it
should have the following relation with the connections
$\bigtriangledown, \bigtriangledown^{'}$. Set

\begin{equation} \label{co1}
\bigtriangledown_{1}:= \big(\begin{matrix}
\bigtriangledown^{'} & 0 \\
0 & \bigtriangledown
\end{matrix}\big),  \  \  \bigtriangledown_{2}:= \big(\begin{matrix}
\bigtriangledown^{'} & T \circ \bigtriangledown - \bigtriangledown^{'} \circ T \\
0 & \bigtriangledown
\end{matrix}\big).
\end{equation}
For the defined connections in (\ref{co1}) on the vector bundle
$(E^{'} \bigoplus E)^{\star}$, $\bigtriangledown_{2}$ should be a
conjugate of $\bigtriangledown_{1}$ by the unipotent matrix
$\big(\begin{matrix}
1 & T \\
0 & 1
\end{matrix}\big)$.

\begin{thm}
With applying Riemann-Hilbert correspondence, one can formulate Connes-Marcolli category (consisting of divergences of a renormalizable theory) in a universal configuration by constructing the universal
category $\mathcal{E}$ of equivalence classes of all flat
equi-singular vector bundles with the the fiber functor given by
$\varphi: \mathcal{E} \longrightarrow \mathcal{V}_{\mathbb{C}},$
$\Theta \longmapsto V$. \cite{CM2, CM4, CM3, CM1, M4}
\end{thm}

We know that there is a representation of a neutral Tannakian category by the
category $\mathfrak{R}_{G^{*}}$ of finite dimensional
representations of the affine group scheme of automorphisms of the
fiber functor of the main category. This fact provides a new reformulation from this specific universal category.


\begin{thm} \label{cat.5}
$\mathcal{E}$ is a neutral Tannakian category. It is equivalent to
the category $\mathfrak{R}_{\mathbb{U}^{*}}$ of finite linear
representations of one special affine group scheme (related to the
universal Hopf algebra of renormalization $H_{\mathbb{U}}$) namely, universal
affine group scheme $\mathbb{U}^{*}$ such that $H_{\mathbb{U}}$ is a
connected graded commutative non-cocommutative Hopf algebra of
finite type. It is the graded dual of the universal enveloping
algebra of the free graded Lie algebra $L_{\mathbb{U}}:=
\bold{F}(1,2,...)_{\bullet}$ generated by elements $e_{-n}$ of
degree $n>0$ (i.e. one generator in each degree). \cite{CM4, CM1}
\end{thm}

\begin{rem} \label{cat.6}
(i) Since the category of filtered vector spaces is not an abelian
category, it is necessary to use the direct sum of bundles and the
above condition on connections.

(ii) The construction of this category over the field $\mathbb{Q}$
is also possible.
\end{rem}

With attention to theorem \ref{geo-dse7} and also given correspondence in theorem
\ref{cat.5}, one can obtain a new prescription from equi-singular
connections such that it will be applied to produce a new universal level for
counterterms, renormalization groups and $\beta-$functions of
theories.

\begin{thm} \label{cat.10}
Let $H$ be a connected graded commutative Hopf algebra of finite
type with the affine group scheme $G$ and $\widetilde{P}^{0}:=B^{0}
\times G^{*}$. Each equivalence class $\omega$ of flat
equi-singular connections on $\widetilde{P}^{0}$ identifies a graded
representation $\rho_{\omega}: \mathbb{U}^{*} \longrightarrow
G^{*}$   (which is identity on $\mathbb{G}_{m}$) and also,
for each graded representation $\rho: \mathbb{U} \longrightarrow G$,
there is one specific class of flat equi-singular connections.
\cite{CM2, CM3}
\end{thm}

Based on the correspondence $\mathcal{R}_{\mathbb{U}^{*}} \simeq \mathcal{E}$, for each object $\Theta=[V, \bigtriangledown]$, there exists a unique representation $\xi_{\Theta}$ of $\mathbb{U}^{*}$ in $V$ such that $\bold{D} \xi_{\Theta} (\gamma_{\mathbb{U}}) \simeq \bigtriangledown$. And also each arbitrary representation $\xi$ of $\mathbb{U}^{*}$ in $V$ determines a unique connection $\bigtriangledown$ (up to equivalence) such that $[V, \bigtriangledown]$ is an object in $\mathcal{E}$. It can be seen that one specific element will be determined from the process namely, the loop {\it universal singular frame}
$\gamma_{\mathbb{U}}$ with values in $\mathbb{U}(\mathbb{C})$ where at this level one hopes to eliminate divergences for generating a finite theory.

Now because of
the equivalence relation between loops (with values in the Lie
group $G(\mathbb{C})$) and elements of the Lie algebra (corresponding to
$G(\mathbb{C})$), with the help of a suitable element of the Lie algebra
$L_{\mathbb{U}}$, one can characterize the universal singular frame.

\begin{lem}
The Lie algebra element corresponding to the universal singular frame is $e=\sum_{n \ge 1} e_{-n}$ (i.e. the sum of the
generators of the Lie algebra).
\end{lem}

Since the universal Hopf algebra of
renormalization is finite type, whenever we pair $e$ with an element
of the Hopf algebra, it will be only a finite sum. Hence $e$ makes
sense.

\begin{thm} \label{cat.9}
(i) $e$ is an element in the completion of $L_{\mathbb{U}}$.

(ii) $e: H_{\mathbb{U}} \longrightarrow \mathbb{K}[t]$ is a linear map. Its
affine group scheme level namely, $ \bold{rg}:
\mathbb{G}_{a}(\mathbb{C}) \longrightarrow \mathbb{U}(\mathbb{C})$
is a morphism that plays an essential role to calculate the
renormalization group.

(iii) The universal singular frame can be reformulated with
$\gamma_{\mathbb{U}}(z,v)=Te^{-\frac{1}{z} \int^{v}_{0} u^{Y}(e)
\frac{du}{u}}$.

(iv) For each loop $\gamma(z)$ in $Loop(G(\mathbb{C}),\mu)$, with
help of the associated representation $\rho: \mathbb{U}
\longrightarrow G$, the universal singular frame
$\gamma_{\mathbb{U}}$ maps to the negative part $\gamma_{-}(z)$ of
the Birkhoff decomposition of $\gamma(z)$ and also, the
renormalization group $\{F_{t}\}_{t}$ in $G(\mathbb{C})$ can be recalculated by
$\rho \circ \bold{rg}$. \cite{CM2, CM3}
\end{thm}

When we apply the universal singular frame in the dimensional
regularization, all divergences will be removed and it means that one
can obtain a finite theory which depends only upon the choice of
local trivialization of the $\mathbb{G}_{m}-$principal bundle $B$.

Positive and negative components of the Birkhoff factorization of the loop
$\gamma_{\mathbb{U}}$ in the pro-unipotent affine group scheme
$\mathbb{U}$ contain a universal meaning. For a given renormalizable theory $\Phi$ these
components, via the identified representations
in the theorem \ref{cat.10}, map to renormalized values and
counterterms, respectively. This fact provides a valuable
concept namely, {\it universal counterterms}.

At last, it is favorable to emphasize the universality of the category $\mathcal{E}$ among other categories determined from physical theories. It means that this category gives us the power of analyzing flat
equi-singular connections for each affine group scheme and this is
the main reason of its universality.

\begin{thm}  \label{cat.7}
Let $H$ be a connected graded commutative Hopf algebra of finite
type with the affine group scheme $G$. Let $\omega$ be a flat
equi-singular connection on $P^{0}$ and $\psi: G \longrightarrow
GL(V)$ a finite dimensional linear graded representation of $G$. We
can correspond an element $\Theta \in Obj(\mathcal{E})$ to the data
$(\omega, \psi)$. The equivalence class of $\omega$ identifies the
same element $\Theta$. \cite{CM2, CM4, CM1}
\end{thm}

\begin{rem} \label{cat.8}
For each flat equi-singular vector bundle $(E,\bigtriangledown)$
(such that $E=B \times V$), the connection $\bigtriangledown$
identifies a flat equi-singular $G_{V}-$valued connection $\omega$.
Indeed, $\bigtriangledown$ is given by adding a $Lie \
G_{V}-$valued one form to the trivial connection. This
correspondence preserves the  mentioned equivalence relation
in the definition \ref{equival-rel}.
\end{rem}

In summary, it is observed that lifting the Connes-Kreiemr perturbative renormalization to the universal configuration can yield the particular Hopf algebra $H_{\mathbb{U}}$. Because of the combinatorial nature of the universal Connes-Kreimer Hopf algebra, it should be reasonable
to search a hidden combinatorial construction in the backbone of this Hopf algebra. This problematic notion is the main topic of the next chapter and its importance will be investigated when we want to generalize the Connes-Marcolli treatment in the study of Dyson-Schwinger equations.


\chapter{\textsf{Universal Hopf algebra of renormalization}}

We discussed that how one can systematically interpret the hidden combinatorics of perturbative renormalization based on the Hopf algebra structure on Feynman diagrams of a pQFT. Furthermore, it was exhibited that the Connes-Kreimer Hopf algebra of rooted trees
(equipped with some decorations which represent primitive 1PI graphs) plays the
role of an available practical model such that with changing labels, it will upgrade for each arbitrary theory. On the other hand, we saw that the universal affine
group scheme $\mathbb{U}$ governs the structure of divergences of
all renormalizable theories and the universality of $H_{\mathbb{U}}$ turns to its independency from all theories.

In this chapter, we are
going to discover the natural combinatorial backbone of $H_{\mathbb{U}}$ by giving a new explicit rooted tree type reformulation from this particular Hopf algebra and its related Lie group. Then with using this new interpretation, first we obtain
rigorous relations between universal Hopf algebra of renormalization and some well-known combinatorial Hopf algebras. Secondly, we expand this new Hall tree formalism to the level of Lie groups and finally, we will describe the universal singular frame based on Hall polynomials such that as the result, new Hall tree scattering formulaes for physical information will be determined.  \cite{S6}


\section{{\it \textsf{Shuffle nature of $H_{\mathbb{U}}$}}}

The Hopf algebra $H_{\mathbb{U}}$ is introduced in the theorem \ref{cat.5} and in particular, one important note
is that as an algebra it is isomorphic to the linear space of
noncommutative polynomials in variables $f_{n}$, $n \in
\mathbb{N}_{>0}$ with the shuffle product. It is a skeleton key for us
to find a relation between this Hopf algebra and rooted trees and because of that at first
we need some more information about shuffle structures.

\begin{defn} \label{shuffle-pro}
Let $V$ be a vector space over the field $\mathbb{K}$ of characteristic zero and
$T(V)= \bigoplus_{n \ge 0} V^{\otimes n}$ be its related tensor algebra. Set
$$S(m,n)=\{ \sigma \in S_{m+n} : \sigma^{-1} (1) < ... < \sigma^{-1}
(m), \ \ \sigma^{-1} (m+1) < ... < \sigma^{-1} (m+n) \}.$$
It is called the set of $(m,n)-$shuffles. For each $x=x_{1}
\otimes ... \otimes x_{m} \in V^{\otimes m}$, $y=y_{1} \otimes ...
\otimes y_{n} \in V^{\otimes n}$ and $\sigma \in S(m,n)$, define
$$\sigma (x \otimes y) = u_{\sigma (1)} \otimes u_{\sigma (2)} \otimes
... \otimes u_{\sigma (m+n)} \in V^{\otimes (m+n)}$$
such that $u_{k} = x_{k}$ for $1\le k\le m$ and $u_{k}=y_{k-m}$ for
$m+1\le k\le m+n$. The shuffle product of $x,y$ is given by
$$x \star y := \sum_{\sigma \in S(m,n)} \sigma (x \otimes y).$$
\end{defn}

\begin{lem} \label{univ1}
$(T(V),\star)$ is a unital commutative associative algebra. \cite{EG1, M6}
\end{lem}

There are some extensions of this product such as quasi-shuffles,
mixable shuffles. Let $A$ be a {\it locally finite set} (i.e. a
disjoint union of finite sets $A_{n}, n \ge 1$). The elements of $A$
are {\it letters} and monomials are called {\it words} such that the
empty word is denoted by $1$. Set $A^{-}:=A \cup \{0\}$.

\begin{defn}
Function $<.,.>: A^{-} \times A^{-} \longrightarrow A^{-}$ is called a Hoffman pairing, if it satisfies following conditions:

- For all $a \in A^{-},$ $<a,0>=0$,

- For all $a,b \in A^{-},$ $<a,b> = <b,a>$,

- For all $a,b,c \in A^{-},$ $<<a,b>,c> = <a,<b,c>>$,

- For all $a,b \in A^{-},$ $<a,b> = 0$ or $|<a,b>| = |a|+|b|$.

A locally finite set $A$ together with a Hoffman pairing $<.,.>$ on $A^{-}$ is
called a Hoffman set.
\end{defn}

\begin{defn} \label{f1}
Let $\mathbb{K}<A>$ be the graded
noncommutative polynomial algebra over $\mathbb{K}$.
The quasi-shuffle product $\star^{-}$ on $\mathbb{K}<A>$ is
defined recursively such that for any word $w$, $1 \star^{-} w = w
\star^{-} 1 = w$ and also for words $w_{1}, w_{2}$ and letters $a,
b$,
$$(aw_{1}) \star^{-} (bw_{2}) = a(w_{1} \star^{-} bw_{2}) + b(aw_{1}
\star^{-} w_{2})+<a,b>(w_{1} \star^{-} w_{2}).$$

\end{defn}

\begin{lem} \label{univ2}
$\mathbb{K}<A>$ together with the product $\star^{-}$ is a graded
commutative algebra such that when $<.,.>=0$, it will be the shuffle
algebra $(T(V),\star)$ where $V$ is a vector space generated by the
set $A$. \cite{EG1, H1, M6}
\end{lem}

Shuffle type products can determine interesting family of Hopf algebras such that it can be possible to reformulate $H_{\mathbb{U}}$ on the basis of this class.
The next theorem gives a complete characterization from this class of Hopf algebras.

\begin{thm} \label{shuffle3} \label{shuffle-cop}

(i) The (quasi-) shuffle product introduces a graded connected commutative non-cocommutative Hopf
algebra structure (of finite type) on $(\mathbb{K}<A>, \star^{-})$
and $(T(V),\star)$.

(ii) There is an isomorphism (as a graded Hopf algebras) between
$(T(V),\star)$ and $(\mathbb{K}<A>, \star^{-})$.

(iii) There is a graded connected Hopf algebra structure (of finite
type) (comes from (quasi-)shuffle product) on the graded dual of
$\mathbb{K}<A>$.

(iv) We can extend the isomorphism in the second part to the graded
dual level. \cite{EG1, H1, M6}
\end{thm}

\begin{proof}
The compatible Hopf algebra structure on the shuffle algebra of
noncommutative polynomials is given by the coproduct
$$
\Delta (w)= \sum_{uv=w} u \otimes v
$$
and the counit
$$
\epsilon (1)=1,  \  \  \  \epsilon (w)=0, \ w \neq 1.
$$

For a given Hoffman pairing $<.,.>$ and any finite sequence $S$ of
elements of the set $A$, with induction define $<S> \in A^{-}$ such
that for any $a \in A$,
$$<a>=a, \  \  \  <a, S>=<a,<S>>.$$
Let $C(n)$ be
the set of compositions of $n$ and $C(n,k)$ be the set of
compositions of $n$ with length $k$. For each word $w=a_{1}...a_{n}$
and composition $I=(i_{1},...,i_{l})$, set
$$
I<w> := <a_{1},...,a_{i_{1}}>
<a_{i_{1}+1},...,a_{i_{1}+i_{2}}>...<a_{i_{1}+...+i_{l-1}+1},...,
a_{n}>.
$$
It means that compositions act on words. Now for any word
$w=a_{1}...a_{n}$, its antipode is given by
$$S(1)=1,$$
$$
S(w)=- \sum_{k=0}^{n-1} S(a_{1}...a_{k}) \star^{-} a_{k+1}...a_{n}=
(-1)^{n} \sum_{I \in C(n)} I<a_{n}...a_{1}>.
$$

The isomorphism between these Hopf algebra structures (compatible
with the shuffle products) is given by morphisms
$$
\tau(w)= \sum_{(i_{1},...,i_{l}) \in C(|w|)} \frac{1}{i_{1}! ...
i_{l}!} (i_{1},...,i_{l}) <w>,
$$

$$
\psi(w)=  \sum_{(i_{1},...,i_{l}) \in C(|w|)}
\frac{(-1)^{|w|-l}}{i_{1} ... i_{l}} (i_{1},...,i_{l}) <w>.
$$

The graded dual $\mathbb{K}<A>^{\star}$ has a basis consisting of
elements $v^{\star}$ (where $v$ is a word on $A$) with the following
pairing such that if $u=v$, then $(u,v^{\star})=1$ and if $u \neq
v$, then $(u,v^{\star})=0$. Its Hopf algebra structure is given by
the concatenation product
$$
conc(u^{\star} \otimes v^{\star})=(uv)^{\star}
$$
and the coproduct
$$
\delta (w^{\star}) = \sum_{u,v} (u \star^{-} v, w^{\star}) u^{\star}
\otimes v^{\star}.
$$

The map

$$
\tau^{\star} (u^{\star}) = \sum_{n \ge 1} \sum_{<a_{1},...,a_{n}>=u}
\frac{1}{n!} (a_{1}...a_{n})^{\star},
$$
is an isomorphism in the dual level and its inverse is given by
$$
\psi^{\star} (u^{\star})= \sum_{n \ge 1} \frac {(-1)^{n-1}}{n}
\sum_{<a_{1},...,a_{n}>=u} (a_{1}...a_{n})^{\star}.
$$

\end{proof}

Because of the relation between Hopf algebras and Lie theory, one can consider shuffle Hopf algebras at the Lie algebra version.

\begin{defn}
Let $\mathcal{L}$ be a Lie algebra over $\mathbb{K}$.
There exists an associative algebra $\mathcal{L}_{0}$ over
$\mathbb{K}$ together with a Lie algebra homomorphism $\phi_{0}:
\mathcal{L} \longrightarrow \mathcal{L}_{0}$ such that for each
couple $(\mathcal{A}, \phi: \mathcal{L} \longrightarrow
\mathcal{A})$ of an algebra and a Lie algebra homomorphism, there is
a unique algebra homomorphism $\phi_{\mathcal{A}}:\mathcal{L}_{0}
\longrightarrow \mathcal{A}$ such that $\phi_{\mathcal{A}} \circ
\phi_{0}=\phi$. $\mathcal{L}_{0}$ is called universal
enveloping algebra of $\mathcal{L}$ and it is unique up to
isomorphism.
\end{defn}

\begin{lem} \label{univ3}

(i) Universal enveloping algebra $\mathcal{L}_{0}$ of the free Lie
algebra $\mathcal{L}(A)$ is a free associative algebra on $A$.

(ii) $\phi_{0}$ is an injective morphism such that $\phi_{0}(\mathcal{L}(A))$ is
the Lie subalgebra of $\mathcal{L}_{0}$ generated by $\phi_{0} \circ
i (A)$. \cite{R1}
\end{lem}

\begin{defn}
The set of Lie polynomials in $\mathbb{K}<A>^{\star}$ is the
smallest sub-vector space of $\mathbb{K}<A>^{\star}$ containing the
set of generators $A^{\star}:=\{ a^{\star} : a \in A \}$ and closed
under the Lie bracket.
\end{defn}

\begin{cor}  \label{lie-poly1}
(i) The set of Lie polynomials in $\mathbb{K}<A>^{\star}$ forms a Lie
algebra. It is the free Lie algebra on $A^{\star}$ such that
$\mathbb{K}<A>^{\star}$ is its universal enveloping algebra.

(ii) In the shuffle product case, the Lie polynomials are exactly
the primitives for $\delta$ and therefore at the level of
(quasi-)shuffle product, the primitives are elements of the form
$\psi^{\star} p$ such that $p$ is a Lie polynomial \cite{H1}.
\end{cor}

Now it is the place to come back to the definition of the universal Hopf algebra of renormalization. It can be seen that
the set $A = \{ f_{n}: n \in \mathbb{N}_{>0} \}$ is a locally finite
set and as an algebra, $H_{\mathbb{U}}$ is isomorphic to $(T(V),\star)$ such
that $V$ is a vector space over $\mathbb{C}$ spanned by the set $A$. Therefore its
Hopf algebra structure is determined with the theorem \ref{shuffle3}. At the Lie
algebra level, we have to go to the dual structure. Corollary \ref{lie-poly1}
shows us that the set of all Lie polynomials in $H_{\mathbb{U}}^{\star}$ is
the free Lie algebra generated by $\{f^{\star}_{n}\}_{n \in
\mathbb{N}_{>0}}$ such that $H_{\mathbb{U}}^{\star}$ is its universal
enveloping algebra and on the other hand, we know that
$H_{\mathbb{U}}^{\star}$ is identified by the universal enveloping of the
free graded Lie algebra $L_{\mathbb{U}}$ generated by $\{e_{-n}\}_{n
\in \mathbb{N}_{>0}}$.

\begin{cor}  \label{order}
This procedure implies to have a reasonable correspondence between generators of the Lie algebra $L_{\mathbb{U}}$ and elements of the set $A^{\star}$.
\end{cor}


\section{{\it \textsf{Rooted tree version}}}

In this part, with attention to the shuffle nature of $H_{\mathbb{U}}$, we introduce a combinatorial version from this specific Hopf algebra in the Connes-Marcolli universal renormalization theory \cite{S6}. We consider completely a new Hall rooted tree reformulation from $H_{\mathbb{U}}$ to obtain interesting relations between this particular Hopf algebra and some well-known combinatorial Hopf algebras introduced in \cite{BF, CL1, H1, H3, H2, H5, H6, H4, LR2, LR, P1, S1, V1}. Moreover, we extend this formalization of $H_{\mathbb{U}}$ to the level of its associated Lie group to display its elements based on formal series of Hall trees. So it provides a new Hall tree type representation from universal singular frame $\gamma_{\mathbb{U}}$ where we will consider the applications of Hall basis and PBW basis to reformulate combinatorially physical information.

\begin{defn}
Defining a partial order $\preceq$ on the set of all rooted
trees $\bold{T}$. We say $t \preceq s$, if $t$ can be obtained from
$s$ by removing some non-root vertices and edges and it implies that
$|t| \le |s|$.
\end{defn}

\begin{defn}
Let $\bold{T(A)}$ ($\bold{F(A)}$) be the set
of all rooted trees (forests) labeled by the set $A$.

(i) For $a \in A$,
$t_{1}, ..., t_{m} \in \bold{T(A)}$ such that $u=t_{1}...t_{m} \in
\bold{F(A)}$, $B^{+}_{a}(u)$ is a labeled rooted tree of degree
$|t_{1}|+...+|t_{m}|+1$ obtained by grafting the roots of $t_{1},
..., t_{m}$ to a new root labeled by $a$. In addition,
$B^{+}_{a}(\mathbb{I})$ is a rooted tree with just one labeled
vertex.

(ii) For $t \in \bold{T(A)}$ and $u \in \bold{F(A)}, $ define a
new element $t \circ u$ such that it is a labeled rooted tree of
degree $|t|+|u|$ given by grafting the roots of labeled rooted trees
in $u$ to the root of $t$.
\end{defn}

\begin{lem} \label{rt-v1}

(i) The operation $\circ$ is not associative.

(ii) $\forall t \in \bold{T(A)},$ $\forall u,v \in \bold{F(A)}:$ $(t
\circ u) \circ v = t \circ (uv) = (t \circ v) \circ u$.

(iii) $t_{1} \circ ... \circ t_{m} \circ u = t_{1} \circ (t_{2}
\circ ... \circ (t_{m} \circ u)), \  \ t^{\circ k}=t \circ ... \circ
t$, $k$ times.

(iv) For each $u \in \bold{F(A)}$, let $per(u)$ be the number of
different permutations of the vertices of a labeled partially
ordered set that representing $u$. Then
$$per(\mathbb{I})=1, \  \  per(B^{+}_{a}(u))=per(u).$$
And if $u=\prod^{m}_{j=1} (t_{j})^{i_{j}}$, then
$$per(u)= \prod^{m}_{j=1} i_{j}! per(t_{j})^{i_{j}}.$$

(v) The bilinear extension of $\circ$ to the linear combinations
of labeled rooted trees (or linear combinations of labeled forests) is
also possible.
\end{lem}

\begin{defn} \label{rt-v2}
A set $\bold{H(T(A))}$ of labeled rooted trees is called {\it Hall
set}, if it has following conditions:

- There is a total order relation $>$ on $\bold{H(T(A))}$.

- If $a \in A$, then $B^{+}_{a}(\mathbb{I}) \in \bold{H(T(A))}$.

- For $a \in A$, $u \in \bold{F(A)} - \{\mathbb{I}\}$ such that $u =
t^{\circ r_{1}}_{1} ... t^{\circ r_{m}}_{m}$, $t_{1},...,t_{m} \in
\bold{H(T(A))}$, $r_{1},...,r_{m} \ge 1$, $t_{1} > ... > t_{m}$,
$$ B^{+}_{a}(u) \in \bold{H(T(A))}  \Longleftrightarrow t_{m} > B^{+}_{a}(t^{\circ r_{1}}_{1} ... t^{\circ r_{m-1}}_{m-1}) \in \bold{H(T(A))}. $$

- If $t = B^{+}_{a}(t^{\circ r_{1}}_{1} ... t^{\circ r_{m}}_{m}) \in
\bold{H(T(A))}$  such that $t_{1},...,t_{m} \in \bold{H(T(A))}$,
$r_{1},...,r_{m} \ge 1$, $a \in A$, then for each $j=1,...,m$,
$t_{j}
> t$.
\end{defn}

\begin{lem}
For each $t \in \bold{H(T(A))}$, $r \ge 1$ and $a
\in A$, it is easy to see that
$$B^{+}_{a}(t^{\circ r}) \in \bold{H(T(A))} \Longleftrightarrow t >
B^{+}_{a}(\mathbb{I}).$$
\end{lem}

\begin{defn}  \label{rt-v3}
For a Hall set $\bold{H(T(A))}$, the set of its forests is given
by
$$
\bold{H(F(A))}:=\{\mathbb{I}\} \cup \{t^{r_{1}}_{1} ...
t^{r_{m}}_{m}: r_{1},...,r_{m} \ge 1, t_{1},...,t_{m} \in
\bold{H(T(A))}, t_{i} \neq t_{j} (i \neq j) \}.
$$
\end{defn}

\begin{lem} \label{rt-v4}
Elements of $\bold{H(F(A))}$ and rooted
trees are in the relation with the map
$$\xi: \bold{H(F(A))} - \{\mathbb{I}\} \longrightarrow \bold{T(A)}$$
$$t^{\circ r_{1}}_{1} ... t^{\circ r_{m}}_{m} \longrightarrow t^{\circ
r_{1}}_{1} \circ (t^{\circ r_{2}}_{2}...t^{\circ r_{m}}_{m}).$$
$\xi$ is injective and its image is the set $\{B^{+}_{a}(u) \in
\bold{T(A)}: u \in  \bold{H(F(A))}, a \in A \}$. \cite{M1}
\end{lem}

\begin{rem}

(i) Hall trees and Hall
forests have no symmetry.

(ii) There is a one to one correspondence
between a Hall set of $A-$labeled rooted trees and a Hall set of
words on $A$. \cite{R1}
\end{rem}

\begin{defn} \label{rt-v5}
For $t \in \bold{H(T(A))}$, there is a standard decomposition
$(t^{1},t^{2}) \in \bold{H(T(A))} \times \bold{H(T(A))}$ such that

- If $|t|=1$, then the decomposition is $t^{1}=t,$ $t^{2}=\mathbb{I}$,

- And if $t= B^{+}_{a}(t^{\circ r_{1}}_{1} ... t^{\circ r_{m}}_{m})$
such that $r_{1},...,r_{m} \ge 1,$ $t_{1},...,t_{m} \in
\bold{H(T(A))}:$  $t_{1} > ... > t_{m},$ $a \in A$, then the
decomposition is given by
$$
t^{1}= B^{+}_{a} (t^{\circ r_{1}}_{1} ... t^{\circ
r_{m-1}}_{m-1}t^{\circ r_{m}-1}_{m}), \  \ t^{2}=t_{m}.
$$

- For a Hall forest $u \in \bold{H(F(A))} - \bold{H(T(A))}$ such
that $u=t^{\circ r_{1}}_{1} ... t^{\circ r_{m}}_{m}$, $t_{1} > ... >
t_{m}$, the decomposition is given by $(u^{1},u^{2}) \in
\bold{H(F(A))} \times \bold{H(T(A))}$ where
$$
u^{1}=t^{\circ r_{1}}_{1} ... t^{\circ r_{m-1}}_{m-1}t^{\circ
r_{m}-1}_{m}, \  \  u^{2}=t_{m}.
$$

\end{defn}

\begin{defn} \label{rt-v6}
For a given map that associates to each word $w$ on $A$ a scalar
$\alpha_{w} \in \mathbb{K}$, define a map $\alpha: \bold{F(A)}
\longrightarrow \mathbb{K}$ such that

- $\mathbb{I} \longmapsto \alpha_{1}$.

- For each $u \in \bold{F(A)}-\{\mathbb{I}\}$, there is a labeled
partially ordered set $(\mathfrak{u}(A), \ge)$ that represents the
forest $u$ such that vertices $x_{1},...,x_{n},...$ of this poset
are labeled by $l(x_{i})=a_{i} \in A$ ($1 \le i$). Let
$>_{\mathfrak{u}(A)}$ be a total order relation on the set of
vertices $\mathfrak{u}(A)$ such that it is an extension of the
partial order relation $\ge$ on $\mathfrak{u}(A)$. For each ordered
sequence $x_{i_{1}} >_{\mathfrak{u}(A)} ...
>_{\mathfrak{u}(A)} x_{i_{n}}$ in $\mathfrak{u}(A)$, its
corresponding word $a_{i_{1}}...a_{i_{n}}$ is denoted by
$w(>_{\mathfrak{u}(A)})$. Set
$$
\alpha (u) := \sum_{>_{\mathfrak{u}(A)}}
\alpha_{w(>_{\mathfrak{u}(A)})},
$$
where the sum is over all total order relations
$>_{\mathfrak{u}(A)}$ (i.e. extensions of the main partial order
relation $\ge$) on the set of vertices of $\mathfrak{u}(A)$.
\end{defn}

\begin{lem}  \label{rt-v7}
Define a map $\pi$ given by
$$\pi: \mathbb{K}[\textbf{T(A)}] \longrightarrow
(\mathbb{K}<A>,\star^{-}),  \    \     \ \pi (u):=
\sum_{>_{\mathfrak{u}(A)}} w(>_{\mathfrak{u}(A)}).$$
One can show that for each $u, v \in \textbf{F(A)}$ and
$a \in A$,

(i) $\pi (B^{+}_{a}(u))=\pi(u)a,$

(ii) $\pi (uv)=\pi(u) \star^{-} \pi(v),$

(iii) $\alpha(u)= \widehat{\alpha}(\pi(u)),$  where $\widehat{\alpha}: (\mathbb{K}<A>,\star^{-}) \longrightarrow
\mathbb{K},$ $\widehat{\alpha} (w)= \alpha_{w}$ is a
$\mathbb{K}-$linear map.
\end{lem}

\begin{defn} \label{f3}
For given maps $\alpha, \beta: \bold{F(A)}  \longrightarrow \mathbb{K}$, one
can define a new map $\alpha \beta: \bold{F(A)}  \longrightarrow
\mathbb{K}$ given by
$$
u \longmapsto \sum_{(\mathfrak{v}(A),\mathfrak{w}(A)) \in
R(\mathfrak{u}(A))} \alpha (v) \beta (w)
$$
such that $\mathfrak{u}(A)$ is a labeled poset representing $u$ and
$\mathfrak{v}(A), \mathfrak{w}(A)$ are labeled partially ordered
subsets of $\mathfrak{u}(A)$ such that the set of all pairs
$(\mathfrak{v}(A),\mathfrak{w}(A))$ with the following conditions is
denoted by $R(\mathfrak{u}(A))$.

- The set of vertices in $\mathfrak{u}(A)$ is the disjoint union of
the set of vertices $\mathfrak{v}(A)$ and $\mathfrak{w}(A)$,

- For each $x,y \in \mathfrak{u}(A)$ such that $x \ge y$, if $x \in
\mathfrak{w}(A)$ then $y \in \mathfrak{w}(A)$.
\end{defn}

\begin{rem} \label{f13}
It can be seen that for each word $w=a_{1}...a_{m}$ on $A$,
$$
(\alpha \beta)_{w} = \alpha_{w} \beta_{1} + \alpha_{1} \beta_{w} +
\sum^{m-1}_{j=1} \alpha_{a_{1}...a_{j}} \beta_{a_{j+1}...a_{m}}.
$$
\end{rem}

\begin{defn}
For the map $\alpha$ given by the definition \ref{rt-v6}, one can introduce an equivalence
relation on $\mathbb{K}[\textbf{T(A)}]$ such that for each $u, v \in
\mathbb{K}[\textbf{T(A)}]$, they are congruent ($u \equiv v$), if
for every map $\widehat{\alpha}: A^{*} \longrightarrow \mathbb{K}$,
$w \mapsto \alpha_{w}$, (such that  $A^{*}$ is the set of all words
on $A$), then we have $\alpha(u) = \alpha(v)$.
\end{defn}

\begin{rem}

(i) $u \equiv v \Longleftrightarrow u-v \in ker \ \pi$.

(ii) $u, v \in \mathbb{K}[\textbf{T(A)}]$, $a \in A$, $u \equiv v
\Longrightarrow B^{+}_{a}(u) \equiv B^{+}_{a}(v)$.

(iii) $\overline{u}, \overline{v} \in \textbf{F(A)}: \overline{u} \equiv
\overline{v} \Longrightarrow u\overline{u} \equiv v\overline{v}$.

(iv) $t \in \textbf{T(A)}, n \ge 1: t^{n} \equiv n! t^{\circ n}$.

(v)  $t \in \textbf{T(A)},$ $i,j \ge 1:$ $t^{\circ i}t^{\circ j}
\equiv \frac{(i+j)!}{i!j!} t^{\circ (i+j)}$.
\end{rem}

\begin{lem}
For $m \ge 2$ and $t_{1},...,t_{m} \in \textbf{T(A)}$, with induction
one can show that
$$t_{1}...t_{m} \equiv \sum^{m}_{i=1} t_{i} \circ \prod_{j \neq i}
t_{j}.$$
There is an algorithm (in finite number of recursion steps) for
rewriting each $u \in \textbf{F(A)}$ as $u \equiv v$ such that $v
\in \mathbb{K} \bold{H(F(A))}$ (i.e. $v$ is a $\mathbb{K}-$linear
combination of Hall forests). \cite{M1}
\end{lem}

\begin{defn} \label{rt-v8}
There is a canonical map $f$ on Hall rooted trees defined by

- $f(a)=a$, if $a \in A$,

- $f(t)=f(t^{1})f(t^{2})$, if $t$ be of
degree $\ge 2$ with the standard decomposition $t=(t^{1},t^{2})$.

The function $f$ is called foliage and for each Hall tree $t$, its
degree $|f(t)|$ is the number of leaves of $t$. The foliage of a
Hall tree is called Hall word.
\end{defn}

\begin{thm}
For each word $w$ on $A$, there is a unique factorization
$w=f(t_{1})...f(t_{n})$ such that $t_{i} \in \bold{H(T(A))}$ and
$t_{1} > ... > t_{n}$. \cite{R1}
\end{thm}

One can show that Hall sets of $A$-labeled rooted
trees can be reconstructed recursively from an arbitrary Hall set of
words on $A$. It means that

\begin{cor}
A Hall set of words on $A$ is the image under the foliage
of a Hall set $\bold{H(T(A))}$ of labeled rooted trees.
\end{cor}

There is an important class of words (i.e. Lyndon words) such that one can deduce Hall sets from them and moreover, this kind of words can store interesting information about shuffle structures. Let us start with a well-known order.

\begin{defn}
Let $A$ be a totally ordered set. The alphabetical ordering
determines a total order on the set of words on $A$ such that for
any nonempty word $v$, put $u < uv$ and also for letters $a < b$ and
words $w_{1}, w_{2}, w_{3}$, put $w_{1}aw_{2} < w_{1}bw_{3}$.
\end{defn}

\begin{defn}
For a given total order set $A$, a non-trivial word $w$ is called Lyndon, if for any non-trivial
factorization $w=uv$, we have $w<v$.
\end{defn}

The first advantage of these words can be seen in their influencing role in making a Hall set.

\begin{thm}
The set of Lyndon words, ordered alphabetically, is a Hall set.  \cite{H1, R1}
\end{thm}

\begin{thm} \label{shuffle-order}
Let $A$ be a locally finite set equipped with a total order relation.
The (quasi-)shuffle algebra $(\mathbb{K}<A>, \star^{-})$ is the free
polynomial algebra on the Lyndon words. \cite{H1}
\end{thm}

It is shown that the universal Hopf algebra of renormalization as an algebra
is defined by the shuffle product on the linear space of
noncommutative polynomials with variables $f_{n}$ ($n \in
\mathbb{N}$). This determines an important order.

\begin{defn} \label{totalorder}
With referring to given correspondence in the result \ref{order}, one
can define a natural total order relation (depending on the degrees of
the generators $e_{-n},$ $(n \in \mathbb{N})$ of the free Lie
algebra $L_{\mathbb{U}}$) on the set $A = \{ f_{n}: n \in
\mathbb{N}_{>0} \}$. It is given by
$$f_{m} > f_{n} \Longleftrightarrow n > m.$$
\end{defn}

Theorem \ref{shuffle-order} shows that $H_{\mathbb{U}}$ (as an algebra) is the
free polynomial algebra of Lyndon words on the set $A$ and therefore one can consider
Hall set of these Lyndon words (ordered alphabetically) such
that its corresponding Hall set of labeled rooted trees is denoted
by $\bold{H(T(A))}_{\mathbb{U}}$. It can be seen obviously that the set of Lyndon words is an influencing factor in determining this
bridge between rooted trees and $H_{\mathbb{U}}$.

It is near to have our interesting rooted tree reformulation.
Let us consider free commutative algebra $\mathbb{K}[\textbf{T(A)}]$
such that the set $\{t^{r_{1}}_{1}...t^{r_{m}}_{m}: t_{1},...,t_{m}
\in \textbf{T(A)}\}$ is a $\mathbb{K}-$basis (as a graded vector
space) where each expression $t^{r_{1}}_{1}...t^{r_{m}}_{m}$ is a
forest.

\begin{defn} \label{rt-v9}
For the forest $u$ with the associated partial order set
$(\mathfrak{u}(A), \ge)$, define a coproduct given by
$$
\Delta (u)= \sum_{(\mathfrak{v}(A),\mathfrak{w}(A)) \in
R(\mathfrak{u}(A))} v \otimes w
$$
such that labeled forests $v,w$ are represented by labeled partially
ordered subsets $\mathfrak{v}(A),\mathfrak{w}(A)$ of
$\mathfrak{u}(A)$.
\end{defn}



\begin{thm} \label{rt-v10}
Coproduct \ref{rt-v9} determines a connected graded commutative Hopf
algebra structure on $\mathbb{K}[\textbf{T(A)}]$ such that the
product in the dual space $\mathbb{K}[\textbf{T(A)}]^{\star}=\{
\alpha:\textbf{T(A)} \longrightarrow^{linear} \mathbb{K} \}$
corresponds to the fixed coproduct namely, dual of the
coalgebra structure and it means that for each $\alpha, \beta \in
\mathbb{K}[\textbf{T(A)}]^{\star}$ and each forest $u$,
$$\alpha \beta (u) = (\alpha \otimes \beta) \Delta(u).$$ \cite{M6, M1}
\end{thm}

\begin{rem}
One can show that $H_{GL}$ (labeled by the set $A$) and
$\mathbb{K}[\textbf{T(A)}]$ are graded dual to each other.
\end{rem}

By using theorem \ref{rt-v10} and operation
$B^{+}_{a}$, one can show that this Hopf algebra has a universal
property.

\begin{thm} \label{rt-v11}
Let $H$ be a commutative Hopf algebra over $\mathbb{K}$
and $\{L_{a}: H \longrightarrow H \}_{a \in A}$ be a family of
$\mathbb{K}-$linear maps such that
$$\cup_{a \in A} Im L_{a} \subset
ker \epsilon_{H}$$
and
$$\Delta_{H}L_{a}(c)=L_{a}(c) \otimes
\mathbb{I}_{H} + (id_{H} \otimes L_{a})\Delta_{H}(c).$$
Then there
exists a unique Hopf algebra homomorphism $\psi_{H}:
\mathbb{K}[\textbf{T(A)}] \longrightarrow H$ such that for each $u
\in \mathbb{K}[\textbf{T(A)}]$ and $a \in A$, we have
$$\psi_{H}(B^{+}_{a}(u))=L_{a}(\psi_{H}(u)).$$ \cite{M6, M1}
\end{thm}

\begin{cor}

Theorem \ref{rt-v11} is a poset version of the combinatorial Connes-Kreimer Hopf algebra and it
means that $H_{CK}$ (labeled by the set $A$) is isomorphic to
$\mathbb{K}[\textbf{T(A)}]$.
\end{cor}

\begin{proof}
It is clearly proved based on the theorem \ref{universality-ck} (universal property of $H_{CK}$).
\end{proof}

\begin{lem}
There is a bijection between the set of non-empty words and the
set of labeled rooted trees without side-branchings.
\end{lem}

\begin{proof}
One can show that the $\mathbb{K}-$linear map $\pi$ is a Hopf
algebra homomorphism and for each $a_{1},...,a_{m} \in A$, we have
$$\pi(B^{+}_{a_{m}}...B^{+}_{a_{2}}(a_{1})) = a_{1}...a_{m}.$$
\end{proof}

\begin{cor}
The map $\pi$ is an epimorphism and for  each $\widehat{\alpha},
\widehat{\beta} \in \mathbb{K}<A>^{\star}$, $u \in
\mathbb{K}[\textbf{T(A)}]$, we have
$$<\widehat{\alpha}\widehat{\beta}, \pi(u)>=<\alpha \beta, u>$$
such that $\alpha, \beta \in \mathbb{K}[\textbf{T(A)}]^{\star}$.
\end{cor}

Everything is ready to introduce a rooted tree version of shuffle type Hopf algebras.

\begin{thm} \label{rt-v13}

(i) The (quasi-)shuffle Hopf algebra $(\mathbb{K}<A>,\star^{-})$ is
isomorphic to the quotient Hopf algebra
$\frac{\mathbb{K}[\textbf{T(A)}]}{I_{\pi}}$ such that $I_{\pi}:=Ker \pi$ is a Hopf ideal in
$\mathbb{K}[\textbf{T(A)}]$ with the generators

$<\{ \prod^{m}_{i=1} t_{i} - \sum^{m}_{i=1} t_{i} \circ \prod_{j
\neq i} t_{j}: m>1, t_{1},...,t_{m} \in \textbf{T(A)} \}> = < \{ t
\circ z + z \circ t - tz :t,z \in  \textbf{T(A)} \} \cup \{ s \circ
t \circ z + s \circ z \circ t - s \circ (tz): t,z,s \in
\textbf{T(A)} \}>= <\{ t \circ z + z \circ t -tz : t,z \in
\textbf{T(A)} \} \cup \{ s \circ (tz) + z \circ (ts) + t \circ (sz)
- tzs: t,z,s \in \textbf{T(A)}\}>.$

(ii) As an $\mathbb{K}-$algebra,
$\frac{\mathbb{K}[\textbf{T(A)}]}{I_{\pi}}$ is freely generated by the set
$$\{t + I_{\pi}: t \in \bold{H(T(A))}\}.$$  \cite{M6,M1}
\end{thm}

With attention to the given operadic picture
from Connes-Kreimer Hopf algebra in the past chapters, next result can be indicated immediately.

\begin{cor} \label{rt-v14}
For each locally finite set $A$ together with a total order
relation, there exist Hopf ideals $J_{1}, J_{2}$ such that
$$(\mathbb{K}<A>,\star^{-}) \cong \frac{\mathbb{K}[\textbf{T(A)}]}{I_{\pi}} \cong \frac{H_{CK}(A)}{J_{1}} \cong \frac{H_{NAP}(A)}{J_{2}}.$$
So universal Hopf algebra of renormalization is isomorphic to a
quotient of the (labeled) incidence Hopf algebra with respect to the
basic set operad NAP.
\end{cor}

\begin{proof}
By theorems \ref{nap-ck}, \ref{rt-v11}, it is enough to set $J_{1}:=I_{\pi}$ and
$J_{2}:=\rho^{-1}{J_{1}}$. In addition, theorem \ref{rt-v13} is in fact a representation of
$H_{U}$ by rooted trees. Because it is enough to replace the set $A$ with
the variables $f_{n}$ such that the identified Lyndon words (with
the shuffle structure of $H_{\mathbb{U}}$) gives us the Hall set
$\bold{H(T(A))}_{\mathbb{U}}$.
\end{proof}

In continue of this part, we use this new redefinition from universal Hopf algebra of renormalization to provide some interesting relations between this Hopf algebra and other combinatorial Hopf algebras.

\begin{prop} \label{diag}
Rooted tree reformulation of $H_{\mathbb{U}}$ determines the
following commutative diagrams of Hopf algebra
homomorphisms.

\begin{gather}
\begin{CD}
NSYM @>{\beta_{1}}>> H_{F}\\
@V{\beta_{3}}VV @V{\beta_{2}}VV\\
SYM @>{\beta_{4}}>> H_{\mathbb{U}}
\end{CD}
\ \ \ \ \
\begin{CD}
SYM @<{\beta_{4}^{\star}}<< U(L_{\mathbb{U}})\\
@V{\beta_{3}^{\star}}VV @V{\beta_{2}^{\star}}VV\\
QSYM @<{\beta_{1}^{\star}}<< H_{\bold{P}}
\end{CD}
\end{gather}
\end{prop}

\begin{proof}

It is easy to see that $SYM \subset QSYM$. As a vector
space, $QSYM$ is generated by the monomial quasi-symmetric functions
$M_{I}$ such that $I=(i_{1},...,i_{k})$ and $M_{I}:=\sum_{n_{1} <
n_{2} < ... < n_{k}} x^{i_{1}}_{n_{1}} ... x^{i_{k}}_{n_{k}}$ and if
we forget order in a composition, then the
generators $m_{\lambda} := \sum_{\bold{c}(I)= \lambda} M_{I}$ of $SYM$
(viewed as a vector space) can be obtained where the map $\bold{c}$ maps compositions to partitions that forgets order.
The Hopf algebra structure on $QSYM$ is determined by the coproduct
\begin{equation}
\Delta(M_{(i_{1},...,i_{k})}) = \sum_{j=0}^{k} M_{(i_{1},...,i_{j})} \otimes M_{(i_{j+1},...,i_{k})}.
\end{equation}
Therefore the power-sum symmetric functions $M_{i}$ are primitives and elementary symmetric functions $M_{(1,...,1)}$ are divided powers.
And also as an algebra, $NSYM$ is the noncommutative polynomials on
the variables $z_{n}$ of degree $n$ such that its Hopf algebra structure can be determined by supposing the generators $z_{n}$ divided powers. \cite{GKLLRT, H2, H5, H6}

On the other hand, for each (planar) rooted tree $t$, one can put different decorations
(with elements of the locally finite set $A=\{f_{n}\}_{n \ge 1}$) on its
vertices. Let $[t]$ be the class of all different possible Hall
(planar) rooted trees with respect to $t$ in $\bold{H(T(A))}_{\mathbb{U}}$ such that $[\mathbb{I}]=\mathbb{I}$.
For the forest $u$, supposing $[u]$ be the class of all different Hall
forests associated to $u$ in $\bold{H(F(A))}_{\mathbb{U}}$. It means that if $u=t_{1}t_{2}...t_{n}$, then $[u]=[t_{1}]...[t_{n}]$.
Now based on the structures of the mentioned Hopf algebras and the given commutative diagrams (6.1) and (6.2) in \cite{H2}, the following diagrams explain a sketch of expected homomorphisms.
\begin{equation}
\xy
(-5,0)*{z_{n}},
(5,0)*{\rightarrow},
(15,0)*{l_{n}},
(-5,-7)*{\downarrow},
(-10,-15)*{m_{\underbrace{(1,...,1)}_{n}}},
(5,-15)*{\rightarrow},
(20,-15)*{\sum_{v \in [l_{n}]}
\pi(v)},
(15,-7)*{\downarrow},
(45,0)*{m_{\underbrace{(1,...,1)}_{n}}},
(60,0)*{\leftarrow},
(70,0)*{e_{-n}},
(70,-7)*{\downarrow},
(45,-7)*{\downarrow},
(70,-15)*{l_{n}},
(45,-15)*{M_{(1,...,1)}},
(60,-15)*{\leftarrow},
\endxy
\end{equation}
Since the Connes-Kreimer coproduct is such that the ladder trees $l_{n}$ are divided powers and the Hopf algebra structure of $H_{F}$ is determined by this coproduct,    therefore by sending the divided powers $z_{n}$ to $l_{n}$, one can define a Hopf algebra homomorphism $\beta_{1}$ from $NSYM$ to $H_{F}$.

Corollary \ref{rt-v14} provides a new picture from $H_{\mathbb{U}}$ based on the labeled Hopf algebra $H_{CK}(A)$ (i.e. definition \ref{rt-v9}). Now for each forest $u=t_{1}...t_{n}$ of labeled planer rooted trees in $H_{F}$, define
\begin{equation}
\beta_{2}(u):= \sum_{v \in [u]} \pi(v).
\end{equation}
The Connes-Kreimer coproduct can be defined based on admissible cuts. It means that for each rooted tree $t$, its coproduct is given by
\begin{equation}
\Delta(t)= \mathbb{I} \otimes t + t \otimes \mathbb{I} + \sum_{c} R_{c}(t) \otimes P_{c}(t)
\end{equation}
such that the sum is over all proper admissible cuts of $t$, $R_{c}(t)$ is a rooted tree with the origin root of $t$ (i.e. that part of the tree which remains connected with the root after applying the cut $c$) and $P_{c}(t)$ is a forest of rooted trees \cite{CK4, EGK1, EGK2}. Both $H_{CK}$ and $H_{F}$ have the same type coproduct and they are connected and graded. So their antipodes are defined naturally from this coproduct.
On the other hand, it can be seen that
\begin{equation}
[\Delta(t)] = [\mathbb{I}][t] + [t][\mathbb{I}] + \sum_{c} [R_{c}(t)] [P_{c}(t)]
\end{equation}
Now since $\pi$ is a Hopf algebra morphism, we have
\begin{equation}
\beta_{2}(t_{1}...t_{n}) = \sum_{s_{1} \in [t_{1}], ..., s_{n} \in [t_{n}]} \pi(s_{1}...s_{n}) = \sum_{s_{1} \in [t_{1}]} \pi(s_{1}) ... \sum_{s_{n} \in [t_{n}]} \pi(s_{n}) = \beta_{2}(t_{1})...\beta_{2}(t_{n}),
\end{equation}
\begin{equation}
\Delta(\beta_{2}(t)) = \Delta(\sum_{s \in [t]} \pi(s)) = \sum_{s \in [t]} \Delta \pi(s)= \sum_{\Delta(s) \in [\Delta(t)]} \pi(\Delta(s))=\beta_{2}(\Delta(t)).
\end{equation}
Therefore $\beta_{2}$ is a Hopf algebra homomorphism.

We know that $m_{(p_{1},...,p_{k})} = \sum_{i_{1},...,i_{k}} x_{i_{1}}^{p_{1}} x_{i_{2}}^{p_{2}}...x_{i_{k}}^{p_{k}}$ and elementary symmetric functions $m_{(1,...,1)}$ are divided powers. On the other hand, by surjective morphism $\pi$, labeled ladder trees are representing divided powers in $H_{\mathbb{U}}$. Therefore the map $\beta_{4}$ that sending each $m_{(1,...,1)}$ to $\sum_{v \in [l_{n}]} \pi(v)$ is a Hopf algebra morphism.

About $\beta_{3}$, it is the abelianization homomorphism that sends each $z_{n}$ to the elementary symmetric function of degree $n$.

For the given diagram at the graded dual level, Hopf algebra structure on $U(L_{\mathbb{U}})$ can be determined by theorem \ref{2}.  The map $\beta_{3}^{\star}$ is just the inclusion map and the morphism $\beta_{1}^{\star}$ is also defined by Hoffman in \cite{H2} such that for each planar rooted tree $t_{J}:=B^{+}(l_{j_{1}},...,l_{j_{k}})$ (where $J=(j_{1},...,j_{k})$), $\beta^{\star}_{1}(t_{J}):=M_{J}$.

For the given balanced bracket representations $c=c_{1}...c_{n}, \  \  c'=c'_{1}...c'_{n}$, the product $c.c'$ in $H_{\bold{P}}$ is given by shuffling the symbols $c_{1}, ..., c_{n}$ into $c'$ and the coproduct structure is given by
\begin{equation}
\Delta(c) = \sum_{i=0}^{n} c_{1}...c_{i} \otimes c_{i+1}...c_{n}.
\end{equation}
Note that Foissy Hopf algebra $H_{F}$ is self-dual and so it is isomorphic to $H_{\bold{P}}$.

For each $n$, the set $\{t \in \beta_{2}^{-1}(w), \  \ deg(w)=n\}$ has only one element $l_{n}$. Now
the other two morphisms are imposed from given morphisms by Hoffman in \cite{H2} in the sense that for each generator $e_{-n}$ of degree $n$,
\begin{equation}
\beta_{2}^{\star}(e_{-n}):= l_{n}, \  \  \beta_{4}^{\star}(e_{-n}):= m_{\underbrace{(1,...,1)}_{n}}.
\end{equation}
The maps $\beta_{2}, \ \beta^{\star}_{2}$ show this note that there is a correspondence $e_{-n} \leftrightarrow \sum f_{k_{1}}...f_{k_{n}}$ between the generators of the Lie algebra $U(L_{\mathbb{U}})$ and Lyndon words related to $H_{U}$. As the result,
one can say that $e_{-n}$s are divided powers that map to divided powers in $H_{\bold{P}}$ and $SYM$, respectively. So $\beta_{2}^{\star}, \ \beta_{4}^{\star}$ are Hopf algebra morphisms.

\end{proof}

\begin{rem}
Holtkamp in \cite{H4} proved that the Foissy Hopf algebra $H_{F}$ is isomorphic to the Hopf algebra $\mathbb{C}[Y_{\infty}]$ of planar binary trees (\cite{LR2}) and moreover, Foissy in \cite{F4} found an isomorphism between $H_{F}$ and photon Hopf algebra $H^{\gamma}$ (\cite{BF}) (related to renormalization). With the help of these facts and proposition \ref{diag}, one can find new homomorphisms between these Hopf algebras and $H_{\mathbb{U}}$.
\end{rem}

Now we want to lift the Zhao's homomorphism given in the definition \ref{zhao} and its
dual to the level of the universal Hopf algebra of renormalization.
It is shown that $\pi$ is a surjective homomorphism from $H_{CK}(A)$ to
$H_{\mathbb{U}}$. On the other hand, $Z^{\star}$ provides a unique surjective
map from $H_{CK}$ to $QSYM$ with the property \ref{zhao3}. For a
word $w$ with length $n$ in $H_{\mathbb{U}}$, there exists a labeled ladder
tree $l^{w}_{n}$ of degree $n$ in $H_{CK}(A)$ such that
$\pi(l^{w}_{n})=w$.

\begin{defn} \label{exten-zhao1}
Define a new map $Z_{u}: H_{\mathbb{U}} \longrightarrow
QSYM$ such that for each element $w \in H_{\mathbb{U}}$, it is given by
$Z_{u}(w):=Z^{\star}(l^{w}_{n})$.
\end{defn}

It can be seen that $Z_{u}$ is a homomorphism of Hopf algebras and
it is unique with respect to the relation \ref{zhao3}. With
the help of theorem \ref{rel-tree-symm} and proposition \ref{diag}, one can define
homomorphisms
\begin{equation}
\theta_{1}: NSYM \longrightarrow H_{\mathbb{U}}, \ \ \theta_{1}:= \beta_{4} \circ \beta_{3} = \beta_{2}
\circ \beta_{1},
\end{equation}
\begin{equation}
\theta_{2}: H_{GL} \longrightarrow H_{\mathbb{U}}, \ \  \theta_{2}:= \beta_{4}
\circ \alpha_{4}^{\star} = \beta_{2} \circ \alpha^{\star}_{2}.
\end{equation}

\begin{lem}
The surjective morphism
$\pi$ induces a new homomorphism $\Xi$ from $H_{CK}$ to $H_{\mathbb{U}}$
such that for each unlabeled forest $u$ in $H_{CK}$, it is defined
by
$$
\Xi(u):= \sum_{v \in [u]} \pi(v).
$$
\end{lem}

\begin{prop}

Rooted tree reformulation of $H_{\mathbb{U}}$ determines the following commutative diagram.

\begin{equation}  \label{hex1}
\begin{split}
\xymatrixcolsep{3pc} \xymatrix{
& SYM  \ar[ld]_{\alpha_{4}} \ar[dd]_{\beta_{4}} \\
H_{CK}  \ar[rd]^{\Xi}  & & H_{GL}  \ar[ul]_{\alpha_{4}^{\star}} \ar[ld]^{\theta_{2}}\\
& H_{\mathbb{U}} \ar[ld]_{Z_{u}}   &\\
QSYM   & & NSYM  \ar[ld]^{\alpha_{3}} \ar[lu]^{\theta_{1}}\\
& SYM \ar[lu]^{Z_{u} \circ \beta_{4}} \ar[uu]^{\beta_{4}} }
\end{split}
\end{equation}
\end{prop}

It is also possible to consider the dual version of the above diagram.

\begin{lem} \label{dualunzh}
Dual of the maps $Z_{u}$ and $\Xi$ can be give by
$$Z_{u}^{\star}: NSYM \longrightarrow U(L_{\mathbb{U}}), \ Z_{u}^{\star}(z_{n}):=e_{-n},
$$
$$\Xi^{\star}: U(L_{\mathbb{U}}) \longrightarrow H_{GL}, \  \Xi^{\star}(e_{-n}):= l_{n}.$$
\end{lem}

\begin{prop}
The dual version
of the diagram (\ref{hex1}) is given by
\begin{equation} \label{hex2}
\begin{split}
\xymatrixcolsep{3pc} \xymatrix{
& SYM  \ar[rd]^{\alpha_{4}} \\
H_{GL} \ar[ur]^{\alpha_{4}^{\star}} & & H_{CK} \\
& U(L_{\mathbb{U}}) \ar[uu]_{\beta_{4}^{\star}}
\ar[dd]^{\beta_{4}^{\star}}
\ar[lu]_{\Xi^{\star}} \ar[rd]_{\theta_{1}^{\star}} \ar[ru]_{\theta_{2}^{\star}} &\\
NSYM \ar[ru]_{Z_{u}^{\star}} \ar[rd]_{(Z_{u} \circ \beta_{4})^{\star}}  & & QSYM \\
& SYM \ar[ru]_{\alpha_{3}^{\star}} }
\end{split}
\end{equation}
\end{prop}

Based on the graded dual relation, there is another procedure to obtain other connections between $H_{U}$ and combinatorial Hopf algebras.

\begin{lem} \label{relation-dual}
Let $H_{1}$ and $H_{2}$ be graded connected locally finite Hopf
algebras which admit inner products $(.,.)_{1}$ and $(.,.)_{2}$,
respectively. If they are dual to each other, then there is a linear
map $\lambda: H_{1} \longrightarrow H_{2}$ such that

(i) $\lambda$ preserves degree,

(ii) For each $h_{1}, h_{2} \in H_{1}:$
$(h_{1},h_{2})_{1}=(\lambda(h_{1}),\lambda(h_{2}))_{2},$

(iii) For each $h_{1}, h_{2},  h_{3} \in H_{1}:$
$$(h_{1}h_{2},h_{3})_{1}=(\lambda(h_{1}) \otimes \lambda(h_{2}), \Delta_{2}(\lambda(h_{3})))_{2},$$
$$(h_{1} \otimes h_{2}, \Delta_{1}(h_{3}))_{1} = (\lambda(h_{1})\lambda(h_{2}), \lambda(h_{3}))_{2}.$$ \cite{H2}
\end{lem}

\begin{rem}  \label{iso}
Lemma \ref{relation-dual} determines an isomorphism $\tau: H_{2}
\longrightarrow H_{1}^{\star}$ such that for each $h_{1} \in H_{1}$
and $h_{2} \in H_{2}$, it is defined by
$$<\tau(h_{2}),h_{1}>:=(h_{2},\lambda(h_{1}))_{2}.$$
\end{rem}

Since $H_{CK}(A)$ and $H_{GL}(A)$ are graded dual to each other,
therefore by \ref{relation-dual} one can find a linear map
$\lambda:H_{CK}(A) \longrightarrow H_{GL}(A)$ with the mentioned
properties such that it defines an isomorphism $\tau_{1}$ from
$H_{GL}(A)$ to $H_{CK}(A)^{\star}$ given by

\begin{equation}
<\tau_{1}(t),s>:=(t, \lambda(s))
\end{equation}
where for rooted trees $t_{1},t_{2}$, if $t_{1}=t_{2}$ then
$(t_{1},t_{2})=|sym(t_{1})|$ and otherwise it will be $0$. For each
word $w$ with length $n$ in $H_{\mathbb{U}}$, there is a labeled ladder tree
$l^{w}_{n}$ in $H_{CK}(A)$ such that $\pi(l^{w}_{n})=w$.

\begin{prop} \label{alpha}
There is a homomorphism of Hopf algebras $F: H_{\mathbb{U}} \longrightarrow H_{GL}(A)$ given by
$$F(w):= \tau_{1}^{-1}((l^{w}_{n})^{\star}).$$
\end{prop}

We know that $H_{\mathbb{U}}$ and $U(L_{\mathbb{U}})$ are graded dual to each other. And
therefore lemma \ref{relation-dual}  introduces a linear map $\theta: H_{\mathbb{U}}  \longrightarrow
U(L_{\mathbb{U}})$ with the mentioned properties. This morphism determines an
isomorphism $\tau_{2}$ from $U(L_{\mathbb{U}})$ to $H_{\mathbb{U}}^{\star}$
such that for each element $x \in U(L_{\mathbb{U}})$ and
word $w \in H_{\mathbb{U}}$, we have
\begin{equation}
<\tau_{2}(x),w>:=(x, \theta(w))
\end{equation}
where $(.,.)$ is the natural pairing on $U(L_{\mathbb{U}})$.

\begin{prop} \label{alpha-dual}
For
a labeled forest $u$ with the corresponding element $\pi(u)$ in $H_{\mathbb{U}}$, based on the natural pairing (given in theorem \ref{shuffle3}), the dual of $F$ is
clarified by
$$ F^{\star}: H_{CK}(A) \longrightarrow  U(L_{\mathbb{U}}),$$
$$F^{\star}(u):= \tau_{2}^{-1}((\pi(u))^{\star}).$$
\end{prop}

It should be remarked that one can extend this rooted tree reformulation from $H_{\mathbb{U}}$ to its related affine group scheme $\mathbb{U}$ where it is done
by the theory of construction of a group from an operad \cite{R2}. Let $P$ be
an augmented set operad and $\mathbb{K}A_{P}= \bigoplus_{n}
\mathbb{K} (P_{n})_{\mathbb{S}_{n}}$ be the direct sum of its
related coinvariant spaces with the completion
$\widehat{\mathbb{K}A_{P}}= \prod_{n} \mathbb{K}
(P_{n})_{\mathbb{S}_{n}}$.

\begin{lem}
There is an associative monoid structure
on $\widehat{\mathbb{K}A_{P}}$. \cite{CL1, S1}
\end{lem}

Let $\mathcal{G}_{P}$ be the set of
all elements of $\widehat{\mathbb{K}A_{P}}$ whose first component is
the unit $\bold{1}$. It is a subgroup of the set of invertible
elements. One can generalize this notion to a categorical level.

\begin{thm}
There is a functor from the
category of augmented operads to the category of groups. \cite{CL1, S1}
\end{thm}

Naturally, one can expect a Hopf algebra based on subgroup $\mathcal{G}_{P}$.

\begin{thm}
For a given operad $P$, we have a commutative Hopf algebra structure on
$\mathbb{K}[\mathcal{G}_{P}]$ given by the set of coinvariants of
the operad. \cite{CL1, S1}
\end{thm}

\begin{proof}
It is a free commutative algebra of functions on
$\mathcal{G}_{P}$ generated by the set $(g_{\alpha})_{\alpha \in
A_{P}}$. Each $f \in \mathcal{G}_{P}$ can be represented by a formal
sum $f = \sum_{\alpha \in A_{P}} g_{\alpha}(f) \alpha$ such that
$g_{\bold{1}}=1$.
\end{proof}

Relation with incidence Hopf algebra is the first essential property of this Hopf algebra. It can be seen that there is a surjective morphism
\begin{equation} \label{group-level1}
\eta: g_{\alpha} \longmapsto \frac{F_{[\alpha]}}{|Aut(\alpha)|}
\end{equation}
from the Hopf algebra $\mathbb{K}[\mathcal{G}_{P}]$ to the incidence
Hopf algebra $H_{P}$. Further, based on this morphism it is observed that at the level of groups,
the Lie group $G_{P}(\mathbb{K})$ of $H_{P}$ is a subgroup
of the group $\mathcal{G}_{P}$. \cite{CL1, V1}

\begin{cor} \label{group-level2}
(i) There is a surjective morphism from the Hopf algebra
$\mathbb{K}[\mathcal{G}_{NAP}]$ to the Connes-Kreimer Hopf algebra
$H_{CK}$ of rooted trees.

(ii) The complex Lie group $G(\mathbb{C})$ of $H_{CK}$ is a
subgroup of $\mathcal{G}_{NAP}$.
\end{cor}

\begin{proof}
With referring to theorem \ref{nap-ck} and (\ref{group-level1}), the map $\rho \circ \eta:
\mathbb{K}[\mathcal{G}_{NAP}] \longrightarrow H_{CK}$ is a
surjective morphism and in terms of groups we will get the second
claim.
\end{proof}

For the operad $NAP$, $\mathcal{G}_{NAP}$ is a group of formal power
series indexed by the set of unlabeled rooted trees and there is
also an explicit picture from the elements of $G_{NAP}(\mathbb{C})$
such that one can lift it to the level of universal Hopf algebra of
renormalization.

\begin{cor}  \label{group-level3}
(i) Group $\mathbb{U}(\mathbb{C})$ is a
subgroup of $\mathcal{G}_{NAP}$ and therefore each of its elements
can be represented by a formal power series indexed with Hall
rooted trees.

(ii) Formal series $K=\sum_{t} g_{t}(K)t$ in $\mathcal{G}_{NAP}$ belongs to
$\mathbb{U}(\mathbb{C})$ if and only if

- $t$ be a Hall tree in $\bold{H(T(A))}_{\mathbb{U}}$, (It does not belong to
the Hopf ideal $I_{\pi}$.)

- And if $t=B^{+}_{f_{n}}(u)$ for some $u=t_{1}...t_{k}$ such that
$t_{1},...,t_{k} \in \bold{H(T(A))}_{\mathbb{U}}$ and $f_{n} \in A$, one has
$$g_{t}(K)=\prod^{k}_{i=1} g_{B^{+}_{f_{n}}(t_{i})}(K).$$
\end{cor}

\begin{proof}

One can extend the morphism (\ref{group-level1}) to the level of the
decorated Hopf algebras and therefore by corollaries \ref{rt-v14} and
\ref{group-level2}, there is a surjective map from
$\mathbb{C}[\mathcal{G}_{NAP}](A)$ to $H_{\mathbb{U}}$. So it shows that in
terms of groups, $\mathbb{U}(\mathbb{C})$ is a subgroup of
$\mathcal{G}_{NAP}$. For the second case, according to the lemma
6.12 in \cite{CL1}, each element of $\mathcal{G}_{NAP}$ is in the
subgroup $G_{NAP}(\mathbb{C})$ if and only if for each tree
$t=B^{+}(t_{1},...,t_{k})$, we have the following condition
$$ |sym(t)|g_{t}(K)=\prod^{k}_{i=1} |sym(B^{+}(t_{i}))|g_{B^{+}(t_{i})}(K).$$
By theorems \ref{nap1}, \ref{nap-ck}, corollaries \ref{rt-v14}, \ref{group-level2} and (\ref{group-level1}),
and since Hall trees have no symmetries, the proof
is completed.
\end{proof}

\begin{rem}
The universal affine group scheme $\mathbb{U}^{*}:=\mathbb{U} \rtimes \mathbb{G}_{m}$ has a manifestation as a motivic Galois group for a category of mixed Tate motives. Suppose $\mathbb{Q}(\zeta_{N})$ be the cyclotomic field of level $N$ and $\mathcal{O}$ be its ring of integers. For $N=3$ or $4$, the motivic Galois group of the category $\mathcal{MT}_{mix}(S_{N})$ of mixed Tate motives where $S_{N}= Spec (\mathcal{O}[\frac{1}{N}])$ is non-canonically of the form $\mathbb{U}^{*}$. Moreover for the scheme $S_{N}=Spec(\mathbb{Z}[i][\frac{1}{2}])$ of $N-$cyclotomic integers, one can find a non-canonical isomorphism between $\mathbb{U}^{*}$ and the motivic Galois group $G_{\mathcal{M}_{T}}(\mathbb{Z}[i][\frac{1}{2}])$  \cite{CM3, CM1, G2, G1}. So with applying corollaries \ref{group-level2} and \ref{group-level3} one can translate this Hall tree formalism to these motivic Galois groups (at the level of groups) to obtain a new Hall tree type representation from their elements.
\end{rem}

And finally, let us have a remark about Lie algebra version. It will help us to characterize PBW basis and Hall basis with respect to the Hall set $\bold{H(T(A))}_{\mathbb{U}}$.

\begin{defn}
Let $H(A)$ be a Hall set and $A^{\star}:=\{ a^{\star}: a \in A \}$.
For each Hall word $w$, its associated Hall polynomial $p_{w}$
in the free Lie algebra $\mathcal{L}(A^{\star})$ is introduced by

- If $f_{k_{j}} \in A$, then $p_{f_{k_{j}}}=f_{k_{j}}^{\star}$,

- If $w$ be a Hall word of length $\ge 2$ such that its
corresponding Hall tree $t_{w}$ has the standard decomposition
$(t_{w_{1}},t_{w_{2}})$, then $p_{w}=[p_{w_{1}},p_{w_{2}}]$.
\end{defn}

\begin{lem}
(i) With induction one can show that each $p_{w}$ is a homogeneous Lie
polynomial of degree equal to the length of $w$ and also, it has the
same partial degree with respect to each letter as $w$.

(ii) For a given Hall set $H$, Hall polynomials form a basis for the free
Lie algebra (viewed as a vector space) and their decreasing products
$p_{f_{k_{1}}}...p_{f_{k_{n}}}$ such that $f_{k_{i}} \in H, \ \
f_{k_{1}} > f_{k_{2}} > ... > f_{k_{n}}$, form a basis for the free associative
algebra (viewed as a vector space) \cite{R1}.
\end{lem}

We found a Hall set $\bold{H(T(A))}_{\mathbb{U}}$ corresponding to $H_{\mathbb{U}}$ such that
$A=\{ f_{n}\}_{n \in \mathbb{N}_{>0}}$ and for each $f_{n}$ its
associated Hall polynomial is given by
\begin{equation}
p_{f_{n}}=e_{-n}.
\end{equation}
The next result completes our Hall tree reconstruction.

\begin{cor}  \label{lie-level}
(i) As a vector space, Hall polynomials associated to the Hall set
$\bold{H(T(A))}_{\mathbb{U}}$ form a basis for the Lie algebra
$L_{\mathbb{U}}$.

(ii) As a vector space, decreasing products of Hall polynomials with
respect to the Hall set
$\bold{H(T(A))}_{\mathbb{U}}$ form a basis for the free algebra $H_{\mathbb{U}}$.
\end{cor}

Therefore in summary, based on the basic set operad NAP on rooted trees, one can provide a complete reconstruction from the specific Hopf algebra $H_{\mathbb{U}}$ in the algebraic perturbative renormalization, its related infinite dimensional complex Lie group and also Lie algebra (with notice to the Milnor-Moore theorem). Now it can be useful to apply this induced version in the Connes-Marcolli's universal treatment such that consequently, new reformulations from physical information such as counterterms can be expected. This project is actually performed by universal singular frame.


\section{{\it \textsf{Universal singular frame}}}

The appearance of the Connes-Marcolli categorification of renormalization theory depends directly upon the Riemann-Hilbert problem such that as the result one can search the application of the theory of motives in describing divergences. Neutral Tannakian formalism of the category of flat equi-singular vector bundles could determine the special loop $\gamma_{\mathbb{U}}$ with values in the universal affine group scheme $\mathbb{U}(\mathbb{C})$. Since now we have enough knowledge about the combinatorics of the universal Hopf algebra of renormalization, so it does make sense to search for a combinatorial nature inside of the universal singular frame.
The explanation of this important frame (free of any divergences) based on Hall basis and PBW basis is the main purpose of this section such that
universality of $\gamma_{\mathbb{U}}$ allows us to modify this new Hall type representation to counterterms of any renormalizable theory. \cite{S6}

Starting with the application of words in geometry.
Let $M$ be a smooth manifold, $C(\mathbb{R}^{+})$ the ring of
real valued piecewise continuous functions on $\mathbb{R}^{+}$ and
$\{X_{a}\}_{a \in A}$ a family of smooth vector fields on $M$.
Suppose $\mathcal{A}$ be the algebra over $C(\mathbb{R}^{+})$ of
linear operators on $C^{\infty}(M)$ generated by the vector fields
$X_{a}$ ($a \in A$). For a family $\{g_{a}\}_{a \in A}$ of elements
in $C(\mathbb{R}^{+})$, set
\begin{equation} \label{f11}
X(x)=\sum_{a \in A} g_{a}(x)X_{a}.
\end{equation}

\begin{lem} \label{f12}
The formal series (\ref{f11}) can be expanded as a series of linear operators in $\mathcal{A}$
of the form $\sum_{w} g_{w}X_{w}$ such that

- $w=a_{1}...a_{m}$ is a word on $A$,

- $X_{1}=Id$ (identity operator), $X_{w}=X_{a_{1}}...X_{a_{m}}$,

- $g_{w} =  \int_{a_{m}} ... \int_{a_{1}} 1_{C(\mathbb{R}^{+})}$
where each $\int_{a_{i}}: C(\mathbb{R}^{+}) \longrightarrow
C(\mathbb{R}^{+}),$ $(1 \le i \le m)$ is a linear endomorphism
defined by
$$\{ \int_{a_{i}}g \}(x):= \int_{0}^{x} g(s)g_{a_{i}}(s)ds.$$
\end{lem}

This technique can be generalized to any arbitrary algebra.

\begin{thm}
For a given associative algebra $\mathcal{A}$ over the
field $\mathbb{K}$ of characteristic zero generated by the elements $\{E_{a}\}_{a
\in A}$, all elements in $\mathcal{A}$ are characterized by formal
series $\sum_{w} \mu_{w}E_{w}$ such that $\mu_{w} \in \mathbb{K}$. \cite{M1, M6, R1}
\end{thm}

\begin{rem}
If $\mathcal{A}$ is a free algebra, then this formal series representation is
unique.
\end{rem}

\begin{defn}
Suppose $\mathfrak{a}$ be the unital Lie algebra generated by the set $\{E_{a}\}_{a \in A}$.
For a given Hall set $\bold{H(T(A))}$ of labeled rooted
trees with the corresponding Hall forest $\bold{H(F(A))}$, one can
assign elements $E(u)$ such that

- $E(\mathbb{I})=\mathfrak{e}$ (The unit of
$\mathfrak{a}$),

- For each Hall tree $t$ with the standard decomposition
$(t^{1},t^{2}) \in \bold{H(T(A))} \times \bold{H(T(A))}$,
$$
E(t)=[E(t^{2}),E(t^{1})]=E(t^{2})E(t^{1}) - E(t^{1})E(t^{2}),
$$

- For each $u \in \bold{H(F(A))} - \bold{H(T(A))}$ with the standard
decomposition $(u^{1},u^{2}) \in \bold{H(F(A))} \times
\bold{H(T(A))}$,
$$
E(u)=E(u^{2})E(u^{1}).
$$
\end{defn}

\begin{lem} \label{rt-phy1}

(i) The Lie algebra $\mathfrak{a}$ is spanned by $\{ E(t): t
\in \bold{H(T(A))}\}$ (as an ordered basis). It is called {\it Hall basis}.

(ii) $\mathcal{A}$ is spanned by $\{ E(u): u \in
\bold{H(F(A))}\}$. It is called {\it PBW basis}. \cite{M6, M1, R1}
\end{lem}

It is the place to give a new understanding from the universal singular frame.

\begin{prop} \label{rt-phy2} \label{f15}
For the locally finite total order set $\{f_{n}\}_{n \in
\mathbb{N}}$, the universal singular frame can be rewritten by
$$\gamma_{U}(-z,v)= \sum_{n \ge 0, k_{j}>0} \alpha^{U}_{f_{k_{1}}f_{k_{2}}...f_{k_{n}}} p_{f_{k_{1}}}...p_{f_{k_{n}}} v^{\sum k_{j}} z^{-n} $$
such that $p_{f_{k_{j}}}$s are Hall polynomials.
\end{prop}

\begin{proof}
From theorem \ref{cat.9}, we know that
$$\gamma_{\mathbb{U}}(z,v)=Te^{-\frac{1}{z} \int^{v}_{0} u^{Y}(e) \frac{du}{u}}.$$
After the application of the time ordered exponential, one can get
$$
\gamma_{\mathbb{U}}(-z,v)= \sum_{n \ge 0, k_{j}>0}
\frac{e_{-k_{1}}...e_{-k_{n}}}{k_{1}(k_{1}+k_{2})...(k_{1}+...+k_{n})}
v^{\sum k_{j}} z^{-n}
$$
such that in this expansion the coefficient of $e_{-k_{1}}...e_{-k_{n}}$ is
calculated by the iterated integral
$$
\int_{0 \le s_{1} \le ... \le s_{n} \le 1}
s_{1}^{k_{1}-1}...s_{n}^{k_{n}-1} ds_{1}...ds_{n}.
$$
On the other hand, theorem \ref{shuffle-order} determines that Hopf algebra $H_{\mathbb{U}}$ is a free polynomial
algebra on Lyndon words on the set $\{f_{n}\}_{n \in
\mathbb{N}_{>0}}$. Consider formal series
$$
E:= f_{k_{1}}+x f_{k_{2}}+x^{2} f_{k_{3}}+...
$$
where
$$
\mu_{k_{j}}(x)=x^{k_{j}-1}.
$$
By the lemma \ref{f12}, for the variables $0 \le s_{1} \le ... \le s_{n} \le
1$, we have
$$
\{\int_{k_{j}} 1\}(s_{j})= \int_{0}^{s_{j}} x^{k_{j}-1}dx.
$$
For each word $f_{k_{1}}f_{k_{2}}...f_{k_{n}}$, we can define the
following well-defined iterated integral
$$
\alpha^{\mathbb{U}}_{f_{k_{1}} f_{k_{2}}...f_{k_{n}} } := \int_{k_{n}} ...
\int_{k_{1}} 1.
$$
It is easy to see that the above integral is agree with the iterated
integral associated to the coefficient of the term
$e_{-k_{1}}...e_{-k_{n}}$. This completes the proof.

\end{proof}

In addition, one can determine uniquely a real
valued map on the set $\bold{F(A)}$.

\begin{defn} \label{un}
Let $A=\{f_{n}\}_{n \in \mathbb{N}}$ be the locally finite total
order set corresponding to the universal Hopf algebra of
renormalization. For the given map in the above proof that associates to
each word $w=f_{k_{1}}f_{k_{2}}...f_{k_{n}}$ a real value
$\alpha^{\mathbb{U}}_{w}$ and based on the definition \ref{rt-v6}, we introduce a new map
$\alpha^{\mathbb{U}}$ on $\bold{F(A)}$ such that

- $\mathbb{I} \longmapsto \alpha^{\mathbb{U}}(\mathbb{I})=1,$

- For each non-empty labeled forest $u$ in $\bold{F(A)}$,
$$ \alpha^{\mathbb{U}}(u)= \sum_{>_{\mathfrak{u}(A)}} \alpha^{\mathbb{U}}_{w(>_{\mathfrak{u}(A)})}. $$
\end{defn}

\begin{lem} \label{uns}
For given labeled rooted trees $t_{1},...,t_{m} \in
\bold{T(A)}$ and $f_{k_{j}} \in A$,
$$
\alpha^{\mathbb{U}}(t_{1}...t_{m})=\alpha^{\mathbb{U}}(t_{1})...\alpha^{\mathbb{U}}(t_{m}), \ \
\alpha^{\mathbb{U}}(B^{+}_{f_{k_{j}}}(t_{1}...t_{m}))=\int_{k_{j}}
\alpha^{\mathbb{U}}(t_{1})...\alpha^{\mathbb{U}}(t_{m}).
$$
The map $\alpha^{\mathbb{U}}$ (determined by the universal singular frame),
with the above properties, can be uniquely identified.
\end{lem}


\begin{lem} \label{betaa}
For the given map in definition
\ref{un} together with the mentioned properties in lemma \ref{uns}, there exists a real valued
map $\beta^{\mathbb{U}}$ on $\bold{F(A)}$ given by
$$ \alpha^{\mathbb{U}}= exp \ \beta^{\mathbb{U}}$$
such that for each $u \in \bold{F(A)} - \bold{T(A)}$,
$\beta^{\mathbb{U}}(u)=0$.
\end{lem}

\begin{proof}
(A sketch of the proof.)
For a given map $\alpha: \bold{F(A)}
\longrightarrow \mathbb{K}$,

If $\alpha(\mathbb{I})=0$, then the
exponential map is defined by
$$exp \ \alpha(\mathbb{I})=1,$$
and for each $u \in \bold{F(A)} - \{\mathbb{I}\}$,
$$exp \ \alpha(u)= \sum_{k=1}^{|u|} \frac{1}{k!} \alpha^{k}(u).$$

And if $\alpha(\mathbb{I})=1$, then the logarithm map is
defined by
$$log \ \alpha(\mathbb{I})=0,$$
and for each $u \in \bold{F(A)} - \{\mathbb{I}\}$,
$$log \ \alpha(u)= \sum_{k=1}^{|u|} \frac{(-1)^{k+1}}{k} (\alpha -
\epsilon)^{k}(u)$$
such that $\epsilon(\mathbb{I})=1$ and for $u \in \bold{F(A)} -
\{\mathbb{I}\}$, $\epsilon(u)=0$. With the help of proposition 7 in \cite{M1}, the proof is completed.
\end{proof}









\begin{prop} \label{53}
$$\sum_{k_{j}>0,n \ge 0} \alpha^{\mathbb{U}}_{f_{k_{1}}f_{k_{2}}...f_{k_{n}}} f_{k_{1}}f_{k_{2}}...f_{k_{n}} =
exp \ (\sum_{t \in \bold{H(T(A))}_{\mathbb{U}}} \beta^{\mathbb{U}}(t)E(t))$$ such that

- $\{E(t): t \in \bold{H(T(A))}_{\mathbb{U}}\}$ (i.e. the set of all Hall
polynomials) is the Hall basis for $L_{\mathbb{U}}$,

- $\{E(u): u \in \bold{H(F(A))}_{\mathbb{U}}\}$ (i.e. the set of decreasing
products of Hall polynomials) is the PBW basis for $H_{\mathbb{U}}$.

\end{prop}

\begin{proof}
One can show that for a given map $\alpha: \bold{F(A)}
\longrightarrow \mathbb{K}$, if $\alpha(\mathbb{I})=0$, then
$$
exp \ (\sum_{w} \alpha_{w}E_{w}) = \sum_{u \in \bold{H(F(A))}} exp \
\alpha(u)E(u),
$$
and if $\alpha(\mathbb{I})=1$, then
$$
 log \ (\sum_{w} \alpha_{w}E_{w})= \sum_{u \in \bold{H(F(A))}} log \ \alpha(u)E(u).
$$
Now by lemma \ref{uns} and with help of the continuous BCH formula (44)
in \cite{M1}, for each word $w=f_{k_{1}}f_{k_{2}}...f_{k_{n}}$ on
the set $A$, one can have
$$
exp \ (\sum_{t \in \bold{H(T(A))}_{\mathbb{U}}} \beta^{\mathbb{U}}(t)E(t)) = \sum_{w}
\alpha^{\mathbb{U}}_{w} w.
$$
Hall basis and PBW basis depended on $H_{\mathbb{U}}$ can be determined by corollary \ref{lie-level} and lemma \ref{rt-phy1}.
\end{proof}

With attention to lemma \ref{rt-phy1}, definition \ref{un} and proposition \ref{rt-phy2}, a Hall
representation from $\gamma_{\mathbb{U}}$ is obtained.

\begin{defn} \label{hall-poly-sing-fra}
The formal series $\sum_{t \in \bold{H(T(A))}_{\mathbb{U}}} \beta^{\mathbb{U}}(t)E(t)$
becomes a {\it Hall polynomial representation} for the universal
singular frame.
\end{defn}

\begin{rem}
There is also another way to show that the universal singular frame
has a rooted tree representation. Corollary \ref{group-level2} and
(\ref{group-level1}) give us a surjective morphism from
$\mathbb{C}[\mathcal{G}_{NAP}](A)$ to the Hopf algebra $H_{\mathbb{U}}$. And
also corollary \ref{group-level3} shows that the infinite dimensional Lie group
$\mathbb{U}(\mathbb{C})$ is a subgroup of $\mathcal{G}_{NAP}$. Since
$\gamma_{\mathbb{U}}$ is a loop with values in
$\mathbb{U}(\mathbb{C})$, therefore for each fixed values $z,v$,
$\gamma_{\mathbb{U}}(z,v)$ should be a formal power series of Hall
rooted trees with the given conditions in corollary \ref{group-level2}.
\end{rem}


Now let us come to conclusion. In \cite{EGK1, EGK2} one can find rooted tree type reformulations of components of the Birkhoff decomposition of a dimensionally regularized Feynman rules character based on Baker-Campbell-Hausdorff (BCH) series and Bogoliubov character such that as the result, scattering formulaes for the renormalization group and the $\beta-$function can be investigated. Now here we can obtain a new Hall rooted tree scattering type formula for counterterms based on the given combinatorial version of $H_{\mathbb{U}}$.

\begin{cor}
One can map the Hall polynomial representation
of the universal singular frame to counterterms of an arbitrary pQFT.
\end{cor}

\begin{proof}
One can show that for each loop $\gamma_{\mu}(z)$ in $Loop(G(\mathbb{C}),\mu)$ (where $\mu$ is the mass parameter), with
help of the graded representation $\xi_{\gamma_{\mu}}: \mathbb{U}
\longrightarrow G$ identified by theorem 1.106 in \cite{CM1}, the universal singular frame
$\gamma_{\mathbb{U}}$ maps to the negative part $\gamma_{-}(z)$ of
the Birkhoff decomposition of $\gamma_{\mu}(z)$. And we know that this minus part does independent of $\mu$ and it determines counterterms. Now it is enough to map Hall trees formulaes given by propositions \ref{f15} and \ref{53} underlying $\xi_{\gamma_{\mu}}$.
\end{proof}

Since Hall set $\bold{H(T(A))}_{\mathbb{U}}$ and its related Hall polynomials determine Hall basis and PBW basis, one can expect to reproduce the BCH type representations of physical information (given in \cite{EGK1, EGK2}) with respect to this Hall tree approach. This possibility reports another reason for the importance of $H_{\mathbb{U}}$ in the study of renormalizable theories.

\begin{prop}
On the basis of formal sums of Hall
trees (Hall forests) and Hall polynomials associated to the
universal Hopf algebra of renormalization, one can introduce new
reformulations from divergences (counterterms)  of each arbitrary renormalizable theory and furthermore, it can be applied to redefine the renormalization group and its infinitesimal generator.
\end{prop}



\chapter{\textsf{Combinatorial Dyson-Schwinger equations and Connes-Marcolli universal treatment}}

The study of Dyson-Schwinger equations (DSEs) could help people as an influenced method of describing unknown non-perturbative circumstances and in fact, it has central role in the development of modern physics. These equations enable us to find an effective conceptional explanation from non-perturbative theory which is the main complicate part of quantum field theory. It seems that a non-perturbative theory can be discovered with solving its corresponding DSE. The significance of this approach for the general identification of quantum field theory is on the increase. \cite{LP1, S13}

In modern physics people concentrate on an analytic
approach to this type of practical equations such that it contains a class of equations related to ill-defined iterated Feynman integrals. More interestingly, Kreimer introduces a new combinatorial reinterpretation from this class of equations such that in his strategy, he systematically applies the combinatorics of renormalization together with Hochschild cohomology theory to obtain perturbative expansions of Hochschild one cocycles depended upon Hopf algebra of Feynman diagrams. This process
justifies combinatorial Dyson-Schwinger equations (DSEs) and so it will be considerable to derive non-perturbative theory from the capsulate renormalization Hopf algebra.  It should be mentioned that in principle, this formalism is strongly connected with the analytic formulation. Because Kreimer could regain analytic DSEs from this combinatorial version by using one particular measure which relates Feynman rules characters (in the Hopf algebraic language) with their corresponding standard forms. \cite{BK1, BK2, K8, K20, K10, K7, K4, K14}

In short, on the one hand, DSEs report non-perturbative phenomena and on the other hand, they can be
formalized on the basis of infinite formal perturbative expansions of one cocycles. So it means that the formalization of quantum field theory underlying Hopf algebra of renormalization leads us widely to clarify a more favorable explanation from non-perturbative quantum field theory.

In this chapter at first we consider
Kreimer's programme in the study of DSEs
based on Hochschild cohomology theory on various Hopf algebras of rooted tress. Then after redefining one cocycles at the Lie algebra
level, we consider
DSEs at the level of the universal Hopf algebra of renormalization (with attention to its described combinatorics). In the next step, we want to show the importance of $H_{\mathbb{U}}$ in the theory of DSEs and this work will be done based on the factorization principal of Feynman diagrams into primitive components. As the result, we will extend reasonably the
universality of $H_{\mathbb{U}}$ (i.e. independency of all renormalizable physical
theories) to non-perturbative theory. And finally, we show that the study of DSEs at the level of the universal Hopf algebra of renormalization hopefully  leads to a new explanation from  this kind of equations in a combinatorial-categorical configuration. This procedure provides one important fact that the universality of the category of equi-singular flat vector bundles can be developed to DSEs and therefore non-perturbative theory. \cite{S7, S5, S14}

Totally, these observations mean that beside analytic and combinatorial techniques, one can expect a new categorical geometric interpretation from this family of equations.


\section{{\it \textsf{Hopf algebraic reformulation of quantum equations of motion based on Hochschild cohomology theory}}}

One crucial property of DSEs is their ability in describing
the self-similarity nature of amplitudes in renormalizable QFTs such that it can help us to introduce a recursive combinatorial version
from these equations. Starting with the physical theory $\Phi$ with the associated Hopf algebra $H_{FG}=H(\Phi)$. With applying the renormalization coproduct, one can indicate a coboundary operator which introduces a Hochschild cohomology theory with respect to the Hopf algebra $H_{FG}$.

\begin{defn} \label{dse7}  \label{dse8}
Define a chain
complex $(C,\bold{b})$ with respect to the coproduct structure of $H$ such
that the set of {\it $n-$cochains} and the {\it coboundary operator} are
determined with
$$
C^{n}:=\{ T: H \longrightarrow H^{\otimes n} : Linear\},
$$
$$
\bold{b}T:= (id \otimes T) \Delta + \sum^{n}_{i=1} (-1)^{i} \Delta_{i} T +
(-1)^{n+1} T \otimes \mathbb{I}
$$
where $T \otimes \mathbb{I}$ is given by $x \longmapsto T(x) \otimes
\mathbb{I}$.
\end{defn}

\begin{lem} \label{dse9}

(i) $\bold{b}^{2} = 0,$

(ii) $H$ is a bicomodule over itself with the right coaction $(id
\otimes \epsilon) \Delta$.

\end{lem}

\begin{prop} \label{dse10}
The cohomology of the complex $(C,\bold{b})$ is denoted by
$HH^{\bullet}_{\epsilon} (H)$ and one can show that for $n \ge 2$, $HH^{\bold{n}}_{\epsilon}(H)$ is trivial. \cite{BK1}
\end{prop}

It should be remarked that this definition of the coboundary operator strongly connected with the Connes-Kreimer coproduct which yields the universality of the pair $(H_{CK},B^{+})$ with respect to Hochschild cohomology theory. Indeed, $\bold{b}$ is defined in the sense that the grafting operator $B^{+}$ can be deduced as a Hochschild one cocycle.
\begin{equation} \label{formula}
\bold{b}T = 0 \Longleftrightarrow \Delta (T) = (id \otimes T) \Delta + T
\otimes \mathbb{I}.
\end{equation}
Moreover, it is remarkable to see that these one cocycles determine generators of $HH^{1}_{\epsilon}(H)$. Thus it is necessary to collect more information about Hochschild one cocycles.

\begin{lem} \label{dse11}
(i) Let $T$ be a generator of $HH^{1}_{\epsilon}(H)$, then
$T(\mathbb{I})$ is a primitive element of the Hopf algebra.

(ii) There is a surjective map $HH^{1}_{\epsilon}(H)
\longrightarrow Prim(H)$, $ T \longmapsto T(\mathbb{I})$.

(iii) We can translate 1-cocycles to the universal enveloping
algebra on the dual side and it means that for example 1-cocycle
$B^{+}:H \longrightarrow H_{lin}$ turns out to the dual map
$(B^{+})^{\star}: \mathcal{L} \longrightarrow U(\mathcal{L})$ where that
is a 1-cocycle in the Lie algebra cohomology. \cite{BK1, K3}
\end{lem}

So it can be seen that primitive elements of the renormalization Hopf algebra and the grafting operator $B^{+}$ can introduce
an important part of the generators of $HH^{1}_{\epsilon}(H)$. In fact, the sum over all primitive graphs (i.e. graphs without any divergent
subgraphs which need renormalization) of a given loop order $n$
defines a 1-cocycle $B^{+}_{n}$ such that every graph is generated
in the range of these 1-cocycles. The importance of cocycles in the reformulation of Green functions
will be observed more clear, if we understand that every
relevant tree or Feynman graph is in the range of a homogeneous
Hochschild 1-cocycle of degree one.

One should notice that insertion into a
primitive graph commutes with the coproduct and therefore it
determines a generator of $HH^{1}_{\epsilon}(H_{CK}(\Phi))$.

\begin{lem} \label{dse12}
In a renormalizable theory $\Phi$, for each $r \in \mathcal{R}_{+}$, we know that
$G_{\phi}^{r}=\phi(\Gamma^{r})$. There are Hochschild 1-cocycles
$B^{+}_{k,r}$ in their expansions such that they can be formulated with
building blocks (i.e. 1PI primitives graphs) of the theory and it means that
$$
B^{+}_{k,r} = \sum_{\gamma \in H(\Phi)^{(1)} \cap H_{lin}(\Phi),
res(\gamma)=r} \frac{1}{|sym(\gamma)|} B^{+}_{\gamma}
$$
where the sum is over all Hopf algebra primitives $\gamma$
contributing to the amplitude $r$ at $k$ loops. \cite{F2, K9, K2}
\end{lem}

In gauge theories, we may have some overlapping sub-divergences with
different external structures. Therefore it is possible to make a
graph $\Gamma$ by inserting one graph into another but in the
coproduct of $\Gamma$ there may be subgraphs and cographs completely
different from those which we used to make $\Gamma$. In this
situation, it is impossible that each $B^{+}_{\Gamma}$ (that
$\Gamma$ is 1PI) be a Hochschild 1-cocycle. Since there may be
graphs appearing on the right hand side of the formula in
(\ref{formula}) which do not appear on the left hand side and for
solving this problem in these theories, there are identities between
graphs. For example: Ward identities in QED and Slavnov-Taylor (ST) identities in
QCD. These identities generate Hopf (co-)ideals in $H_{CK}(QED)$ and $H_{CK}(QCD)$, respectively where with working on the related
quotient Hopf algebras, we will obtain Hochschild 1-cocyles.
\cite{K4, K14, K15, KY2, V3, V6, Y1}

\begin{lem}
The locality of counterterms and the finiteness of
renormalized Green functions are another applications of
Hochschild cohomology theory in QFT. \cite{BK2, CK3, K10, K2}
\end{lem}

Another important influence of Hochschild cohomology in quantum field theory
can be found in rewriting quantum equations of motion in terms of Hopf algebra
primitives and elements in
\begin{equation}
H_{lin}(\Phi) \cap \{ker Aug^{(2)} / ker
Aug^{(1)}\}
\end{equation}
such that it can be applied to characterize DSEs.

\begin{defn} \label{dse13} \label{f8}
For a given renormalizable theory $\Phi$, let $H$ be its associated
free commutative connected graded Hopf algebra and
$(B^{+}_{\gamma_{n}})_{n \in \mathbb{N}}$ be a collection of
Hochschild 1-cocycles (related to the primitive 1PI graphs in $H$).
A class of combinatorial Dyson-Schwinger equations in $H[[ \alpha]]$
has the form
$$
X = \mathbb{I} + \sum_{n \ge 1} \alpha^{n} \omega_{n}
B^{+}_{\gamma_{n}} (X^{n+1})
$$
such that $\omega_{n} \in \mathbb{K}$ and $\alpha$ is a constant.
\end{defn}

This family of equations provides
a new source of Hopf subalgebras such that with working on rooted trees, some interesting relations between Hopf subalgebras of rooted trees (decorated
by primitive 1PI Feynman graphs) and Dyson-Schwinger equations (at
the combinatorial level) can be found.

\begin{thm} \label{dse14}
Each combinatorial Dyson-Schwinger equation DSE has a unique
solution (given by $c=(c_{n})_{n \in \mathbb{N}}, \  c_{n} \in H$)
such that it generates a Hopf subalgebra of $H$. \cite{BK1, BK2, K3}
\end{thm}

\begin{proof}

The elements $c_{n}$ are determined by
$$c_{0}=\mathbb{I},$$
$$ c_{n}= \sum ^{n}_{m=1} \omega_{m} B^{+}_{\gamma_{m}} (\sum _{k_{1}+...+k_{m+1}=n-m, k_{i} \ge 0} c_{k_{1}}...c_{k_{m+1}} ). $$
And so the unique solution is given by $X = \sum_{n \ge 0}
\alpha^{n} c_{n}$. The related Hopf subalgebra structure from this
unique solution is given by the following coproduct
$$\Delta (c_{n}) = \sum^{n}_{k=0} P^{n}_{k} \otimes c_{k}$$
such that $P^{n}_{k} := \sum_{l_{1}+...+l_{k+1}=n-k} c_{l_{1}} ...
c_{l_{k+1}}$ are homogeneous polynomials of degree $n-k$ in $c_{l}$
($l \ge n$).
\end{proof}

\begin{rem}
The independency of the coproduct from the scalars $\omega_{k}$ determines an isomorphism between the induced Hopf subalgebras by all DSEs
of this class. It means that all Hopf subalgebras which come from a
fixed general class of DSEs are isomorphic.
\end{rem}

\begin{lem}
For a connected graded Hopf algebra $H$ and an element $a \in H$,
set
$$
val(a):=max \{ n \in \mathbb{N}: a \in \bigoplus_{k \ge n} H_{k} \}.
$$
It defines a distance on $H$ such that for each $a, b \in H$,
$$
d(a,b):=2^{-val(a-b)}
$$
where its related topology is called $n$-adic topology.
\end{lem}

\begin{lem}
The completion of $H$ with this topology is denoted by
$\overline{H}=\prod^{\infty}_{n=0} H_{n}$ such that its elements are
written in the form $\sum a_{n}$ where $a_{n} \in H_{n}$. One can
show that $\overline{H}$ has a Hopf algebra structure originally based on
$H$ and in fact, the solution of a DSE belongs to this
completion. \cite{F1, F3}
\end{lem}

Since Hopf algebra of Feynman diagrams of a given
theory fundamentally defined by the insertion of graphs into each other
(namely, pre-Lie operation $\star$), therefore one can have a new description from
Hochschild one cocycles in terms of this operator. In continue we consider this notion.

\begin{defn}
For the Lie algebra $\mathfrak{g}$, let $U(\mathfrak{g})$ be its
universal enveloping algebra. For a given $\mathfrak{g}-$module $M$
(with the Lie algebra morphism $\mathfrak{g} \longrightarrow
End_{\mathbb{K}}(M)$), one can define a $U(\mathfrak{g})-$bimodule
structure given by $M^{ad}=M$ with left and right
$U(\mathfrak{g})-$actions: $X.m=X(m),$ $m.X=0$.
\end{defn}

\begin{defn}
Suppose $C^{n}_{Lie} (\mathfrak{g},M) :=
Hom(\bigwedge^{n} \mathfrak{g},M)$ be
the set of all alternating $n-$linear maps $f(X_{1},...,X_{n})$ on
$\mathfrak{g}$ with values in $M$. The Chevalley-Eilenberg
complex is defined by
$$
M \longrightarrow^{\delta} C^{1}(\mathfrak{g},M) \longrightarrow^{\delta} C^{2}(\mathfrak{g},M) \longrightarrow ...
$$
such that
$$
\delta (f)(X_{1},...,X_{n+1}):=
$$
$$ \sum_{i < j} (-1)^{i+j} f([X_{i},X_{j}],...,\widehat{X_{i}},...,\widehat{X_{j}},...,X_{n+1}) +
\sum_{i} (-1)^{i}X_{i}.f(X_{1},...,\widehat{X_{i}},...,X_{n+1}).$$
Cohomology of the chain complex $(C^{n}_{Lie} (\mathfrak{g},M),
\delta)$ is called Lie algebra cohomology.
\end{defn}

\begin{thm} \label{f10}
There is an
isomorphism between Lie algebra (co)homology and Hochschild
(co)homology. In other words,
$$
H^{n}_{Lie} (\mathfrak{g},M) \cong HH^{n}(U(\mathfrak{g}),M).
$$
\cite{K5}
\end{thm}

Now let $\mathcal{L}$ be the Lie algebra
(which comes from the pre-Lie operation $\star$) of Feynman diagrams
of a given renormalizable theory. Because of the importance of one cocycles just we work on the case $n=1$ and calculate the Lie algebra cohomology
$H^{1}_{Lie}(\mathcal{L}, U(\mathcal{L}))$. We have

\begin{equation}
C^{1}_{Lie} (\mathcal{L}, U(\mathcal{L})) = \{ f:\mathcal{L} \longrightarrow U(\mathcal{L}): Linear, Alternating\},
\end{equation}

\begin{equation}
\delta f(X_{1},X_{2}) = -f([X_{1},X_{2}]) - X_{1}f(X_{2}) + X_{2}f(X_{1})
\end{equation}
such that $X_{1},X_{2}$ are disjoint unions of 1PI graphs. Moreover
\begin{equation}
\delta f = 0 \Longleftrightarrow f([X_{1},X_{2}]) = X_{2}f(X_{1}) - X_{1}f(X_{2})
\end{equation}
where it means that
$$ f(X_{1} \star X_{2} - X_{2} \star X_{1}) = f(\sum_{\Gamma_{1}:1PI} n(X_{1},X_{2},\Gamma_{1}) \Gamma_{1} -
\sum_{\Gamma_{2}:1PI} n(X_{2},X_{1},\Gamma_{2}) \Gamma_{2})$$
\begin{equation}
\Longrightarrow \sum_{\Gamma_{1}, \Gamma_{2}:1PI} (n(X_{1},X_{2},\Gamma_{1}) -
n(X_{2},X_{1},\Gamma_{2})) f(\Gamma_{1} - \Gamma_{2}) = X_{2}f(X_{1}) - X_{1}f(X_{2})
\end{equation}
such that $\Gamma_{1}/X_{2} = X_{1},$ $\Gamma_{2}/X_{1} = X_{2}$. It is an explicit condition for one cocycles at the Lie algebra level.

\begin{prop} \label{29}
Let $H$ be the Hopf algebra of Feynman diagrams of the theory $\Phi$
and $L: H \longrightarrow H_{lin}$ a linear map with its dual
$L^{\star}: \mathcal{L} \longrightarrow U(\mathcal{L})$. Then
$$ \Delta(L)=(id \otimes L) \Delta + L \otimes \mathbb{I} \Longleftrightarrow$$

$$ \sum_{\Gamma_{1}, \Gamma_{2}:1PI} (n(X_{1},X_{2},\Gamma_{1}) - n(X_{2},X_{1},\Gamma_{2})) L^{\star}(\Gamma_{1} -
\Gamma_{2}) = X_{2}.L^{\star}(X_{1}) - X_{1}.L^{\star}(X_{2}) $$
where $\Gamma_{1}, \Gamma_{2}$ are 1PI graphs such that $\Gamma_{1}/X_{2} = X_{1}, \   \Gamma_{2}/X_{1} = X_{2}$.
\end{prop}

\begin{cor} \label{30}
For the Connes-Kreimer Hopf algebra of rooted trees, we have
$$H^{n}_{Lie} (\mathcal{L}_{CK},U(\mathcal{L}_{CK})) \simeq HH^{n}(U(\mathcal{L}_{CK}),U(\mathcal{L}_{CK})) \simeq HH^{n}(H_{GL},H_{GL}).$$
\end{cor}

Result \ref{30} shows that when we want to identify one cocycles for the
Hopf algebra $H_{CK}$ (at the Lie algebra level), it is enough to
find Hochschild one cocycles on the Hopf algebra $H_{GL}$.

It is possible to do above procedure for the universal Hopf algebra of renormalization to characterize its corresponding one cocycles at the Lie algebra level.
It was shown that $H_{\mathbb{U}} \simeq
\frac{\mathbb{K}[\textbf{T(A)}]}{I_{\pi}}$ such that $A=(f_{n})_{n \in
\mathbb{N}}$. Therefore for each element $u$ in
$\frac{\mathbb{K}[\textbf{T(A)}]}{I_{\pi}}$, there exist $t \in
\bold{H(T(A))}_{\mathbb{U}}$ and $i \in I_{\pi}$ such that
\begin{equation}
u=t+i.
\end{equation}
And furthermore there exist $t_{1}, t_{2}, z_{1}, z_{2}, s \in
\textbf{T(A)}$ such that
\begin{equation}
u = t + t_{1} \circ z_{1} + z_{1} \circ t_{1} - t_{1}z_{1} + s \circ (t_{2}z_{2}) + z_{2} \circ (t_{2}s) + t_{2} \circ (sz_{2}) - t_{2}z_{2}s.
\end{equation}
It is easy to see that $u$ is a linear combination of rooted trees
and therefore its coproduct should be a linear combination of
coproducts. It means that when each component of the above sum is
primitive, $u$ is primitive. For example, coproduct of
$t_{1} \circ z_{1}$ contains non-trivial terms that induced by
admissible cuts on rooted trees $t_{1}, z_{1}$ and also admissible
cuts on $t_{1} \circ z_{1}$. If $t_{1} \circ z_{1}$ is primitive,
then all of these terms should be canceled.

\begin{lem} \label{31}
For a given primitive element $u \in \frac{\mathbb{K}[\textbf{T(A)}]}{I_{\pi}}$,
$t_{1}, t_{2}, z_{1}, z_{2}, s$ are empty tree.
\end{lem}

\begin{cor} \label{32}
$u=t+i \in \frac{\mathbb{K}[\textbf{T(A)}]}{I_{\pi}}$ is primitive iff $i$ be the empty tree and $t$ primitive.
\end{cor}

With attention to the Hall set corresponding to $H_{\mathbb{U}}$ and the related Lyndon words, a new characterization from primitives at this level can be given.

\begin{lem} \label{33}
Let $\bold{H(T(A))}_{\mathbb{U}}$ be the corresponding Hall set to the Lyndon
words on the locally finite total order set $A=(f_{n})_{n \in
\mathbb{N}}$.  $t \in \bold{H(T(A))}_{\mathbb{U}}$ is primitive iff $t=B^{+}_{a}(\mathbb{I})$ for some $a \in A$.
\end{lem}

It means that $H_{\mathbb{U}}$ has just primitive elements of degree zero
(i.e. a vertex labeled by one arbitrary generator $f_{n}$) and its
related one cocycle is denoted by $B^{+}_{f_{n}}$. In fact, lemma \ref{33}
is a representation of this note that a (Lyndon) word $w$ is
primitive if and only if $w=a$ for some $a \in A$.
This story introduces generators (i.e. one cocycles)
of $HH^{1}_{\epsilon}(H_{\mathbb{U}})$ corresponding to primitive elements
of this Hopf algebra.
On the other hand, theorem \ref{f10} makes clear the Hochschild cohomology of $H_{\mathbb{U}}$ at the Lie algebra level and it means that
\begin{equation}
H^{n}_{Lie}(L_{\mathbb{U}},U(L_{\mathbb{U}})) \cong HH^{n}(U(L_{\mathbb{U}}),U(L_{\mathbb{U}})).
\end{equation}

\begin{cor}
Lie polynomials are
exactly the primitives in the graded dual of $H_{\mathbb{U}}$ and therefore
for each Lie polynomial, one can identify its corresponding one
cocycle at the level of the Lie algebra cohomology.
\end{cor}


\section{{\it \textsf{Universal Hopf algebra of renormalization and factorization problem}}}

If we look at to the structure of the universal affine group scheme $\mathbb{U}$ and its corresponding Hopf algebra $H_{\mathbb{U}}$, then it can be understood that this Hopf algebra is free from any physical dependency. In other words, it is independent of all renormalizable theories and this property separates this specific Hopf algebra from other renormalization Hopf algebras of Feynman diagrams of theories.  In this part, with attention to the combinatorics of Dyson-Schwinger equations and on the basis of factorization problem of Feynman graphs, we are going to improve the independency of the specific Hopf algebra $H_{\mathbb{U}}$ to the level of non-perturbative study.  This property leads us to remove the uniqueness problem of factorization and also, it plays an essential role to consider the behavior of the Connes-Marcolli categorical approach with respect to these equations where as the consequence, we will explain in the next section that the Connes-Marcolli's universal category $\mathcal{E}$ could preserve its universality at the level of Dyson-Schwinger equations.

Kreimer discovered a very trickly interaction between Dyson-Schwinger equations and Euler products and in this process he applied one important concept in physics namely, factorization. Factorization of Feynman diagrams into primitive components is in fact a brilliant highway for physicist to transfer information from perturbative theory (as a well-defined world) to non-perturbative theory (as an ill-defined world). This conceptional translation can be explained by Dyson-Schwinger equations and we know that they are equations which formally solved
in an infinite series of graphs. On the other hand, it was shown that
Feynman diagrams can be decomposed to primitive graphs with bidegree one in a recursive procedure such that the extension of this mechanism
to the level of DSEs is known as one important problem. Fortunately, with introducing a new technical shuffle type product on 1PI Feynman graphs, one can find the answer of this question. One can rewrite the solution of a DSE based on this new shuffle type product. The surprising note is that Euler factorization and
Riemann $\zeta-$function play a large role in this process. \cite{K8, K9, K20, K10, K7, K2}

With help of the given rooted tree reformulation of the universal Hopf
algebra of renormalization and based on the combinatorial approach in the study of DSEs, we are going to consider this family of equations
at the level of $H_{\mathbb{U}}$ and after that with respect to the uniqueness problem of
factorization of Feynman diagrams, we shall see that how one can extend the
universal property of this Hopf algebra to non-perturbative theory. This result can provide remarkable requirements to find a new geometric interpretation from DSEs underlying a categorical configuration. \cite{S7, S5}

\begin{defn}
For each labeled rooted tree $t$, the finite value
$$
w(t):=\sum_{v \in t^{[0]}}
|dec(v)|
$$
is called decoration weight of $t$.
\end{defn}

\begin{thm} \label{dse-un1}
For a given equation DSE in the Hopf algebra $H_{CK}$, there is an explicit
presentation from the generators $c_{n},$ ($n \in \mathbb{N}$)
(identified with the unique solution of DSE) at this level.
We have
$$
c_{0}=\mathbb{I},  \   \ c_{n}= \sum_{t, w(t)=n} \frac{t}{|sym(t)|}
\prod_{v \in t^{[0]}} \rho_{v}
$$
such that
$$
\rho_{v} = \omega_{|dec(v)|}
\frac{(|dec(v)|+1)!}{(|dec(v)|+1-fer(v))!},
$$
for the case $fert(v)
\le |dec(v)|+1$ and
$\rho_{v} =0,$ for otherwise. \cite{BK1}
\end{thm}

Corollary \ref{rt-v14} allows us to lift this theorem to the level of $H_{\mathbb{U}}$.

\begin{prop}  \label{dse-un2}
For a given combinatorial equation DSE in $H_{\mathbb{U}}$, its unique
solution $c=(c_{n})_{n}$ is determined  by
$$c_{0}=I_{\pi},$$
$$c_{n}= \sum_{t \in \bold{H(T(A))}_{\mathbb{U}}, t \notin I_{\pi}, w(t)=n} t \prod_{v \in t^{[0]}} \rho_{v} + I_{\pi}.$$
\end{prop}

\begin{proof}
It is proved by proposition \ref{dse14}, theorem \ref{dse-un1} and this note that Hall
trees (forests) have no symmetries.
\end{proof}

Let $\mathbb{K}[[h]]$ be the ring of formal series in one variable
over $\mathbb{K}$. Foissy in \cite{F1, F3} considers some interesting
classes of DSEs (given by elements of $\mathbb{K}[[h]]$) in Hopf
algebras of rooted trees and also, he classifies systematically their associated
Hopf subalgebras. One can lift his results to the level of labeled
rooted trees and hence $H_{\mathbb{U}}$.

\begin{defn} \label{dse-un3}
The composition of formal series gives a group structure on the set
$$G:=\{ h + \sum_{n \ge 1} a_{n}h^{n+1} \in \mathbb{K}[[h]]\}.$$
Hopf algebra of functions on the opposite of the group $G$ is called
{\it Faa di Bruno Hopf algebra}. It is a connected graded
commutative non-cocommutative Hopf algebra and denoted by $H_{FdB}$
such that for each $f \in H_{FdB}$ and $P,Q \in G$, its coproduct is
given by
$$\Delta (f) (P \otimes Q) = f (Q \circ P).$$
\end{defn}

\begin{rem}
One can show that $H_{FdB}$ is the polynomial ring in variables
$Y_{i}$ ($i \in \mathbb{N}$), where
$$Y_{i}: G \longrightarrow \mathbb{K}, \  \ h+ \sum_{n \ge 1} a_{n}
h^{n+1} \longmapsto a_{i}.$$
\end{rem}

\begin{prop} \label{dse-un4}
Let $\mathbb{K}[[h]]_{1}$ be the set of elements in
$\mathbb{K}[[h]]$ with constant term $1$ and $P \in \mathbb{K}[[h]]_{1}$.

(i) Dyson-Schwinger equation $X_{P}=B^{+}_{f_{n}}(P(X_{P}))$ in
$H_{\mathbb{U}}[[h]]$ has a unique solution with the associated Hopf
subalgebra $H^{\alpha, \beta}_{\mathbb{U}}(P)$ iff there exists $
(\alpha,\beta) \in \mathbb{K}^{2}$ such that
$$(1-\alpha\beta h)
P^{'}(h)=\alpha P(h), \  \   P(0)=1$$

(ii) For $\beta \neq -1$, $H^{1, \beta}_{\mathbb{U}}(P)$ is isomorphic to the
quotient Hopf algebra $ \frac{H_{FdB}}{I_{\pi}}$.

(iii) $H^{1,-1}_{\mathbb{U}}(P)$ is isomorphic to a quotient Hopf algebra
$\frac{SYM}{J}$ of Hopf algebra of symmetric functions.

(iv) $H^{0,1}_{\mathbb{U}}(P)$ is isomorphic to the quotient Hopf algebra
$\frac{\mathbb{K}[\bullet]}{I_{\pi}}$.
\end{prop}

\begin{proof}
It is proved based on the rooted tree version of $H_{\mathbb{U}}$ and the
given results in sections 3, 4, 5 in \cite{F1}. For the third part,
there is a homomorphism $\theta: SYM \longrightarrow H_{CK}$ that
sends each generator $m_{\underbrace{(1,...,1)}_{n}}$ to the ladder
tree $l_{n}$. Set $J:=
\theta^{-1}(I_{\pi})$.
\end{proof}

Moreover, Hoffman suggests a new procedure to study DSEs with
translating equations to a quotient of noncommutative version of
the Connes-Kreimer Hopf algebra namely, Foissy Hopf algebra. One can improve his main result to our interesting level.

\begin{prop} \label{dse-un5}
The unique solution of the equation
$$
X= I_{\pi} + B_{f_{n}}^{+}(X^{p})
$$
in $H_{\mathbb{U}}$ where $p \in \mathbb{R}$ is determined by
$$t_{n}= \sum_{t \in \bold{H(T(A))}_{\mathbb{U}}} e(t)C_{p}(t)t$$
such that $e(t)$ is the number of Hall planar rooted trees $s$ such that
$\alpha_{2}(s)=t$ (defined in theorem \ref{main10}).
\end{prop}

\begin{proof}
The unique solution of DSE is
given by a formal sum
$$
X= I_{\pi}+t_{1}+t_{2}+...
$$
such that $t_{n}$ is a Hall tree in $\bold{H(T(A))}_{\mathbb{U}}$ with degree
$n$. Set
$$
\widetilde{X}:=t_{1}+t_{2}+...
$$
Equation DSE can be changed to the form
$$
\widetilde{X}=B_{f_{n}}^{+}((I_{\pi}+\widetilde{X})^{p}).
$$
Since the operator $B_{f_{n}}^{+}$ increases degree, it is easy to
see that

$$
t_{n+1}=B_{f_{n}}^{+}(\{ I_{\pi} +(\begin{matrix}
 p \\
 1
\end{matrix})\widetilde{X}+ (\begin{matrix}
 p \\
 2
\end{matrix}) \widetilde{X}^{2}+...\}_{n})
$$
such that $\{\}_{n}$ is the component of degree $n$. There is a
natural homomorphism $\alpha_{2}: H_{F} \longrightarrow H_{CK}$ that maps
each planar rooted tree to its corresponding rooted tree without
notice to the order in products. One can lift this homomorphism to
the level of labeled rooted trees and apply it to study the given
DSE at the level of the quotient Hopf algebra $\frac{H_{F}(A)}{I_{\pi}}$.
Let
$$
I_{\pi}+\widetilde{Y} = I_{\pi}+s_{1}+s_{2}+...
$$
be its solution in this new level such that $s_{n}$ is a Hall planar
rooted  tree of degree $n$ in $\bold{H(T(A))}_{\mathbb{U}}$. Induction shows that
$$
s_{n}= \sum_{t \in \bold{H(P_{n-1}(A))}_{\mathbb{U}}} C_{p}(t)t
$$
such that
$$
C_{p}(t)=\prod_{v \in \overline{V}(t)} \big(\begin{matrix}
 p \\
 c(v)
\end{matrix})
$$
where $c(v)$ is the number of leaves of $v$, $\overline{V}(t)$ is
the set of  vertices of $t$ with $c(v) \neq 0$ and
$\bold{H(P_{n-1}(A))}_{\mathbb{U}}$ is the Hall subset of
$\bold{H(T(A))}_{\mathbb{U}}$ generated by planar rooted trees of degree $n$.
It is observed that for each Hall planar rooted tree $s$,
$C_{p}(s)=C_{p}(\alpha_{2}(s))$. Since
$\alpha_{2}(\widetilde{Y})=\widetilde{X}$, according to the theorem 6.2 in
\cite{H2}, one can obtain a clear presentation from the unique
answer of DSE.
\end{proof}

Now it is the time to consider the attractive relation between Dyson-Schwinger equations and Euler products underlying the factorization problem.

\begin{defn}
The analytic continuation of the sum $\sum_{n} \frac{1}{n^{s}}$ is
called Riemann $\zeta-$function. If $\mathfrak{R}(s)>1$, then it
has an Euler product over all prime numbers given by
$\zeta(s)=\prod_{p} \frac{1}{1-p^{-s}}$.
\end{defn}

\begin{defn}
For a sequence of primitive 1PI graphs
$J=(\gamma_{1},...,\gamma_{k})$ in the renormalizable theory $\Phi$,
a Feynman graph $\Gamma$ is called compatible with $J$ (i.e.
$\Gamma \sim J$), if
$$<Z_{\gamma_{1}} \otimes ... \otimes Z_{\gamma_{k}},
\Delta^{k-1}(\Gamma)>=1.$$
\end{defn}

\begin{lem}
Let $n_{\Gamma}$ be the number of compatible sequences with
$\Gamma$. Define a product on 1PI graphs given
by
$$\Gamma_{1} \uplus \Gamma_{2}:= \sum_{I_{1} \sim \Gamma_{1}, I_{2} \sim
\Gamma_{2}} \sum_{\Gamma \sim I_{1} \star I_{2}}
\frac{1}{n_{\Gamma}} \Gamma.$$
It is a commutative associative product.
\end{lem}

\begin{lem}
Define a relation
$$\Gamma_{1} \le \Gamma_{2} \Longleftrightarrow
<Z_{\Gamma_{1}}^{-},\Gamma_{2}> \neq 0.$$
It determines a partial order relation based on subgraphs.
\end{lem}

\begin{rem}
(i) One can rewrite the Connes-Kreimer coproduct by
$$\Delta(\Gamma)= \sum_{\Gamma_{1},\Gamma_{2}}
\zeta(\Gamma_{1},\Gamma_{2}) \Gamma_{1} \otimes \Gamma_{2}$$
where if $\Gamma_{1} \le \Gamma_{2}$, then
$\zeta(\Gamma_{1},\Gamma_{2})=1$ and  otherwise
$\zeta(\Gamma_{1},\Gamma_{2})=0$.

(ii) The product $\uplus$ is a
generalization of the shuffle product $\star$ and it means that  $\star$ appropriates
for totally ordered sequences whenever $\uplus$ is just for partial order
relation (on subgraphs). \cite{K8, K9}
\end{rem}

With the help of defined shuffle type product on Feynman diagrams, Kreimer shows that Euler products exist in solutions of DSEs.

\begin{lem} \label{dse-un6} \label{ast-eu}
Unique solution of the equation
$$
X= \mathbb{I} + \sum_{\gamma} \alpha^{k_{\gamma}} B^{+}_{\gamma}
(X^{k_{\gamma}})
$$
(such that $k_{\gamma}$ is the degree of $\gamma$) has an
$\uplus-$Euler product given by
$$
 X = \prod^{\uplus}_{\gamma} \frac{1}{1- \alpha^{k_{\gamma}} \gamma}.
$$
\cite{K8, K9, K10}
\end{lem}

\begin{rem}
The uniqueness of this factorization in
gauge theories is lost and for removal this problem, we should work
on the quotient Hopf algebras. \cite{K9, K10, V2, V3, V6}
\end{rem}

Now we know that uniqueness is the main problem of factorization and in continue we want to focus on this lack to show the advantage of the universal Hopf algebra of renormalization.

\begin{thm}
Consider set $H_{pr}$ of all sequences $(p_{1},...,p_{k})$ of prime numbers such that the empty
sequence is denoted by $1$ and define a map $B^{+}_{p}$ such that its application on a sequence
$J=(p_{1},...,p_{k})$ is the new sequence $(p,p_{1},...,p_{k})$.
There is a Hopf algebra structure on $H_{pr}$ such that its compatible coproduct with the shuffle product is
determined by
$$\Delta(B^{+}_{p}(J)) = B^{+}_{p}(J) \otimes 1 + [id \otimes
B^{+}_{p}] \Delta(J),$$
$$\Delta (1) = 1 \otimes 1, $$
$$\Delta ((p)) = (p) \otimes 1 + 1 \otimes (p).$$
$$B^{+}_{p_{1}} (J_{1}) \star B^{+}_{p_{2}} (J_{2}) =
B^{+}_{p_{1}}(J_{1} \star B^{+}_{p_{2}}(J_{2})) +
B^{+}_{p_{2}}(B^{+}_{p_{1}}(J_{1}) \star J_{2}).$$ \cite{K4}
\end{thm}

\begin{rem}
It is natural to think that the operator $B^{+}_{p}$ is a Hochschild
one cocycle and it means that with this operator one can introduce Dyson-Schwinger equations at the level of $H_{pr}$.
\end{rem}

Consider Dyson-Schwinger type equation
\begin{equation}\label{dse-un7}
X(\alpha)= 1 + \sum_{p} \alpha B^{+}_{p}[X(\alpha)]
\end{equation}
in $H_{pr}$. It has a decomposition given by
\begin{equation} \label{sta-eu}
X(\alpha) = \prod_{p}^{\star} \frac{1}{1 -\alpha (p)}.
\end{equation}

Because of the shuffle nature of this Hopf algebra, one can suggest to apply universal Hopf algebra of
renormalization. It is observed that for
finding the Euler factorization, one should define a new product on
Feynman diagrams of a theory and on the other hand, the
uniqueness of this factorization is not available in general. Now we see that with working at the level of $H_{\mathbb{U}}$, this indicated problem can be solved.

\begin{prop} \label{dse-un8} \label{uniq} \label{dse-un9}
There exists an Euler factorization in the universal Hopf algebra of renormalization.
\end{prop}

\begin{proof}
There are two shuffle structures to apply
in $H_{\mathbb{U}}$ for receiving factorization. One of them namely,
$\star$ is exactly the same as the shuffle product on $H_{\mathbb{U}}$. One can show that

(i) The product of $H_{\mathbb{U}}$ is integral,

(ii) There is a combinatorial $\star-$Euler product (comes from a
class of DSEs) in the universal Hopf algebra of renormalization.

For the second claim, it is clear that for each
variable $f_{n}$ ($n \in \mathbb{N}_{>0}$) in $H_{\mathbb{U}}$,
$B^{+}_{f_{n}}$ is a Hochschild one cocycle. Consider equation
$$
X(\alpha)= 1 + \sum_{f_{n}} \alpha B^{+}_{f_{n}}[X(\alpha)]
$$
in $H_{\mathbb{U}}$. By replacing prime
numbers with these variables in (\ref{sta-eu}) and also
with notice to the shuffle structure in $H_{\mathbb{U}}$, one can get the decomposition
$$X(\alpha) = \prod_{f_{n}}^{\star} \frac{1}{1 -\alpha (f_{n})}$$
such that $(f_{n})$ is a word with length one.

\end{proof}

It was shown that an extension of the shuffle product can be applied to
obtain a factorization for the formal series of Feynman diagrams of
the solution of a DSE. For the word $w=f_{k_{1}}...f_{k_{n}}$ in
$H_{\mathbb{U}}$, a word $v$ is called compatible with $w$  (i.e. $v \sim
w$), if
\begin{equation}
<p_{f_{k_{1}}} \otimes ... \otimes p_{f_{k_{n}}},
\Delta^{k_{n}-1}(v)>=1
\end{equation}
such that $p_{f_{n}} \equiv e_{-n}$s are Hall polynomials of
primitive elements $f_{n}$s. (One should stress that Hall
polynomials form a basis (at the vector space level) for the Lie
algebra). Therefore
\begin{equation}
v \sim w \Longleftrightarrow  <e_{-k_{1}} \otimes ... \otimes
e_{-k_{n}}, \Delta^{k_{n}-1}(v)>=1.
\end{equation}
Let $n_{w}$ be the number of compatible words with $w$. It determines our interesting product
\begin{equation}
w_{1} \uplus w_{2}:= \sum_{v_{1} \sim w_{1}, v_{2} \sim w_{2}} \sum_{w
\sim v_{1} \star v_{2}} \frac{1}{n_{w}} w.
\end{equation}
Now consider Dyson-Schwinger equation
\begin{equation}
X= 1 + \sum_{f_{n}} \alpha^{n} B^{+}_{f_{n}} (X^{n})
\end{equation}
in the universal Hopf algebra of renormalization. Its unique solution has an $\uplus-$Euler
product given by
\begin{equation}
X = \prod^{\uplus}_{f_{n}} \frac{1}{1- \alpha^{n} f_{n}}.
\end{equation}

\begin{cor}
There is a unique factorization into the Euler product in the universal
Hopf algebra of renormalization.
\end{cor}

\begin{proof}
With attention to the given coproduct in theorem \ref{shuffle3} and also proposition
\ref{uniq}, it can be seen that
$$
v \sim w \Longleftrightarrow v=w \Longrightarrow n_{w}=1
\Longrightarrow w_{1} \uplus w_{2} = w_{1} \star w_{2}.
$$
Therefore  $\uplus-$Euler product and $\star-$Euler product in the universal Hopf
algebra of renormalization are the same. On the other hand,
 we know that each word has a unique representation with a decreasing
decomposition to Hall polynomials and moreover, corollary
\ref{lie-level} shows that these elements determine a basis at the
vector space level for the free algebra $H_{\mathbb{U}}$.
\end{proof}

At last, let us consider the possibility of defining the Riemann $\zeta-$function in $H_{\mathbb{U}}$.

\begin{prop} \label{dse-un11} \label{sta-eu0}
The Riemann $\zeta-$function can be reproduced from a class of DSEs in the universal Hopf algebra of renormalization.
\end{prop}

\begin{proof}
One can define an injective homomorphism from $H_{prime}$ to $H_{\mathbb{U}}$ given
by
$$
J=(p_{1},...,p_{n}) \longmapsto w_{J}=f_{p_{1}}...f_{p_{n}}.
$$
There is an interesting notion (in \cite{K8}) to
obtain Riemann $\zeta-$function from a class of DSEs in $H_{prime}$
such that one can lift it to the level of $H_{\mathbb{U}}$. Consider the
equation
$$
X(\alpha)= 1 + \sum_{p} \alpha B^{+}_{f_{p}}[X(\alpha)]
$$
in $H_{\mathbb{U}}$ such that the sum is on prime numbers. Choose a
homomorphism $\phi_{s}$ of $H_{\mathbb{U}}$ given by
$$
\phi_{s}(w)=\frac{1}{|w|!}pr(w)^{-s}
$$
such that for the word $w=f_{k_{1}}...f_{k_{n}}$,
$pr(w):=k_{1}...k_{n}$. It is observed that
$$
lim_{\alpha \longrightarrow 1} \phi_{s}[X(\alpha)] = \zeta(s).
$$
The Euler product of the above DSE is given by
$$
X(\alpha) = \prod_{p}^{\star} \frac{1}{1 -\alpha (f_{p})}
$$
and one can see that
$$
\phi_{s} (\prod_{p}^{\star} \frac{1}{1 -\alpha (f_{p})})
= \prod_{p} \frac{1}{1-p^{-s}} = \zeta(s).
$$
\end{proof}

Roughly speaking, this explained procedure yields one essential reason to generalize the concept of universality of $H_{\mathbb{U}}$ in non-perturbative theory.

\begin{cor}
Universal Hopf algebra of renormalization can preserve its independency from physical theories
at the level of Dyson-Schwinger equations and therefore
non-perturbative theory.
\end{cor}


\section{{\it \textsf{Categorical configuration in the study of DSEs}}}

We discussed that how combinatorial Dyson-Schwinger equations can determine a systematic formalism to consider non-perturbative theory based on the
Connes-Kreimer perturbative renormalization. It means that this Hopf algebraic reinterpretation can lead to an extremely practical strategy to discover some unknown parts of the theory of quantum fields. In this process we understood that from each DSE one can associate a Hopf subalgebra of Feynman diagrams of a fixed theory. In this part we want to work on these Hopf subalgebras and introduce a new framework to study Dyson-Schwinger equations based on the Connes-Marcolli's universal approach.
This purpose can be derived by finding an interrelationship
between these equations and objects of the category of flat
equi-singular vector bundles where it has an essential universal property in
the mathematical treatment of the perturbative renormalization.
In this generalization one can characterize a new family of equations (i.e. {\it universal Dyson-Schwinger equations}) such that it establishes a new procedure to calculate Feynman integrals in the sense that at first, one can find the solution of an equation at the simplified universal level (i.e. DSEs in $H_{\mathbb{U}}$) and then with using graded representations (introduced by Connes-Marcolli theory), we will project this solution to the level of an arbitrary renormalizable theory.
In addition, by this
way, one can find an arterial road from combinatorial DSEs to the universal
category $\mathcal{E}$ where as the consequence, it mentions a new geometric interpretation from Dyson-Schwinger equations underlying the Riemann-Hilbert correspondence.
So it makes possible to expand the universality of the category $\mathcal{E}$ to the level of
these equations. \cite{S7, S5}

The universality of the category $\mathcal{E}$ is determined by its relationship with categories connected with renormalizable theories and it was shown in theorem \ref{cat.5} that this neutral Tannakian category enables to cover categories of all renormalizable theories as full subcategories and moreover, it determines the specific affine group scheme $\mathbb{U}^{*}$. Here we are going to apply Hopf subalgebra structures depended on Dyson-Schwinger equations to find a relation between these equations and objects of $\mathcal{E}$.

Let us consider the Dyson-Schwinger equation
DSE in $H[[\alpha]]$ with the associated Hopf
subalgebra $H_{c}$ and affine group scheme $G_{c}$ and let $\mathfrak{g}_{c}$ be its corresponding
Lie algebra.

\begin{lem}
Generators of the Lie algebra $\mathfrak{g}_{c}$
are linear maps
$$Z_{n}: H_{c} \longrightarrow \mathbb{C}, \  Z_{n}(c_{l})=
\delta_{n,l}.
$$
\end{lem}

\begin{lem}
One can make a trivial principal $G_{c}-$bundle $P_{c}= B \times
G_{c}$ over the base space $B= \bold{\Delta} \times
\mathbb{G}_{m}(\mathbb{C})$ such that its restriction on $B^{0}$ is
denoted by $P^{0}_{c}=B^{0} \times G_{c}$.
\end{lem}

\begin{cor}
(i) Equivalence relation given by definition \ref{equival-rel} provides a bijective correspondence between minus
parts of the Birkhoff decomposition of loops (with values in
$G_{c}$) and elements of the Lie algebra $\mathfrak{g}_{c}$.

(ii) Theorem \ref{geo-dse7} guarantees the existence of the classes of
equi-singular flat connections with respect to the elements of this
Lie algebra. Therefore a classification of equi-singular flat
connections on the vector bundle connected with the combinatorial
equation DSE in the theory $\Phi$ will be determined.
\end{cor}

\begin{cor} \label{geo-dse10}
For a given equation DSE in the theory $\Phi$, equivalence classes
of flat equi-singular $G_{c}-$connections on $P^{0}_{c}$ are represented
by elements of the Lie algebra $\mathfrak{g}_{c}$ and also, each
element of this Lie algebra identifies one specific class of
equi-singular $G_{c}-$connections. This process is done independent
of the choice of a local regular section $\sigma: \bold{\Delta}
\longrightarrow B$ with $\sigma(0)=y_{0}$.
\end{cor}

\begin{prop}
Let $c=(c_{n})_{n \in \mathbb{N}}$ be the unique solution of DSE. Result
\ref{geo-dse10} shows that for each $k\in \mathbb{N}$ there exists a
unique class of flat equi-singular connections $\omega^{k}_{c}$ on
$P^{0}_{c}$ such that
$$\omega^{k}_{c} \sim \bold{D} \gamma_{Z_{k}},$$
$$
\gamma_{Z_{k}}(z,v)= Te^{-\frac{1}{z} \int^{v}_{0}
{u^{Y}(Z_{k})}\frac{du}{u}}, \  u=tv, \  t \in [0,1].
$$
\end{prop}

Let $V^{l}$ be an arbitrary $l-$dimensional vector space generated
by some elements of $(c_{n})_{n \in \mathbb{N}}$ and
$\psi^{l}_{c}:G_{c} \longrightarrow Gl(V^{l})$ be a graded
representation. By theorem \ref{cat.10}, the pair
$(\omega^{k}_{c},\psi^{l}_{c})$ identifies an element from the
category of flat equi-singular vector bundles $\mathcal{E}$.

\begin{cor} \label{geo-dse11}
For a given Dyson-Schwinger equation DSE in the theory $\Phi$, a
family of objects of the category $\mathcal{E}$ will be determined.
\end{cor}

When we consider the Riemann-Hilbert correspondence underlying the Connes-Kreimer
theory of perturbative renormalization, the universality of
the category $\mathcal{E}$ is characterized by this fact that
$\mathcal{E}$ carries a geometric representation from all renormalizable
theories as subcategories. Now this notion can be expanded to the level of combinatorial DSEs.

\begin{defn} \label{geo-dse12}
For a fixed equation DSE in the theory $\Phi$, one can define a
subcategory $\mathcal{E}^{\Phi}_{c}$ of $\mathcal{E}$ such that its
objects are introduced by corollary \ref{geo-dse11}.
\end{defn}

\begin{prop}
For each given equation DSE, there are classes
of flat equi-singular $G_{c}$-connections such that they introduce a category. Instead of
working on this category, one can go to a universal framework and concentrate on a full
subcategory of $\mathcal{E}$ of those flat equi-singular vector
bundles that are equivalent to the finite dimensional linear
representations of $G_{c}^{*}$. It provides this fact that
$\mathcal{E}^{\Phi}_{c}$ has power to store a geometric description from
DSE.
\end{prop}

With attention to the algebro-geometric dictionary (in the
minimal subtraction scheme in dimensional regularization
\cite{CM2}), we know that each loop $\gamma_{\mu}$ in the space of diffeographisms can associate
a homomorphism $\phi_{\gamma_{\mu}}: H \longrightarrow \mathbb{C}$ and then
we can perform Birkhoff decomposition at the level of these
homomorphisms to obtain physical information.

\begin{lem}
There is a surjective map from the affine
group scheme $G$ to the affine group scheme
$G_{c}$.
\end{lem}

\begin{proof}
We know that $H_{c}$ is a Hopf subalgebra of $H$. With
restriction one can map each element $\phi$ in the complex Lie group
scheme $G(\mathbb{C})$ to its corresponding element $\phi_{c} \in G_{c}(\mathbb{C})$.
On the other hand, there is an injection from $H_{c}$ to $H$ such
that it determines an epimorphism from $Spec(H)$ to $Spec(H_{c})$.
\end{proof}

\begin{prop} \label{geo-dse13}
Objects of the category $\mathcal{E}^{\Phi}_{c}$ store some parts of
physical information of the theory with respect to the
Dyson-Schwinger equation DSE.
\end{prop}

This fact shows that with help the of objects of the category of
flat equi-singular vector bundles, a geometrically analysis from all
of the combinatorial DSEs in a given theory can be investigated and
since according to \cite{K4} these equations address
non-perturbative circumstances, therefore category $\mathcal{E}$ can
preserve its universal property at this new level.

It was shown that Hopf algebras of renormalizable physical theories
and $H_{\mathbb{U}}$ have the same combinatorial source (namely, rooted trees). By applying the next fact, one can find an interesting idea to compare
Dyson-Schwinger equations at the level of renormalizable physical
theories with their corresponding at the level of the universal Hopf
algebra of renormalization.


We know that with the help of Hochschild one cocycles (identified by
primitive elements of the Hopf algebra), one can characterize
combinatorial DSEs. Fix an equation DSE in $H$ with the associated
Hopf subalgebra $H_{c}$. According to theorem \ref{cat.10}, for the
equivalence class of flat equi-singular connections $\omega$ on
$\widetilde{P}^{0}$ one can identify a graded representation
$\rho_{\omega}$. Let $\omega_{c}$ be the flat equi-singular
connection on $\widetilde{P}^{0}_{c}:=B^{0} \times G^{*}_{c}$ corresponding to $\omega$ with
the associated graded representation $\rho_{\omega_{c}}$.
On the other hand, one can consider DSE in a decorated version of the Connes-Kreimer Hopf
algebra of rooted trees. Theorem \ref{dse14} provides an explicit reformulation from
generators of the Hopf algebra $H_{c}$ such that by proposition \ref{dse-un2},
we can lift these generators to the level of the rooted tree
representation of $H_{\mathbb{U}}$. This process determines a new equation
$DSE_{u}$ and a Hopf subalgebra $H_{u}$ of $H_{\mathbb{U}}$ with the related
affine group scheme $\mathbb{U}_{c}$.

\begin{lem}
For a fixed equi-singular flat connection $\omega$, one can provide a graded representation
$\rho^{c}_{\omega_{c}}:\mathbb{U}_{c}^{*} \longrightarrow
G_{c}^{*}$ such that it is a lift of the representation
$\rho_{\omega}$ and characterized with $\rho_{\omega_{c}}$. In summary, we have the
following commutative diagram.

\begin{equation} \label{geo-dse15}
\xymatrix{ \mathbb{U}^{*} \ar@/_/[ddr]_{}
\ar@/^/[drr]^{\rho_{\omega}}
  \ar@/^/[dr]|-{\rho_{\omega_{c}}}         \\
& G_{c}^{*} \ar@{<.}[d]|-{\rho^{c}_{\omega_{c}}} \ar@{<-}[r]|-{}
                    & G^{*}         \\
& \mathbb{U}_{c}^{*}                                  }
\end{equation}
\end{lem}

\begin{prop} \label{geo-dse16}
The morphism $\rho^{c}_{\omega_{c}}$ induced by
representations $\rho_{\omega}$,
$\rho_{\omega_{c}}$ provides the concept of {\it
universal Dyson-Schwinger equations} and it means that $DSE_{u}$
maps to DSE (or DSE lifts to $DSE_{u}$) under the representation $\rho^{c}_{\omega_{c}}$.
\end{prop}

When $\mathbb{K}=\mathbb{Q}$, the fiber functor $\varphi:
\mathcal{E}_{\mathbb{Q}} \longrightarrow \mathcal{V}_{\mathbb{Q}}$
is given by $\varphi = \bigoplus \varphi_{n}$ such that for each
element $\Theta$ of the category,
\begin{equation} \label{fiberfunctor}
\varphi_{n} (\Theta):= Hom(\mathbb{Q}(n),Gr^{W}_{-n}(\Theta)).
\end{equation}
For each $n$, $\mathbb{Q}(n) = [V,\bigtriangledown] $ is an object
in $\mathcal{E}_{\mathbb{Q}}$ such that $V$ is an one dimensional
$\mathbb{Z}-$graded $\mathbb{Q}-$vector space placed in degree $n$
and $\bigtriangledown=d$ (i.e. ordinary differentiation in one
variable). We explained that how one can identify some elements of this
category with objects given in corollary \ref{geo-dse11} and now it would be remarkable to see that elements $\mathbb{Q}(n)$ are
represented with respect to a given DSE and for this work, the class
of trivial connections should be calculated. Because proposition
1.101 in \cite{CM1} provides this fact that the connection
$\bigtriangledown$ identifies a connection $\omega$ where
$\bigtriangledown=d+\omega$. In other words, we have to find an
element in the Lie algebra $\mathfrak{g}_{c}$ such that its
corresponding equi-singular connection is equivalent to $0$ and one
can show that $Z_{0}$ plays this role. We have
\begin{equation}
\theta_{-t}(Z_{k})=e^{-tk}Z_{k}
\end{equation}
\begin{equation}
k=0 \Longrightarrow \theta_{-t}(Z_{0})=Z_{0}:H_{c} \longrightarrow \mathbb{C}, \  \  Z_{0}(c_{k})=\delta_{0,k}.
\end{equation}
By theorem 1.60 in \cite{CM1}, for
the element $\beta = Z_{0}$ in $\mathfrak{g}_{c}$, we have
\begin{equation}
\gamma^{c}_{-}(z)=Te^{-\frac{Z_{0}}{z}\int^{\infty}_{0} dt}.
\end{equation}
On the other hand, we know that the expression $Te^{\int^{u}_{0}
\alpha(t)dt}$, such that $\alpha(t)=1$ for $t=t_{0}$ and
$\alpha(t)=0$ for otherwise, is the value $g(u)$ of the solution for
the equation
\begin{equation} \label{geo-dse17}
dg(u)=g(u)\alpha(u)du, \ \  g(0)=1.
\end{equation}
Therefore by (\ref{geo-dse17}), it can be seen that
\begin{equation}
g(t)= const. \Longrightarrow \bold{D}g=0 \Longrightarrow \omega \sim 0.
\end{equation}

\begin{cor} \label{geo-dse18}
One can represent elements $\mathbb{Q}(n)$ with a given equation DSE.
\end{cor}

\begin{proof} Let $c=(c_{n})_{n \in \mathbb{N}}$ be the unique
solution of a fixed equation DSE and $V$ be the one dimensional
$\mathbb{Q}-$vector space generated by $c_{n}$ placed in degree $n$.
It is observed that $\omega_{0} \sim D \gamma^{c}_{Z_{0}}$ such that
$\gamma^{c}_{Z_{0}}$ is the constant loop and it means that
$\omega_{0}=0$. Since $\bigtriangledown=d+\omega_{0}$, the proof is
complete.
\end{proof}

One should note that there are different choices to define the
vector space $V$ in the representation of $\mathbb{Q}(n)$ with a
fixed DSE but all of them belong to the isomorphism class of
one dimensional $\mathbb{Z}-$graded $\mathbb{Q}-$vector spaces. For
a given equation DSE, result \ref{geo-dse18} shows that the equation
(\ref{fiberfunctor}) can play the role of the fiber functor for the full abelian
tensor category $\mathcal{E}^{\Phi}_{c}$.


\chapter{\textsf{Conclusion and future improvements}}

Here we try to have an overview from the whole structure of this work and then with attention to the given results, we will introduce some major problems which help to advance theory of integrable systems and theoretical understanding of non-perturbative phenomena underlying the Connes-Kreimer-Marcolli postulates.


\section{{\it \textsf{Overview}}}

Roughly speaking, this research attempts to establish new progresses in the study of quantum field theory underlying renormalization Hopf algebra and in particular, we
focused on three important problems namely, theory of quantum integrable systems, combinatorial representations of universal Hopf algebra and universal singular frame and  categorification of non-perturbative studies (with respect to the Kreimer's approach).

The first purpose was focused on integrable systems. Firstly, with the help of non-commutative differential forms, we considered an essential problem in quantum field theory namely, quantum integrable systems. The key point in our chosen approach is summarized in the deformation of algebras based on Nijenhuis operators where these maps are induced from regularization or renormalization schemes. Indeed, multiplicativity of renormalization reports about a hidden algebraic nature inside of the BPHZ method namely, Rota-Baxter structure such that on the basis of this property we could introduce a new family of quantum Hamiltonian systems depended on renormalization or regularization schemes. Secondly, we studied motion integrals related to Feynman rules characters and then it was observed that how Connes-Kreimer renormalization group makes possible to obtain an infinite dimensional integrable quantum Hamiltonian system. Thirdly, based on Bogoliubov character and BCH formula, we found a new class of fixed point equations related to motion integrals where it yields to search more geometrical meanings inside of physical parameters.

The second purpose was concentrated on very special object in the Connes-Marcolli universal interpretation of renormalizable quantum field theories. We analyzed universal Hopf algebra of renormalization and also with the help of its shuffle nature, we reproduced it with Hall rooted trees. This new description provided a strong tool to consider the relation between $H_{\mathbb{U}}$ and some other combinatorial Hopf algebras. Moreover, we developed this rooted tree reformulation to the level of the universal affine group scheme and its related Lie algebra. From this story, we investigated that how theory of operads and poset theory (i.e. Hall rooted trees and Hall polynomials) can be entered into the study of quantum field theory. According to the universal property of $H_{\mathbb{U}}$ among all renormalizable physical theories and based on its large role in finding an isomorphism between universal category of flat equi-singular vector bundles and the category of finite dimensional linear representations of universal affine group scheme $\mathbb{U}^{*}$, it was explained that how we can apply this rooted tree version to obtain a Hall tree-Hall polynomial representation from universal singular frame.

The third purpose can be summarized in working on a geometric studying of non-perturbative theory. This project was done in two steps such that in the first stage, the extension of independency of the universal Hopf algebra of renormalization to the level of Dyson-Schwinger equations was considered where in this process, we studied DSEs at the level of $H_{\mathbb{U}}$. Furthermore, with attention to factorization of Feynman diagrams (determined with solutions of Dyson-Schwiger equations), we showed that this factorization uniquely exists in $H_{\mathbb{U}}$ and also, one can reformulate the Riemann $\zeta-$function with DSEs at the level of this particular Hopf algebra. In the next stage, with the help of generated Hopf subalgebras by DSEs, we showed that the category of equi-singular flat bundles can recover all of these equations. Generalization of the categorical interpretation (in the study of renormalizable theories) to Kreimer's non-perturbative modeling and introducing the concept of {\it universal Dyson-Schwinger equations} were immediate observations in this direction.


\section{{\it \textsf{Other integrable systems}}}

Renormalization group determined the compatibility of the introduced integrable Hamlitonian systems with the Connes-Kreimer perturbative renormalization.
It should be interesting to find another integrable Hamiltonian systems. For instance with deforming the algebra $L(H,A)$ underlying Nijenhuis maps (determined with regularization schemes), one can have this chance. On the other hand, finding a physical theory such that components of Birkhoff factorization of its associated Feynman rules character play the role of motion integrals for the character, can be very important question. We want to know that is there any specific algebro-geometric property in this kind of theories?


\section{{\it \textsf{Relation between motion integrals and DSEs}}}

Working on this question can help us to develop our approach to integrable systems to the level of non-perturbative theory and further, it leads to discover a new description from quantum motions based on Nijenhuis type Poisson brackets. It seems that with this notion one can introduce a new interrelationship between theory of Rota-Baxter algebras and Dyson-Schwinger equations.


\section{{\it \textsf{More about universal DSEs}}}

Here we introduced universal Dyson-Schwinger equations and then we just showed that with using the universal nature of the universal singular frame, one can lift DSEs in different physical theories to the level of $DSE_{u}$ (i.e. as the process of comparison). There are some essential questions about the importance of this class of equations. For instance, Can we apply these equations in computations as kind of a simplified toy model? What type of physical information can be analyzed explicitly with these equations?


\newpage


\chapter*{\textsf{Index}}

\small
\begin{center}
\textbf{\textit{A}}
\end{center}
Admissible cut \  \ 12, 13, 72, 86 \\
Alphabetical order \  \  69 \\
Amplitude \  \  1, 23, 26, 27, 28, 29, 82, 83 \\
Antipode \  \  7, 8, 13, 33, 65, 72 \\
Augmentation ideal \  \  9, 24, 25, 29 \\
Augmented modified quasi-shuffle product  \  \  39 \\
Augmented operad  \  \  17, 75 \\
\begin{center}
\textbf{\textit{B}}
\end{center}
Balanced bracket representation  \  \  14, 15, 72 \\
BCH series  \  \  53, 54, 55, 80, 96 \\
Beta function  \  \  2, 10, 29, 32, 49, 53, 56, 61, 80 \\
Bialgebra  \  \  6, 7, 8, 51 \\
Birkhoff decomposition  \  \  2, 13, 20, 29, 30, 31, 32, 33, 34, 49, 57, 58, 59, 61, 80, 93, 94 \\
Bogoliubov  \ \  1, 28, 31, 33, 34, 53, 55, 80, 96 \\
BPHZ  \  \  1, 2, 12, 20, 28, 31, 32, 33, 47, 52, 53, 57, 59, 96 \\
\begin{center}
\textbf{\textit{C}}
\end{center}
Casimir function  \  \  52 \\
Coinvariant  \  \  75 \\
Connected filtered  \  \  8 \\
Connected graded  \  \  7, 9, 10, 13, 14, 15, 16, 23, 28, 57, 61, 62, 64, 69, 74, 83, 84, 87 \\
Connes-Kreimer Hopf algebra  \  \  3, 11, 13, 14, 17, 19, 23, 24, 25, 28, 33, 40, 53, 55, 62, 63, 70, 75, 85, 88, 94 \\
Connes-Marcolli  \  \  3, 10, 19, 57, 61, 62, 66, 77, 81, 86, 92, 93, 96 \\
Connes-Moscovici Hopf algebra  \  \  15 \\
Conservation law  \  \  21 \\
Convolution product  \  \  6, 8, 13, 29, 33, 54 \\
Counterterm  \  \  1, 2, 3, 10, 20, 24, 28, 29, 31, 33, 34, 57, 59, 61, 62, 77, 80, 83 \\
\begin{center}
\textbf{\textit{D}}
\end{center}
deRham complex  \  \  42 \\
Deformed algebra  \  \  40, 51, 54 \\
Dimensional Regularization  \  \  1, 2, 3, 20, 29, 30, 32, 33, 34, 47, 48, 57, 59, 62, 94 \\
Diffeographism  \  \  31, 32 \\
Dimensionally regularized Feynman rules character  \  \ 2, 31, 33, 47, 48, 49, 50, 51, 52, 80 \\
Dyson-Schwinger equation   \  \  3, 15, 24, 62, 81, 83, 86, 88, 89, 90, 91, 92, 93, 94, 95, 97 \\
\begin{center}
\textbf{\textit{E}}
\end{center}
Elimination  \  \  1, 13  \\
Equi-singular  \  \  2, 3, 4, 33, 57, 58, 59, 60, 61, 62, 77, 82, 92, 93, 94, 95, 96, 97 \\
External edge  \  \  21, 22, 26, 27, 28, 83 \\
\begin{center}
\textbf{\textit{F}}
\end{center}
Faa di Bruno Hopf algebra  \  \  87 \\
Feynman diagram  \  \  1, 2, 3, 4, 8, 10, 11, 20, 21, 22, 23, 24, 25, 26, 27, 28, 29, 32, 33, 34, 37, 40, 48, 49, 51, 63, 81, 83, 84, 85, 86, 87, 89, 90, 91, 92, 97 \\
Fibre functor  \  \  60, 61, 95 \\
\begin{center}
\textbf{\textit{G}}
\end{center}
Gluon  \  \  3 \\
Graded representation  \  \  61, 62, 80, 93, 94 \\
Grafting  \  \  11, 19, 25, 66, 82 \\
Green function  \  \  26, 27, 83 \\
Grossman-Larson Hopf algebra  \  \  14 \\
\begin{center}
\textbf{\textit{H}}
\end{center}
Hall forest  \  \  67, 68, 71, 77, 80 \\
Hall polynomial  \  \  63, 76, 77, 78, 79, 80, 91, 96, 97 \\
Hall set  \  \  67, 68, 69, 71, 76, 77, 80, 86 \\
Hamiltonian derivation (vector field)  \  \  36, 40, 42, 43 \\
Hochschild one cocycle  \  \  11, 14, 27, 81, 82, 83, 84, 85, 90, 91, 94  \\
Hoffman pairing  \  \  64 \\
\begin{center}
\textbf{\textit{I}}
\end{center}
Idempotent  \  \  31, 38, 40, 41, 43, 44, 46, 48, 49, 50, 53 \\
Incidence Hopf algebra  \  \  17, 18, 19, 71, 75 \\
Infinitesimal generator  \  \  2, 10, 29, 32, 33, 34, 56, 80 \\
Insertion  \  \  13, 21, 23, 27, 83, 84  \\
Integral of motion (motion integral)  \  \  34, 36, 43, 45, 46, 47, 48, 49, 50, 51, 52, 53, 54, 55, 56, 58, 96, 97 \\
Integrable system  \  \  2, 3, 30, 33, 34, 35, 36, 37, 39, 40, 45, 46, 50, 51, 52, 56, 96, 97 \\
Internal edge  \  \  8, 21, 22, 23, 28 \\
\begin{center}
\textbf{\textit{J}}
\end{center}
Jacobi identity  \  \  37, 42, 43 \\
\begin{center}
\textbf{\textit{L}}
\end{center}
Ladder tree  \  \  13, 16, 17, 53, 54, 72, 73, 74, 88 \\
Leibniz  \  \  41, 43 \\
Letter  \  \  38, 64, 69, 76 \\
Lie algebra cohomology  \  \  82, 84, 85, 86 \\
Lie polynomial  \  \  65, 66, 76, 86 \\
Logarithmic  \  \  9, 27, 58 \\
Loop algebra  \  \  52 \\
Lyndon word  \  \  17, 69, 71, 73, 78, 86 \\
\begin{center}
\textbf{\textit{M}}
\end{center}
Milnor-Moore theorem  \  \  9, 10, 23, 77 \\
Minimal subtraction  \  \  1, 2, 20, 29, 30, 32, 33, 34, 47, 48, 57, 59, 94 \\
Modified quasi-shuffle product  \  \  39 \\
(Modified) Yang-Baxter equation  \  \  37 \\
Monodromy  \  \  20, 58 \\
Multiplicativity  \  \  2, 29, 34, 96 \\
\begin{center}
\textbf{\textit{N}}
\end{center}
n-adic topology  \  \  84 \\
Newton's second law  \  \  35, 36 \\
Neutral  \  \  3, 60, 61, 77, 93 \\
Nijenhuis algebra (tensor)  \  \  35, 37, 38, 39, 40, 41, 44, 46, 47, 51, 97 \\
Non-degenerate  \  \  42, 43, 44, 45 \\
Non-perturbative  \  \  3, 4, 81, 82, 86, 87, 92, 94, 96, 97 \\
\begin{center}
\textbf{\textit{O}}
\end{center}
One particle irreducible (1PI) graph  \  \  21, 22, 23, 24, 25, 26, 27, 28, 31, 48, 49, 50, 63, 83, 85, 86, 89 \\
\begin{center}
\textbf{\textit{P}}
\end{center}
Poisson  \  \  34, 36, 37, 42, 43, 44, 45, 53, 97 \\
Photon  \  \ 3 \\
Pro-unipotent  \  \  20, 31, 32 \\
\begin{center}
\textbf{\textit{Q}}
\end{center}
Quantum Hamiltonian system  \  \  3, 37, 39, 46, 54, 96 \\
Quark  \  \  3 \\
Quasi-shuffle product \  \  39, 64, 66, 69, 70 \\
Quotient graph  \  \  22 \\
\begin{center}
\textbf{\textit{R}}
\end{center}
Renormalization  \  \  1, 2, 3, 4, 5, 8, 11, 12, 20, 21, 23, 25, 26, 28, 29, 30, 32, 33, 34, 37, 40, 41, 43, 44, 46, 47, 48, 49, 52, 54, 57, 59, 62, 63, 66, 73, 77, 81, 82, 92, 93, 96, 97 \\
Renormalization bundle  \  \  33, 59 \\
Renormalization group  \  \  2, 3, 10, 29, 32, 33, 34, 37, 39, 46, 47, 49, 50, 51, 52, 53, 56, 61, 80, 96, 97 \\
Renormalized value  \  \  2, 10, 20, 29, 31, 32, 33, 34, 62 \\
Residue  \  \  23 \\
Riemann-Hilbert  \  \  2, 3, 20, 28, 29, 31, 34, 60, 61, 77, 93 \\
Riemann zeta function  \  \ 87, 89, 92, 93, 97 \\
Rooted tree  \  \  3, 4, 11, 12, 13, 14, 15, 16, 17, 18, 19, 21, 23, 24, 25, 28, 53, 54, 56, 63, 66, 67, 68, 69, 70, 71, 72, 73, 74, 75, 76, 77, 79, 80, 83, 85, 86, 87, 88, 89, 94, 96, 97 \\
Rota-Baxter  \  \  2, 3, 29, 30, 31, 33, 34, 37, 38, 39, 40, 43, 47, 52, 53, 54, 96, 97 \\
\begin{center}
\textbf{\textit{S}}
\end{center}
(Quasi-)Symmetric function  \  \  14, 15, 16, 71, 72, 88 \\
Set operad  \  \  17, 18, 19, 71, 75, 77 \\
Slavnov-Taylor identity  \  \  83 \\
Standard decomposition  \  \  67, 68, 76, 78 \\
Superficially divergency  \  \  23, 26, 27, 31, 48, 49, 50 \\
Symplectic structure (form)  \  \  35, 36, 37, 40, 41, 43, 44, 45, 46, 47, 52 \\
Symplectomorphism  \  \  36 \\
\begin{center}
\textbf{\textit{T}}
\end{center}
Tadpole  \  \  26 \\
Tannakian  \  \ 3, 60, 61, 77, 93 \\
Time ordered exponential  \  \  2, 58, 78 \\
Trivial connection  \  \  60, 62, 95 \\
\begin{center}
\textbf{\textit{U}}
\end{center}
Universal affine group scheme  \  \  61, 63, 76, 77, 78, 86, 96, 97 \\
Universal counterterm  \  \ 62 \\
Universal Dyson-Schwinger equation \  \  92, 95, 97  \\
Universal enveloping algebra  \  \  5, 7, 14, 23, 31, 61, 65, 66, 82, 84 \\
Universal Hopf algebra of renormalization  \  \  3, 4, 15, 16, 17, 61, 63, 66, 69, 71, 73, 75, 77, 79, 80, 81, 82, 85, 86, 87, 90, 91, 92, 94, 96, 97 \\
Universal Rota-Baxter algebra  \  \  39 \\
Universal Nijenhuis algebra  \  \  39, 40, 41, 44 \\
Universal singular frame  \  \  4, 57, 61, 62, 63, 66, 79, 80, 96, 97 \\
\begin{center}
\textbf{\textit{V}}
\end{center}
Vacuum  \  \  26 \\
Virtual particle  \  \  28 \\
\begin{center}
\textbf{\textit{W}}
\end{center}
W-connection  \  \  60 \\
Ward identity  \  \ 83 \\
\begin{center}
\textbf{\textit{Y}}
\end{center}
Yang-Baxter equation \  \  34, 37, 38 \\
\begin{center}
\textbf{\textit{Z}}
\end{center}
Zhao's homomorphism  \  \  16, 17, 73 \\

\end{document}